\documentclass[ngerman,english,official]{IISthesis}

\usepackage{graphicx}           %
\usepackage[usenames]{color}
\usepackage[toc, section=section]{glossaries}
\usepackage[automake]{glossaries-extra}

\usepackage{amsmath}
\RequirePackage[hidelinks]{hyperref}
\usepackage{xspace}
\usepackage{textcomp,lmodern}
\usepackage{multirow}
\usepackage{rotating}
\usepackage{lscape}
\usepackage{array} 
\renewcommand{\arraystretch}{1.10}
\usepackage[nottoc]{tocbibind}

\usepackage{amsmath,amssymb,amsfonts}
\RequirePackage{algorithmicx}
\RequirePackage{algorithm}
\RequirePackage{algpseudocode}
\usepackage{booktabs}
\usepackage{threeparttable}
\usepackage[table,xcdraw]{xcolor}
\usepackage{acronym}
\usepackage[detect-all, per-mode=symbol, range-units=single, range-phrase=~to~]{siunitx}
\usepackage[shortcuts]{extdash}
\usepackage{fixfoot}
\usepackage{url}
\usepackage{cite}
\usepackage{textcomp}
\usepackage{multicol}
\hypersetup{breaklinks=true}
\usepackage{soul}
\usepackage{makecell}
\usepackage[shortlabels]{enumitem}
\usepackage{wrapfig}
\usepackage{rotating}
\usepackage{adjustbox}
\usepackage{tabularx}
\usepackage{float} 

\usepackage{bibentry} 

\usepackage{pifont}
\usepackage{makecell}
\usepackage{adjustbox} %
\usepackage[multiple]{footmisc}
\usepackage{refcount}
\usepackage[noabbrev,capitalise]{cleveref}
\usepackage{microtype}
\usepackage{multirow}

\usepackage{tikz}
\usepackage{calc}
\usepackage{eso-pic}
\usepackage{ragged2e}
\usepackage[utf8]{inputenc}
\usepackage[T1]{fontenc}
\usepackage{orcidlink}
\usepackage{subcaption}
\usepackage{listings}
\usepackage{arydshln}
\usepackage{pdflscape}
\usepackage{afterpage}
\usepackage[graphicx]{realboxes}
\usepackage{placeins}

\definecolor{ieee-bright-dblue-100}{rgb}{0.0, 0.3828, 0.6055}
\definecolor{ieee-bright-dblue-80}{rgb}{0.0, 0.4883, 0.6797}
\definecolor{ieee-bright-dblue-60}{rgb}{0.3633, 0.6094, 0.7617}
\definecolor{ieee-bright-dblue-40}{rgb}{0.5898, 0.7383, 0.8398}
\definecolor{ieee-bright-dblue-20}{rgb}{0.8906, 0.8984, 0.9219}
\definecolor{ieee-bright-red-100}{rgb}{0.7266, 0.0469, 0.1836}
\definecolor{ieee-bright-red-80}{rgb}{0.832, 0.3164, 0.3281}
\definecolor{ieee-bright-red-60}{rgb}{0.8906, 0.4922, 0.4805}
\definecolor{ieee-bright-red-40}{rgb}{0.9336, 0.6562, 0.6406}
\definecolor{ieee-bright-red-20}{rgb}{0.9688, 0.8203, 0.8125}
\definecolor{ieee-bright-orange-100}{rgb}{0.9961, 0.6367, 0.0}
\definecolor{ieee-bright-orange-80}{rgb}{0.9844, 0.6953, 0.3125}
\definecolor{ieee-bright-orange-60}{rgb}{0.9883, 0.7695, 0.4844}
\definecolor{ieee-bright-orange-40}{rgb}{0.9922, 0.8359, 0.6562}
\definecolor{ieee-bright-orange-20}{rgb}{0.9961, 0.9219, 0.8164}
\definecolor{ieee-bright-yellow-100}{rgb}{0.9961, 0.8164, 0.0}
\definecolor{ieee-bright-yellow-80}{rgb}{0.9961, 0.8477, 0.2148}
\definecolor{ieee-bright-yellow-60}{rgb}{0.9961, 0.875, 0.4492}
\definecolor{ieee-bright-yellow-40}{rgb}{0.9961, 0.9062, 0.6328}
\definecolor{ieee-bright-yellow-20}{rgb}{0.9961, 0.9531, 0.8125}
\definecolor{ieee-bright-lgreen-100}{rgb}{0.4688, 0.7422, 0.125}
\definecolor{ieee-bright-lgreen-80}{rgb}{0.5742, 0.7852, 0.332}
\definecolor{ieee-bright-lgreen-60}{rgb}{0.6875, 0.8398, 0.5039}
\definecolor{ieee-bright-lgreen-40}{rgb}{0.793, 0.8906, 0.6641}
\definecolor{ieee-bright-lgreen-20}{rgb}{0.8945, 0.9414, 0.8281}
\definecolor{ieee-bright-dgreen-100}{rgb}{0.0, 0.5156, 0.2383}
\definecolor{ieee-bright-dgreen-80}{rgb}{0.1641, 0.6055, 0.3867}
\definecolor{ieee-bright-dgreen-60}{rgb}{0.3906, 0.6953, 0.5234}
\definecolor{ieee-bright-dgreen-40}{rgb}{0.6094, 0.8008, 0.6719}
\definecolor{ieee-bright-dgreen-20}{rgb}{0.8047, 0.8945, 0.8359}
\definecolor{ieee-bright-purple-100}{rgb}{0.5938, 0.1133, 0.5898}
\definecolor{ieee-bright-purple-80}{rgb}{0.6992, 0.3281, 0.668}
\definecolor{ieee-bright-purple-60}{rgb}{0.7812, 0.4961, 0.7461}
\definecolor{ieee-bright-purple-40}{rgb}{0.8555, 0.6602, 0.8281}
\definecolor{ieee-bright-purple-20}{rgb}{0.9219, 0.8281, 0.9023}
\definecolor{ieee-bright-lblue-100}{rgb}{0.0, 0.6094, 0.6484}
\definecolor{ieee-bright-lblue-80}{rgb}{0.0, 0.6797, 0.7188}
\definecolor{ieee-bright-lblue-60}{rgb}{0.2109, 0.75, 0.7812}
\definecolor{ieee-bright-lblue-40}{rgb}{0.5469, 0.8242, 0.8438}
\definecolor{ieee-bright-lblue-20}{rgb}{0.7695, 0.918, 0.9219}
\definecolor{ieee-bright-cyan-100}{rgb}{0.0, 0.707, 0.8828}
\definecolor{ieee-bright-cyan-80}{rgb}{0.0, 0.7227, 0.9453}
\definecolor{ieee-bright-cyan-60}{rgb}{0.2656, 0.7812, 0.957}
\definecolor{ieee-bright-cyan-40}{rgb}{0.5547, 0.8438, 0.9688}
\definecolor{ieee-bright-cyan-20}{rgb}{0.7773, 0.9141, 0.9805}
\definecolor{ieee-bright-white-100}{rgb}{0.9961, 0.9961, 0.9961}
\definecolor{ieee-bright-white-80}{rgb}{0.9961, 0.9961, 0.9961}
\definecolor{ieee-bright-white-60}{rgb}{0.9961, 0.9961, 0.9961}
\definecolor{ieee-bright-white-40}{rgb}{0.9961, 0.9961, 0.9961}
\definecolor{ieee-bright-white-20}{rgb}{0.9961, 0.9961, 0.9961}
\definecolor{ieee-dark-red-100}{rgb}{0.5234, 0.1211, 0.2539}
\definecolor{ieee-dark-red-80}{rgb}{0.6445, 0.2812, 0.3828}
\definecolor{ieee-dark-red-60}{rgb}{0.7422, 0.4727, 0.5234}
\definecolor{ieee-dark-red-40}{rgb}{0.832, 0.6445, 0.6758}
\definecolor{ieee-dark-red-20}{rgb}{0.918, 0.8203, 0.832}
\definecolor{ieee-dark-orange-100}{rgb}{0.9062, 0.4648, 0.1328}
\definecolor{ieee-dark-orange-80}{rgb}{0.9648, 0.5664, 0.3164}
\definecolor{ieee-dark-orange-60}{rgb}{0.9766, 0.6758, 0.4805}
\definecolor{ieee-dark-orange-40}{rgb}{0.9844, 0.7773, 0.6523}
\definecolor{ieee-dark-orange-20}{rgb}{0.9922, 0.8789, 0.8125}
\definecolor{ieee-dark-yellow-100}{rgb}{0.9961, 0.7773, 0.1719}
\definecolor{ieee-dark-yellow-80}{rgb}{0.9961, 0.8086, 0.375}
\definecolor{ieee-dark-yellow-60}{rgb}{0.9961, 0.875, 0.4492}
\definecolor{ieee-dark-yellow-40}{rgb}{0.9961, 0.8984, 0.6875}
\definecolor{ieee-dark-yellow-20}{rgb}{0.9961, 0.9453, 0.8438}
\definecolor{ieee-dark-lgreen-100}{rgb}{0.3945, 0.5508, 0.0938}
\definecolor{ieee-dark-lgreen-80}{rgb}{0.5078, 0.6289, 0.293}
\definecolor{ieee-dark-lgreen-60}{rgb}{0.6367, 0.7188, 0.4688}
\definecolor{ieee-dark-lgreen-40}{rgb}{0.7539, 0.8047, 0.6367}
\definecolor{ieee-dark-lgreen-20}{rgb}{0.875, 0.9023, 0.8125}
\definecolor{ieee-dark-dgreen-100}{rgb}{0.0, 0.3867, 0.2539}
\definecolor{ieee-dark-dgreen-80}{rgb}{0.1836, 0.5, 0.3906}
\definecolor{ieee-dark-dgreen-60}{rgb}{0.3984, 0.6172, 0.5273}
\definecolor{ieee-dark-dgreen-40}{rgb}{0.5938, 0.7422, 0.6758}
\definecolor{ieee-dark-dgreen-20}{rgb}{0.793, 0.8711, 0.8359}
\definecolor{ieee-dark-purple-100}{rgb}{0.4648, 0.1445, 0.5117}
\definecolor{ieee-dark-purple-80}{rgb}{0.5898, 0.3242, 0.6016}
\definecolor{ieee-dark-purple-60}{rgb}{0.6914, 0.4883, 0.6953}
\definecolor{ieee-dark-purple-40}{rgb}{0.7969, 0.6523, 0.793}
\definecolor{ieee-dark-purple-20}{rgb}{0.8945, 0.8203, 0.8945}
\definecolor{ieee-dark-cyan-100}{rgb}{0.0, 0.4492, 0.4648}
\definecolor{ieee-dark-cyan-80}{rgb}{0.0, 0.5469, 0.5664}
\definecolor{ieee-dark-cyan-60}{rgb}{0.3047, 0.6602, 0.668}
\definecolor{ieee-dark-cyan-40}{rgb}{0.5586, 0.7695, 0.7734}
\definecolor{ieee-dark-cyan-20}{rgb}{0.7734, 0.8789, 0.8789}
\definecolor{ieee-dark-dblue-100}{rgb}{0.0, 0.1562, 0.332}
\definecolor{ieee-dark-dblue-80}{rgb}{0.1797, 0.3008, 0.4609}
\definecolor{ieee-dark-dblue-60}{rgb}{0.3828, 0.4609, 0.5859}
\definecolor{ieee-dark-dblue-40}{rgb}{0.5781, 0.6289, 0.7188}
\definecolor{ieee-dark-dblue-20}{rgb}{0.7852, 0.8047, 0.8555}
\definecolor{ieee-dark-grey-100}{rgb}{0.457, 0.4688, 0.4805}
\definecolor{ieee-dark-grey-80}{rgb}{0.5625, 0.5625, 0.5742}
\definecolor{ieee-dark-grey-60}{rgb}{0.6641, 0.6641, 0.6758}
\definecolor{ieee-dark-grey-40}{rgb}{0.7734, 0.7695, 0.7773}
\definecolor{ieee-dark-grey-20}{rgb}{0.8789, 0.8828, 0.8828}
\definecolor{ieee-dark-black-100}{rgb}{0.0, 0.0, 0.0}
\definecolor{ieee-dark-black-80}{rgb}{0.3438, 0.3477, 0.3555}
\definecolor{ieee-dark-black-60}{rgb}{0.5, 0.5078, 0.5195}
\definecolor{ieee-dark-black-40}{rgb}{0.6523, 0.6602, 0.6719}
\definecolor{ieee-dark-black-20}{rgb}{0.8164, 0.8242, 0.8281}

\newcommand{\kGE}{\ensuremath{\,}k\GE}  %

\newcommand{\GHz}{\ensuremath{\,}GHz\xspace}

\newcommand{\etal}{\emph{et al.}}
\newcommand{\cmark}{\ding{51}}%
\newcommand{\xmark}{\ding{55}}%

\newcommand{\tus}{{\textunderscore}}

\newcommand{\artistic}{ArtistIC}
\newcommand{\dc}{{Synopsys} {Design} {Compiler} {NXT} 2023.12}
\newcommand{\dutctl}{DUTCTL}
\newcommand{\gfs}{{GlobalFoundries'}}
\newcommand{\gftech}{{GF12LP+}}
\newcommand{\gf}{{GlobalFoundries}}
\newcommand{\imagemagick}{ImageMagick}
\newcommand{\klayout}{KLayout}
\newcommand{\pt}{{Synopsys} {PrimeTime} 2022.03}
\newcommand{\riscv}{\teb{\mbox{RISC-V}}}
\newcommand{\axipack}{AXI-Pack}

\newcommand{\carfield}{{Carfield}}
\newcommand{\cheshire}{{Cheshire}}
\newcommand{\controlpulp}{{ControlPULP}}
\newcommand{\pulp}{\gls{pulp}}
\newcommand{\mempool}{{MemPool}}
\newcommand{\occamy}{{Occamy}}
\newcommand{\pulpopen}{{PULP-open}}

\newcommand{\idma}{iDMA}
\newcommand{\idmae}{{\idma} engine}
\newcommand{\iDmaE}{{\idma} Engine}
\newcommand{\idmaes}{{\idma} engines}
\newcommand{\dma}{\gls{dma}}

\newcommand{\dmaes}{\gls{dma} engines}
\newcommand{\dmae}{\gls{dma} engine}
\newcommand{\DmaE}{DMA Engine}
\newcommand{\Dmaes}{\Gls{dma} engines}

\newcommand{\dmaa}{\gls{dma} architecture}

\newcommand{\fe}{front-end}
\newcommand{\fes}{{\fe}s}
\newcommand{\me}{mid-end}
\newcommand{\mes}{{\me}s}
\newcommand{\be}{back-end}
\newcommand{\bes}{{\be}s}
\newcommand{\Fe}{Front-end}
\newcommand{\Fes}{Front-ends}
\newcommand{\Me}{Mid-end}
\newcommand{\Mes}{Mid-ends}
\newcommand{\Be}{Back-end}

\newcommand{\axirealm}{{AXI-REALM}}
\newcommand{\irealm}{i{realm}}
\newcommand{\erealm}{e{realm}}

\setabbreviationstyle[acronym]{long-short}
\newacronym{ac}{AC}{application-class}
\newacronym{aces}{ACES}{Autonomy, Connectivity, Electrification, and Shared mobility}
\newacronym{adas}{ADAS}{advanced driver-assistance system}
\newacronym{admm}{ADMM}{alternating direction method of multipliers}
\newacronym{aes}{AES}{autonomous embodied system}
\newacronym{agi}{AGI}{artificial general intelligence}
\newacronym{ai}{AI}{artificial intelligence}
\newacronym{alap}{ALAP}{as-late-as possible}
\newacronym{alu}{ALU}{arithmetic logic unit}
\newacronym{amba}{AMBA}{advanced microcontroller bus architecture}
\newacronym{amd}{AMD}{approximate minimum degree}
\newacronym{aod}{AoD}{always-on domain}
\newacronym{aot}{AOT}{ahead of time}
\newacronym{aox-01}{AOX-01}{AetherOre eXecutable version 01}
\newacronym{ap}{AP}{application-class processor}
\newacronym{api}{API}{application programming interface}
\newacronym{apu}{APU}{Application Processing Unit}
\newacronym{asap}{ASAP}{as-soon-as possible}
\newacronym{asic}{ASIC}{application-specific integrated circuit}
\newacronym{asip}{ASIP}{application-specific instruction-set processor}
\newacronym{ate}{ATE}{automated test equipment}
\newacronym{avsbus}{AVSBUS}{adaptive voltage scaling}
\newacronym{axi4}{AXI4}{advanced eXtensible interface 4}
\newacronym{axirealm}{AXI-REALM}{AXI real-time regulation and traffic monitoring}
\newacronym{axis}{AXIS}{advanced eXtensible interface stream}
\newacronym{bcsr}{BCSR}{blocked compressed sparse rows}
\newacronym{blas}{BLAS}{Basic Linear Algebra Subprograms}
\newacronym{bmc}{BMC}{baseboard management controller}
\newacronym{bram}{BRAM}{block RAM}
\newacronym{bs}{BS}{backward substitution}
\newacronym{c2c}{C2C}{chip-to-chip}
\newacronym{ca}{CA}{command/address}
\newacronym{casr}{CSR}{configuration and status register}
\newacronym{cgra}{CGRA}{coarse-grained reconfigurable array}
\newacronym{ci}{CI}{continuous integration}
\newacronym{cisr}{CISR}{cooperating indexed stream registers}
\newacronym{clic}{CLIC}{core-local interrupt controller}
\newacronym{clint}{CLINT}{core-local interruptor}
\newacronym{cmos}{CMOS}{complementary metal-oxide-semiconductor}
\newacronym{cots}{COTS}{commercial off-the-shelf}
\newacronym{cppc}{CPPC}{collaborative processor performance control}
\newacronym{cps}{CPS}{cyber-physical system}
\newacronym[longplural={central processing units}]{cpu}{CPU}{central processing unit}
\newacronym{crtes}{CRTES}{critical real-time embedded system}
\newacronym{csc}{CSC}{compressed sparse column}
\newacronym{csf}{CSF}{compressed sparse fiber}
\newacronym{csr}{CSR}{compressed sparse rows}
\newacronym{csrmm}{CsrMM}{CSR matrix-matrix multiplication}
\newacronym{csrmv}{CsrMV}{CSR matrix-vector multiplication}
\newacronym{d2d}{D2D}{die-to-die}
\newacronym{dag}{DAG}{directed acyclic graph}
\newacronym{db}{DB}{data bus}
\newacronym{dbgm}{DBGM}{debug module}
\newacronym{dcls}{DCLS}{dual-core lockstep}
\newacronym{dcspm}{DCSPM}{dynamically configurable scratchpad memory}
\newacronym{dcu}{DCU}{domain control unit}
\newacronym{ddr}{DDR}{double data rate}
\newacronym{deepc}{DeePC}{data-enabled predictive control}
\newacronym{dfs}{DFS}{dynamic frequency scaling}
\newacronym{dl}{DL}{deep learning}
\newacronym{dlt}{DLT}{data layout transform}
\newacronym{dm}{DM}{data mover}
\newacronym{dma}{DMA}{direct memory access}
\newacronym{dmac}{DMAC}{direct memory access controller}
\newacronym{dmae}{DMAE}{direct memory access engine}
\newacronym{dmp}{DMP}{discrete model pruning}
\newacronym{dos}{DoS}{denial of service}
\newacronym{dpllc}{DPLLC}{dynamically partitionable last-level cache}
\newacronym{dram}{DRAM}{dynamic random access memory}
\newacronym{drc}{DRC}{design rule check}
\newacronym{ds}{DS}{domain-specific}
\newacronym{dsa}{DSA}{domain-specific accelerator}
\newacronym{dsp}{DSP}{digital signal processing}
\newacronym{dtm}{DTM}{dynamic thermal management}
\newacronym{dut}{DUT}{device under test}
\newacronym{dvfs}{DVFS}{dynamic voltage and frequency scaling}
\newacronym{dvs}{DVS}{dynamic voltage scaling}
\newacronym{e2e}{E2E}{end-to-end}
\newacronym{ecc}{ECC}{error correcting code}
\newacronym{ecu}{ECU}{electronic control unit}
\newacronym{eda}{EDA}{electronic design automation}
\newacronym{edram}{eDRAM}{embedded DRAM}
\newacronym{epi}{EPI}{European Processor Initiative}
\newacronym{esr}{ESR}{egress SR}
\newacronym{eth}{ETH}{Ethernet}
\newacronym{etm}{ETM}{energy and thermal management}
\newacronym{fame}{FAME}{FPGA architecture model execution}
\newacronym{fdsoi}{FD-SOI}{fully-depleted silicon-on-insulator}
\newacronym{fe}{FE}{forward elimination}
\newacronym{fem}{FEM}{finite element analysis}
\newacronym{fg}{FG}{fast gradient}
\newacronym{fifo}{FIFO}{first in, first out}
\newacronym{fil}{FIL}{FPGA in the loop}
\newacronym{fireore}{Fireore}{fiercely immature, risky, empirically obscure, radical easteregg}
\newacronym{fiwlcsp}{FI-WLCSP}{fan-in wafer-level chip-scale packaging}
\newacronym{fll}{FLL}{frequency locked loop}
\newacronym{flop}{FLOP}{floating-point operation}
\newacronym{flops}{FLOPS}{floating-point operations per second}
\newacronym{fma}{FMA}{fused multiply-add}
\newacronym{foss}{FOSS}{free and open-source}
\newacronym{fp}{FP}{floating-point}
\newacronym{fpga}{FPGA}{field-programmable gate array}
\newacronym{fpu}{FPU}{floating point unit}
\newacronym{frep}{FREP}{floating-point repetition}
\newacronym{fsbl}{FSBL}{first-stage bootLoader}
\newacronym{fsm}{FSM}{finite state machine}
\newacronym{fu}{FU}{functional unit}
\newacronym{fub}{FUB}{functional unit block}
\newacronym{gcn}{GCN}{graph convolutional network}
\newacronym{gdsii}{GDSII}{Graphic Data System II}
\newacronym{gemm}{GEMM}{general matrix multiply}
\newacronym{gp}{GP}{general-purpose}
\newacronym{gpos}{GPOS}{general-purpose operating system}
\newacronym{gpp}{GPP}{general purpose processor}
\newacronym{gpt}{GPT}{Globally Unique Identifier Partition Table}
\newacronym{gpu}{GPU}{graphics processing unit}
\newacronym{grli}{GRLI}{globally regular, locally irregular}
\newacronym{gsm}{GSM}{Graduate Student Member}
\newacronym{guipt}{GPT}{Globally Unique Identifier Partition Table}
\newacronym{ha}{HA}{hardware accelerator}
\newacronym[longplural={hardware abstraction layers}]{hal}{HAL}{hardware abstraction layer}
\newacronym{hbm}{HBM}{high-bandwidth memory}
\newacronym{hbm2e}{HBM2E}{high-bandwidth memory 2E}
\newacronym{heicps}{He-iCPS}{heterogeneous integrated cyber-physical system}
\newacronym[longplural={heterogeneous systems on chip}]{hesoc}{HeSoC}{heterogeneous system on chip}
\newacronym{hil}{HIL}{hardware-in-the-loop}
\newacronym{hlc}{HLC}{high-level controller}
\newacronym{hls}{HLS}{high-level synthesis}
\newacronym{hmr}{HMR}{hybrid modular redundancy}
\newacronym{hpc}{HPC}{high-performance computing}
\newacronym{hw}{HW}{hardware}
\newacronym{hwrot}{HWRoT}{hardware root of trust}
\newacronym{hypt}{ParSPL}{parallel sparsity-pattern-leveraging triangular linear system solver}
\newacronym{ibmocc}{IBM OCC}{IBM on-chip controller}
\newacronym{ic}{IC}{integrated circuit}
\newacronym{idma}{iDMA}{intelligent DMA}
\newacronym{ihls}{iHLS}{IP-based high-level synthesis}
\newacronym{ima}{IMA}{integrated modular avionics}
\newacronym{io}{I/O}{input/output}
\newacronym{iommu}{IOMMU}{IO memory management unit}
\newacronym{iot}{IoT}{internet of things}
\newacronym{iotlb}{IOTLB}{IO translation lookaside buffer}
\newacronym[longplural={intellectual properties}]{ip}{IP}{intellectual property}
\newacronym{ipc}{IPC}{instructions per cycle}
\newacronym{ipu}{IPU}{integer processing unit}
\newacronym{irq}{IRQ}{interrupt request}
\newacronym{isa}{ISA}{instruction set architecture}
\newacronym{isr}{ISR}{indexed stream register}
\newacronym{issr}{ISSR}{indirection stream semantic register}
\newacronym{isut}{ISUT}{integrated system under test}
\newacronym{jid}{jID}{job ID}
\newacronym{kkt}{KKT}{Karush-Kuhn-Tucker}
\newacronym{l1}{L1}{level-one}
\newacronym{l2}{L2}{level-two}
\newacronym{l3}{L3}{level-three}
\newacronym{la}{LA}{linear algebra}
\newacronym{led}{LED}{light-emitting diode}
\newacronym{llc}{LLC}{last-level cache}
\newacronym{llctl}{LLC}{low-level controller}
\newacronym{llm}{LLM}{large language model}
\newacronym{lns}{LNS}{logarithmic number system}
\newacronym{lpddr}{LPDDR}{low-power double data rate}
\newacronym{lqr}{LQR}{linear quadratic regulator}
\newacronym{lti}{LTI}{linear time-invariant}
\newacronym{mac}{MAC}{multiply-accumulate}
\newacronym{macp}{MCP}{manageability control processor}
\newacronym{mcm}{MCM}{multi-chip module}
\newacronym{mcp}{MCP}{multi-core processor}
\newacronym{mcs}{MCS}{mixed-criticality system}
\newacronym{mctp}{MCTP}{management component transport protocol}
\newacronym{mcu}{MCU}{microcontroller unit}
\newacronym{mil}{MIL}{model in the loop}
\newacronym{mimo}{MIMO}{multi-input multi-output}
\newacronym{ml}{ML}{machine learning}
\newacronym{mmu}{MMU}{memory management unit}
\newacronym{mpam}{MPAM}{memory system resource partitioning and monitoring}
\newacronym{mpc}{MPC}{model predictive control}
\newacronym[longplural={multi-processor systems on chip}]{mpsoc}{MPSoC}{multi-processor system on chip}
\newacronym{mqtt}{MQTT}{message queuing telemetry transport}
\newacronym{mtunit}{M\&R unit}{monitoring and regulation unit}
\newacronym{nd}{N-D}{N-dimensional}
\newacronym{nn}{NN}{neural network}
\newacronym[longplural={networks-on-chip}]{noc}{NoC}{network-on-chip}
\newacronym{nsrrp}{NSRRP}{non-stallable request-response protocol}
\newacronym{numa}{NUMA}{non-uniform memory architecture}
\newacronym{obi}{OBI}{open bus interface}
\newacronym{occ}{OCC}{on-chip controller}
\newacronym{ocm}{OCM}{on-chip memory}
\newacronym{ooc}{OOC}{out-of-context}
\newacronym{oom}{OoM}{order of magnitude}
\newacronym{ooo}{OoO}{out-of-order}
\newacronym{os}{OS}{operating system}
\newacronym{oseda}{OSEDA}{open-source electronic design automation}
\newacronym{osh}{OSH}{open-source hardware}
\newacronym{ospm}{OSPM}{operating system-directed configuration and power management}
\newacronym{osqp}{OSQP}{operator-splitting quadratic programming}
\newacronym{p2p}{P2P}{point-to-point}
\newacronym{pcb}{PCB}{printed circuit board}
\newacronym{pcf}{PCF}{power control firmware}
\newacronym{pcg}{PCG}{preconditioned conjugate gradient}
\newacronym{pcie}{PCIe}{Peripheral Component Interconnect Express}
\newacronym{pcs}{PCS}{power controller system}
\newacronym{pcu}{PCU}{power control unit}
\newacronym{pdk}{PDK}{process design kit}
\newacronym{pe}{PE}{processing element}
\newacronym{pes}{PS}{per-system}
\newacronym{pfct}{PFCT}{periodic frequency control task}
\newacronym{pgs}{PGS}{Processing System}
\newacronym{phy}{PHY}{physical layer}
\newacronym{pid}{PID}{proportional integral derivative}
\newacronym{pil}{PIL}{processor in the loop}
\newacronym{pl}{PL}{Programmable Logic}
\newacronym{pldm}{PLDM}{Platform Level Data Model}
\newacronym{plic}{PLIC}{platform-level interrupt controller}
\newacronym{pll}{PLL}{phase locked loop}
\newacronym{pmbus}{PMBUS}{Power Management Bus}
\newacronym{pmca}{PMCA}{programmable many-core accelerator}
\newacronym{pmp}{MMU}{physical memory protection unit}
\newacronym{pnr}{P\&R}{place and route}
\newacronym{ppa}{PPA}{power-performance-area}
\newacronym{ps}{PS}{per-system}
\newacronym[longplural={page table entries}]{pte}{PTE}{page table entry}
\newacronym{ptw}{PTW}{page table walker}
\newacronym{pu}{PU}{per-unit}
\newacronym{pulp}{PULP}{parallel ultra-low power}
\newacronym{pur}{PUR}{per unit and region}
\newacronym{pvct}{PVCT}{periodic voltage control task}
\newacronym{pvt}{PVT}{process-voltage-temperature}
\newacronym{qnn}{QNN}{quantized neural networks}
\newacronym{qos}{QoS}{quality of service}
\newacronym{qp}{QP}{quadratic programming}
\newacronym{rac}{RAC}{runtime active control}
\newacronym{rf}{RF}{register file}
\newacronym{rhs}{RHS}{right-hand-side}
\newacronym{rl}{RL}{reinforcement learning}
\newacronym{rmse}{RMSE}{root mean square error}
\newacronym{rob}{ROB}{reorder buffer}
\newacronym{rpc}{RPC}{reduced pin count}
\newacronym{rpcdram}{RPC DRAM}{reduced-pin-count DRAM}
\newacronym{rr}{RR}{round-robin}
\newacronym{rtl}{RTL}{register transfer level}
\newacronym{rtos}{RTOS}{real-time OS}
\newacronym{rtu}{RTU}{Real Time Unit}
\newacronym{rtunit}{REALM unit}{real-time regulation and traffic monitoring unit}
\newacronym{sbc}{SBC}{single board computer}
\newacronym{scarabaeus}{Scarabaeus}{Specifically Crafted Acronym Referencing an Ariane in a Brilliantly and Artistically Engineered Unprecedented SoC}
\newacronym{scmi}{SCMI}{System Control and Management Interface}
\newacronym{scp}{SCP}{System Control Processor}
\newacronym{scpi}{SCPI}{standard commands for programmable instruments}
\newacronym{sdma}{sDMA}{sensor DMA}
\newacronym{sdr}{SDR}{single data rate}
\newacronym{sdst}{SDST}{software-defined SoC testing}
\newacronym{sdv}{SDV}{software-defined vehicles}
\newacronym{shv}{SHV}{selective hardware vectoring}
\newacronym{sil}{SIL}{safety integrity level}
\newacronym{silcontrol}{SIL}{software in the loop}
\newacronym{silsafety}{SIL}{safety integrity level}
\newacronym{simd}{SIMD}{single instruction, multiple data}
\newacronym[longplural={systems in package}]{sip}{SiP}{system in package}
\newacronym{sl}{SL}{serial link}
\newacronym{slength}{SL}{streaming length}
\newacronym{sm}{SM}{streaming multiprocessor}
\newacronym{smmu}{sMMU}{streaming memory management unit}
\newacronym{smp}{SMP}{symmetric multiprocessing}
\newacronym{smu}{SMU}{System Management Unit}
\newacronym{soa}{SoA}{state-of-the-art}
\newacronym[longplural={systems-on-chip}]{soc}{SoC}{system-on-chip}
\newacronym{soi}{SOI}{silicon-on-insulator}
\newacronym{spm}{SPM}{scratchpad memory}
\newacronym{spmm}{SpMM}{sparse matrix-matrix multiply}
\newacronym{spmspm}{SpMSpM}{sparse-sparse matrix multiply}
\newacronym{spmv}{SpMV}{sparse matrix-vector multiply}
\newacronym{sptrsv}{SpTRSV}{sparse triangular linear system solver}
\newacronym{spvv}{SpVV}{sparse vector-vector multiplication}
\newacronym{sr}{SR}{stream register}
\newacronym[longplural={static random access memories}]{sram}{SRAM}{static random access memory}
\newacronym{ssr}{SSR}{stream semantic register}
\newacronym{sssr}{SSSR}{sparse stream semantic register}
\newacronym{stm}{STM}{static thermal management}
\newacronym{su}{SU}{streaming unit}
\newacronym{sut}{SUT}{system under test}
\newacronym{sv}{SV}{SystemVerilog}
\newacronym{sw}{SW}{software}
\newacronym{swapc}{SWaP-C}{space, weight, power, and cost}
\newacronym{tcdm}{TCDM}{tightly coupled data memory}
\newacronym{tcls}{TCLS}{triple-core lockstep}
\newacronym{tdma}{TDMA}{time division multiple access}
\newacronym{tdp}{TPD}{thermal design power}
\newacronym{tid}{tID}{transaction ID}
\newacronym{tlb}{TLB}{translation lookaside buffer}
\newacronym{tluh}{TL-UH}{TileLink Uncached Heavyweight}
\newacronym{tlul}{TL-UL}{TileLink Uncached Lightweight}
\newacronym{tmu}{TMU}{transaction monitor unit}
\newacronym{uav}{UAV}{unmanned aerial vehicles}
\newacronym{ulp}{ULP}{ultra-low-power}
\newacronym{vlsi}{VLSI}{very large scale integration}
\newacronym{vlsu}{VLSU}{vector load-store unit}
\newacronym{vm}{VM}{virtual memory}
\newacronym{vrm}{VRM}{voltage regulator module}
\newacronym{vv}{V\&V}{validation and verification}
\newacronym{wcdt}{WCDT}{worst-case detection time}
\newacronym{wcet}{WCET}{worst-case execution time}
\newacronym{zif}{ZIF}{zero insertion force}

\DeclareSIUnit[quantity-product= ]\percent{\%}
\DeclareSIUnit\bit{bit}
\DeclareSIUnit\dpi{dpi}
\DeclareSIUnit\flops{FLOPS}
\DeclareSIUnit\flop{FLOP}
\DeclareSIUnit\gateeq{GE}
\DeclareSIUnit\GE{GE}
\DeclareSIUnit\kGE{\kilo\GE}
\DeclareSIUnit\MGE{\mega\GE}
\DeclareSIUnit\ops{OPS}
\DeclareSIUnit\pixel{px}
\DeclareSIUnit{\x}{\!\times}
\DeclareSIUnit\ff{FF}
\DeclareSIUnit\lut{LUT}
\crefname{lstlisting}{Listing}{Listings}
\Crefname{lstlisting}{Listing}{Listings}

\newcommand{\glsf}[1]{\glsreset{#1}\gls{#1}}
\newcommand{\glsplf}[1]{\glsreset{#1}\glspl{#1}}
\newcommand{\Glsf}[1]{\glsreset{#1}\Gls{#1}}

\newcommand{\glsu}[1]{\glsunset{#1}\gls{#1}}

\newlength\myheight
\newlength\mydepth
\settototalheight\myheight{Xygp}
\newcommand*\circnum[1]{\tikz[baseline=(char.base)]{%
            \node[white,shape=circle,fill=ieee-dark-black-100,draw,inner sep=1pt] (char) {\color{ieee-bright-white-100}\sffamily #1};}}

\newenvironment{publications}{\begin{itemize}[align=parleft]\footnotesize}{\end{itemize}}
\newcommand{\publication}[1]{\item[\textbf{\cite{#1}}] \bibentry{#1}}

\nobibliography* %

\newcommand{\rott}[1]{\rotatebox[origin=c]{90}{#1}}
\newcommand{\divl}{\noalign{\smallskip} \arrayrulecolor{gray}\cline{2-15} \noalign{\smallskip}}

\newcommand{\res}[4]{\makecell[cl]{\SI{#1}{#3}{#4} \\ $\mathcal{O}$\textit{(#2)}}}

\newcommand{\resh}[4]{\SI{#1}{#3}{#4}}

\newcommand{\resn}{\textcolor{gray}{\textit{0}}}

\newcommand{\hol}[1]{\makecell[cc]{\textbf{#1}}}
\newcommand{\htl}[2]{\makecell[cc]{\textbf{#1} \\ \textbf{#2}}}

\newcommand{\did}[2]{\vspace{0.05cm}\makecell[cc]{{#1} \\ \emph{#2}}\vspace{0.05cm}}
\newcommand{\dis}[1]{\vspace{0.05cm}\makecell[cc]{\emph{#1}}\vspace{0.05cm}}
\newcommand{\dtw}[2]{\vspace{0.05cm}\makecell[cc]{\textbf{#1} \\ \emph{#2}}\vspace{0.05cm}}

\newcommand{\rot}[1]{\rotatebox[origin=c]{90}{#1}}
\newcommand{\tilt}[1]{\hspace{-1cm}\rotatebox[origin=c]{32}{#1}}
\newcommand{\noc}{\textcolor{ieee-dark-grey-40}{0}}

\newcommand{\na}{\textcolor{ieee-dark-grey-40}{N.A.}}
\newcommand{\nad}{\textcolor{ieee-dark-grey-40}{-}}
\newcommand{\dl}[2]{\makecell[cc]{#1 \\ #2}}
\newcommand{\dll}[2]{\makecell[cl]{#1 \\ #2}}
\newcommand{\dlb}[2]{\makecell[cc]{\textbf{#1} \\ \textbf{#2}}}
\newcommand{\tl}[3]{\makecell[cc]{#1 \\ #2 \\ #3}}
\newcommand{\tll}[3]{\makecell[cl]{#1 \\ #2 \\ #3}}
\newcommand{\tlb}[3]{\makecell[cc]{\textbf{#1} \\ \textbf{#2} \\ \textbf{#3}}}
\def\thetitle{Development of an Energy-Efficient and Real-Time Data Movement Strategy for Next-Generation Heterogeneous Mixed-Criticality Systems}

\IfFileExists{identifier.tex}{
    \def\reviewpass{v1.0.0-48ffc2d995b0f07f}

}{
    \def\reviewpass{Overleaf Version - Do not distribute!}
}

\ifx\showrevision\undefined
\else
    \AddToShipoutPictureFG{%
        \put(%
            8mm,%
            \paperheight-1.2cm%
            ){\vtop{{\null}\makebox[0pt][c]{%
                \rotatebox[origin=c]{90}{%
                    \large\textcolor{red!75}{\reviewpass}%
                }%
            }}%
        }%
    }
    \AddToShipoutPictureFG{%
        \put(%
            \paperwidth-6mm,%
            \paperheight-1.2cm%
            ){\vtop{{\null}\makebox[0pt][c]{%
                \rotatebox[origin=c]{90}{%
                    \large\textcolor{red!30}{ETH Zurich - Unpublished - Confidential - Draft - Copyright Thomas 2025}%
                }%
            }}%
        }%
    }
\fi

\ifx\showrevision\undefined
    \newcommand{\todo}[1]{{#1}}
\else
    \newcommand{\todo}[1]{{\textcolor{red}{#1}}}
\fi

\ifx\showrebuttal\undefined
    
\else
    
\fi

\ifx\showrebuttal\undefined
    \newcommand{\revdel}[1]{}
\else
    \newcommand{\revdel}[1]{\textcolor{ieee-bright-red-100}{\st{#1}}}
\fi

\ifx\showrebuttal\undefined
    
\else
    
\fi

\ifx\showrebuttal\undefined
    \newcommand{\revprg}[1]{}
\else
    \newcommand{\revprg}[1]{\hspace{-0.5ex}\textcolor{ieee-bright-red-100}{\scalebox{.2}[1.5]{$\blacksquare$}}\hspace{-0.5ex}}
\fi

\ifx\showlandscape\undefined
    \newenvironment{sidetab}{\begin{sidewaystable}}{\end{sidewaystable}}
    \newenvironment{sidefig}{\begin{sidewaysfigure}}{\end{sidewaysfigure}}
\else
    \newenvironment{sidetab}{\begin{landscape}\centering\begin{table}[p]}{\end{table}\end{landscape}}
    \newenvironment{sidefig}{\begin{landscape}\centering\begin{figure}[p]}{\end{figure}\end{landscape}}
\fi

\ifx\showfeedback\undefined
    \newcommand{\lb}[1]{#1}
    \newcommand{\alc}[1]{#1}
    \newcommand{\teb}[1]{#1}
    \newcommand{\tebsr}[1]{#1}
\else
    \newcommand{\lb}[1]{{\textcolor{ieee-bright-lblue-100}{#1}}}
    \newcommand{\alc}[1]{{\textcolor{ieee-bright-dgreen-100}{#1}}}
    \newcommand{\teb}[1]{{\textcolor{ieee-bright-red-100}{#1}}}
    \newcommand{\tebsr}[1]{{\textcolor{ieee-bright-purple-100}{#1}}}
\fi

\makeglossaries
\renewcommand{\glossarysection}[2][]{}

\setglossarystyle{long}
\renewcommand{\glsnamefont}[1]{\textbf{#1}}
\renewcommand{\glspostdescription}{}                %
\renewcommand{\glsnumberformat}[1]{\hspace{1.5mm}(p.$\,$\glshypernumber{#1})}  %

\hypersetup{
  pdftitle = {\thetitle},
  pdfauthor = {Thomas Emanuel Benz},
  pdfsubject = {Ph.D. dissertation, ETH Zurich, Zurich, Switzerland},
  pdfkeywords = {DMA, DMA Architecture, RISC-V, Real-Time, Mixed-Criticality, Embedded Systems, HPC},
  pdfstartview = {Fit},     %
  bookmarksopen = {true},
  bookmarksnumbered = {true},
  bookmarksopenlevel = {0}, %
  colorlinks = {false},  %
  pdfstartpage = 1          %
}

\makeatletter
\def\bstctlcite{\@ifnextchar[{\@bstctlcite}{\@bstctlcite[@auxout]}}
\def\bstctlcite[#1]#2{\@bsphack
  \@for\@citeb:=#2\do{%
    \edef\@citeb{\expandafter\@firstofone\@citeb}%
    \if@filesw\immediate\write\csname #1\endcsname{\string\citation{\@citeb}}\fi}%
  \@esphack}
\makeatother

\makeatletter
  \def\theHALC@line{\thealgorithm-\theALC@line}
  \def\theHALC@rem{\thealgorithm-\theALC@rem}
\makeatother

\begin{document}

\newcolumntype{R}{>{\raggedleft\arraybackslash}X}

\title{\thetitle}
\author{Thomas Emanuel\ Benz}

\dissnum{\teb{31645}}
\acdegree{DOCTOR OF SCIENCES \\ (Dr. sc. ETH Zurich)}
\dateofbirth{April~19th,~1994}
\examiner{Prof.\ Dr.\ Luca\ Benini}
\coexaminer{Prof.\ Dr.\ Alessandro\ Capotondi}
\thesisyear{2025}

\volume{\todo{-}}
\isbn{\todo{?-?????-???-?}}         %
\isbnLong{\todo{???-?-?????-???-?}} %
\published{2025}

\frontmatter

\maketitle       %

\selectlanguage{english}
\clearpage
\cleardoublepage
\newcommand\narrowstyle{\SetTracking{encoding=*}{-60}\lsstyle}
\thispagestyle{empty}
\vfill

\begin{center}

    \parbox{10.5cm}{%
        \begin{justify}
            \itshape%
            The worthwhile problems are the ones you can really solve or help solve, the ones you can really contribute something to.... No problem is too small or too trivial if we can really do something about it. \par\bigskip
        \end{justify}
        \raggedleft{Richard Feynman}\par%
    }

    \vspace{2.5cm}

    \parbox{10.5cm}{%
        \begin{justify}
            \itshape%
            Efficiency is doing things right; effectiveness is doing the right things. \par\bigskip
        \end{justify}
        \raggedleft{Peter Drucker}\par%
    }

    \vspace{2.5cm}

    \parbox{10.5cm}{%
        \begin{justify}
            \itshape%
            Better three hours too soon than a minute too late. \par
        \end{justify}
        \raggedleft{William Shakespeare}\par%
    }

    \vspace{2.5cm}

    \parbox{10.5cm}{%
        \begin{justify}
            \itshape%
            I am Doug Dimmadome, owner of the Dimmsdale Dimmadome!\par\bigskip
        \end{justify}
        \raggedleft{Doug Dimmadome \\ The Fairly OddParents}\par%
    }

\end{center}
\vfill\vfill
\clearpage

\clearpage
\chapter{Acknowledgments}
\label{chap:acknowledgments}

Working towards my Ph.D degree has been the most intense part of my education so far.
It not only provided me with a broad knowledge base and an immense set of practical and technical skills but also helped me grow personally and professionally through countless challenges.
This unbelievable journey I made over the past five years would, of course, not have been possible without the support of many people I now want to thank.

First and foremost, I want to thank my supervisor, Prof. Dr. Luca Benini, for guiding and supporting me along my academic journey by constantly providing me with critical but constructive feedback.
Luca, the opportunities you provided me during the last five years are truly unique; the unequaled scientific research, technical work, and teaching opportunities were fantastic for me to refine my skill set and to grow as a researcher, teacher, and person.
I further want to thank you for your immense patience with me, the amount of trust issued, and the level of freedom granted while I was exploring countless side quests not directly related to my core research, but proved still invaluable to my personal development or to the advancement of the entire research group.
I highly doubt that I would have found an equally supportive research group that could have provided me with the same opportunities elsewhere.

Second, I want to thank my second supervisor, Prof. Dr. Torsten Hoefler, for your support along my journey.
I am especially thankful for your valuable feedback on my first journal manuscript, allowing me to submit a much more polished draft.

Third, I owe my deepest gratitude to my co-examiner, Prof. Dr. Alessandro Capotondi for his invaluable and insightful feedback, allowing me to improve and finalize my thesis\tebsr{ }\teb{and for driving an engaging discussion during my defense}.

Further, I want to thank Dr. Frank Kagan Gürkaynak for leading the Microelectronics Design Center (DZ), shielding us from tedious (European) bureaucracy, and creating and supporting a fantastic work environment for us to foster.
Frank, I know your role here is not easy, but this group would not exist in today's thriving form without your constant support. Thank you.
Special thanks go to the entire DZ for providing us with an excellently maintained EDA environment, a broad set of cutting-edge PDKs, and an unbelievably vast knowledge in all facets of IC design. Beat Muheim, Alfonso Blanco Fontao, Zerun Jiang, and \teb{Dr. Arianna Rubino}, thank you for your tireless work.

One of the fundamental pillars of our group's success is the fertile collaboration with our colleagues at the University of Bologna.
I am especially grateful to Prof. \teb{Dr.} Angelo Garofalo, Prof. \teb{Dr.} Davide Rossi, Prof. \teb{Dr.} Andrea Bartolini, \teb{Dr. Luca Valente}, Simone Manoni, and Chaoqun Liang.

I am deeply indebted to Prof. em. Dr. Hubert Kaeslin for writing two such excellent books on VLSI design and for getting me into ASIC design.

A large thank you to IIS's IT team for providing such an excellent IT infrastructure for us to complete our work.
Christoph Wicki, Mateo Juric, and Adam Feigin, thank you.

During my Ph.D. I was able to manufacture countless prototypes.
These would not have been possible without the constructive support and tireless work of ETH's \teb{technical staff}.
I am deeply grateful to Hansjörg Gisler, Thomas Quanbrough, Thomas Kleier, Aldo Rossi, Michael Lerjen, and Silvio Scherr.

I want to thank Irina Rau for her many cheerful words in phases of great despair and for constantly radiating an aura of positivity.

I am very grateful to Dr. Luca Aloatti for introducing me at an early stage during my studies to the importance of free and open-source hardware and systems.

This endeavor would not have been possible without Paul Scheffler.
Paul, I am deeply grateful for the last eleven years of knowing you both as a friend and as a colleague.
I've enjoyed our countless academic and recreational (and sometimes heated) discussions alongside my journey towards all my scientific qualifications to date.
Your organization, excellent scientific and technical methodology and skills, and your pristine eye for detail are truly inspiring.

I was fortunate enough to spend most of my time during my studies in J85, a truly magical place with many oddities and facilities other offices could not even imagine.
I am extremely grateful to my J85 colleagues, Paul Scheffler, Nils Wistoff, Matteo Perotti, and Philippe Sauter, for having many fruitful technical discussions and for making this office feel like home.

During my studies, I had the fortune to work alongside many bright minds on similar journeys.
I am deeply indebted to Alessandro Ottaviano, Michael Rogemoser, \teb{Jannis Schönleber,} Christopher Reinwardt, Enrico Zelioli, Sergio Mazzola, Robert Balas, Lorenzo Leone, Tim Fischer, Luca Colagrande, Luca Bertaccini, \teb{Victor Jung, Viviane Potocnik, Cyril Koenig,} Cristian Cioflan, \teb{Yichao Zhang, and Chi Zhang} for so many fruitful collaborations.

My thesis builds on the foundation laid by previous generations of Ph.D. students working at IIS.
Many thanks to Dr. Florian Zaruba, Dr. Fabian Schiki, Dr. Andreas Kurth, \teb{Dr. Pirmin Vogel,} Dr. Bjoern Forsberg, Dr. Gianna Paulin, Dr. Georg Rutishauser, Dr. Manuel Eggimann, Dr. Samuel Riedel, Dr. Matheus Cavalcante, \teb{\tebsr{Dr.} Davide Schiavone, \tebsr{Dr.} Moritz Scherer, \tebsr{Dr.} Alfio Di Mauro,} and Wolfgang Rönninger.

I am also grateful to Armin P. Barth, Thomas Notter, Kurt Doppler, and Ruedi \teb{Georg} Vogt for awakening my interest in physics, mathematics, computer science, natural sciences and for excellently preparing me for my studies at ETH Zürich.

Many thanks to Prof. Dr. Alberto Colotti, Dr. Quentin Lohmayer, Prof. em. Dr. Elsbeth Stern, Prof. Dr. Hanspeter Hochreutener, and Dr. Beatrice Trummer for guiding me in becoming a better teacher and for strengthening my didactical skill set.

Besides oneself, the journey towards a Ph.D. degree also takes a toll on family and friends.

Words cannot express my gratitude towards my parents, Ursula Ramona Benz-Wullschleger and Peter Johann Benz, for constantly supporting me since the day I was born, for often putting my interests above theirs, and for suffering through countless educational crises along the journey towards my degree.

My sincere thanks go to Ursula Elisabeth Trüb and Jürg Friedrich Trüb for so many joyful and happy Friday afternoons spent together.
Ursula and Jürg, thank you for sparking my interest in electronics and computers and for supporting me during so many of my earlier projects.

I want to thank Edith Benz for providing me some distraction from my Ph.D. studies by posing me one or another technical challenge to solve.

I further want to thank Rolf Wullschleger for supporting me alongside my journey.
I know a car workshop is not a playground, yet you have provided me with a unique place to explore and learn about cars and technical machines.

I would like to express my deepest gratitude to Karla Madžarević. Karla, thanks for all the love, unwavering support, endless encouragement, and limitless patience in the face of long working hours, countless night shifts, and many weekends spent working. This unbelievable journey would not have been possible without you.
A special thanks goes to Jasminka and Karlo Madžarević for so lovingly accepting me into their family.

Thanks should also go to all my friends for helping me forget work for a couple of hours and for many therapeutic conversations far from electronics, science, and my thesis.
I in particular want to thank David Perels, \teb{Lorenz Becker-Sander, Gani Aliguzhinov, Gian Marti, Thomas Kramer,} Sheila Peterhans, Seline Frei, Raphael Kuhn, and Juri Nowak.

Further, I want to thank Martin Stössel for being very patient and understanding with me and my life as a Ph.D. student.
I am further thankful for teaching me excellently how to keep a cool head in stressful situations, how to prioritize important tasks under stress, and the importance of always being one step ahead of the machine you are operating.

Last but by no means least, many thanks to Sherlock Olaf, Prof. Moriarty, Indira Yuna Aimé, and Cookie for providing me company during so many hours of my Ph.D. journey.

\ifx\showcats\undefined
    \newpage
    \null
    \thispagestyle{empty}
    \newpage
\else
    \begin{figure}[t]
        \centering%
        \includegraphics[width=\columnwidth]{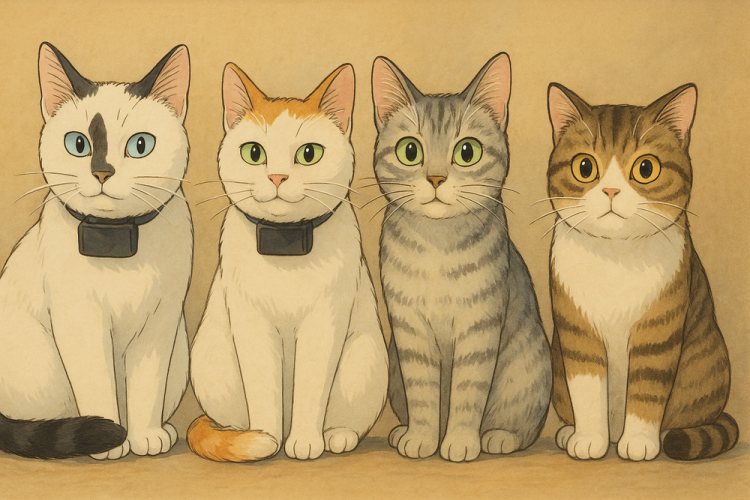}%
        \caption*{%
            An AI-generated render, using OpenAI's GPT-4o model, of Sherlock Olaf, Prof. Moriarty, Indira Yuna Aimé, and Cookie (left to right). %
        }%
        \label{fig:ack:cats}%
    \end{figure}
\fi

\selectlanguage{ngerman}
\clearpage
\chapter{Danksagung}
\label{chap:danksagung}

Die Arbeit an meiner Doktorarbeit war der intensivste Teil meiner bisherigen Ausbildung.
Sie hat mir nicht nur ein breites Wissen und eine Vielzahl praktischer und technischer Fähigkeiten vermittelt, sondern mir auch durch unzählige Herausforderungen geholfen, mich persönlich und beruflich weiterzuentwickeln.
Diese unglaubliche Reise, die ich in den letzten fünf Jahren unternommen habe, wäre natürlich ohne die Unterstützung vieler Menschen, denen ich nun danken möchte, nicht möglich gewesen.

Zuallererst möchte ich meinem Betreuer, Prof. Dr. Luca Benini, dafür danken, dass er mich auf meinem akademischen Weg begleitet und unterstützt hat, indem er mir stets kritisches, aber konstruktives Feedback gegeben hat.
Luca, die Möglichkeiten, die du mir in den letzten fünf Jahren geboten hast, sind wirklich einzigartig und die unvergleichlichen wissenschaftlichen Forschungs-, und Lehrmöglichkeiten waren fantastisch für mich, um meine Fähigkeiten zu verfeinern und mich als Forscher, Lehrer und Mensch weiterzuentwickeln.
Ausserdem möchte ich dir für deine immense Geduld mit mir, das mir entgegengebrachte Vertrauen und die Freiheit danken, die du mir gewährt hast, während ich unzählige Nebenprojekte verfolgt habe, die zwar nicht direkt mit dem Kern meiner Forschung zu tun hatten, sich aber dennoch als unschätzbar wertvoll für meine persönliche Entwicklung und den Fortschritt der gesamten Forschungsgruppe erwiesen haben.
Ich bezweifle sehr, dass ich anderswo eine ebenso unterstützende Forschungsgruppe gefunden hätte, die mir die gleichen Möglichkeiten geboten hätte.

Zweitens möchte ich meinem \teb{Zweitbetreuer}, Prof. Dr. Torsten Hoefler, für seine Unterstützung während meiner gesamten Laufbahn danken.
Ich bin besonders dankbar für sein wertvolles Feedback zu meinem ersten Artikel in einer Fachzeitschrift, wodurch ich einen wesentlich ausgefeilteren Entwurf einreichen konnte.

Drittens bin ich meinem Koexaminator, Prof. Dr. Alessandro Capotondi, zu tiefstem Dank verpflichtet für sein unschätzbares und aufschlussreiches Feedback, das mir half, meine Dissertation zu verbessern und fertigzustellen \teb{und für die anregende Diskussion während meiner Verteidigung.}

Ausserdem möchte ich Dr. Frank Kagan Gürkaynak dafür danken, dass er das Microelectronics Design Center (DZ) leitet, uns vor mühsamer Bürokratie schützt und ein fantastisches Arbeitsumfeld für uns geschaffen und unterstützt hat.
Frank, ich weiss, dass deine Aufgabe hier nicht einfach ist, aber ohne deine ständige Unterstützung würde diese Gruppe heute nicht in ihrer gedeihend Form existieren. Vielen Dank.
Ein besonderer Dank geht an das gesamte DZ für die Bereitstellung einer hervorragend gepflegten EDA-Umgebung, einer breiten Palette an modernsten PDKs und einem unglaublich umfangreichen Wissen in allen Bereichen des IC-Designs.
Beat Muheim, Alfonso Blanco Fontao, Zerun Jiang und \teb{Dr. Arianna Rubino}, vielen Dank für eure unermüdliche Arbeit.

Eine der grundlegenden Säulen für den Erfolg unserer Gruppe ist die fruchtbare Zusammenarbeit mit unseren Kollegen an der Universität Bologna.
Mein besonderer Dank gilt Prof. \teb{Dr.} Angelo Garofalo, Prof. \teb{Dr.} Davide Rossi, Prof. \teb{Dr.} Andrea Bartolini, \teb{Dr. Luca Valente}, Simone Manoni und Chaoqun Liang.

Ich bin Prof. em. Dr. Hubert Kaeslin zu tiefstem Dank verpflichtet, dass er zwei so hervorragende Bücher über VLSI-Design geschrieben und mich in das ASIC-Design eingeführt hat.

Ein grosses Dankeschön geht an das IT-Team des IIS, das uns eine hervorragende IT-Infrastruktur zur Verfügung gestellt hat, damit wir unsere Arbeit abschliessen konnten.
Christoph Wicki, Mateo Juric und Adam Feigin, vielen Dank.

Während meiner Promotion konnte ich unzählige Prototypen herstellen.
Ohne die konstruktive Unterstützung und die unermüdliche Arbeit der \teb{technischen} Mitarbeiter der ETH wäre dies nicht möglich gewesen.
Ich bin Hansjörg Gisler, Thomas Quanbrough, Thomas Kleier, Aldo Rossi, Michael Lerjen und Silvio Scherr zutiefst dankbar.

Ich möchte Irina Rau für ihre vielen aufmunternden Worte in Phasen grosser Verzweiflung und dafür danken, dass sie stets eine positive Ausstrahlung hat.

Ich bin Dr. Luca Aloatti sehr dankbar, dass er mir schon früh während meines Studiums die Bedeutung von freier und quelloffener Hardware und Systemen nähergebracht hat.

Ohne Paul Scheffler wäre dieses Unterfangen nicht möglich gewesen.
Paul, ich bin zutiefst dankbar für die letzten elf Jahre, in denen ich dich sowohl als Freund als auch als Kollegen kennenlernen durfte.
Ich habe unsere unzähligen akademischen und unterhaltsamen (und manchmal hitzigen) Diskussionen während meiner bisherigen wissenschaftlichen Laufbahn sehr genossen. Deine Organisation, deine exzellenten wissenschaftlichen und technischen Methoden und Fähigkeiten sowie dein unfehlbares Auge fürs Detail sind wirklich inspirierend.

Ich hatte das Glück, den grössten Teil meiner Studienzeit in J85 zu verbringen, einem wahrhaft magischen Ort mit vielen Kuriositäten und Einrichtungen, die andere Büros sich nicht einmal vorstellen können.
Ich bin meinen Kollegen von J85, Paul Scheffler, Nils Wistoff, Matteo Perotti und Philippe Sauter, sehr dankbar für die vielen fruchtbaren technischen Diskussionen und dafür, dass ich mich in diesem Büro wie zu Hause gefühlt habe.

Während meines Studiums hatte ich das Glück, mit vielen klugen Köpfen zusammenzuarbeiten, die sich auf einer ähnlichen Reise befanden.
Ich bin Alessandro Ottaviano, Michael Rogemoser, \teb{Jannis Schönleber,} Christopher Reinwardt, Enrico Zelioli, Sergio Mazzola, Robert Balas, Lorenzo Leone, Tim Fischer, Luca Colagrande, Luca Bertaccini, \teb{Victor Jung, Viviane Potocnik, Cyril Koenig,} Cristian Cioflan\teb{\tebsr{,} Yichao Zhang und Chi Zhang} für die vielen fruchtbaren Kooperationen zu tiefstem Dank verpflichtet.

Meine Dissertation baut auf den Grundlagen auf, die von früheren Doktoranden am IIS gelegt wurden.
Mein herzlicher Dank gilt Dr. Florian Zaruba, Dr. Fabian Schiki, Dr. Andreas Kurth, \teb{Dr. Pirmin Vogel,} Dr. Bjoern Forsberg, Dr. Gianna Paulin, Dr. Georg Rutishauser, Dr. Manuel Eggimann, Dr. Samuel Riedel, Dr. Matheus Cavalcante, \teb{\tebsr{Dr.} Davide Schiavone, \tebsr{Dr.} Moritz Scherer, \tebsr{Dr.} Alfio Di Mauro} und Wolfgang Rönninger.

Ich bin auch Armin P. Barth, Thomas Notter, Kurt Doppler und Ruedi \teb{Georg} Vogt dankbar dafür, dass sie mein Interesse an Physik, Mathematik, Informatik und Naturwissenschaften geweckt und mich hervorragend auf mein Studium an der ETH Zürich vorbereitet haben.

Vielen Dank an Prof. Dr. Alberto Colotti, Dr. Quentin Lohmayer, Prof. em. Dr. Elsbeth Stern, Prof. Dr. Hanspeter Hochreutener und Dr. Beatrice Trummer, die mich dabei begleitet haben, ein besserer Lehrer zu werden, und meine didaktischen Fähigkeiten gestärkt haben.

Neben einem selbst fordert der Weg zum Doktorgrad auch von Familie und Freunden seinen Tribut.

Ich kann meine Dankbarkeit gegenüber meinen Eltern, Ursula Ramona Benz-Wullschleger und Peter Johann Benz, gar nicht in Worte fassen. Sie haben mich seit meiner Geburt stets unterstützt, oft meine Interessen über ihre eigenen gestellt und unzähligen Bildungskrisen auf meinem Weg zum Abschluss mitgemacht.

Mein aufrichtiger Dank gilt Ursula Elisabeth Trüb und Jürg Friedrich Trüb für so viele fröhliche und glückliche Freitagnachmittage, die wir gemeinsam verbracht haben.
Ursula und Jürg, ich danke euch dafür, dass ihr mein Interesse an Elektronik und Computern geweckt und mich bei so vielen meiner früheren Projekte unterstützt habt.

Ich möchte Edith Benz dafür danken, dass sie mir durch die Lösung der einen oder anderen technischen Herausforderung eine willkommene Abwechslung zu meinem Doktoratsstudium geboten hat.

Ausserdem möchte ich Rolf Wullschleger dafür danken, dass er mich auf meinem Weg unterstützt hat.
Ich weiss, dass eine Autowerkstatt kein Kinderspielplatz ist, aber du hast mir einen einzigartigen Ort geboten, an dem ich Autos und technische Maschinen erkunden und mehr über sie lernen konnte.

Meine tiefste Dankbarkeit gilt Karla Madžarević.
Karla, danke für all die Liebe, die unerschütterliche Unterstützung, die unendliche Ermutigung und die grenzenlose Geduld angesichts der langen Arbeitszeiten, der unzähligen Nachtschichten und der vielen Wochenenden, die ich mit Arbeiten verbracht habe.
Ohne dich wäre diese unglaubliche Reise nicht möglich gewesen.
Ein besonderer Dank geht an Jasminka und Karlo Madžarević, die mich so liebevoll in ihre Familie aufgenommen haben.

Mein Dank gilt auch all meinen Freunden, die mir geholfen haben, die Arbeit für ein paar Stunden zu vergessen, und für viele therapeutische Gespräche fernab von Elektronik, Wissenschaft und meiner Dissertation.
Insbesondere möchte ich David Perels, \teb{Lorenz Becker-Sander, Gani Aliguzhinov, Gian Marti, Thomas Kramer,} Sheila Peterhans, Seline Frei, Raphael Kuhn und Juri Nowak danken.

Ausserdem möchte ich Martin Stössel dafür danken, dass er mir und meinem Leben als Doktorand gegenüber so geduldig und verständnisvoll war.
Ich bin ihm auch dankbar dafür, dass er mir auf hervorragende Weise beigebracht hat, wie man in Stresssituationen einen kühlen Kopf bewahrt, wie man unter Druck wichtige Aufgaben priorisiert und wie wichtig es ist, der Maschine, die man bedient, immer einen Schritt voraus zu sein.

Zu guter Letzt möchte ich mich ganz herzlich bei Sherlock Olaf, Prof. Moriarty, Indira Yuna Aimé und Cookie bedanken, die mir während meiner langjährigen Promotion Gesellschaft geleistet haben.

\ifx\showcats\undefined
    \newpage
    \null
    \thispagestyle{empty}
    \newpage
\else
    \begin{figure}[t]
        \centering%
        \includegraphics[width=\columnwidth]{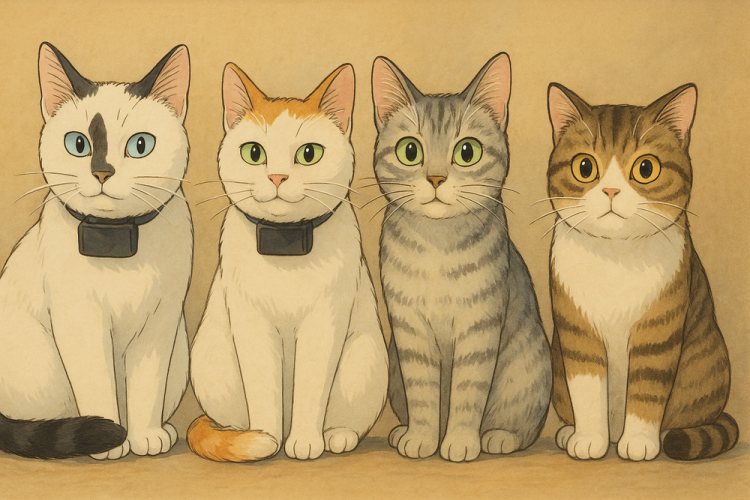}%
        \caption*{%
        Eine KI-generierte Darstellung mit dem GPT-4o-Modell von OpenAI von Sherlock Olaf, Prof. Moriarty, Indira Yuna Aimé und Cookie (von links nach rechts). %
        }%
        \label{fig:ack:cats:ger}%
    \end{figure}
\fi

\selectlanguage{english}
\clearpage
\chapter{Abstract}
\label{chap:abstract}

Industrial domains such as automotive, robotics, and aerospace are rapidly evolving to satisfy the increasing demand for machine-learning-driven \gls{aces}.
This paradigm shift inherently and significantly increases the requirement for onboard computing performance and high-performance communication infrastructure.
At the same time, {Moore’s Law} and {Dennard Scaling} are grinding to a halt, in turn, driving computing systems to larger scales and higher levels of heterogeneity and specialization, through application-specific hardware accelerators, instead of relying on technological scaling only.
Approaching \gls{aces} requires this substantial amount of compute at an increasingly high energy-efficiency, since most use cases are fundamentally resource-bound.

This increase in compute performance and heterogeneity goes hand in hand with a growing demand for high memory bandwidth and capacity as the driving applications grow in complexity, operating on huge and progressively irregular data sets and further requiring a steady influx of sensor data, increasing pressure both on on-chip and off-chip interconnect systems.
Further, \gls{aces} combines real-time time-critical with general compute tasks on the same physical platform, sharing communication, storage, and micro-architectural resources.
These heterogeneous \glspl{mcs} place additional pressure on the interconnect, demanding minimal contention between the different criticality levels to sustain a high degree of predictability.
Fulfilling the performance and energy-efficiency requirements across a wide range of industrial applications requires a carefully co-designed process of the memory system with the use cases as well as the compute units and accelerators.

Firstly, this thesis tackles efficient, agile, and high-performance data movement in heterogeneous systems by presenting a modular and highly customizable \gls{dma} architecture, \emph{\idma}, serving the diverse needs of today's heterogeneous platforms.
Secondly, we introduce \emph{\axirealm}, a modular interconnect extension, tackling the predictability problem arising in heterogeneous \gls{mcs} running real-time-critical applications in the presence of \gls{dma} transfers originating from domain-specific accelerators.

In multiple case studies, we show the applicability and the benefits of {\idma} in silicon-implemented systems and across the entire interconnect hierarchy, including an application-grade Linux-capable \gls{soc}, a high-performance general-purpose compute accelerator, and an automotive-grade \gls{mcs}.
In the latter, we present the benefits of {\axirealm} restoring performance of time-critical applications running on \glspl{mcs} in the presence of contention caused by domain-specific accelerator \gls{dma} engines.

\selectlanguage{ngerman}
\glsresetall
\clearpage
\chapter{Zusammenfassung}
\label{chap:zusammenfassung}

Industrielle Bereiche wie Automobilbau, Robotik und Luft- und Raumfahrt entwickeln sich rasant weiter, um der steigenden Nachfrage nach  Autonomie, Konnektivität, Elektrifizierung und gemeinsamer Mobilität (ACES), gesteuert durch machinelles Lernen, gerecht zu werden.
Dieser Paradigmenwechsel erhöht naturgemäss und erheblich die Anforderungen an die Rechenleistung an Bord und an eine leistungsstarke Kommunikationsinfrastruktur.
Gleichzeitig kommen Moores Gesetz und Dennards Skalierung zum Stillstand, was wiederum dazu führt, dass Computersysteme durch anwendungsspezifische Hardwarebeschleuniger grösser und heterogener werden und sich stärker spezialisieren, anstatt sich nur auf technologische Skalierung zu verlassen.
Die Umsetzung von ACES erfordert diese erhebliche Rechenleistung bei einer immer höheren Energieeffizienz, da die meisten Anwendungsfälle grundsätzlich ressourcengebunden sind.
Diese Steigerung der Rechenleistung und Heterogenität geht Hand in Hand mit einem wachsenden Bedarf an hoher Speicherbandbreite und -kapazität, da die treibenden Anwendungen immer komplexer werden, mit riesigen und zunehmend unregelmässigen Datensätzen arbeiten und darüber hinaus einen stetigen Zufluss von Sensordaten erfordern, was den Druck auf Kommunikationsnetzwerke, innerhalb desselben integrierten Schaltkreises oder bei dessen Kommunikation mit der Aussenwelt, erhöht.
Darüber hinaus kombiniert ACES zeitkritische Echtzeitaufgaben mit allgemeinen Rechenaufgaben auf derselben physischen Plattform und teilt sich dabei Kommunikations-, Speicher- und Mikroarchitekturressourcen.
Diese heterogenen Systeme mit unterschiedlichen Kritikalitätsstufen (MCS) üben zusätzlichen Druck auf die Verbindung aus und erfordern eine minimale Konkurrenz zwischen den verschiedenen Kritikalitätsstufen, um ein hohes Mass an Vorhersagbarkeit aufrechtzuerhalten.
Um die Leistungs- und Energieeffizienzanforderungen in einem breiten Spektrum industrieller Anwendungen zu erfüllen, ist ein sorgfältig abgestimmter Prozess des Speichersystems mit den Anwendungsfällen sowie den Recheneinheiten und Beschleunigern erforderlich.

Erstens befasst sich diese Arbeit mit der effizienten, agilen und leistungsstarken Datenübertragung in heterogenen Systemen, indem sie eine modulare und hochgradig anpassbare Architektur für direkte Speicherzugriffseinheiten (DMA) namens iDMA vorstellt, die den vielfältigen Anforderungen heutiger heterogener Plattformen gerecht wird.
Zweitens stellen wir AXI-REALM vor, eine modulare Kommunikationsnetzwerkserweiterung, die das Problem der Vorhersagbarkeit angeht, das in heterogenen MCS auftritt, die zeitkritische Anwendungen ausführen, wenn DMA-Übertragungen von domänenspezifischen Beschleunigern stattfinden.

In mehreren Fallstudien zeigen wir die Anwendbarkeit und die Vorteile von iDMA in siliziumimplementierten Systemen und über die gesamte Kommunikationsnetzwerkshierarchie hinweg, einschliesslich eines Linux-fähigen Ein-Chip-System (SoC), eines hochleistungsfähigen Allzwecksrechenbeschleunigers und eines MCS in Automobilqualität.
Im letzteren Fall präsentieren wir die Vorteile von AXI-REALM, das die Leistung zeitkritischer Anwendungen wiederherstellt, die auf MCSs laufen, wenn es zu Konflikten durch direkte Speicherzugriffseinheiten der domänenspezifischen Beschleuniger kommt.

\glsresetall
\clearpage
\chapter{Zämäfassig}
\label{chap:zaemaefassig}

Industrielli Beriich wie de Automobilbau, d Robotik, und Luft- und Ruhmfahrt entwickled sich \teb{rasch} wiiter um mit de steigende Nachfrog nach Autonomie, Konnektivität, Elektrifizierig und gmeinsamer Mobilität (ACES), gstüüred dur maschinells Lerne, grecht z werde.
Dä Paradigmewechsel erhöht naturgmäss und au erheblich d Aforderige ad Recheleischtige an Board und ane leistigsstarchi Kommunikationsinfrastruktur.
Gliichziitig chömmed im Moores sis Gsetz und em Dennard sini Skalierig zumne Halt, was wederum dezu fürt, dass Computersystem dur awendigsspezifischi Hardwarebeschlüniger grösser und heterogener werded und sich stärcher spezialisiered, anstatt sich nur uf die technologischi Skalirig z verloh.
D Umsetzig vo ACES erfordered die erheblichi Recheleischtig binere immer höchere Energieffizienz, da die meischte Awändigsfäll grundsätzlich ressourcegebunde sind.
Die Steigerig vo de Recheleischtig und de Hereogenität gohd Hand in Hand mit emne wachsende Bedarf a höcher Speicherbandbreiti und -kapazität, da die tribende Awändige immer komplexer werded, mit risigne und zuhnemned unregelmässigere Datesätz schaffed und drüberhinus en stetige Zuefluss vo Sensordate erfordered, was de Druck ufd Kommunikationsnetzwerk, innerhalb vom gliche integrierte Schaltchreis oder bi desse Kommunikation mit de Ussewelt, erhöht.
Drüberhinus kombiniert ACES ziitkritischi Ächtziitufgabe mit allgemerine Recheufgabe uf dergliche physische Plattform und teilt sich debie Kommunikations-, Spicher-, und Mikroproarchitekturressource.
Die heterogene System mit unterschiedliche Kritikalitätsstufene (MCS) üebed zuesätzlich Druck ufd Verbindig us und erfordered nur minimali konkurrenz zwische de verschidene Kritikalitätsstufene, um es hochs Mass a Vorhersagbarkeit ufrechtserhalte.
Um die Leistigs- und Energieeffizienzaforderige imne breite Spektrum industrielle Awändige z erfülle, isch en sorgfälltig abgstimmte Prozess vom Spichersystem mit de Awändigsfäll sowie de Recheeinheite und Beschlüniger erforderlich.

\teb{Zerscht} befasst sich die Arbet mit de effiziente, aglie und leischtigsstarche Dateüberträgig i heterogene Systeme, indem sie e modulari und hochgradig ahpassbari Architektur für direkti Speicherzuegriffseinheite (DMA) names iDMA vorstellt, die de vielfälltige Aforderige hütiger heterogene Plattformene grecht wird.
Als zweits stelled mer AXI-REALM vor, e modulari Kommunikationsnetzwerkserwiiterig, die s Problem vo der Vorhersagbarkeit agohd, weles in heterogene MCS uftritt, die ziitkritischi Awändige usführed, wenn DMA-Überträgige vo domänespezifische Beschlüuniger stattfinded.

I mehrere Fallstudie zeigemer d Anwendbarkeit und d Vorteil v iDMA i siliziumimplementierte System und über die gsamti Kommunikationsnetzwerkhierarchie hinweg, einschliesslich vomne Linux-fähigen Eis-Chip-System (SoC), vomne hochleistigsfähige Allzweckrechebeschlüniger und vomne MCS in Automobilqualität.
Im letschtere Fall präsentiered mer d Vorteil vo AXI-REALM, welles d Leischtig vo ziitkritischer Awendige wederherstellt, die uf MCSs laufed, wenns zu Konflikte usglöst dur direkte Speicherzuegriffseinheite vo de domänespezifische Beschlüniger chunnt.

\selectlanguage{english}
\clearpage
\cleardoublepage
\thispagestyle{empty}
\null

{\scriptsize
In reference to IEEE copyrighted material which is used with permission in this thesis, the IEEE does not endorse any of ETH Zurich's products or services.
Internal or personal use of this material is permitted.
If interested in reprinting/republishing IEEE copyrighted material for advertising or promotional purposes or for creating new collective works for resale or redistribution, please go to \url{http://www.ieee.org/publications_standards/publications/rights/rights_link.html} to learn how to obtain a license from RightsLink.
If applicable, University Microfilms and/or ProQuest Library, or the Archives of Canada may supply single copies of the dissertation.
}

\newpage

\tableofcontents  %

\mainmatter
\glsresetall

\chapter{Introduction}
\label{chap:introduction}

In this chapter, we motivate this thesis and present the fundamental problems and challenges tackled throughout the following chapters.

\section{Motivation}
\label{chap:introduction:motivation}

Humankind's century-long strive for complete automation has reached an unprecedented level thanks to recent developments in the field of \gls{ai}.
Generative models such as {OpenAI}'s \emph{o3}~\cite{openai2025introducingopen} even achieved human-like results on general intelligence benchmarks~\cite{bennett2024anaisystemhasre}.

With such potential in automation, the recent entry of \gls{ai} into people's everyday life has become inevitable, as consumers value extensive digitization, e.g., through software-enabled features, leading to a fundamental paradigm shift in many industrial domains, including automotive, robotics, and aerospace\teb{,} towards \gls{aces}~\cite{srcmaptmicroelectr, burkackygettingreadyfor, mutschlerautomotiveoemsf, jiang2023towardshardreal, kasarapu2025performanceande, fletcherthecaseforanend}.
To provide the customer with these needs, connected cyber-physical systems must continuously update software, engage with a wide variety of digital ecosystems, and provide high-bandwidth and low-latency access to on- and off-board data~\cite{burkackygettingreadyfor}.

The widespread use of \gls{ai} and the trend to \gls{aces} both significantly increase the desire for onboard computing performance and communication bandwidth, which in turn raises their energy demand~\cite{iea2025aiissettodrives}, yet their use in automotive, robotics, and aerospace applications limits their power envelope~\cite{ntabeni2024devicelevelener} pushing architectures and systems towards an increase in energy efficiency.

\begin{figure}
    \centering
    \includegraphics[width=\linewidth]{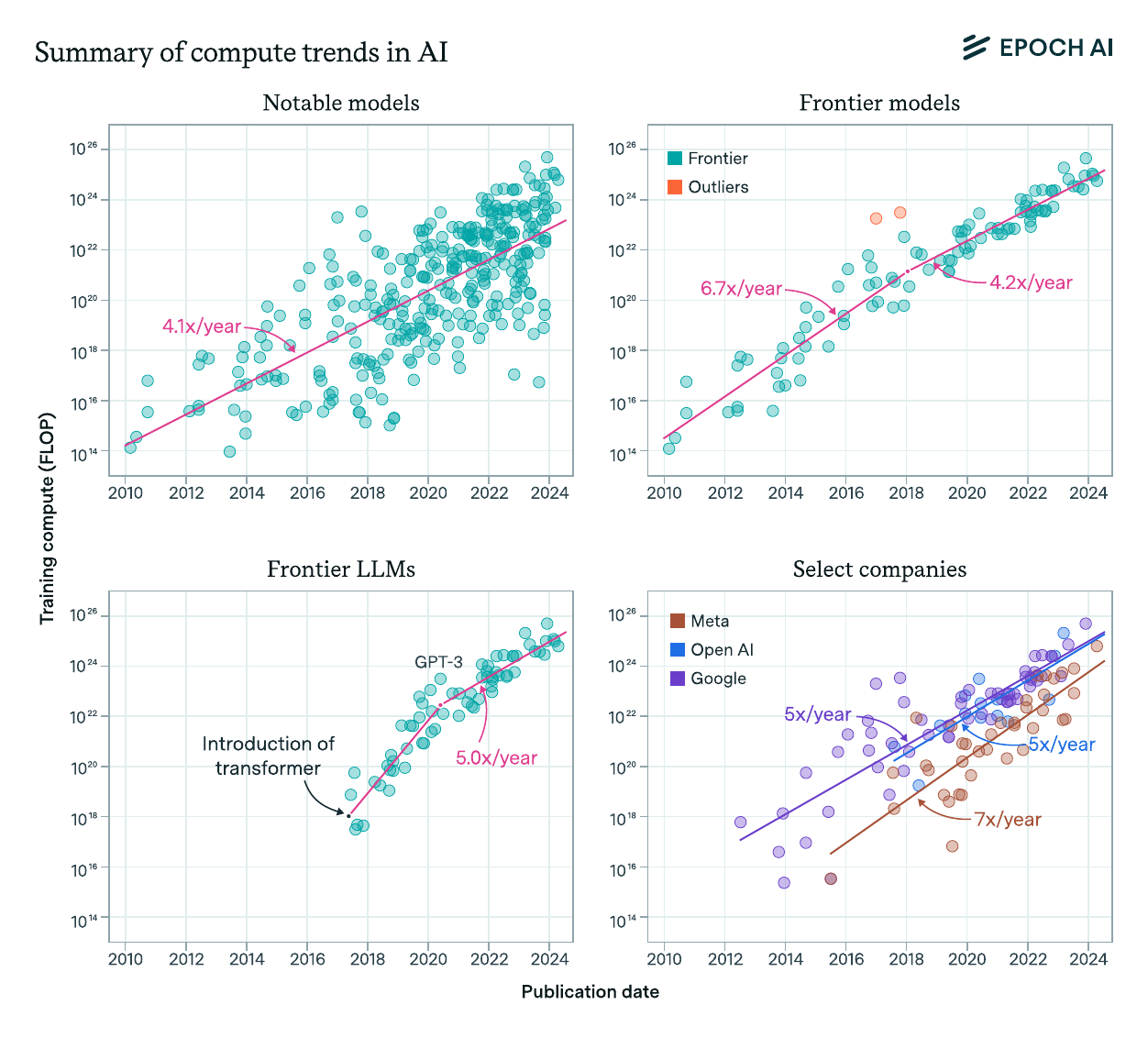}
    \caption{\alc{Training compute and thus number of parameters of \gls{ml} models and their trends.
    Image taken from~\cite{sevilla2024trainingcompute}.}}
    \label{fig:introduction:paramtrends}
\end{figure}

The high level of automation of cyber-physical systems requires complex, more diverse, and multi-modal \gls{ml} models with an ever-increasing number of parameters~\cite{srcmaptmicroelectr, burkackygettingreadyfor, mutschlerautomotiveoemsf, villalobos2022trendsintrainin, villalobos2024willwerunoutofd, villalobos2022machinelearning}.
As can be seen in \Cref{fig:introduction:paramtrends}, \alc{recent years showed exponential growth in required training compute leading to parameter counts of hundreds of billions~\cite{villalobos2022machinelearning, sevilla2024trainingcompute}.}
Coupled with an increased number of sensors required to perceive the environment~\cite{pandharipande2023sensingandmachi}, more and more pressure is placed on both the \gls{io} and the on-chip memory system~\cite{lee2015decoupleddirect} to fetch the required input data from the environment, to process internally, and to finally actuate the environment.

\begin{figure}
    \centering
    \begin{subcaptionblock}{0.815\linewidth}%
        \centering%
        \includegraphics[width=\linewidth]{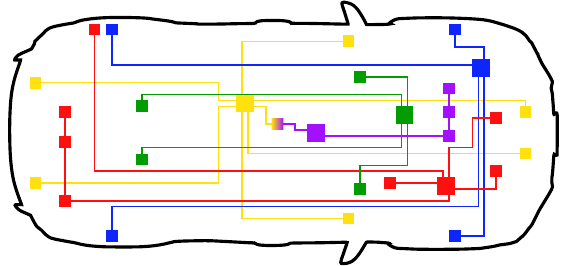}%
        \vspace{-1mm}
        \caption{Traditional flat architecture}%
        \label{fig:introduction:architectures_flat}%
        \vspace{2mm}
    \end{subcaptionblock}\hfill
    \begin{subcaptionblock}{0.815\linewidth}%
        \centering%
        \includegraphics[width=\linewidth]{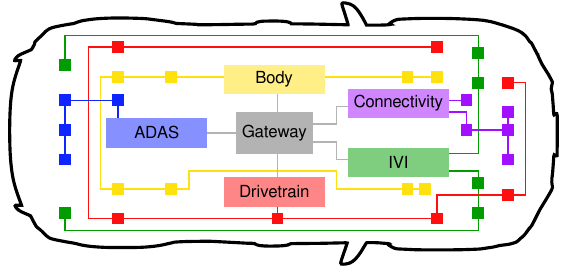}%
        \vspace{-1mm}
        \caption{Domain architecture}%
        \label{fig:introduction:architectures_domain}%
        \vspace{2mm}
    \end{subcaptionblock}\hfill
    \begin{subcaptionblock}{0.815\linewidth}
        \centering%
        \includegraphics[width=\linewidth]{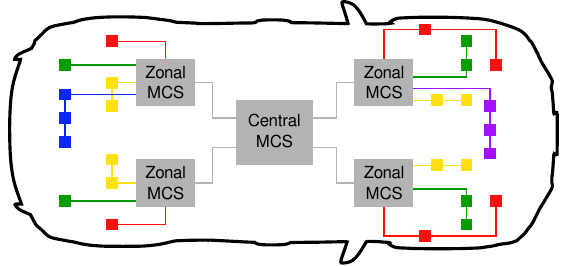}%
        \vspace{-1mm}
        \caption{Zonal architecture}%
        \label{fig:introduction:architectures_zone}%
    \end{subcaptionblock}
    \caption{
        \alc{Transition from a flat or distributed \gls{ecu} architecture to a structured domain \teb{and finally a} zonal architecture with \gls{mcs} nodes.}
    }
    \label{fig:introduction:architectures}
\end{figure}

In the automotive domain, the current trend pushes towards \gls{sdv}~\cite{mutschlerautomotiveoemsf} transforming hardware-defined cars into software-defined transportation platforms, unifying classical \glspl{mcu} with high-level automation functions like \gls{adas}, or \gls{ai}-based self-driving capabilities and immersive on-board entertainment functionality~\cite{fletcherthecaseforanend, mclellan2023whatiszonalarch}.

Traditionally, cars rely on hundreds of embedded real-time \glspl{ecu}, distributed throughout the vehicle.
This architecture cannot meet the growing compute and communication demands and complicates cable harness management, impacting \gls{swapc}~\cite{fletcherthecaseforanend, lim2025aframeworkforde, pinto2019virtualizationo}.
Hence, integrated, interconnected \emph{zonal} and \emph{domain} architectures are becoming the preferred replacements for discrete \glspl{ecu}~\cite{mclellan2023whatiszonalarch}, as they deliver the flexibility and compute capability required for \gls{aces} mobility and the \gls{swapc} problem by simplifying the connectivity~\cite{jang2023designofahybrid, pinto2019virtualizationo}, see \Cref{fig:introduction:architectures}.

To serve both the ever-increasing demand for core compute and memory throughput, as well as enabling \emph{zonal} architectures, cyber-physical systems need to be adapted to embody highly heterogeneous \alc{\glsplf{mcs}}~\cite{esper2018anindustrialvie, burns2017asurveyofresear}.
These architectures usually combine general-purpose and domain-specific sub-systems with diverse real-time and specialized computing requirements coupled through a high-performance interconnect fabric, both executing workloads concurrently on the same silicon die, sharing communication, storage, and micro-architectural resources~\cite{majumder2020partaaarealtime, sá2022afirstlookatris}.
Some subsystems handle hard safety- and time-critical workloads, such as engine, brake, and cruise control~\cite{fletcherthecaseforanend, pagani2019abandwidthreser, gray2023axiicrttowardsa}, while others run less time-critical but computationally demanding tasks like perception pipelines, infotainment, communication, and commodity applications~\cite{fletcherthecaseforanend, ditty2022nvidiaorinsyste, talpes2020computesolution}.
\alc{For example, Tesla's \emph{Full Self-Driving Chip}~\cite{contributorsfsdchiptesla, talpes2020computesolution} and NVIDIA's Jetson \emph{Orion} platform~\cite{ditty2022nvidiaorinsyste} are prominent examples of heterogeneous \glspl{mcs} used in automotive and robotics applications.}

However, in heterogeneous \glspl{mcs}, this process is complicated by the increased interference generated by multiple compute and real-time domains contending for shared memory resources on the same platform~\alc{\cite{pagani2019abandwidthreser, carletti2025takingacloserlo}}.

To summarize, cyber-physical systems in multiple industrial domains are forced to evolve due to an increased level of automation requested by consumers.
The increased heterogeneity to meet energy-efficiency demands, the requirement to transition to \glspl{mcs} to support zonal architectures, and the accelerating growth of machine-learning models increase pressure on the memory interconnect.
While already causing limitations today, without novel and lightweight architectural solutions, the memory architecture will become the bottleneck for tomorrow's cyber-physical systems.

\alc{My thesis aims to tackle these memory architecture bottlenecks by developing a scalable and flexible \gls{dma} architecture to maximize the efficiency of data transfers, while supporting multiple industry on-chip communication protocols tailoring it to the needs of heterogeneous platforms, and by proposing a light-weight interconnect extension enabling real-time interconnect guarantees in \gls{mcs}.}

In the following \Cref{chap:introduction:dmaengines}, we provide a solid background on efficient data movement.
Then, we overview a selection of commonly used industry-grade non-coherent on-chip protocols in \Cref{chap:introduction:protocols}.
Finally, we provide a detailed overview of this thesis' content in \Cref{chap:introduction:overview}, provide a summary of the
contributions in \Cref{chap:introduction:contributions}, and conclude with a list of publications acting as the base of this work in \Cref{chap:introduction:publications}.

\newpage
\section{Efficient Data Movement: DMA Engines}
\label{chap:introduction:dmaengines}

\begin{figure}
    \centering
    \includegraphics[width=0.85\linewidth]{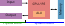}
    \caption{
        Abstract representation of a computing system;
        a \gls{pe} including an \gls{alu} sits at the center accessing both the physical world through \gls{io} and memory.
    }
    \label{fig:introduction:cpusys}
\end{figure}

Since the advent of computing more than a century and a half ago~\cite{menabrea1843sketchoftheanal}, including the transition to electronic computers~\cite{burks1946preliminarydisc}, the vast majority of digital computer architectures share the same fundamental blueprint.
The data to be processed is received from the \emph{physical world} through an \emph{input device} and stored in \emph{memory}.
Computation on the data is done in the \emph{\glsf{cpu}} or \emph{\glsf{pe}} by fetching both instructions and data from memory.
Results are communicated with the physical world through an \emph{output device}.
Traditionally, data is moved to and from the physical world through the \gls{io} devices using the \gls{pe}, see \Cref{fig:introduction:cpusys}.

Using the \gls{pe} to move data either from \glspl{io} or memory is inherently inefficient as this procedurally simple task burdens the processor unnecessarily and prevents it from performing \emph{useful} compute.
This is especially true for \gls{io} devices as they generally run at a lower transmission speed compared to the \gls{pe}'s internal data access, requiring either slow accesses synchronized through polling, backpressure, or interrupts.

\begin{figure}
    \centering
    \includegraphics[width=\linewidth]{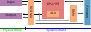}
    \caption{
        Abstract representation on a computing system extended with \gls{dma} units.
        A \gls{io} \gls{dma} accelerates accesses to the physical world, whereas a memory \gls{dma} can be used to copy and reorganize data in memory.
    }
    \label{fig:introduction:dmasys}
\end{figure}

Linear data movement can be accelerated through a specialized \gls{dma} unit~\cite{ma2019mtdmaadmacontro}.
Depending on their position in the computing architecture and their primary role, we distinguish between two different classes of \gls{dma} units:
\emph{memory} and \emph{\gls{io}} units.
As can be seen in \Cref{fig:introduction:dmasys}\teb{.}
\teb{T}he former is placed close to the system's memory and the latter between the \gls{io} devices and the memory.

\paragraph*{\textbf{Memory Units}}

They are placed close to the system's memory, allowing for fast access when initializing, moving, or reorganizing data.
This class of \gls{dma} engines is particularly important in hierarchical memory systems with explicitly managed scratchpad memories and/or off-chip storage~\cite{pullini2019mrwolfanenergyp}; data needs to be transferred from large, but slow memories to small, but fast memories located closely to the \glspl{pe}.

\paragraph*{\textbf{IO Units}}

Placed between \gls{io} and memory, they are specialized to interact with slow peripherals without stalling or slowing down either the core or the memory system.
\Gls{io} units are constructed simpler and thus with a smaller resource footprint than memory units, as they do not need to be optimized for throughput nor latency~\cite{pullini2017textmudmaanauto}.

\alc{\vspace{0.5cm}}

\begin{figure}
    \centering
    \includegraphics[width=0.8\linewidth]{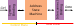}
    \caption{
        Abstract view of the internals of a \gls{dma} unit.
        \emph{Control} and \emph{data} \emph{\teb{planes}} can be differentiated; the former includes a system binding allowing the \gls{pe} to interface the unit, the latter includes the interconnect binding over which the unit accesses memory and/or \gls{io}.
        At the center, an address generator \teb{enumerates} the memory region to be copied.
    }
    \label{fig:introduction:dmacomp}
\end{figure}

Internally, \gls{dma} engines consist of two parts: the \emph{control} and \emph{data} \emph{\teb{planes}}, as presented in \Cref{fig:introduction:dmacomp}.
The control \teb{plane} handles the interface with the \gls{pe}, called \emph{system binding}, and the orchestrating functions of the engine, e.g., the emission of interrupts or events once a \emph{\gls{dma} job} is complete.

An internal \emph{address generator} enumerates the individual memory transfers to be completed sequentially to perform the data movement operation commanded; basic units support linear transfers only~\cite{ma2019mtdmaadmacontro}.
More advanced units enable more complex transfer patterns~\cite{rossi2014ultralowlatency, fjeldtvedt2019cubedmaoptimizi}.

The data \teb{plane} executes the data transfer using a memory interconnect binding that fits the memory architecture of the system.
In traditional IBM-based system, data movement is usually done using an expansion bus protocol, e.g., \gls{pcie}, whereas on-chip communication protocols, see \Cref{chap:introduction:protocols}, are used on \glspl{soc}.

\alc{To summarize, \gls{dma} units can be split into a control and a data \teb{plane}; the former defines how the unit is interfacing the system, the latter which protocol is used to copy a deterministic stream of bytes.}
\alc{We work on this principle in \Cref{chap:idma}, where we modularize our architecture along this split of \teb{planes}.}

\newpage
\section{On-Chip Interconnect Protocols}
\label{chap:introduction:protocols}

\afterpage{%
\begin{sidetab}
    \centering
    \caption{%
        Common non-coherent on-chip protocols and their key characteristics. All protocols share \emph{byte addressability} and a \emph{ready-valid} handshake.%
    }%
    \label{tab:intro:protocols}
    \renewcommand*{\arraystretch}{1.1}
    \resizebox{\linewidth}{!}{%
    \begin{threeparttable}
        \begin{tabular}{@{}llllll@{}} \toprule
            \textbf{Protocol} &
            \textbf{\makecell[cl]{Request \\ Channel}} &
            \textbf{\makecell[cl]{Response \\ Channel}} &
            \textbf{\makecell[cl]{Key Characteristics}} &
            \textbf{Applications} &
            \textbf{Burst Support} \\
            \midrule
            \emph{AMBA AXI4+ATOP}~\cite{arm2023ambaaxiandacepr, kurth2022anopensourcepla} &
            \emph{AW~\tnote{a}, W~\tnote{a}, AR~\tnote{b}} &
            \emph{B~\tnote{a}, R~\tnote{b}} &
            \makecell[cl]{High-Performance \\ High-Speed \\ Fully Decoupled \\ Max. Piplineability\vspace{1mm}} &
            \makecell[cl]{HPC Interconnect} &
            \makecell[cl]{\emph{256 beats} \\ or \emph{\SI{4}{\kilo\byte}~\tnote{c}}} \\
            \emph{AMBA AXI4 Lite}~\cite{arm2023ambaaxiandacepr} &
            \emph{AW~\tnote{a}, W~\tnote{a}, AR~\tnote{b}} &
            \emph{B~\tnote{a}, R~\tnote{b}} &
            \makecell[cl]{High-Speed \\ Fully Decoupled \\ Max. Piplineability\vspace{1mm}} &
            \makecell[cl]{Peripheral \& \\ Configuration \\ Interconnect} &
            no \\
            \emph{AMBA AXI4 Stream}~\cite{arm2021ambaaxistreampr} &
            \emph{T~\tnote{d}} &
            \emph{T~\tnote{d}} &
            \makecell[cl]{High-Speed \\ Versatile \\ Point-to-Point\vspace{1mm}} &
            Data Streams &
            unlimited \\
            \emph{OpenHW OBI}~\cite{silicon2020obi1} &
            \emph{D} &
            \emph{R} &
            \makecell[cl]{Tightly Coupled \\ Low-Power} &
            \makecell[cl]{Tightly Coupled \\ Direct Memory \\ (TCDM) \& \\ \glspl{mcs} Interconnect} &
            no \\
            \emph{SiFive TileLink UH}~\cite{sifivesifivetilelinks} &
            \emph{A} &
            \emph{R} &
            \makecell[cl]{High-Performance \\ Scalable \\ Tightly Coupled \\ Piplineable\vspace{1mm}} &
            \makecell[cl]{HPC Interconnect} &
            powers of two \\
            \emph{SiFive TileLink UL}~\cite{sifivesifivetilelinks} &
            \emph{A} &
            \emph{R} &
            \makecell[cl]{Scalable \\ Tightly Coupled \\ Piplineable\vspace{1mm}} &
            \makecell[cl]{Peripheral \& \\ Configuration \\ Interconnect} &
            no \\
            \bottomrule
        \end{tabular}
        \addtolength{\tabcolsep}{-3pt}
        \begin{tablenotes}[para, flushleft]
            \teb{\item[a] write}
            \teb{\item[b] read}
            \item[c] whichever is reached first
            \item[d] symmetrical \emph{RX/TX} channels
        \end{tablenotes}
    \end{threeparttable}
    }
\end{sidetab}
}

There are multiple commonly used non-coherent on-chip protocols currently available with an open-source specification~\cite{arm2023ambaaxiandacepr, arm2021ambaaxistreampr, silicon2020obi1, sifivesifivetilelinks, opencoreswishbonesystemo}.
\Cref{tab:intro:protocols} presents and summarizes a selection of on-chip protocols chosen for their presence in research systems or for their importance to the content presented in this thesis.

\subsection{Focus: Non-Coherent Protocols}

To simplify application development on multicore systems, it is assumed all \glspl{pe} see the exact same state of the system's memory~\cite{dubois1982effectsofcachec}.
This assumption holds trivially if all cores are connected to a single large memory.
With the \emph{Memory Wall} or the \emph{Processor-Memory Gap}~\cite{burger1996memorybandwidth, wilkes2001thememorygapand, hennessy2011computerarchite} increasingly growing, a fast and large unified memory becomes increasingly impossible to design and implement.
Multi-level \gls{pe}-coupled cache architectures promise to fill this gap but at a cost of requiring implicit synchronization of their contents to provide a unified and coherent memory view.

Specialized on-chip protocols have been developed to carry this synchronization traffic alongside the regular data stream~\cite{cavalcante2020designofanopens, molka2009memoryperforman, arm2023ambaaxiandacepr}.

This thesis focuses on explicitly managed memory architectures where data copy operations are orchestrated by the user and carried out through highly optimized \gls{dma} engines for their predictability, and high energy efficiency.
While playing a significant role in application-grade systems, coherency does not play a major role in such systems, since coherency provides few benefits in explicitly managed systems, while coherent protocols impose a significant overhead in resource utilization and energy footprint, heightened latency, and a drop in bandwidth~\cite{cavalcante2020designofanopens}.

\subsection{Protocols Discussed}

\paragraph*{\textbf{AMBA AXI4}}

This protocol supports high-performance and high-speed communication between manager and subordinate devices~\cite{arm2023ambaaxiandacepr}.
Its five-channel fully handshaked latency-tolerant architecture effectively decouples read and write data as well as transfer requests, data streams, and responses, maximizing pipelinability, minimizing stalls, and facilitating a dataflow-oriented implementation.
\Gls{axi4} supports bursts up to 256 elements in length or \SI{4}{\kilo\byte} in size, enabling movement of large chunks of contiguous data with only one request and response beat.
Through the use of \glspl{tid}, \gls{axi4} supports transfer reordering and multiple outstanding transactions.

Implementing the full capabilities of \gls{axi4} requires a relatively complex manager and subordinate design.
For simpler devices, e.g., configuration registers, a simplified subset of the specification, \gls{axi4} Lite, can be selected.
While still featuring the five channels, support for bursts and transfer reordering is removed to simplify the design of the interconnect and the devices.

\paragraph*{\textbf{AMBA AXI4 Stream}}

This protocol targets streaming applications, where versatile, high-speed, unidirectional, point-to-point connections between transmitters and receivers are required~\cite{arm2021ambaaxistreampr}.
Although primarily specified as a point-to-point protocol, interconnects can be employed to connect multiple \gls{axi4} Stream components together.
\gls{axi4} Stream only describes the data transfer mechanism, but not the data's meaning, allowing \gls{axi4} Stream to carry a variety of high-level protocols.

\paragraph*{\textbf{OpenHW OBI}}

Featuring two double-handshaked channels targeting single-cycle, tightly coupled memory connections.
\Gls{obi} is optimized for low-power \glspl{mcu} and can be used as the main interconnect towards fixed-latency \gls{tcdm} or to connect to variable-latency peripheral devices.
The \gls{obi} subordinate device is very easy to implement, allowing trivial connection to \gls{sram} banks and making it a great protocol to connect configuration registers.

\paragraph*{\textbf{SiFive TileLink Uncached}}

With two two double-handshaked channels, \gls{tluh} implements a tightly coupled, scalable, pipelinable, and stateless bus protocol optimized to be used as the main interconnect in \glspl{soc}.
\Gls{tluh} is considered a high-performance protocol thanks to the support of out-of-order completion, various hints, atomic operations, and burst accesses.

\gls{tlul} supports only read and write operations and is thus primarily used to access configuration space.

\vspace{0.5cm}

The number of used protocols clearly indicates that there is no one optimal and universal protocol to rule them all.
Although these protocols differ significantly in their specifications (number of channels, signal names, capabilities), at its core, all protocols share ready-valid-handshaked interfaces and the facility to read and write data on byte-addressed granularity.

With such a zoo of protocols existing, heterogeneous \glspl{soc} will necessarily feature more than one interconnect protocol, thus requiring protocol adaptation.
Although possible due to the general support for backpressure and byte-addressability, protocol adapters incur a considerable resource overhead and are likely to introduce a latency overhead and drop in utilization.
In explicitly memory-managed systems, it is thus better to create a multi-protocol \gls{dma} engine directly capable of translating between different on-chip protocols, see \Cref{chap:idma}.
\newpage
\section{Thesis Overview and Outline}
\label{chap:introduction:overview}

\begin{figure}
    \centering
    \includegraphics[width=0.96\linewidth]{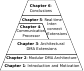}
    \caption{
        Overview of this thesis's chapters.
        \alc{\Cref{chap:idma} with {\idma} builds the \teb{\emph{foundation}} of this thesis.}
        \alc{\Cref{chap:dmaext} defines important \gls{dma} extension which are developed into a communication processor in \Cref{chap:comcpu}\teb{, supporting our goal of a} energy-efficient real-time data movement strategy presented in \Cref{chap:realm}.}
    }
    \label{fig:introduction:overview}
\end{figure}

In this section, we will summarize and motivate the structure of this thesis chapter by chapter.
The overall organization is given in \Cref{fig:introduction:overview}.
A substantial part of the content discussed in this thesis has been published in peer-reviewed conferences and as journal papers.
A comprehensive listing of publications contributing to this thesis can be found in \Cref{chap:introduction:publications}.

\paragraph*{\textbf{Chapter 2}}
We first design a modular and highly versatile \gls{dma} architecture, called {\idma}, which can be used in applications ranging from simple and dedicated \glsu{io}-\glspl{dma} controlling autonomous data transfer for peripherals to high-performance multi-channel engines that provide massive throughput at near-ideal bus utilization with minimal latency overheads.
With the key insight in mind that most commonly used on-chip interconnect protocols share a byte-addressed data stream, {\idma} is designed to support five industry-grade protocols, translating seamlessly between them, facilitating {\idma}'s integration in heterogeneous \glspl{soc}.

\paragraph*{\textbf{Chapter 3}}
We integrate {\idma} in a wide range of systems, ranging from \gls{ulp} to \gls{hpc} use cases, covering a large set of application scenarios and proving the applicability of the architecture.
As part of these integration efforts, we develop three different system bindings for {\idma}:
a simple configuration-register-based interface, a Linux-capable transfer-descriptor-based scheme, and an agile {\riscv}-compliant instruction-based {\fe}.
To accelerate commonly occurring transfer patterns, we develop a set of \gls{dma} extensions facilitating data orchestration by handling \alc{\gls{nd}} tensor transfers and scheduling of real-time transfers directly in hardware. We further include a standardized interface in {\idma}'s data path to integrate in-stream operations.
Finally, we develop a stream-optimized \gls{mmu} to be integrated into {\idma}, enabling the \gls{dma} to autonomously handle \gls{vm} in conjunction with an application-grade host.

\paragraph*{\textbf{Chapter 4}}
We combine {\idma} with a tiny \SI{20}{\kilo\gateeq} {\riscv} processor tightly coupled through our instruction-based system binding, creating an agile communication processor capable of complex data scheduling and orchestration.
We integrate this communication processor into a compute-cluster-based accelerator architecture with a focus on an optimal connection between the {\dma} engine and the cluster-local scratchpad memory.
This cluster architecture is then scaled out to a 432-core dual-chiplet manycore system featuring fourteen {\idma} engines and \SI{16}{\gibi\byte} of \gls{hbm2e} connected through a hierarchical crossbar-based point-to-point interconnect.

\paragraph*{\textbf{Chapter 5}}
With an efficient \gls{dma} engine developed and its integration into multiple systems studied, we shift focus towards the host by investigating real-time data transfers in heterogeneous \glspl{mcs}, which become increasingly important with the advent of \gls{ai}-driven applications in automotive and aeronautical use cases.
We develop {\axirealm}, a lightweight and interconnect-agnostic helper-module-based architecture to module the amount of data managers can inject into the shared interconnect.
This is especially important, as {\idma}-equipped accelerators are tuned for maximal bandwidth efficiency through long data bursts, which can, if kept unregulated, lead to adverse performance of time-critical applications running on the host cores.
We evaluate the effectiveness of {\axirealm} by integrating it into an open-source \gls{mcs} research platform and evaluating using real-world workloads.

\paragraph*{\textbf{Chapter 6}}
In this last chapter, we conclude the thesis, summarizing the key achievements and outlining a roadmap to continue this research and development.

\newpage
\section{Summary of Contributions}
\label{chap:introduction:contributions}

This thesis provides the following contributions:

\begin{enumerate}
    \item Specifying, designing, and implementing a modular and highly parametrizable {\dmaa}, called {\idma}, targeting a wide range of systems from \gls{ulp} to \gls{hpc} systems \alc{by natively supporting multiple industry-grade on-chip protocols.}
    \item Developing adapted integrations for {\idma} to suit multiple systems;
    this includes specifying a low-overhead configuration-register based programming interface for \gls{ulp} system, combining a transfer-descriptor-based {\fe} to form an autonomous and Linux-compatible \gls{dmac} , and extending the {\riscv} instruction set with light-weight \gls{dma} extensions tighly coupling the {\dmae} to the \glspl{pe}.
    \item Creating multiple {\dma} extensions to accelerate various aspects of commonly occurring transfer patterns.
    This includes accelerating \alc{\gls{nd}} tensor transfers, a flexible \gls{vm} extension, and automatically scheduling real-time memory transfers.
    \item Combining {\idma} with a light-weight \SI{20}{\kilo\gateeq} {\riscv} core through the developed \gls{dma} instruction extension forming a capable \emph{communication processor} in charge of scheduling and orchestrating complex memory transfers in a compute-cluster-based \gls{hpc} system.
    Proofing the viability and performance of the memory architecture by scaling it to an open-source 432-core {\riscv} manycore system.
    \item Designing and implementing a light-weight and topology-agnostic on-chip interconnect extension to provide real-time guarantees in high-performance heterogeneous \glspl{mcs} by modulating the ingress of managers into the interconnect fabric.
    Integrating this real-time unit with a subordinate protection mechanism into an open-source \gls{mcs} platform, demonstrating its performance using real-world applications.

\end{enumerate}

This thesis discusses some elements developed in collaboration with Alessandro Ottaviano~\cite{ottaviano2025designofheterog} and Paul Scheffler~\cite{scheffler2025accelerationofs}.
Their respective theses cover these elements partially with a different focus.

\newpage
\section{List of Publications}
\label{chap:introduction:publications}

\subsection{Previous Publications}

The foundation for this thesis was laid completing my Master's Thesis.

\begin{publications}
    \publication{benz2020snitchscaleouto}
\end{publications}

\subsection{Core Publications}

The following publications are completely encompassed within this thesis.

\begin{publications}
    \publication{benz2024ahighperformanc}
    \publication{benz2025adirectmemoryac}
    \publication{benz2024axirealmalightw}
    \publication{benz2025axirealmsafemod}
\end{publications}

\subsection{Related Publications}

The following publications containing my contributions were partially covered by the content of this thesis.

\begin{publications}
    \publication{di2021pspinahighperfo}
    \publication{di2021ariscvinnetwork}
    \publication{benz2021a10coresocwith2}
    \publication{kurth2022anopensourcepla}
    \publication{ottaviano2023cheshirealightw}
    \publication{rogenmoser2024sentrycorearisc}
    \publication{ottaviano2025controlpulpleta}
    \publication{khalilov2024osmosisenabling}
    \publication{paulin2024occamya432core2}
    \publication{liang2024agigabitdmaenha}
    \publication{liang2025towardsreliable}
    \publication{fischer2025floonoca645gbsl}
    \publication{garofalo2025areliabletimepr}
    \publication{scheffler2025occamya432cored}
    \publication{al2025carfieldanopens}
\end{publications}

\subsection{Unrelated Publications}

I further contributed to the following publications while working towards my degree; while not being directly covered by the contents of this thesis, they might reveal additional insights.

\begin{publications}
    \publication{mazzola2022adatadrivenappr}
    \publication{jain2023patronocparalle}
    \publication{zhang2023axipacknearmemo}
    \publication{benz2023iguanaanendtoen}
    \publication{sauter2024basiliskachievi}
    \publication{sauter2024insightsfrombas}
    \publication{scheffler2024basiliskanendto}
    \publication{zhang2024nearmemoryparal}
    \publication{mazzola2025datadrivenpower}
    \publication{benz2024dutctlaflexible}
    \publication{benz2025artisticanopens}
    \publication{sauter2025crocanendtoendo}
    \publication{sauter2025basiliska34mm²e}
    \publication{zhang2025flatattentionda}
\end{publications}

\chapter{Modular DMA Architecture}
\label{chap:idma}

\section{Introduction}
\label{chap:idma:introduction}

\alc{\Dmaes} form the communication backbone of many contemporary computers~\cite{ma2019mtdmaadmacontro}. %
They concurrently move data at high throughput while hiding memory latency and minimizing processor load, freeing the latter to do useful compute. %
This function becomes increasingly critical with the trend towards physically larger systems~\cite{choquette2021nvidiaa100tenso} and ever-increasing memory bandwidths~\cite{lee2022futurescalingof}. %
With Moore's Law slowing, 2.5D and 3D integration are required to satisfy future applications' computational and memory needs, leading to wider and higher-bandwidth memory systems and longer access latencies~\cite{blythe2021xehpcpontevecch, wang2021hamhotspotaware}. %

Without {\dmaes}, \glspl{pe} need to read and write data from and to remote memory, often relying on deep cache hierarchies to mitigate performance and energy overheads.
In this thesis, we focus on explicitly managed memory hierarchies, where copies across the hierarchy are handled by {\dmaes}. %
We refer the interested reader to \cite{branco2022cacherelatedhar, nagarajan2020aprimeronmemory, jain2019cachereplacemen, balasubramonian2011multicorecacheh} for excellent surveys on cache-based memory systems. %
Caches and {\dmaes} often coexist in modern computing systems as they address different application needs. %
Dedicated {\dmaes} are introduced to efficiently and autonomously move data for workloads where memory access is predictable, weakly data-dependent, and made in fairly large chunks, decoupling memory accesses from execution and helping maximize \gls{pe} time spent on useful compute. %

When integrating {\dmaes}, three main design challenges must be tackled: the control-plane interface to the \glspl{pe}, the intrinsic data movement capabilities of the engine, and the on-chip protocols supported in the data plane. %
The sheer number of {\dmaes} present in literature and available as commercial products explains why these choices are usually fixed at design time. %
The increased heterogeneity in today's accelerator-rich computing environments leads to even more diverse requirements for {\dmaes}. %
Different on-chip protocols, programming models, and application profiles lead to a large variety of different {\glsf{dma}} units used in modern \glspl{soc}, hindering integration and verification efforts. %

In this \teb{chapter}, we tackle these issues by presenting a modular and highly parametric {\dmaa} called \emph{\glsf{idma}}, which is composed of three distinct parts: the \emph{\fe} handling \gls{pe} interaction, the \emph{\me} managing the engine's lower-level data movement capabilities and facilitating complex data-shuffling patterns, and the \emph{\be} implementing one or more on-chip protocol interfaces. %
We call concrete implementations of our {\idma} architecture \emph{\idmaes}.
All module boundaries are standardized to facilitate the substitution of individual parts, allowing for the same {\dmaa} to be used across a wide range of systems and applications, reducing management, verification, and maintenance efforts. %
While this chapter focuses on the general architecture of {\idma}, \Cref{chap:dmaext} presents a wide range of system integrations, {\dma} extensions, and use cases served by {\idma} in great detail.
To foster the use of our {\dmaa}, we provide the synthesizable \gls{rtl} description of {\idma}, silicon-proven in various instances, and the system bindings are available free and open-source under a libre Apache-based license\,\footnote{\texttt{\url{https://github.com/pulp-platform/iDMA}}}.

In more detail, this chapter presents the following contributions to \gls{soa} {\dma} research: %
\begin{itemize}
    \item We specify a \emph{modular}, \emph{parametric} {\dmaa} composed of interchangeable parts, allowing {\idma} to accommodate and benefit any system. %
    \item We optimize {\idma} to minimize hardware buffering through a highly agile, read-write decoupled, dataflow-oriented transport engine that maximizes bus utilization in any context.  %
    Our architecture incurs no idle time between transactions, even when adapting between different on-chip protocols, and incurs only \emph{two} cycles of initial latency to launch a \teb{\gls{nd}} affine transfer. %
    \item We propose and implement a two-stage transfer acceleration scheme: the {\mes} manage (distribute,  repeat, and modify) transfers while an in-stream \emph{acceleration port} enables configurable in-flight operation on the data being transferred. %
    \item Thanks to our modular \teb{system binding}\tebsr{s} ({\fes}) and our support for multiple industry-standard on-chip protocols ({\bes}), {\idma} can be used in a wide range of contexts, from \gls{ulp} to \gls{hpc} systems. %
    A lightweight \gls{fifo} port allows easy integration into accelerators and peripherals without passing through a complex on-chip protocol.
    This interface can also be used to initialize memory given various data patterns. %
    \item We thoroughly characterize our architecture in area, timing, and latency by creating area and timing models with less than \SI{9}{\percent} mean error and an analytical latency model, easing instantiation in third-party designs and accelerating system prototyping. %
    \item We use synthetic workloads to show that {\idma} achieves high bus utilization in ultra-deep memory systems with affordable area growth. %
    Our architecture perfectly hides latency in systems with memory hierarchies hundreds of stages deep. %
    It reaches full bus utilization on transfers as small as \SI{16}{\byte} while occupying an area footprint of less than \SI{25}{\kilo\GE}\,%
    \footnote{%
    Gate equivalent (\si{\GE}) is a technology-independent figure of merit measuring circuit complexity. %
    A \si{\GE} represents the area of a two-input, minimum-strength {NAND} gate. %
    } %
    in a 32-\si{\bit} configuration. %
\end{itemize}

\newpage
\section{Architecture}
\label{chap:idma:architecture}

\begin{figure}[t]
    \centering%
    \includegraphics[width=\textwidth]{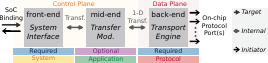}%
    \caption{%
        Schematic of {\idma}: Our engines are split into three parts: at least one {\fe}, one or multiple optional {\mes}, and at least one {\be}. %
    }%
    \label{fig:arch:arch}%
\end{figure}

Unlike \glsf{soa} {\dmaes}, we propose a modular and highly parametric {\dmae} architecture composed of three distinct parts, as shown in \Cref{fig:arch:arch}. %
The \emph{\fe} defines the interface through which the processor cores control the {\dmae}, corresponding to the \emph{control plane}.
The \emph{\be} or \emph{data plane} implements the on-chip network manager port(s) through which the {\dmae} moves data. %
Complex and capable on-chip protocols like \gls{axi4}~\cite{arm2023ambaaxiandacepr}, to move data through the system, and simpler core-local protocols like \gls{obi}~\cite{silicon2020obi1} to connect to \gls{pe}-local memories, are supported by the {\be}.
The \emph{\me}, connecting the front- and {\be}, slices complex transfer descriptors provided by the {\fe} (e.g., when transferring \teb{\gls{nd}} tensors) into one or multiple simple 1-D transfer descriptors for the {\be} to process.
Additionally, multiple {\mes} may be chained to enable complex transfer processing steps. %
\alc{In a case study in \Cref{chap:dmaext:realtime} we show this chaining mechanism by connecting a real-time and a {\teb{3-D} tensor} {\me} to efficiently and autonomously gather sensor data.} %
We discuss the available {\fes}, the corresponding system bindings, and transfer acceleration through {\mes} in detail in \Cref{chap:dmaext}.

\begin{figure}
    \centering%
    \includegraphics[width=\textwidth]{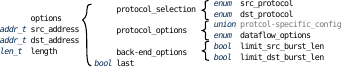}%
    \caption{%
        Outline of the \emph{\teb{1-D} transfer descriptor} (exchanged between mid- and {\be}).
        \alc{The \emph{options} field contains protocol-specific configuration independently of the chosen configuration to keep the {\be}'s interface interchangeable.}
    }%
    \label{fig:arch:arch-request}%
\end{figure}

To ensure compatibility between these three different parts, we specify their interfaces.
From the {\fe} or the last {\me}, the {\be} accepts a \emph{\teb{1-D} transfer descriptor} specifying a \emph{source address}, a \emph{destination address}, \emph{transfer length}, \emph{protocol}, and \emph{{\be} options}, as seen in \Cref{fig:arch:arch-request}. %
{\Mes} receive bundles of {\me} configuration information and a \emph{\teb{1-D} transfer descriptor}. %
A {\me} will strip its configuration information while modifying the \emph{\teb{1-D} transfer descriptor}. %
All interfaces between front-, mid-, and {\bes} feature \emph{ready-valid} handshaking and can thus be pipelined.

\subsection{\Be}
\label{sec:idma:architecture:backend}

\begin{figure}
    \centering%
    \includegraphics[width=\textwidth]{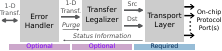}%
    \caption{%
        The internal architecture of the {\be}. %
        The \emph{transport layer} handles the actual copying of the data on the on-chip protocol, supported by the optional \emph{transfer legalizer} and the \emph{error handler}.
    }%
    \label{fig:arch:arch-backend}%
\end{figure}

For given on-chip protocols, a {\be} implements efficient \emph{in-order} \emph{\teb{1-D} arbitrary-length} transfers. %
\emph{Multichannel} {\dmaes} featuring multiple ports into the memory system, can be built by connecting multiple {\bes} to a single front- or {\me}, requiring a similar area to a true multichannel {\be} but less verification and design complexity. %
Arbitration between the individual {\bes} can either be done explicitly by choosing the executing {\be} through software or by an \emph{arbitration} {\me} using round-robin or address-based distribution schemes. %
Depending on the application, explicitly managed or automatic distribution schemes might be more beneficial.
In a \gls{ulp} or a system with a highly non-uniform memory architecture, explicitly managed channels will deliver better results than round-robin or address-based distribution schemes.
Similar hardware will be required in a true multichannel {\be} to distribute transactions to available channels. %

The {\be} comprises three parts, see \Cref{fig:arch:arch-backend}. The \emph{error handler} communicates and reacts to failing transfers, the \emph{transfer legalizer} reshapes incoming transfers to meet protocol requirements, and the \emph{transport layer} handles data movement and possibly in-cycle switches between protocol-specific data plane ports. %
Of these three units, only the transport layer is mandatory. %

\paragraph*{\textbf{Error Handler}}

An \emph{error handler} may be included if the system or application requires protocol error reporting or handling. %
Our current error handler can either \emph{continue}, \emph{abort}, or \emph{replay} erroneous transfers. %
Replaying erroneous transfers allows complex \alc{\gls{nd}} transfers to continue in case of errors in single {\be} iterations without the need to abort and restart the entire transfer. %

When an error occurs, the {\be} pauses the transfer processing and passes the offending transfer's legalized burst base address to its {\fe}.  %
The \glspl{pe} can then specify through the {\fe} which of the three possible actions the error handler should take to resolve the situation. %

We discuss the error handler in more detail in \Cref{chap:idma:architecture:errorhandling}.

\paragraph*{\textbf{Transfer Legalizer}}

\begin{figure}[t]
    \centering%
    \includegraphics[width=\textwidth]{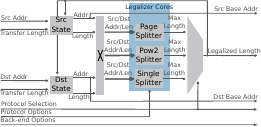}%
    \caption{%
        The internal architecture of the \emph{transfer legalizer}. %
        Any given transfer can be legalized except for zero-length transactions: They may optionally be rejected. %
    }%
    \label{fig:arch:arch-legalizer}%
\end{figure}

Shown in  \Cref{fig:arch:arch-legalizer}, the transfer legalizer accepts a 1-D transfer and legalizes it to be supported by the specific on-chip protocol(s) in use. %
Transfer information is stored internally and modular \emph{legalizer cores} determine the transfer's maximum legal length supported given user constraints and the protocols' properties. %
For protocols that do not support bursts, the legalizer decomposes transfers into individual bus-sized accesses. %
Otherwise, splitting happens at page boundaries, the maximum burst length supported by the protocol, or user-specified burst length limitations. %
The source and destination protocols' requirements are considered to guarantee only legal transfers are emitted. %
In area-constrained designs, the transfer legalizer may be omitted; legal transfers must be guaranteed in software. %

\paragraph*{\textbf{Transport Layer}}
\label{sec:arch_transport-layer}

\begin{figure}
    \centering%
    \includegraphics[width=\textwidth]{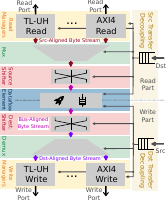}%
    \caption{%
        The architecture of the \emph{transport layer}. %
        One or multiple \emph{read manager(s)} feed a stream of bytes into the \emph{source shifter}, the \emph{data flow element}, the \emph{destination shifter}, and finally, into one or multiple \emph{write manager(s)}. %
        \protect\settototalheight\myheight{Xygp}\settodepth\mydepth{Xygp}\hspace{-2pt}\raisebox{-\mydepth}{%
        \includegraphics[height=\myheight]{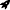}}\hspace{-8pt}\protect~~~denotes the \emph{in-stream accelerator}. %
    }%
    \label{fig:arch:arch-transport}%
\end{figure}

The parametric and modular \emph{transport layer} implements the protocol characteristics for legalized transfers and decouples read and write operations, maximizing the bus utilization of any transfers. %
It uses \emph{read} and \emph{write managers} to handle protocol-specific operations, allowing it to internally operate only on generic byte streams, as shown in \Cref{fig:arch:arch-transport}. %
This enables our {\idma} to easily support multiple on-chip protocols and multiple ports of the same protocol. %
The number and type of protocol ports available in the engine must be set at compile time, whereas the given protocol port a transaction uses can be selected during run-time through the {\fe}. %

Both the \emph{multiplexer} and \emph{demultiplexer} within the transport layer (see \Cref{fig:arch:arch-transport}) can be modified to support multiple concurrent input and/or output streams to support\teb{,} among others\teb{,} reliable data transfers (reading/writing the same data from multiple memories) or vector operations.%

The read and write stages of the transport layer are decoupled from the legalizer by \gls{fifo} buffers, allowing a configurable number of outstanding transfers. %
A \emph{dataflow element} decouples the read and write stages, ensuring that only protocol-legal back pressure is applied to the memory system at each end, coalesces transfers, and cuts long timing paths to increase the engine's maximum operating frequency.
Fully buffered operation may be required depending on the system and the memory endpoints; in this case, the small \gls{fifo} buffer in the dataflow element may be replaced with an \glsunset{sram}\gls{sram}-based buffer, allowing entire transfers to be stored. %
On the other hand; if all endpoints are single-cycle memories (e.g., \gls{tcdm} over \gls{obi}), the internal \gls{fifo} can be omitted to reduce the area footprint at the cost of no longer cutting the data path through the transport layer. %
We provide more insights regarding buffer sizing in \Cref{chap:idma:architecture:buffer}.

Two shifters --- one at each end of the dataflow element --- align the byte stream to bus boundaries.
To reduce the area footprint, the shifters can be merged into one single shifter at the ingress of the dataflow element at a cost of having a destination-aligned over a bus-aligned data stream in the dataflow element.
\emph{In-stream accelerators}, e.g., transposition units, allowing operations performed on the data stream during data movement, may be integrated into the dataflow element, augmenting the buffer in the transport layer. %
Our dataflow-oriented architecture allows us to switch between multiple read managers, write managers, and in-stream accelerators \emph{in-cycle}, allowing our engine to asymptotically reach perfect bus utilization even when the used protocols or acceleration schemes change regularly. %

\paragraph*{\textbf{Protocol Managers}}
\label{sec:arch:protocol-managers}

\begin{table}
    \centering
    \caption{%
        Available on-chip protocols and their key characteristics. All protocols share \emph{byte addressability} and a \emph{ready-valid} handshake.%
    }%
    \label{tab:arch:managers}
    \renewcommand*{\arraystretch}{1.6}
    \resizebox{\linewidth}{!}{
    \begin{threeparttable}
        \addtolength{\tabcolsep}{1pt}
        \begin{tabular}{@{}lllllll@{}} \toprule
            \textbf{Protocol} &
            \textbf{Version} &
            \textbf{\makecell[cl]{Request \\ Channel}} &
            \textbf{\makecell[cl]{Response \\ Channel}} &
            \textbf{Bursts} \\
            \midrule
            \emph{AMBA AXI4+ATOP}~\cite{arm2023ambaaxiandacepr} &
            \emph{H.c} &
            \emph{AW~\tnote{a}, W~\tnote{a}, AR~\tnote{b}} &
            \emph{B~\tnote{a}, R~\tnote{b}} &
            \makecell[cl]{\emph{256 beats} \\ or \emph{\SI{4}{\kilo\byte}~\tnote{c}}} \\
            \emph{AMBA AXI4 Lite}~\cite{arm2023ambaaxiandacepr} &
            \emph{H.c} &
            \emph{AW~\tnote{a}, W~\tnote{a}, AR~\tnote{b}} &
            \emph{B~\tnote{a}, R~\tnote{b}} &
            no \\
            \emph{AMBA AXI4 Stream}~\cite{arm2021ambaaxistreampr} &
            \emph{B} &
            \emph{T~\tnote{d}} &
            \emph{T~\tnote{d}} &
            unlimited \\
            \emph{OpenHW OBI}~\cite{silicon2020obi1} &
            \emph{v1.5.0} &
            \emph{D} &
            \emph{R} &
            no \\
            \emph{SiFive TileLink}~\cite{sifivesifivetilelinks}~\tnote{e} &
            \emph{v1.8.1} &
            \emph{A} &
            \emph{R} &
            \makecell[cl]{UH: \\ \emph{power of two}} \\
            \emph{Fifo/Init}~\tnote{f} &
            N.A. &
            \emph{REQ} &
            \emph{RSP} &
            unlimited \\
            \bottomrule
        \end{tabular}
        \addtolength{\tabcolsep}{-3pt}
        \begin{tablenotes}[para, flushleft]
            \teb{\item[a]} write
            \teb{\item[b]} read
            \item[c] whichever is reached first
            \item[d] symmetrical \emph{RX/TX} channels
            \item[e] \emph{TL-UL} \& \emph{TL-UH} supported
            \item[f] simple double-handshaked \gls{fifo} protocol
        \end{tablenotes}
    \end{threeparttable}
    }
\end{table}

The transport layer abstracts the interfacing on-chip protocols through \emph{read} and \emph{write managers} with standardized interfaces, allowing for true multi-protocol capabilities. %
This modularization eases adapting {\idma} to different on-chip protocols and verifying its proper operation as the bulk of the {\dmae} does not require any changes.

\emph{Read managers} receive the current read transfer's base address, transfer length, and protocol-specific configuration information as inputs. %
They then emit a \emph{read-aligned} stream of data bytes to the downstream transport layer. %
\emph{Write managers} receive the write transfer information and the \emph{write-aligned} stream of data bytes from the upstream  transport layer to be emitted over their on-chip protocol's manager port. %

\Cref{tab:arch:managers} provides a complete list of implemented protocols; more details can be found in \Cref{chap:introduction:protocols}.%

The \emph{Fifo/Init} protocol provides a doubly handshaked streaming interface to easily connect custom hardware blocks without having to go through an on-chip communication protocol.
It can be used to emit a configurable stream of either the \emph{same repeated value}, \emph{incrementing values}, or a \emph{pseudorandom sequence}.
This enables our engine to accelerate memory initialization.

\subsection{Buffer Considerations}
\label{chap:idma:architecture:buffer}

Buffer sizing is an elemental part of configuring our {\dmaa} to achieve the full performance potential of the unit.
Three distinct types of buffers are present in {\idma}, each of them requiring careful tuning.

\subsubsection{Job Buffer}
\label{idma:arch:jobbuffer}

\emph{Job buffers} can be placed between the engine's \teb{\fe} and the {\be} or first {\me} to allow the \glspl{pe} to launch multiple non-blocking transfers.
Operation without this buffer is possible but results in every \gls{dma} transfer being \emph{blocking} as only one transfer can be issued by the {\fe} and worked on by the engine.
A new transaction can only be accepted once the {\be} completes the operation, \alc{which is marked by a handshake on the response channel.}

Depending on the system and the application, a job buffer of two to sixteen elements is recommended.

\subsubsection{Meta Buffer}

\emph{Meta buffers} are placed throughout the {\idma}'s architecture to hold transfer-specific information, including addresses, transfer lengths, and configurations, among others.
While not explicitly configurable through an exposed parameter, these buffers linearly scale with the number of outstanding transfers the {\dmae} supports.

To achieve a high bus utilization, the number of outstanding transfers should be chosen large enough to hide the latency of the desired memory endpoint when employing the targeted burst length.
In a well-designed system, expected values range from two to up to 64 outstanding transfers; to justify the latter, a deep memory system and an application issuing regularly short bursts is required.

\begin{figure}
    \centering%
    \includegraphics[width=\textwidth]{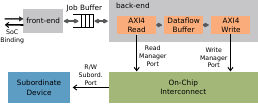}%
    \caption{%
        \alc{Simple system containing an {\idma}; the subordinate device can either handle one read or write burst at the same time.}
        \alc{Should the endpoint accept the write burst first and the {\be}-internal dataflow buffer is not large enough to hold the full burst, a deadlock occurs as the data cannot be read from the endpoint concurrently.}
    }%
    \label{fig:arch:bufferchart}%
\end{figure}

\subsubsection{Dataflow Buffer}

The \emph{dataflow buffer} within the dataflow element, see \Cref{sec:arch_transport-layer}, logically and physically decouples the read and write stage of {\idma}.
Unlike the other two types of buffers previously discussed, incorrectly sizing the dataflow buffer not only leads to subpar performance but can result in memory interconnect deadlocks for certain combinations of systems, protocols, and endpoints.

Ideally, the dataflow buffer should just be configured for logical and physical decoupling as well as coalescing of the read and write streams, storing at most three words of the \gls{dma} transfer.
This ensures maximized performance while keeping the area and resource footprint minimal.
For most systems, this configuration does not result in any adverse effects and should thus be used.

Burst-based protocols, like \Gls{axi4}, usually arbitrate their transfers on the granularity of an individual burst~\cite{kurth2022anopensourcepla} instead of a single data beat.
Should an endpoint only support either serving reads or writes at the same time, the burst-based protocol dictates that it first complete the read or the write transaction before serving the other.

In conjunction with a minimally sized dataflow buffer, this could lead to a deadlock scenario.
Let us assume a reorganization operation employing \SI{4}{\kilo\byte} bursts on a single endpoint is commanded to {\idma}\alc{, see \Cref{fig:arch:bufferchart}.}
The {\dmae} will issue both read and write operations towards the endpoint over the interconnect, which is free to reorder read and write operations freely.
Should the write operation be accepted first by the endpoint, a deadlock will be inevitable as the memory endpoint will not provide any read data unless the write burst operation is complete and the \gls{dma} cannot issue any writes before receiving the read data.

As part of this thesis, \teb{we develop} multiple mitigation mechanisms to prevent deadlocks from occurring.

\paragraph*{\textbf{Serialized and Debursted Operation}}

The transfer legalizer supports debursting the transfers to any power-of-two granularity.
Without any bursts emitted by {\idma}, transfer reordering by the interconnect does allow the endpoint to handle the memory accesses interleaved.

\paragraph*{\textbf{R-AW-Coupled Mode}}

In \alc{newer} versions of {\idma}, write requests can be stalled until the corresponding read data arrives in the dataflow element and is thus ready to be written.
In conjunction with the debursted operation, this ensures deadlock-free operation.

\paragraph*{\textbf{Leveraging AXI-REALM}}

As described in \Cref{chap:realm}, AXI-REALM can be used to fragment a data stream and to prevent \emph{ahead-of-time} blocking of subordinate resources.
These two features together allow for the complete prevention of the deadlock scenario described above.

\paragraph*{\textbf{R-W-Decoupled Endpoints}}

If we have full control over the memory endpoints in the system, they can easily be enhanced to handle reads and writes interleaved instead of blocking the read until the completion of the write burst or vice versa.
We provide an enhanced \emph{axi\_to\_mem\_interleaved} module as part of the \emph{axi4} repository\,\footnote{\url{github.com/pulp-platform/axi}}.

\paragraph*{\textbf{Fully Buffered Operation}}

As a final option, the \gls{fifo}-based buffer within the dataflow element can be replaced with a larger \gls{sram}-based buffer designed to hold at least one maximum-sized burst.
This should be considered as a last resort as buffer sizes in excess of \SI{4}{\kilo\byte} will be required, usually implemented through expensive and slow dual-port \glspl{sram}.

\subsection{Multi-Protocol Customization}
\label{chap:idma:architecture:mario}

\begin{figure}
    \centering%
    \includegraphics[width=0.8\textwidth]{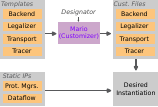}%
    \caption{%
        Our script, \emph{Mario}, customizes and properly connects the \gls{sv} template files of {\idma} to adapt it to specified protocol configurations.
    }%
    \label{fig:arch:mario}%
\end{figure}

As presented earlier in this section, one of the advantages of {\idma} over \gls{soa} is its modular architecture achieved through the use of highly parameteric \alc{\glsf{sv}} code and the modularity of the legalizer cores and protocol managers.
Although \Gls{sv} has well-developed capabilities of handling code parameterization, we could not identify an elegant way of handling {\idma}'s multi-protocol use case with its varying interface signature.
We decided to use a \alc{\emph{Mako}}\teb{-}template-based configuration approach, shown in \Cref{fig:arch:mario}, splitting the source into two parts:
a static \gls{ip} codebase built around parameterized \gls{sv} units and
a their template-based wrapper files customized by a script called \emph{Mario}, instantiating and connecting the proper \glspl{ip} and creating the interface connections.
Unlike other approaches, \emph{Mario} was designed with code introspection in mind, emitting nicely formatted, human-readable and modifiable code.
Mario\,\footnote{Named after Super Mario, the world's best-known plumber.} does not generate \gls{sv} code, it simply connects or plumbs the fundamental \glspl{ip} together using the available template code.

Mario assembles the desired configuration according to a unique identifier directly describing its protocol capabilities.
Designators are constructed as a sequence of protocols and protocol capabilities.
A capability is either described as read-only, \emph{r}, write-only, \emph{w}, or full-access, \emph{rw}.
Each available protocol has an identification string describing it, e.g., \emph{axi4} for \gls{axi4} or \emph{axis} for \gls{axi4} Stream.
The designator is handed to Mario through the build environment and is used to prefix the customized \gls{sv} code to prevent namespace conflicts of the different configurations.

As part of {\idma}'s provided \gls{ci} framework, a fundamental set of designators is created spanning all currently available protocols.
Next to the customized source code, Mario further emits a testbench for each configuration option used in \gls{ci} to verify {\idma}'s proper operation.

\newpage

\subsection{Error Handling}
\label{chap:idma:architecture:errorhandling}

Integrated in complex heterogeneous \glspl{soc}, it is an invalid assumption that {\idma} will never encounter a failing memory transfer.
Such errors can occur should the user or the operating system access an invalid memory region or if a subordinate device encounters an unexpected failure.
Although uncommon, such errors need to be handled efficiently.

\begin{figure}
    \centering%
    \includegraphics[width=0.8\textwidth]{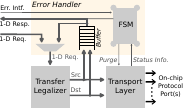}%
    \caption{%
        Internal architecture of {\idma}'s \emph{error handler}.
        A \emph{buffer} holds the emitted transfer information to be reported or replayed in the case of an error.
    }%
    \label{fig:arch:error-handler}%
\end{figure}

As can be seen in \Cref{fig:arch:error-handler} and previously outlined in \Cref{sec:idma:architecture:backend}, the error handler is an optional part of the {\be}.
It introduces two additional doubly handshaked signals to the {\be} facing towards the {\fe}: the \emph{response} and the \emph{error handler} interfaces.
The response interface informs the {\fe} whether a transfer completed successfully or resulted in an internal, e.g., zero-length transfer, or an external error, e.g., decode or subordinate error.
The error handler interface is used to communicate the intent of how to handle the issue.
It further intersects the request channel carrying the 1-D transfer information from the {\fe} to the legalizer within the {\be} to relaunch previously failed transfers.

\subsubsection{Error Operation}

Once an error is detected through one of the protocol ports, the transport layer informs the error handler that the current transfer failed.
The entire {\be} halts operation, and a failing response is issued.
It is ample for the software running on the \gls{pe} to understand the exact location of the error that occurred; this is especially true for \teb{\gls{nd}} tensor transfers.
The error handler thus has a \emph{transfer buffer} holding the address of the legalized transfers currently in-flight.
This address information, bundled with the type of error, is then handed to {\fe}.

Depending on the application's or platform's needs, one of three possible options can be selected to handle the error \alc{at runtime}.

\paragraph*{\textbf{Abort}}

The internal \gls{fsm} of the error handler allows the failed transfer in the dataflow element to finish, setting the write strobe to all zeros through the \emph{poison} signal to save energy by omitting \teb{writing} wrong data.
The current transfer is finished, as many protocols do not allow terminating a burst prematurely.
The state of the legalizer is flushed, bringing the {\dmae} into an idle state.
After completion, new transfers are allowed to be started again.

\paragraph*{\textbf{Continue}}

Like in the \emph{abort} case, the current transfer is completed without effects through issuing the poison signals.
Following this, normal operation is just resumed, effectively ignoring the error.
After completing the rest of the transfer successfully, the software may relaunch the failed part of the transfer manually.

\paragraph*{\textbf{Replay}}

It is the most complex handling mode currently implemented in {\idma}.
Like before, the currently failing transfer is completed without an effect, followed by a completion of all the outstanding transfers in the transport layer.
Once the transport layer becomes idle, the failed transfer is fetched from the transfer buffer and automatically relaunched.
It is assumed that the relaunched transfers are now complete and successful; if not, a new error handling operation can be launched, this time only allowing abort and continue.

\newpage
\section{Architectural Results}
\label{chap:idma:archresults}

\afterpage{%
\begin{sidetab}
    \caption{%
    Area decomposition of the {\dmae} configuration used in the \teb{\pulpopen} cluster, see \Cref{chap:dmaext:tensor:pulpopen}. %
    The \emph{base} area is always required, the contribution of each protocol added is shown. %
    If the area contribution is non-zero, the parameter influencing the value is provided using the \emph{big-$\mathcal{O}$} notation. %
    The area contribution scales linearly with the data width (\textit{DW}) if no scaling is provided. %
    }
    \label{tab:results:area}
    \renewcommand*{\arraystretch}{2.2}
    \resizebox{\linewidth}{!}{%
    \begin{threeparttable}
        \begin{tabular}{@{}l!{\color{black}\vrule}l!{\color{black}\vrule}llllllllllllll@{}} \toprule
            \multicolumn{3}{l}{Units} &
            Base &
            \multicolumn{2}{c}{\makecell[cc]{\teb{\gls{axi4}} \\ read \hspace{0.5cm} write}} &
            \multicolumn{2}{c}{\makecell[cc]{\teb{\gls{axi4}} Lite\\ read \hspace{0.5cm} write}} &
            \multicolumn{2}{c}{\makecell[cc]{\teb{\gls{axi4}} Stream\\ read \hspace{0.5cm} write}} &
            \multicolumn{2}{c}{\makecell[cc]{\teb{\gls{obi}} \\ read \hspace{0.5cm} write}} &
            \multicolumn{2}{c}{\makecell[cc]{\teb{\gls{tluh}} \\ read \hspace{0.5cm} write}} &
            Init\teb{/\gls{fifo}} \\
            \midrule
            \multirow{8}{*}{\rott{Backend}} &
            \rott{Decoupling} &
            - &
            \res{3.7}{NAx}{\kilo\GE}{~\tnote{a}} &
            \res{1.4}{NAx}{\kilo\GE}{} &
            \res{1.4}{NAx}{\kilo\GE}{} &
            \res{310}{NAx}{\GE}{} &
            \res{310}{NAx}{\GE}{} &
            \res{310}{NAx}{\GE}{} &
            \res{310}{NAx}{\GE}{} &
            \res{310}{NAx}{\GE}{} &
            \res{310}{NAx}{\GE}{} &
            \res{310}{NAx}{\GE}{} &
            \res{310}{NAx}{\GE}{} &
            \resn \\ \divl
            &
            \multirow{4}{*}{\rott{Legalizer}} &
            State &
            \res{1.5}{AW}{\kilo\GE}{~\tnote{b}} &
            \res{710}{AW}{\GE}{~\tnote{c}} &
            \res{710}{AW}{\GE}{~\tnote{c}} &
            \res{200}{AW}{\GE}{~\tnote{c}} &
            \res{200}{AW}{\GE}{~\tnote{c}} &
            \res{180}{AW}{\GE}{~\tnote{c}} &
            \res{180}{AW}{\GE}{~\tnote{c}} &
            \res{180}{AW}{\GE}{~\tnote{c}} &
            \res{180}{AW}{\GE}{~\tnote{c}} &
            \res{215}{AW}{\GE}{~\tnote{c}} &
            \res{215}{AW}{\GE}{~\tnote{c}} &
            \res{21}{AW}{\GE}{} \\
            &
            &
            Page Split &
            \resn &
            \res{95}{1}{\GE}{} &
            \res{105}{1}{\GE}{} &
            \res{7}{1}{\GE}{} &
            \res{8}{1}{\GE}{} &
            \resn &
            \resn &
            \res{5}{1}{\GE}{} &
            \res{5}{1}{\GE}{} &
            \resn &
            \resn &
            \resn \\
            &
            &
            Pow2 Split &
            \resn &
            \resn &
            \resn &
            \resn &
            \resn &
            \resn &
            \resn &
            \resn &
            \resn &
            \res{20}{1}{\GE}{} &
            \res{20}{1}{\GE}{} &
            \resn \\ \divl
            &
            \multirow{4}{*}{\rott{Transport Layer}} &
            Dataflow Element &
            \resh{1.3}{DW}{\kilo\GE}{~\tnote{d}} &
            \resn &
            \resn &
            \resn &
            \resn &
            \resn &
            \resn &
            \resn &
            \resn &
            \resn &
            \resn &
            \resn \\
            &
            &
            \makecell[cl]{Contribution \\ of Each Read/ \\ Write Manager \\ Respectively}&
            \resh{70}{DW}{\GE}{} &
            \resh{190}{DW}{\GE}{} &
            \resh{30}{DW}{\GE}{} &
            \resh{60}{DW}{\GE}{} &
            \resh{60}{DW}{\GE}{} &
            \resh{60}{DW}{\GE}{} &
            \resh{60}{DW}{\GE}{} &
            \resh{60}{DW}{\GE}{} &
            \resh{35}{DW}{\GE}{} &
            \resh{230}{DW}{\GE}{} &
            \resh{150}{DW}{\GE}{} &
            \resh{55}{DW}{\GE}{} \\
            &
            &
            Shifter/Muxing &
            \resh{120}{DW}{\GE}{} &
            \resh{250}{DW}{\GE}{~\tnote{c}} &
            \resh{250}{DW}{\GE}{~\tnote{c}} &
            \resh{75}{DW}{\GE}{~\tnote{c}} &
            \resh{75}{DW}{\GE}{~\tnote{c}} &
            \resh{180}{DW}{\GE}{~\tnote{c}} &
            \resh{180}{DW}{\GE}{~\tnote{c}} &
            \resh{170}{DW}{\GE}{~\tnote{c}} &
            \resh{170}{DW}{\GE}{~\tnote{c}} &
            \resh{65}{DW}{\GE}{~\tnote{c}} &
            \resh{65}{DW}{\GE}{~\tnote{c}} &
            \resn \\
            \bottomrule
        \end{tabular}
        \begin{tablenotes}[para, flushleft]
            \item[a] \emph{NAx}: Number of outstanding transfers supported. Used configuration: \textbf{16}.
            \item[b] \emph{AW}: Address width. Used configuration: \textbf{32-\si{\bit}}.
            \item[c] If multiple protocols are used, only the maximum is taken.
            \item[d] \emph{DW}: Data Width. Used configuration: \textbf{32-\si{\bit}}.
        \end{tablenotes}
    \end{threeparttable}
    }
\end{sidetab}
}

To deepen the insight into {\idma} and highlight its versatility, we provide \glsu{ip}-level implementation results in this section. %
We first present area and timing models characterizing the influence of parametrization on our architecture, enabling quick and accurate estimations when integrating engines into new systems. %
We then use these models to show that {\idma}'s area and timing scale well for any reasonable parameterization. %
Finally, we present latency results for our {\be} and discuss our engine's performance in three sample memory systems. %
For implementation experiments, we use {\gfs} {\emph{\gftech}} technology with a 13-metal stack and 7.5-track standard cell library in the typical process corner. %
We synthesize our designs using {\dc} in topological mode to account for place-and-route constraints, congestion, and physical phenomena. %

\subsection{Area Model}

\begin{figure}[p]
   \begin{subcaptionblock}{\linewidth}
       \centering%
       \includegraphics[width=\linewidth]{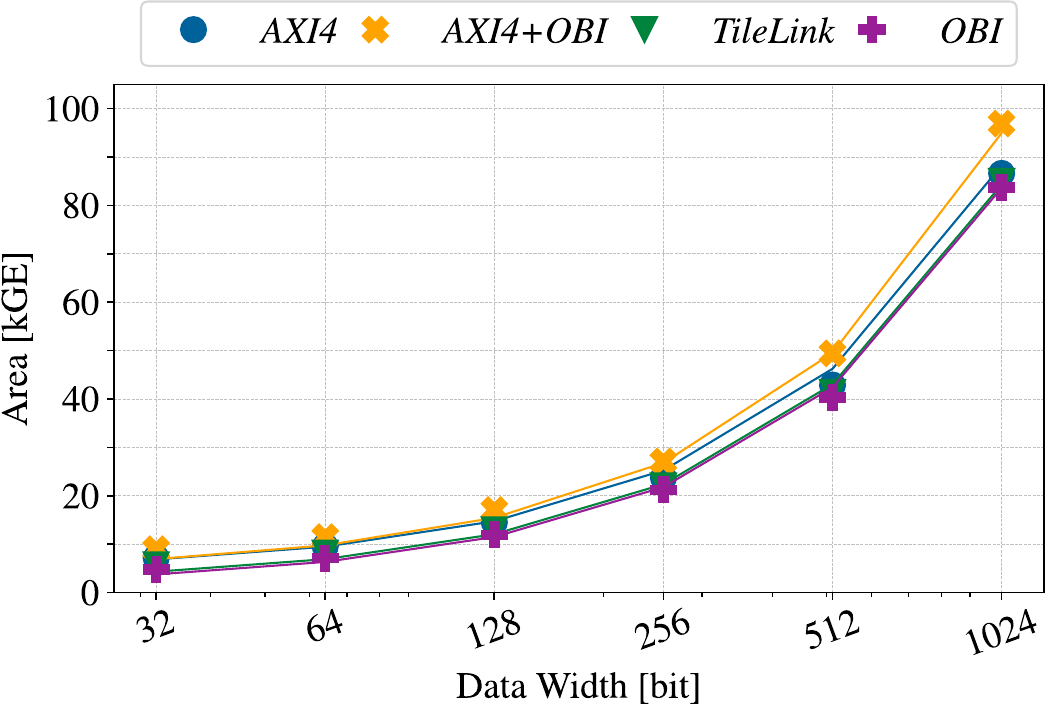}%
       \vspace{0.17cm}
       \label{fig:results:area:dw}%
   \end{subcaptionblock}
   \centering
   \begin{subcaptionblock}{0.512\linewidth}%
       \centering%
       \includegraphics[width=\linewidth]{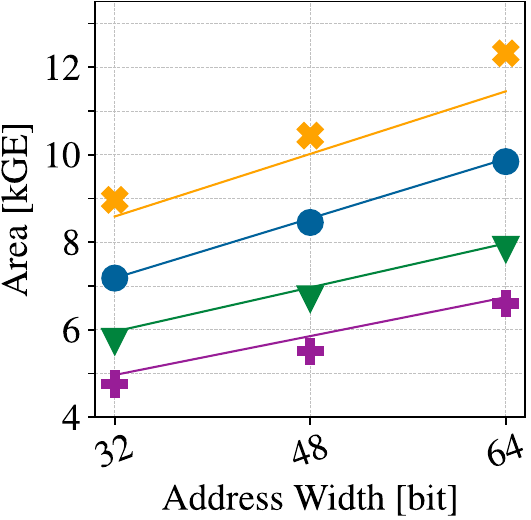}%
       \label{fig:results:area:aw}%
   \end{subcaptionblock}
   \hspace{0.15cm}
   \begin{subcaptionblock}{0.451\linewidth}
       \centering%
       \includegraphics[width=\linewidth]{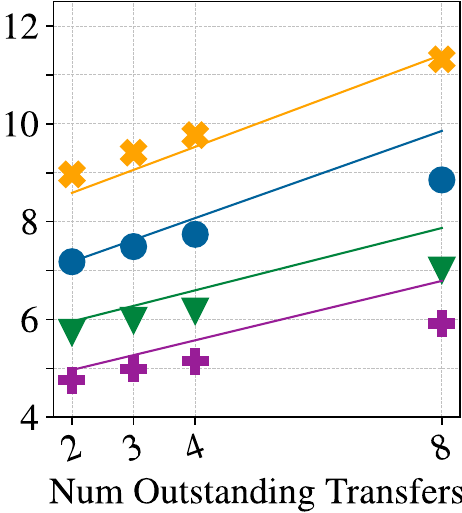}%
       \label{fig:results:area:num-ax}%
   \end{subcaptionblock}\hfill
   \vspace{0.2cm}
   \caption{ %
      Area scaling of a {\be} base configuration (\emph{32-\si{\bit}} address and data width, \emph{two} outstanding transactions). %
      Markers represent the measurement points and lines the fitted model. %
   } %
   \label{fig:results:area}
\end{figure}

We focus our evaluation and modeling effort on our {\idma} {\be}, as both the {\fe} and {\me} are very application- and platform-specific and can only be properly evaluated in-system. %
Our area model was evaluated at \SI{1}{\giga\hertz} using the typical process corner in {\gftech}. %

For each of the {\be}'s major area contributors listed in \Cref{tab:results:area}, we fit a set of linear models using \emph{non-negative least squares}. %
For each parametrization, our models take a vector containing the number of ports of each protocol as an input. %
This set of models allows us to estimate the area decomposition of the base hardware of the {\be} and the area contributions of any additional protocol port, given a particular parameterization, with an average error of less than \SI{4}{\percent}. %
For example, \Cref{tab:results:area} shows the modeled area decomposition for our \emph{base} configuration of 32-\si{\bit} address width, 32-\si{\bit} data width, and two outstanding transfers. %

A second step is required to estimate area contributions to the {\be} depending on the parameterization, the number, and the type of ports. %
We created a second \emph{param} model estimating the influence of the three main parameters, \emph{area width (AW)}, \emph{data width (DW)}, and the \emph{number of outstanding transactions (NAx)}, on the {\be}'s area contributions. %
We can estimate the area composition of the {\be} with an average error of less than \SI{9}{\percent}, given both the parameterization and the used read/write protocol ports as input. %

We provide a qualitative understanding of the influence of parameterization on area by listing the parameter with the strongest correlation using \emph{big-$\mathcal{O}$} notation in \Cref{tab:results:area}. %

To outline the accuracy of our modeling approach, we show the area scaling of four of our {\idmaes} for different protocol configurations, depending on the three main parameters, starting from the \emph{base} configuration. %
The subplots of \Cref{fig:results:area} present the change in area when one of the three main parameters is modified and the output of our two linear models combined. %
The combined area model tracks the parameter-dependent area development with an average error of less than \SI{9}{\percent}. %
In those cases where the model deviates, the modeled area is overestimated, providing a safe upper bound for the {\be} area. %

\subsection{Timing Model}

\begin{figure}
    \begin{subcaptionblock}{\linewidth}
        \centering%
        \includegraphics[width=\linewidth]{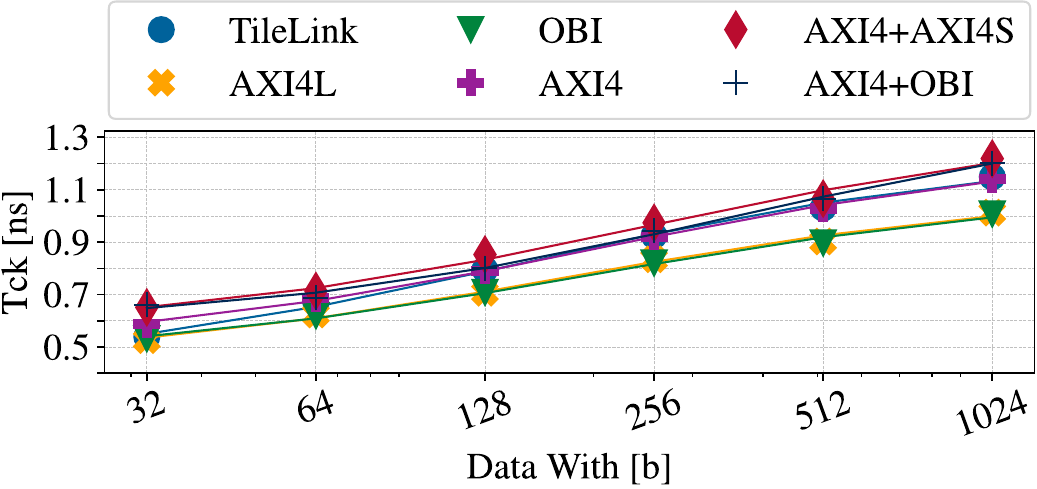}%
        \vspace{0.15cm}
        \label{fig:results:tck:data-width}%
    \end{subcaptionblock}
    \centering
    \begin{subcaptionblock}{0.52\linewidth}%
        \centering%
        \includegraphics[width=\linewidth]{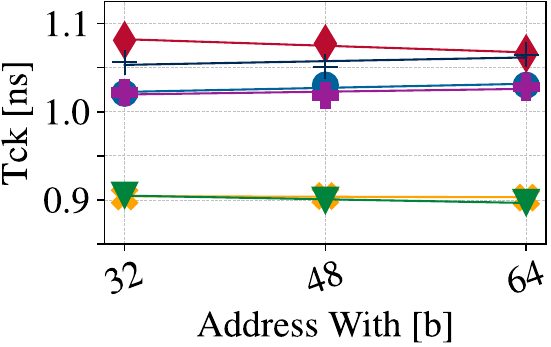}%
        \label{fig:results:tck:addr-width}%
    \end{subcaptionblock}
    \hspace{0.15cm}
    \begin{subcaptionblock}{0.445\linewidth}
        \centering%
        \includegraphics[width=\linewidth]{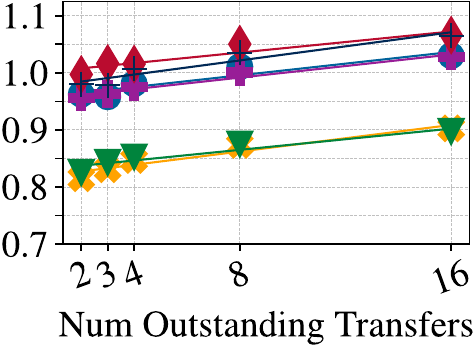}%
        \label{fig:results:tck:num-ax}
    \end{subcaptionblock}\hfill
    \vspace{0.2cm}
    \caption{%
        \alc{Clock period} scaling of a {\be} base configuration (\emph{32-\si{\bit}} address and data width, \emph{two} outstanding transactions).
        \alc{Lower is better.}
        \teb{\emph{AXI4L} denotes \gls{axi4} Lite, \emph{AXI4S}, \gls{axi4} Stream.}
    }
    \label{fig:results:tck}
\end{figure}

We again focus our timing analysis on the {\be}, as the {\fe} should be analyzed in-system and {\mes} may be isolated from the {\idmae}'s timing by cutting timing paths between front-, mid-, and back-ends. %
Our investigation shows a \emph{multiplicative inverse} dependency between the longest path in \emph{\si{\nano\second}} and our main parameters. %
We use the \emph{base} configuration of the {\be} to evaluate our timing model by sweeping our three main parameters. %
The tracking of our model is presented in \Cref{fig:results:tck} for six representative configurations ranging from simple \gls{obi} to complex multi-protocol configurations involving \gls{axi4}. %
Our timing model achieves an average error of less than \SI{4}{\percent}. %

The results divide our {\bes} into two groups: simpler protocols, \emph{OBI} and \emph{AXI Lite}, run faster as they require less complex legalization logic, whereas more complex protocols require deeper logic and thus run slower. %
Engines supporting multiple protocols and ports also run slower due to additional arbitration logic in their data path. %
\emph{Data width} has a powerful impact on {\idmae}'s speed, mainly due to wider shifters required to align the data. %
The additional slowdown at larger data widths can be explained by physical routing and placement congestion of the increasingly large buffer in the dataflow element. %
\emph{Address width} has little effect on the critical path as it does not pass through the legalizer cores, whose timing is most notably affected by address width. %
Increasing the number of \emph{outstanding transactions} sub-linearly degrades timing due to more complex \gls{fifo} management logic required to orchestrate them. %

\subsection{Latency Model}

Our {\idma} {\bes} have a fixed latency of \emph{two} cycles from receiving a 1-D transfer from the {\fe} or the last {\me} to the read request at a protocol port. %
Notably, this is \emph{independent} of the protocol selection, the number of protocol ports, and the three main {\idma} parameters. %
This rule only has one exception: in a {\be} without hardware legalization support, the latency reduces to \emph{one} cycle. %
Generally, each {\me} presented in \Cref{chap:dmaext} requires \emph{one} additional cycle of latency.
We note, however, that the \emph{tensor{\tus}ND} {\me} (\Cref{chap:dmaext:tensor}) can be configured to have \emph{zero} cycles of latency, meaning that even for an \alc{\gls{nd}} transfer, we can ensure that the first read request is issued \emph{two} cycles after the transfer arrives at the {\me} from the {\fe}. %

\subsection{Standalone Performance}

\begin{figure}[p]
    \begin{subcaptionblock}{\linewidth}
        \centering%
        \includegraphics[width=\linewidth]{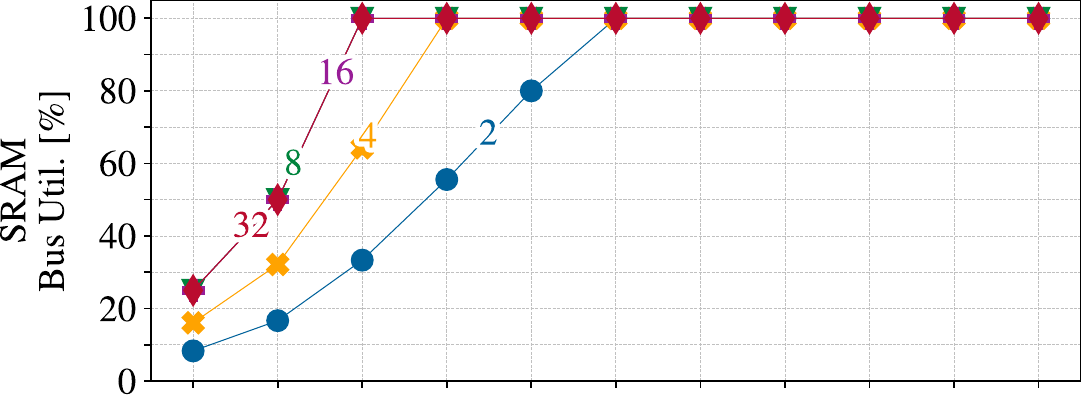}%
        \vspace{0.14cm}
        \label{fig:results:perf:pulp}%
    \end{subcaptionblock}
    \centering
    \begin{subcaptionblock}{\linewidth}%
        \centering%
        \includegraphics[width=\linewidth]{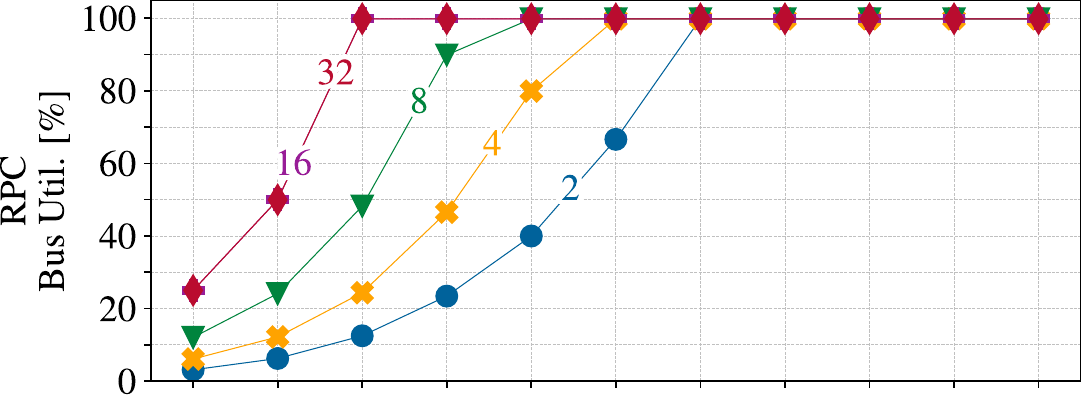}%
        \vspace{0.14cm}
        \label{fig:results:perf:rpc}%
    \end{subcaptionblock}
    \begin{subcaptionblock}{\linewidth}
        \centering%
        \includegraphics[width=\linewidth]{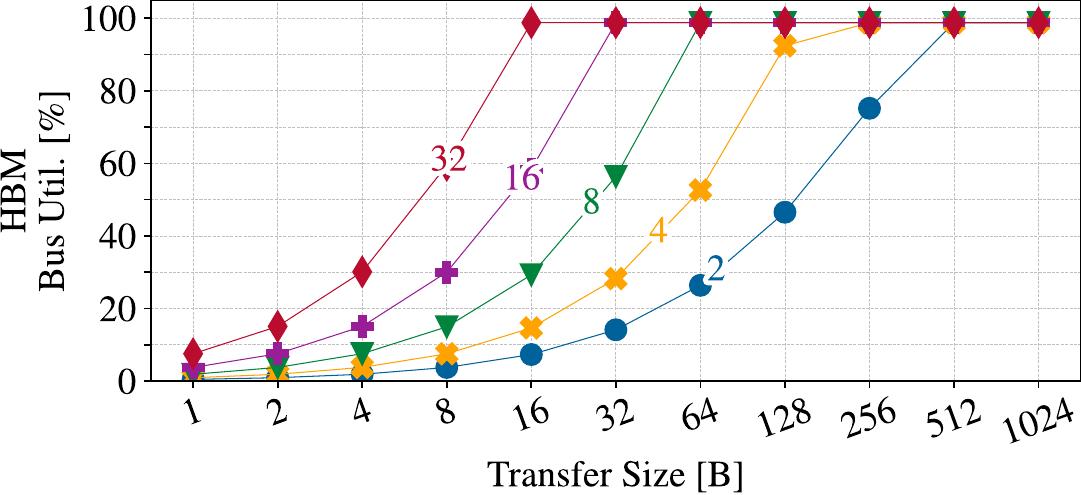}%
        \vspace{0.14cm}
        \label{fig:results:perf:hbm}%
    \end{subcaptionblock}\hfill
    \vspace{-0.6cm}
    \caption{%
        \alc{Bus utilization of our {\idmae} in the base configuration (\emph{32-\si{\bit}} address and data width) with varying amounts of outstanding transactions in three different memory systems; \emph{SRAM}, \emph{RPC DRAM}~\cite{etron2024256mbhighbandwi}, \emph{HBM}~\cite{zaruba2020a4096coreriscvc}.} %
    }
    \label{fig:results:perf}
\end{figure}

We evaluated the out-of-context performance of an {\idmae} in the \emph{base} configuration copying a \emph{\SI{64}{\kibi\byte}} transfer fragmented in individual transfer sizes between \SI{1}{\byte} and \SI{1}{\kibi\byte} in three different memory system models. %
The analysis is protocol-agnostic as all implemented protocols support a similar outstanding transaction mechanism; we thus use \emph{AXI4} in this subsection without loss of generality. %
The three memory systems used in our evaluation differ in access cycle latency and number of outstanding transfers. %

\emph{SRAM} represents the \gls{l2} memory found in the \teb{\pulpopen} system (\Cref{chap:dmaext:tensor:pulpopen}) with three cycles of latency and eight outstanding transfers. %
\teb{\emph{RPC-DRAM}} uses the characteristics of an open-source AXI4 controller for the \gls{rpc} DRAM technology~\cite{etron2024256mbhighbandwi,etron2022reducedpincount} run at \SI{933}{\mega\hertz}, around thirteen cycles of latency and support for sixteen outstanding transactions. %
\teb{\emph{HBM}} models an industry-grade \gls{hbm}~\cite{zaruba2020a4096coreriscvc} interface with a latency in the order of 100 cycles and supporting the tracking of more than 64 outstanding transfers. %

In shallow memory systems, the {\idmae} reaches almost perfect bus utilization, copying single bus-sized data transfers required to track as low as eight outstanding transactions. %
More outstanding requests are required in deeper memory systems to sustain this perfect utilization. %
Any transfers smaller than the bus width will inevitably lead to a \emph{substantial} drop in utilization, meaning that unaligned transfers inherently limit the maximum possible bus utilization our engines can achieve.
Nevertheless, our fully decoupled data-flow-oriented architecture \emph{maximizes the utilization of the bus} even in these scenarios. %

\Cref{fig:results:perf} shows that even in very deep systems with hundreds of cycles of latency, our engine can achieve almost perfect utilization for a relatively small transfer granularity of four times the bus width. %
This agility in handling transfers allows us to copy \teb{\gls{nd}} tensors with a narrow inner dimension efficiently. %

The cost of supporting such fine-granular transfers is an increased architectural size of the engines' decoupling buffers. %
As shown in \Cref{fig:results:area:num-ax}, these scale linearly in the number of outstanding transactions to be supported, growing by roughly \SI{400}{\GE} for each added buffer stage. %
In our base configuration, supporting 32 outstanding transfers keeps the engine area below \SI{25}{\kilo\GE}. %

\subsection{Energy Efficiency}

With {\idma}'s area-optimized design, the capability of selecting the simplest on-chip protocol, and the effort to minimize buffer area, we can curtail the area footprint and thus the static power consumption of our {\idmaes}. %

{\idma}'s decoupled, agile, data-flow-oriented architecture is explicitly designed to handle transfers efficiently while maximizing bus utilization, limiting the unit's active time to a minimum. %
Coupled with our minimal area footprint and low-buffer design, this directly minimizes energy consumption. %
Furthermore, the engine's efficiency allows \emph{run-to-completion} operating modes where we maximize the interconnect's periods of inactivity between transfers, allowing efficient clock gating of the {\idmae} and the interconnect, further increasing energy efficiency. %

\newpage
\section{Related Work}
\label{chap:idma:relatedwork}

\afterpage{%
\begin{sidetab}
    \scriptsize{%
        \centering
        \caption{%
            Comparison of {\idma} to the \gls{soa}.
        }%
        \vspace{-0.25cm}
        \label{tab:res:soac}
        \renewcommand*{\arraystretch}{2.3}
        \setlength\extrarowheight{-16pt}
        \resizebox{0.982
        \linewidth}{!}{%
        \begin{threeparttable}
            \begin{tabular}{@{}ccccccccc@{}}
                \toprule
                \hol{\teb{\DmaE}} &
                \hol{Application} &
                \hol{Technology} &
                \htl{Supported}{Protocols} &
                \htl{Transfer}{Type} &
                \htl{Programming}{Model} &
                \htl{Stream}{Mod. Cap.} &
                \htl{Modularity}{Configurability} &
                \hol{Area} \\
                \midrule
                \did{{CubeDMA}~\cite{fjeldtvedt2019cubedmaoptimizi}}{Fjeldtvedt~\etal} &
                Hyperspectral Imaging &
                FPGA &
                \makecell[cc]{\teb{\gls{axi4}}~\tnote{a} \\ \teb{\gls{axi4}} Stream~\tnote{b}} &
                3-D &
                Register File &
                None &
                Limited Configurability &
                \makecell[cc]{2162~LUTs \\ 1796~FFs} \\
                \did{{RDMA}~\cite{paraskevas2018virtualizedmult}}{Paraskevas~\etal} &
                HPC &
                FPGA &
                \teb{\gls{axi4}} &
                Linear &
                Transfer Descriptors &
                None &
                None &
                N.A. \\
                \did{{cDMA}~\cite{rhu2018compressingdmae}}{Rhu~\etal} &
                DNN &
                ASIC~\tnote{c} &
                N.A. &
                Linear &
                None &
                \makecell[cc]{Compression \& \\ Decompression} &
                No &
                \makecell[cc]{\SI{0.21}{\milli\metre\squared}~\tnote{d} \\ \SI{420}{\kilo\GE}~\tnote{e}} \\
                \dis{{Rossi~\etal~\cite{rossi2014ultralowlatency}}} &
                ULP &
                Tech. - Indep. &
                STBus~\tnote{f}, \teb{\gls{obi}}~\tnote{f}~\tnote{g} &
                Linear &
                Per-PE Register File &
                None &
                Limited Configurability &
                \makecell[cc]{$\approx$ \SI{0.04}{\milli\metre\squared} \\ $\approx$ \SI{82}{\kilo\GE}} \\
                \did{{MT-DMA}~\cite{ma2019mtdmaadmacontro}}{Ma~\etal} &
                \makecell[cc]{Scientific \\ Computing} &
                ASIC &
                Custom &
                \makecell[cc]{2-D \\ Arb. Strides} &
                Transfer Descriptors &
                \makecell[cc]{Block \\ Transp.} &
                None &
                \makecell[cc]{\SI{1.07}{\milli\metre\squared} \\ \SI{1.5}{\mega\GE}~\tnote{h}} \\
                \did{{FastVDMA}~\cite{2019antmicrorelease}}{Antmicro} &
                General-\teb{P}urpose &
                Tech. - Indep. &
                \makecell[cc]{\teb{\gls{axi4}}~\tnote{f}~\tnote{i} \\ \teb{\gls{axi4}}-Stream~\tnote{f}~\tnote{i} \\ Wishbone~\tnote{f}~\tnote{i}} &
                Linear &
                Register File &
                None &
                \makecell[cc]{Protocol \\ Selectable} &
                455~Slices~\tnote{j} \\
                \did{{DMA-330}~\cite{arm2012corelinkdma330d}}{ARM} &
                General-\teb{P}urpose &
                Tech. - Indep. &
                \makecell[cc]{AXI3~\tnote{k} \\ Peripheral Intf.~\tnote{l}} &
                \makecell[cc]{2-D \\ Scatter-Gather } &
                \makecell[cc]{Custom \\ Instructions } &
                None &
                Yes, But Non-\teb{M}odular &
                N.A. \\
                \did{{AXI DMA v7.1}~\cite{xilinx2022axidmav71logico}}{Xilinx} &
                General-\teb{P}urpose &
                \makecell[cc]{FPGA \\ Xilinx-only} &
                \makecell[cc]{\teb{\gls{axi4}}~\tnote{f} \\ \teb{\gls{axi4}}-Stream~\tnote{f}} &
                \makecell[cc]{optional 2-D \\ Scatter-Gather } &
                Transfer Descriptors &
                None &
                Yes, But Non-\teb{M}odular &
                \makecell[cc]{2745~LUTs~\tnote{m} \\ 4738~FFs~\tnote{m} \\ \SI{216}{\kilo\bit}~BRAM~\tnote{m}} \\
                \did{{vDMA AXI}~\cite{rambusdmaaxiipcontrol}}{RAMBUS} &
                General-\teb{P}urpose &
                Tech. - Indep. &
                AXI3/4 &
                \makecell[cc]{2-D \\ Scatter-Gather } &
                Transfer Descriptors &
                None &
                Yes, But Non-\teb{M}odular &
                N.A. \\
                \did{{DW\_axi\_dmac}~\cite{synopsysdesignwareipsol}}{Synopsys} &
                General-\teb{P}urpose &
                Tech. - Indep. &
                \makecell[cc]{AXI3/4 \\ Peripheral Intf.~\tnote{l}} &
                \makecell[cc]{2-D \\ Scatter-Gather } &
                Transfer Descriptors &
                None &
                Yes, But Non-\teb{M}odular &
                N.A. \\
                \did{{Dmaengine}~\cite{stmicroelectronicsdmaengineovervi}}{STMicroelectronics} &
                MCU &
                STM32 &
                \makecell[cc]{STBus~\tnote{n} \\ Peripheral Intf.~\tnote{n}} &
                Linear &
                \makecell[cc]{Custom \\ Instructions } &
                None &
                No &
                N.A. \\
                \did{{DDMA}~\cite{designddmamultichanne}}{DCD-SEMI} &
                MCU &
                Tech.- Indep.~\tnote{n} &
                Custom 32-bit~\tnote{o} &
                Linear, fixed &
                Register File &
                None &
                No &
                N.A. \\
                \did{{{\textmu}DMA}~\cite{pullini2017textmudmaanauto}}{Pullini~\etal} &
                MCU &
                Tech. - Indep. &
                \makecell[cc]{\teb{\gls{obi}}~\tnote{f}~\tnote{g} \\ RX/TX Channels~\tnote{f}~\tnote{p}} &
                Linear &
                Register File &
                None &
                Yes, But Non-\teb{M}odular &
                \SI{15.4}{\kilo\GE} \\
                \dis{Morales~\etal~\cite{morales2019alowareadirectm}} &
                IoT &
                Tech. - Indep. &
                AHB~\tnote{f}, Perif. Intf.~\tnote{f} &
                Linear &
                Register File &
                None &
                Yes, But Non-\teb{M}odular &
                \SI{3.2}{\kilo\GE} \\
                \dis{Su~\etal~\cite{su2011aprocessordmaba}} &
                General-\teb{P}urpose &
                \makecell[cc]{Tech.- \\ Indep.~\tnote{n}} &
                \teb{\gls{axi4}} &
                Linear~\tnote{n} &
                Register File &
                None &
                No &
                N.A. \\
                \did{{VDMA}~\cite{nandan2014highperformance}}{Nandan~\etal} &
                Video &
                Custom &
                N.A. &
                \makecell[cc]{2-D \\ Arb. Strides} &
                Integrated &
                None &
                No &
                N.A. \\
                \dis{Comisky~\etal~\cite{comisky2000ascalablehighpe}} &
                MCU &
                N.A. &
                TR Bus &
                Linear~\tnote{n} &
                Register File &
                None &
                No &
                N.A. \\
                \midrule
                \dtw{\idma}{Architecture} &
                \makecell[cc]{Extreme-\teb{E}dge ULP, \\ Datacenter HPC, \\ Application-\teb{G}rade} &
                Tech. - Indep. &
                \makecell[cc]{\teb{\gls{axi4}}, \teb{\gls{axi4}} Lite, \\ \teb{\gls{axi4}}-Stream, \\ \teb{\gls{tlul}}, \teb{\gls{tluh}}, \\ \teb{\gls{obi}}} &
                \makecell[cc]{Optional \alc{\gls{nd}} \\ Arb. Strides \\ Scatter-Gather} &
                \makecell[cc]{Register File, \\ Transfer Descriptors, \\ RISC-V ISA Ext., \\ Custom} &
                \makecell[cc]{Memory Init., \\ In-\teb{S}tream \\ Accelerator} &
                \makecell[cc]{Configurable \\ and Modular} &
                $\geq$ \SI{2}{\kilo\GE} \\
                \arrayrulecolor{black!30}\bottomrule
            \end{tabular}
            \begin{tablenotes}[para, flushleft]
                \item[a] read-only
                \item[b] write-only
                \item[c] FreePDK45
                \item[d] \SI{28}{\nano\metre} node
                \item[e] assuming \SI{0.5}{\micro\metre\squared} per \SI{1}{\GE}
                \item[f] cross-protocol operation only
                \item[g] pre-1.0 version
                \item[h] assuming \SI{0.7}{\micro\metre\squared} per \SI{1}{\GE}
                \item[i] \emph{one read-only} and \emph{one write-only} protocol selectable
                \item[j] \SI{32}{\bit}, \teb{\gls{axi4}} read, \teb{\gls{axi4}}-Stream write
                \item[k] one manager port, main interface
                \item[l] optional
                \item[m] \emph{UltraScale\_mm2s\_64DW\_1\_100 (xcku040, ffva1156, 1)}
                \item[m] to the best of our knowledge
                \item[o] wrapper for APB, AHB, \teb{\gls{axi4}} Lite available
                \item[p] very similar to \teb{\gls{obi}}
            \end{tablenotes}
        \end{threeparttable}
        }
    }
\end{sidetab}
}

We compare {\idma} to an extensive selection of commercial {\dma} solutions and {\dmaes} used in research platforms; an overview is shown in \Cref{tab:res:soac}. %

In contrast to this work, existing {\dmaes} are designed for a given system, a family of systems, or even a specific application on a system. %
These engines lack modularity and cannot be readily retargeted to a different system. %
To the best of our knowledge, our work is the first fully modular and universal {\dmaa}. %
Moreover, most of the {\dmaes} in our comparison are closed-source designs and thus not accessible to the research community, hindering or even preventing benchmarking and quantitative comparisons. %

We identify two general categories of {\dmaes}: large high-bandwidth engines specialized in efficient memory transfers and low-footprint engines designed for accessing peripherals efficiently. %
Ma~\etal~\cite{ma2019mtdmaadmacontro}, Paraskevas~\etal~\cite{paraskevas2018virtualizedmult}, and Rossi~\etal~\cite{rossi2014ultralowlatency} present high-performance {\dmaes} ranging in size from \SI{82}{\kilo\GE} to over \SI{1.5}{\mega\GE}. %
On the contrary, {\dmaes} designed for accelerating accesses to chip peripherals, as shown in the works of Pullini~\etal~\cite{pullini2017textmudmaanauto} and Morales~\etal~\cite{morales2019alowareadirectm} trade off performance for area efficiency by minimizing buffer space~\cite{pullini2017textmudmaanauto} and supporting only simpler on-chip protocols like AHB~\cite{morales2019alowareadirectm} or OBI~\cite{pullini2017textmudmaanauto}. %
Our {\idma} can be parameterized to achieve peak performance as a high-bandwidth engine in \gls{hpc} systems, see \Cref{chap:comcpu} and \Cref{chap:comcpu}, as well as to require less area ($<$\SI{2}{\kilo\GE}) than the ultra-lightweight design of Pullini~\etal's \emph{{\textmu}DMA}~\cite{pullini2017textmudmaanauto}. %

\gls{soc} {\dmaes}, e.g., Fjeldtvedt~\etal~\cite{fjeldtvedt2019cubedmaoptimizi}, Rossi~\etal~\cite{rossi2014ultralowlatency}, and Morales~\etal~\cite{morales2019alowareadirectm} feature a fixed configuration of on-chip protocol(s). %
Some controllers allow selectively adding a simple interface to connect to peripherals. %
The {\dma} \glspl{ip} from Synopsys~\cite{synopsysdesignwareipsol} and ARM~\cite{arm2012corelinkdma330d} are two examples. %
We identify one exception to this rule: \emph{FastVDMA} from {Antmicro}~\cite{2019antmicrorelease} can be configured to select one read- and one write-only port from a selection of three protocols. %
\emph{FastVDMA} only supports unidirectional data flow from one read to the other write port. %
\emph{Inter-port} operation, meaning copying data from one port and storing it using the same port, is not supported, which can limit its usability. %
{\idma} allows the selective addition of one or multiple read or write interface ports from a list of currently five industry-standard on-chip protocols. %
If configured, our engines allow bidirectional data movement in \emph{inter-port} and \emph{intra-port} operations. %
Thanks to the standardization of interfaces and the separation of data movement from protocol handling, new on-chip protocols can be added quickly by implementing \emph{at most} three modules, each only a couple of hundred \si{\GE}s of complexity. %

\emph{FastVDMA}~\cite{2019antmicrorelease} shows basic modularity by allowing users to select one read and one write protocol from a list of three on-chip protocols. %
Its modularity is thus limited to the {\be}, and there is neither any facility to change between different programming interfaces nor a way of easily adding more complex \teb{\gls{nd}} affine stream movement support. %
As presented in this work, our approach tackles these limitations by specifying and implementing the first fully modular, parametric, universal {\dmaa}. %

\newpage
\section{Summary and Conclusion}
\label{chap:idma:conclusion}

We present {\idma}, a modular, highly parametric {\dmaa} composed of three parts ({\fe}, {\me}, and {\be}), allowing our engines to be customized to suit a wide range of systems, platforms, and applications. %

We evaluate the area, timing, latency, and performance of {\idma}, resulting in area and timing models that allow us to estimate the synthesized area and timing characteristics of any parameterization within \SI{9}{\percent} of the actual result. %

Our architecture enables the creation of both ultra-small {\idmaes} incurring less than \SI{2}{\kilo\gateeq}, as well as large high-performance {\idmaes} running at over \SI{1}{\giga\hertz} on a \SI{12}{\nano\metre} node. %
Our {\bes} incur only two cycles of latency from accepting an \teb{1-D} transfer descriptor to the first read request being issued on the engine's protocol port. %
They show high agility, even in ultra-deep memory systems. %
Flexibility and parameterization allow us to create configurations that achieve asymptotically full bus utilization and can fully hide latency in arbitrary deep memory systems while incurring less than \SI{400}{\gateeq} per trackable outstanding transfer. %
In a \SI{32}{\bit} system, our {\idmaes} achieve almost perfect bus utilization for \SI{16}{\byte}-long transfers when accessing an endpoint with 100 cycles of latency. %
The synthesizable \gls{rtl} description of {\idma} is available free and open-source. %

\chapter{Architectural DMA Extensions}
\label{chap:dmaext}

\section{Introduction}
\label{chap:dmaext:introduction}

In \Cref{chap:idma}, we discussed the importance of efficiently moving linear streams of data in today's high-performance heterogeneous \glspl{soc}~\cite{ma2019mtdmaadmacontro}, and we proposed {\idma}, a modular and parameterized {\dma} architecture to fill this need.
This chapter presents {\idma} integrated in a wide range of systems and enhanced with multiple extensions, further accelerating \gls{dma} transfers and increasing their efficiency.

\begin{table}
    \centering
    \caption{%
        Identifiers and descriptions of {\fes} employed in the use cases. %
        {\Fes} in \textcolor{ieee-dark-grey-100}{gray} are available but not further discussed in this thesis. %
    }%
    \resizebox{1\columnwidth}{!}{%
    \renewcommand*{\arraystretch}{1.0}
    \begin{tabular}{@{}lll@{}}\toprule
        \textbf{\Fe} &
        \textbf{Description} &
        \textbf{Conf.} \\ \midrule
        \textcolor{ieee-dark-grey-100}{\emph{reg\tus32}} &
        \multirow{5}{*}{%
            \makecell[tl]{%
                Core-private register-based \\
                configuration interface for \gls{ulp}-systems%
            }%
        } &
        \textcolor{ieee-dark-grey-100}{32-\si{\bit}, 1-D} \\
        \textcolor{ieee-dark-grey-100}{\emph{reg\tus32\tus2d}} &                                                              &
        \textcolor{ieee-dark-grey-100}{32-\si{\bit}, 2-D} \\
        \emph{reg\tus32\tus3d} &
        & 
        32-\si{\bit}, 3-D \\
        \textcolor{ieee-dark-grey-100}{\emph{reg\tus64}} &
        &
        \textcolor{ieee-dark-grey-100}{64-\si{\bit}, 1-D} \\
        \textcolor{ieee-dark-grey-100}{\emph{reg\tus64\tus2d}} &
        &
        \textcolor{ieee-dark-grey-100}{64-\si{\bit}, 2-D} \\%
        \arrayrulecolor{ieee-dark-grey-100}\midrule
        \emph{reg\tus32\tus{rt}\tus3d}&
        \makecell[tl]{%
            Core-private register-based system \\
            binding supporting our real-time {\me}%
        } %
        &
        32-\si{\bit}, 3-D \\
        \arrayrulecolor{ieee-dark-grey-100}\midrule
        \emph{desc\tus64} &
        \makecell[tl]{%
            Transfer-descriptor-based interface \\%
            designed for 64-\si{\bit} systems compatible \\%
            with the Linux \gls{dma} interface%
        }&
        64-bit, 1-D \\
        \arrayrulecolor{ieee-dark-grey-100}\midrule
        \emph{inst\tus64} &
        \makecell[tl]{%
            Interface decoding a custom \idma\\%
            \emph{RISC-V} instructions used in \gls{hpc} systems%
        } &
        64-\si{\bit}, 2-D \\
        \arrayrulecolor{black}\bottomrule
    \end{tabular}
    }
    \label{tab:arch:fe}
\end{table}

To enable efficient {\dma} transfers integrated in a system, an efficient and low-latency connection between the \glspl{pe} and the {\dmaes} is of essential importance.
{\dmaes} can be grouped according to their system binding: register, transfer-descriptor, and instruction-based. %
Engines requiring a high degree of agility~\cite{fjeldtvedt2019cubedmaoptimizi, rossi2014ultralowlatency} or featuring a small footprint~\cite{morales2019alowareadirectm, pullini2017textmudmaanauto, 2019antmicrorelease} tend to use a register-based interface. %
\Glspl{pe} write the transfer information in a dedicated register space and use a read or write operation to a special register location to launch the transfer. %
In more memory-compute-decoupled systems~\cite{paraskevas2018virtualizedmult} or manycore environments~\cite{paraskevas2018virtualizedmult, ma2019mtdmaadmacontro} transfer descriptors prevail. %
In some \gls{mcu} platforms~\cite{stmicroelectronicsdmaengineovervi, arm2012corelinkdma330d} {\dmaes} are programmed using a custom instruction stream. %
Generally, {\dmaes} only feature one programming interface with some exceptions: both Xilinx's \emph{AXI DMA v7.1} and Synopsys' \emph{DW\_axi\_dmac} support, next to their primary transfer-descriptors-based interface, a register-based interface usable with only a reduced subset of the engines' features~\cite{xilinx2022axidmav71logico, synopsysdesignwareipsol}. %
This chapter will combine these three system bindings with {\idma}, showing its capabilities as a universal {\dmaa}.
With our standardized interfaces, any custom binding can be implemented, fully tailoring the engine to the system it is attached to.
We provide an overview of the available bindings for {\idma} in \Cref{tab:arch:fe}.

Many {\dmaes} support transfers with more than one addressing dimension. %
\teb{2-D} transfers are commonly accelerated in hardware~\cite{ma2019mtdmaadmacontro,  arm2012corelinkdma330d, xilinx2022axidmav71logico}. %
Fjeldtvedt~{\etal}'s \emph{CubeDMA} can even handle \teb{3-D} transfers. %
Any higher-dimensional transfer is handled in software either by repetitively launching simpler transfers~\cite{fjeldtvedt2019cubedmaoptimizi}~\cite{rossi2014ultralowlatency} or by employing transfer descriptor chaining~\cite{ma2019mtdmaadmacontro, xilinx2022axidmav71logico}. %
In \Cref{chap:dmaext:tensor}, we present our \emph{tensor{\tus}ND} {\me}, which can execute arbitrary high-dimensional transfers in hardware. %
Our \emph{desc\tus64} {\fe}, presented in \teb{\Cref{chap:dmaext:linux}}, supports descriptor chaining to handle arbitrarily shaped transfers without putting any load on the \gls{pe}. %
Our flexible architecture can easily accelerate special transfer patterns required by a specific application: once programmed, we present our novel \emph{rt{\tus}3D} {\me} in \Cref{chap:dmaext:realtime}, which can autonomously fetch strided sensor data without involving any \gls{pe}. %
Rhu~\etal~\cite{rhu2018compressingdmae} and Ma~\etal~\cite{ma2019mtdmaadmacontro} present {\dmaes} able to modify the data while it is copied. %
Although their work proposes a solution for their respective application space, none of these engines present a standardized interface to exchange \emph{stream acceleration modules} between platforms easily. %
Additionally, both engines, especially \emph{MT-DMA}~\alc{\cite{ma2019mtdmaadmacontro}}, impose substantial area overhead, limiting their applicability in \gls{ulp} designs. %
Our engines feature a well-defined interface accessing the byte stream while data is copied, allowing us to \teb{e.g.,} include both buffered and buffer-less matrix transposition units \teb{as well as \glspl{alu}} in {\idma}. %

In heterogeneous systems, physically addressed accelerators are combined with host systems running \glspl{os} \teb{employing} virtual memory~\cite{garofalo2025areliabletimepr, al2025carfieldanopens}.
Thanks to the modularity of {\idma}, we can present a stream-optimized \gls{mmu}, called \emph{sMMU} in \Cref{chap:dmaext:vm}, allowing the accelerator to do the address translation directly within the {\dmae}.

\teb{To summarize this chapter.}
\alc{\Cref{chap:dmaext:linux} introduces a Linux-capable transfer-descriptor-based {\fe} which we combine with an {\idma} {\be}, creating a \glsf{dmac}. %
\Cref{chap:dmaext:vm} introduces virtual memory support. %
In \Cref{chap:dmaext:tensor} we introduce a low-overhead configuration-register-based {\fe} and our \gls{nd} tensor extension accelerating edge-\gls{ml} inference workloads. %
\Cref{chap:dmaext:multichannel} introduces how we coordinate multiple {\idma} {\bes} in a manycore system. %
\Cref{chap:dmaext:iodma} presents a generalized methodology on how to combine peripherals with {\idma}. %
In \Cref{chap:dmaext:realtime} we presents an extension, allowing us to efficiently schedule sensor data accesses.
Finally, \Cref{chap:dmaext:instrction} gives a peek into \Cref{chap:comcpu} and in \Cref{chap:dmaext:conclusion} we conclude the current chapter. %
}

\newpage
\section{Linux Support and Extensions}
\label{chap:dmaext:linux}

\subsection{Introduction}

Modern computing systems are rapidly increasing in complexity and scale to combat the slowdown of Dennard scaling and to satisfy the ever-increasing need for more computing performance and memory, driven by \gls{ml} and big data workloads~\cite{frazelle2021chipmeasuringco}. %
Kumar~\teb{\etal} highlight the importance of irregular memory accesses in sparse data structures when dealing with large-scale graph applications~\cite{kumar2016efficientimplem}. %
Today's systems require high-performance interconnects with components that efficiently move the data required to supply their compute units. %
Using such specialized \alc{\glsplf{dmac}\,\footnote{\alc{We explicitly use DMAC for Linux-capable \gls{dma} units. This includes an {\idma} {\be} and our transfer-descriptor-based {\fe} \emph{(desc\tus64)}.}}} is a well-established method to transfer data independently of processors, thereby promising to achieve high throughput while the processor is free to perform computationally useful work~\cite{su2011aprocessordmaba, ma2009designandimplem, comisky2000ascalablehighpe, chen2010dmaengineforrep, xilinx2022axidmav71logico, synopsysdesignwareipsol}. %
With a shift towards more heterogeneous architectures, as well as smaller datatypes~\cite{frazelle2021chipmeasuringco, jang2022encorecompressi}, more diverse transfers are emerging, requiring greater flexibility and less overhead in the programmability of \glspl{dmac}. %

Highly flexible programming hardware interfaces and \teb{\glspl{hal}} for \glspl{dmac} are usually based on \emph{descriptors}, data structures stored in shared memory that hold the information of a transfer. %
Descriptors have multiple advantages compared to simpler register-based programming interfaces, which are widely used in embedded and \gls{mcu} applications~\cite{rossi2014ultralowlatency}. %
Descriptors massively reduce the requirement for dedicated configuration memory space by storing the transfer specification in general-purpose memory segments, thus eliminating the need for configuration space replication in multicore applications to enforce atomic transfer launching~\cite{rossi2014ultralowlatency}. %
Descriptors can be chained as a linked list, enabling the automatic launch of subsequent transfers. %
This allows \teb{\gls{nd}} affine and fully arbitrary and irregular workloads to be processed~\cite{fjeldtvedt2019cubedmaoptimizi}. %
A concrete example of a \gls{dmac} is the \emph{LogiCORE IP DMA}, an \gls{axi4} \gls{dma} soft \gls{ip} by \emph{Xilinx}~\cite{xilinx2022axidmav71logico}. It is a high-bandwidth \gls{dmac} with a descriptor-based programming interface and is designed to transfer data between a memory-mapped \gls{axi4} interface and an \gls{axi4}-Stream target device. %

Applying this descriptor configuration model to fine-grained, irregular transfers leads to long chains of individual descriptors, requiring the \gls{dmac} unit to handle a large amount of data when executing the transfer. %
Excessive descriptor size degrades the throughput of such fine-grained transfers, as the \gls{dmac} may require multiple cycles to fetch the descriptors. %
Furthermore, larger descriptor sizes result in more significant resource utilization and power overhead required by buffering logic. %
\emph{{Synopsys'}} \emph{DesignWare} \gls{axi4} \gls{dma} controller presents a parametrizable and high-performance \gls{dmac} solution, using 64-byte-long descriptors in \emph{scatter-gather mode}~\cite{synopsysdesignwareipsol}. %
Paraskevas~\teb{\etal} describe an efficient 32-byte-long descriptor format without prefetching capabilities~\cite{paraskevas2018virtualizedmult}. %
In their work, descriptors are stored in dedicated pages of the core's on-chip scratchpad memory. %
Ma~\teb{\etal} describe a five-entry-long descriptor format supporting chaining for \teb{\gls{nd}} data transfers~\cite{ma2019anefficientdire}. %
To ensure efficient fetching of the descriptors, they are stored in a dedicated \gls{dmac}-internal \emph{parameter {RAM}}. %
As descriptors are usually handled in sequence~\cite{arm2023ambaaxiandacepr}, requesting the next descriptor once the prior is read, full \gls{dmac} utilization is only reached if the described transfer is long enough to hide the latency of fetching a descriptor. %
This can no longer be guaranteed for fine-grained transfers in a non-ideal memory system, limiting the maximum achievable \gls{dmac} utilization for such transfers. %
We tackle these issues by introducing a \gls{dmac} with a minimal descriptor format, as well as a low-overhead prefetching mechanism; our contributions are:

\begin{itemize}
    \item %
    A lightweight, minimal, and efficient descriptor format holding only the essential information required to describe a transfer. %
    Our format supports chaining and provides a mechanism to track transfer completion, increasing \gls{dmac} utilization by \SI{3.9}{\x} for {64-\si{\byte}} transfers compared to the \emph{LogiCORE IP DMA}~\cite{xilinx2022axidmav71logico}.
    \item %
    Implementing our descriptor format, as well as speculative descriptor prefetching, together with {\idma} presented in \Cref{chap:idma}, creates a fully parametrizable, synthesizable, and technology-independent \gls{dmac}. %
    \newpage
    \item %
    Evaluating our \gls{dmac} \gls{ooc} regarding its performance, area requirements, and timing in a \SI{12}{\nano\metre} node. %
    \alc{Our \gls{axi4}-compliant} \gls{dmac} can achieve near-ideal performance while exceeding clock frequencies of \SI{1.5}{\giga\hertz} and requiring only \SI{49.5}{\kilo\gateeq}. %
    \item %
    Integrating the resulting \gls{dmac} into {\cheshire}~\cite{ottaviano2023cheshirealightw}, our 64-\si{\bit} Linux-capable \gls{soc} platform, we achieve an improvement in terms of latency by \SI{1.66}{\x} compared to the \emph{LogiCORE IP DMA} \gls{axi4} \gls{dma}~\cite{xilinx2022axidmav71logico}, while requiring \SI{11}{\percent} fewer lookup tables, \SI{23}{\percent} fewer flip-flops, and no block {RAM}s.
\end{itemize}

\subsection{Architecture}

With efficient data transfer being an essential requirement for \gls{ml} workloads, we make use of our {\idma} architecture presented in \Cref{chap:idma}. %
This fully open-source \gls{dma} engine directly interfaces with \teb{our} \teb{AMBA} \gls{axi4} memory system and is capable of asymptotically utilizing the available bandwidth. %
Furthermore, \teb{\idma} is optimized for low area utilization, low transfer launch latencies, and high clock frequencies.
\teb{The {\idma} architecture} does not directly provide a programming interface to the system. %
In the following section, we describe a descriptor-based programming interface implemented as an {\idma} {\fe}, called \emph{desc\_64}. %
The \emph{desc\_64} {\fe} and {\idma} {\be} together form the \gls{dmac}, as shown in \Cref{fig:frontend_overview}. %

\subsection{DMA {\Fe} Design}\label{sec:arch:dma-ctrl}

\begin{figure}[t]
    \centering
    \includegraphics[width=0.85\linewidth]{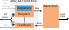}
    \caption{Overview of the \gls{dmac}, containing request logic with internal registers for configuration and read logic to fetch the descriptor, and feedback logic to update the system once the \teb{\idma} {\be} completes the transfer.}
    \label{fig:frontend_overview}
\end{figure}

To configure a \gls{dma} transfer, the {desc\_64} {\fe} exposes a memory-mapped \teb{\gls{casr}}, which accepts an address pointing to a \gls{dma} transfer descriptor in shared memory, described in \Cref{sec:arch:descr}. %
Once this pointer is written into the \teb{\gls{casr}}, the {\fe} requests the descriptor from memory through the read channel of an \gls{axi4} manager port, shown as the request logic in \Cref{fig:frontend_overview}. %
This manager port is configurable in both \gls{axi4} address width, allowing descriptors to be located in any memory location, and \gls{axi4} data width, ranging from \SIrange{16}{512}{bits}. %
Using this port, {desc\_64} retrieves the necessary information for a generic linear memory transfer: source address pointer, destination address pointer, transfer length, and configuration. %
Once fetched, {desc\_64} forwards the information to the {\be}, which executes the transfer. %
To improve performance, both the \teb{\gls{casr}} and the connection to the {\be} implement a queue. %
This allows multiple transfers to be enqueued, maximizing the utilization of the {\be}.

Once the {\be} has completed a transfer, {desc\_64} reports completion back to the system, shown as the feedback logic in \Cref{fig:frontend_overview}. %
For each transfer, the corresponding descriptor is modified to indicate its completion, and an interrupt is signaled if configured. %

\subsection{DMAC Transfer Descriptor}
\label{sec:arch:descr}

Our \gls{dmac} descriptor contains the information necessary to fully describe a linear memory transfer: a 64-\si{\bit} source and destination address, and the length of the transfer. %
The transfer length is stored in an unsigned 32-\si{\bit} field, allowing individual transfers of up to \SI{4}{\gibi\byte} in size. %
Longer transfers can easily be achieved by chaining together multiple descriptors. %

A \emph{config} field in the descriptor holds configuration information for both the \teb{\idma} front- and {\be}. %
For the former, different \gls{irq} options can be set, while for the latter, various \gls{axi4}-related parameters are configurable. %
A complete descriptor structure can be found in \Cref{lst:descriptor}. %
Apart from information describing the transfer, the descriptor contains a pointer to the next descriptor to be processed, enabling \emph{descriptor chaining}. %
This allows our \gls{dmac} to process a \emph{linked list} or \emph{chain of descriptors} in memory without involving the \gls{cpu}. %
The last descriptor in a chain carries all ones (equals to -1) in the \emph{next} field; we call this value \emph{end-of-chain}. This value was chosen as no descriptor can fit at the corresponding address. %
Descriptor chaining allows the construction of arbitrary and irregular transfers from simple linear transfers. %
When designing the descriptor format, we minimized its size while keeping it a multiple of the \gls{axi4} bus width. %
The former has two benefits; it not only reduces the required bandwidth of the memory subsystem when storing and fetching descriptors, but also the overall memory footprint to describe a given transfer. %
The latter allows us to fetch the 256-\si{\bit} descriptors and chains thereof without losing utilization in memory systems with widths up to \SI{256}{\bit}. %
In systems featuring a 512-\si{\bit} infrastructure, such as a wide range of \emph{Xilinx Zynq UltraScale+ MPSoCs}, two full descriptors could be fetched in one cycle. %

\begin{lstlisting}[language=C, caption={Descriptor Layout}, label={lst:descriptor}]
struct descriptor {
  u32 length;
  u32 config;
  u64 next;
  u64 source;
  u64 destination;
}
\end{lstlisting}

To compare, the \emph{LogiCORE IP DMA}~\cite{xilinx2022axidmav71logico} uses a descriptor format of thirteen 32-\si{\bit} words or \SI{416}{\bit}, of which usually only \SI{256}{\bit} are read. %
Its \gls{axi4} manager interface used to fetch descriptors is limited to a data width of \SI{32}{\bit}, leading to a descriptor read latency of at least eight to thirteen cycles. %
In contrast, our \gls{dmac} may read a descriptor in four cycles in a comparable 64-bit system. %

\subsection{Speculative Descriptor Prefetching}\label{sec:arch:prefetch}

To compensate for memory latency, we employ \emph{speculative descriptor prefetching}. %
Once a descriptor address is written to the \teb{\gls{casr}}, we not only request the first descriptor over {desc\_64}'s manager interface but send up to a configurable amount of requests with sequential addresses. %
The number of descriptors speculatively requested is configured using the \emph{prefetching} compile-time parameter, zero deactivating the prefetching logic, as can be seen in \Cref{tab:param}. %
Once a descriptor arrives at the \gls{dma} {\fe}, we compare the \emph{next} field of this descriptor with the speculatively requested address. %
On a match, the speculative address is committed and one speculation slot is freed up. %
\alc{Transfers are only forwarded to the {\be} once the speculatively fetched descriptor is confirmed through a positive match ensuring data integrity.}
Should a misprediction occur, we discard all descriptor addresses in the \emph{speculation slots} and start to fetch from the correct \emph{next} address while ignoring the incoming data that was mispredicted. %
Care was taken not to introduce any latency in the case of mispredictions \alc{compared to prefetching disabled}.
Assuming there is space in the \emph{speculation slots}, the proper request is issued over the \gls{axi4} manager interface in the same cycle the \gls{dma} {\fe} receives the \emph{next} field. %
This is the same latency we observe in the case of prefetching disabled. %
Thus, the only performance degradation that may occur is caused by minimal additional contention in the memory system due to fetching data that is directly discarded. %

\subsection{{\cheshire} integration}\label{sec:arch:soc}

\begin{figure}
    \centering
    \includegraphics[width=0.9\linewidth]{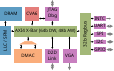}
    \caption{%
        Integration of our \gls{dmac} into the {\cheshire}~\cite{ottaviano2023cheshirealightw}.
        The two manager interfaces, after arbitration,  as well as the subordinate configuration port of the \gls{dmac}, are connected to the \gls{soc}'s interconnect. The \gls{irq} line is connected to the platform's \gls{plic}.
    }
    \label{fig:integration_soc}
\end{figure}

Heterogeneous systems often rely on a high-performance 64-\si{\bit} memory system due to compatibility with the central host processor. %
While parametrizable, our implementation configures the \gls{dmac} for such a memory system, using \SI{64}{\bit} both for address and data width of the \gls{axi4} bus in accordance with {\cheshire}~\cite{ottaviano2023cheshirealightw}, we integrate our \gls{dmac} into. %
An overview of the resulting system can be seen in \Cref{fig:integration_soc}: The two manager interfaces of our \gls{dmac}, as well as the subordinate configuration interface, are connected to the memory system of the \gls{soc}. %
We occupy one new \gls{irq} channel at the systems' \gls{plic}, which is used to signal transfer completion when configured. %
For lightweight in-system progress reporting, we repurpose a transfer descriptor by overwriting its first 8 bytes with \emph{all ones} after the transfer is completed. %
This allows us to forego raising an interrupt after each linear transfer is completed, thus making interrupt notification optional. %

\subsection{Linux Driver}

To ensure simple integration into existing environments, we provide a sample \emph{Linux} driver with an accompanying device-tree file. %
The \gls{dma} subsystem of the Linux kernel exposes a broad \gls{api}~\cite{thelinuxkerneld}, of which we implement the \emph{memcpy} interface. %

For a \gls{dma} client to request a data transfer, the application requests the driver to prepare the \emph{memcpy} transfer. %
This is done by allocating one or more chained descriptors and populating \emph{source}, \emph{destination}, \emph{length}, and \emph{config} fields. %
For more complex transfers, multiple descriptors are allocated and chained. %
Transfer completion notifications are only needed after a (possibly multi-descriptor) transfer is completed. Therefore, \gls{irq} signaling is only enabled in the last descriptor of a transfer.
Should a transfer consist of more than one descriptor, then only the last has \gls{irq} signaling enabled.
As a second step, the client commits to specific transfers, which results in the driver chaining them in a \gls{fifo} fashion to a new chain. %
It is checked if any transfers are already in progress; if so, the driver attempts to commit the transfers to the end of the running chain. Otherwise, they are queued for submission.
Third, the client requests to submit all committed transfers to the hardware. %
The driver checks whether fewer than the maximum number of allowed chains are already running on the \gls{dmac}; if so, it schedules the new chain with a write to the \gls{dmac}'s \teb{\gls{casr}}, otherwise, the transfers are stored to be scheduled later. %
Finally, on transfer completion, the \gls{dmac} raises an \gls{irq}. %
This leads to a call of the \emph{interrupt handler}, which schedules any completion callbacks the client has registered, updates the number of active chains if the transfer was the last of a chain, and schedules stored transfers. %

\begin{table}
    \caption{The compile-time parameters used in our \gls{ooc} experiments.}
    \begin{center}
    \renewcommand{\arraystretch}{1.1} %
    \resizebox{\columnwidth}{!}{
    \begin{tabular}{@{}ccc@{}}
    \hline
    \textbf{Configuration} & Descriptors In-flight & Prefetching \\
    \hline
    \emph{LogiCORE IP DMA}~\cite{xilinx2022axidmav71logico} & 4 & \textit{N.A.} \\
    \emph{base} & 4 & Disabled (0) \\
    \emph{speculation} & 4 & 4 \\
    \emph{scaled} & 24 & 24 \\
    \hline
    \end{tabular}
    }
    \label{tab:param}
    \end{center}
\end{table}

\subsection{Results}
\label{chap:dmaext:linux:results}

\subsubsection{\gls{ooc} Results}
\label{chap:dmaext:linux:results:ooc}

\begin{figure}
    \centering
    \includegraphics[width=0.8\linewidth]{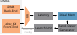}
    \caption{The \gls{ooc} testbench setup; the \gls{dmac} has its two \gls{axi4} manager interfaces connected to a fair round-robin arbiter (RR), which in turn is connected to a latency-configurable memory system. %
    \alc{Contention between \gls{dma} data transfers and reading descriptors is thus modeled.} %
    Descriptors are loaded into the memory using backdoor access and are launched via the \gls{dmac}'s subordinate configuration interface.}
    \label{fig:ooc-testbench}
\end{figure}

We evaluate our \gls{dmac} with our optimized descriptor format \gls{ooc} and integrated into {\cheshire}. %
For the \gls{ooc} evaluation, we attach our \gls{dmac} to a configurable memory system to assess its performance, as can be seen in \Cref{fig:ooc-testbench}. %
We then present area and timing results from synthesizing our controller out-of-context using {\gfs} {\gftech} node. %
We then show both performance and implementation results of our \gls{dmac} integrated into {\cheshire}~\cite{ottaviano2023cheshirealightw} implemented on a \emph{Diligent Genesys 2} \gls{fpga}~\cite{genesys2referen}. %

To evaluate the standalone performance of our \gls{dmac}, we created a testbench environment consisting of a \emph{latency-configurable} memory system and a \emph{launch unit} to set up and execute random streams of descriptors. %
To simulate a real system, both of our \gls{dmac}'s \gls{axi4} manager ports are connected to the same memory system using a fair round-robin arbiter ({RR}), as shown in \Cref{fig:ooc-testbench}. %
To stay aligned with {\cheshire}, we set the address and data width of our \gls{ooc} testbench to \SI{64}{\bit}. %
The randomness of the descriptions can be closely controlled, allowing us to emulate different transfer characteristics. %
The corresponding descriptors are immediately preloaded into our simulation memory using a backdoor, while the actual launch of the transfers is controlled using our \gls{dmac}'s \teb{\gls{casr}} interface. %
The bus utilization is measured at the {\be}'s \gls{axi4} \emph{manager} interface; only \emph{useful} payload traffic contributes to utilization. %
We only report \emph{steady state} bus utilization, suppressing any possible cold-start phenomena. %

In our analysis, we assessed three distinct memory system configurations reflecting different use cases:
\begin{enumerate}
    \item \emph{Ideal Memory:} We configure our simulation memory to have one cycle latency, emulating an SRAM-based main memory. %
    \item \emph{DDR3 Main Memory:} Replicating the conditions found on the \emph{Diligent Genesys 2} \gls{fpga}~\cite{genesys2referen} when accessing DDR3 off-chip memory, we include a configuration with \emph{thirteen} cycles latency. %
    \item \emph{Ultra-deep Memory:} Representing a large \gls{noc} system found in a modern \gls{soc}, we include a configuration with a latency of \emph{one hundred} cycles. %
\end{enumerate}

To ensure a fair comparison against the \emph{LogiCORE IP DMA}, we include a \emph{base} configuration closely matching the \emph{LogiCORE IP DMA}'s default configuration. %
In our evaluation, we included two additional configurations; one enabling \emph{speculation} while closely resembling the \emph{base} configuration, and a \emph{scaled} configuration setting both the number of \emph{descriptors} \alc{outstanding} and the \emph{prefetching} to \emph{24}. %
We summarize the respective parameter configurations in \Cref{tab:param}. %

\begin{table}
    \caption{Area requirements at the maximum clock frequency of the \gls{dmac} and its main sub-components; {desc\_64} and the {\idma} {\be}. Clock frequencies are achieved in typical conditions.}
    \begin{center}
    \renewcommand{\arraystretch}{1.1} %
    \resizebox{\linewidth}{!}{
    \begin{tabular}{@{}cccc|c@{}}
    \hline
    \textbf{Config.} & \textbf{{desc\_64} {\fe}} & \textbf{{\Be}} & \textbf{\gls{dmac}} & \textbf{Achievable Freq.} \\
    \hline
    \emph{base}        & \SI{25.8}{\kilo\gateeq}  & \SI{15.4}{\kilo\gateeq} & \SI{41.2}{\kilo\gateeq}  & \SI{1.71}{\giga\hertz} \\
    \emph{speculation} & \SI{34.8}{\kilo\gateeq}  & \SI{14.7}{\kilo\gateeq} & \SI{49.5}{\kilo\gateeq}  & \SI{1.44}{\giga\hertz} \\
    \emph{scaled}      & \SI{151.1}{\kilo\gateeq} & \SI{37.3}{\kilo\gateeq} & \SI{188.4}{\kilo\gateeq} & \SI{1.23}{\giga\hertz} \\
    \hline
    \end{tabular}
    }
    \label{tab:area_ooc}
    \end{center}
\end{table}

\begin{figure}[p]
    \begin{subcaptionblock}{\linewidth}
        \centering%
        \includegraphics[width=\linewidth]{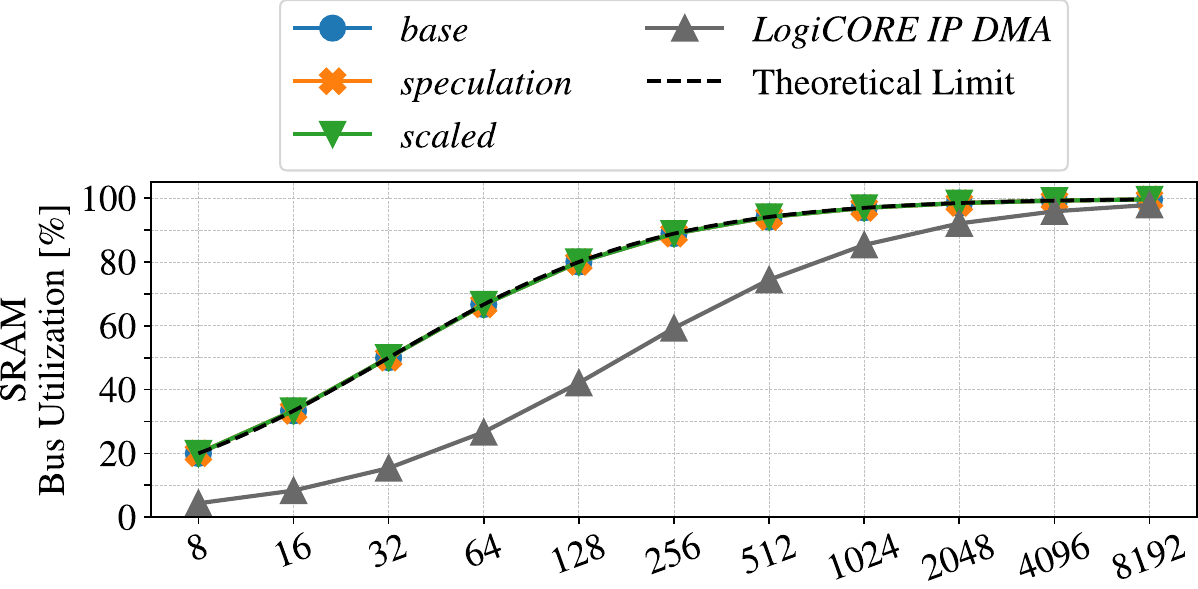}%
        \vspace{0.05cm}
        \label{fig:ooc:0}
    \end{subcaptionblock}
    \centering
    \begin{subcaptionblock}{\linewidth}%
        \centering%
        \includegraphics[width=\linewidth]{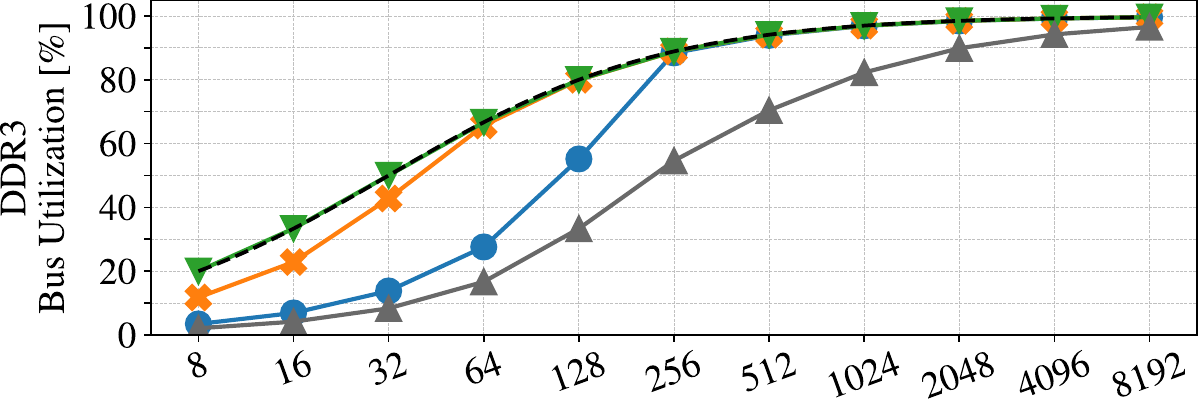}%
        \vspace{0.05cm}
        \label{fig:ooc:13}
    \end{subcaptionblock}
    \begin{subcaptionblock}{\linewidth}
        \centering%
        \includegraphics[width=\linewidth]{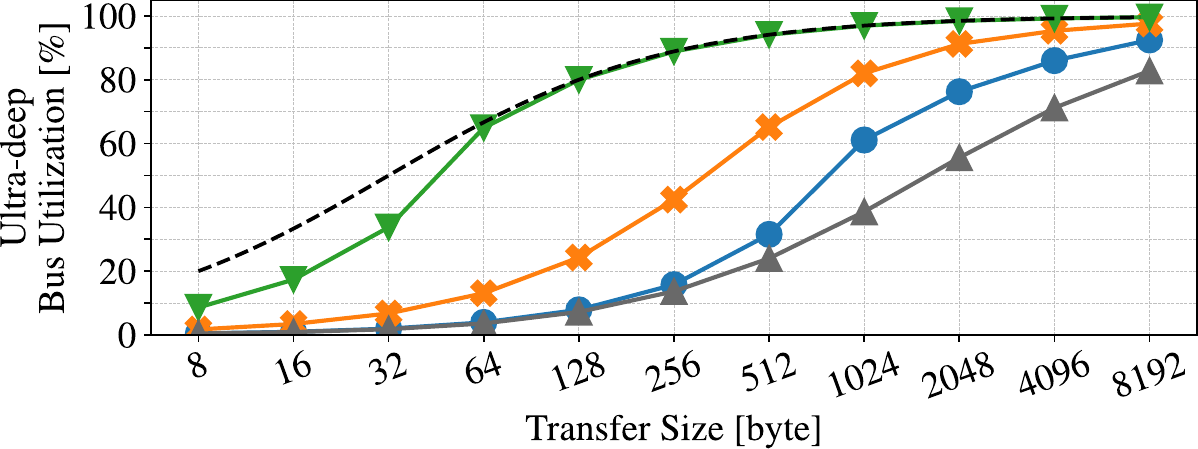}%
        \vspace{0.05cm}
        \label{fig:ooc:100}
    \end{subcaptionblock}\hfill
    \vspace{-0.5cm}
    \caption{
        \Gls{dmac} \emph{steady-state} bus utilization given a prefetch hit rate of \SI{100}{\percent} in memory systems featuring various latencies.
        \alc{\emph{Ideal}, \emph{DDR3}, \emph{Ultra-deep} memory systems featuring \emph{1}, \emph{13}, \emph{100} cycles of latency, respectively.}
    }
    \label{fig:ooc}
\end{figure}

\begin{figure}
    \centering
    \includegraphics[width=0.95\linewidth]{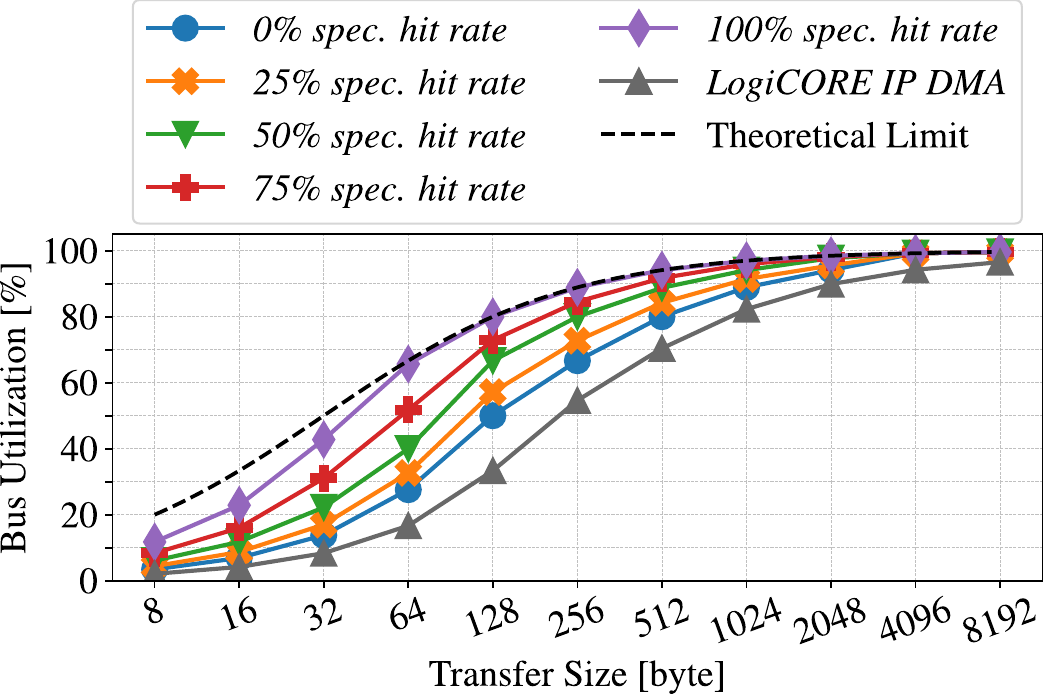}
    \caption{\Gls{dmac} steady-state bus utilization in the case of the DDR3 main memory with speculation misses; \emph{speculation} configuration.}
    \label{fig:cache_misses}
\end{figure}

As access to main memory is shared between the \teb{\idma} \teb{front-} and \teb{\be}, the bus utilization, as defined above, cannot reach \SI{100}{\percent} %
The transfer of the payload will be interrupted by descriptor transfers, limiting the \emph{ideal bus utilization}, $\bar{u}$ -- see \Cref{eq:ideal-util} -- where $n$ is the transfer size in byte. %

\begin{equation}\label{eq:ideal-util}
\bar{u} = \frac{n}{n + 32}
\end{equation}

Descriptor misprediction, in the case of speculative prefetching enabled, further limits the ideal utilization, as it inflates the number of additional bytes fetched by the \teb{\idma} \teb{\fe} per transfer. %
In this case, the \emph{ideal bus utilization} $\bar{u}$ is given by:

\begin{equation}\label{eq:ideal-util-pred}
\bar{u} = \frac{n}{n + 32(1 + m)}
\end{equation}

In \Cref{eq:ideal-util-pred} $n$ denotes the transfer size and $m$ the average number of wrongfully fetched descriptors for each transfer. %
In very shallow or ideal memory systems, our \emph{base} configuration already achieves ideal steady-state utilization for any bus-aligned transfer size, as shown in~\Cref{fig:ooc:0}. %
At transfer sizes of \SI{64}{\byte} -- a typical cache line size in many memory architectures -- we improve the utilization by \SI{2.5}{\x} compared to the \emph{LogiCORE IP DMA}. %
When using the \emph{Genesys 2 DDR3} latency configuration, we achieve ideal steady-state utilization at \SI{256}{\byte} without and \SI{64}{\byte} with prefetching enabled, as can be seen in~\Cref{fig:ooc:13}. %
This increases the utilization by up to \SI{1.7}{\x} and \SI{3.9}{\x}\!, respectively, compared to the \emph{LogiCORE IP DMA}. %
Finally, we show that our \gls{dmac} can be configured to still achieve near-ideal steady state utilization even in ultra-deep memory systems. As can be seen in \Cref{fig:ooc:100}, the \emph{scaled} configuration achieves ideal utilization starting from \SI{128}{\byte}.
Varying prefetching hit rates of \SIrange{75}{0}{\percent}, our achievable increase in bus utilization compared to the \emph{LogiCORE IP DMA} still ranges from \SI{1.65}{\x} to \SI{3.1}{\x} at \SI{64}{\byte}, see \Cref{fig:cache_misses}. %

We evaluate the timing and resource requirements of our \gls{dmac} in the various configurations presented in \Cref{tab:param} by synthesizing our work in \emph{{GlobalFoundies'} {GF12LP+} FinFET} technology using \emph{Synopsys' Design Compiler NXT} in \emph{topological} mode. %
All results are presented in the typical corner of the library at \SI{25}{\celsius} at \SI{0.8}{\volt}, in \Cref{tab:area_ooc}. %
Our \emph{base} configuration requires an area of \SI{41.2}{\kilo\gateeq}, achieving a maximum clock frequency of \SI{1.71}{\giga\hertz}. %
Enabling prefetching adds \SI{8.3}{\kilo\gateeq} while reducing the achievable maximum clock frequency to \SI{1.44}{\giga\hertz}. %
We synthesized our design in numerous configurations, creating a model of the circuit area as a function of the parameters in \Cref{tab:param}. %
The design's area in \emph{\si{\kilo\gateeq}} is described by: $A = 20.30 + 5.28 \* d + 1.94 \* s$, where $d$ denotes the number of descriptors in flight and $s$ the number of speculatively launched descriptors. %
The total area is linear in $d$ and $s$, allowing the hardware to be easily scaled to larger sizes. %
The \emph{scaled} configuration requires a total of \SI{188.4}{\kilo\gateeq} achieving \SI{1.23}{\giga\hertz}. %
Comparing these numbers to \teb{CVA6~\cite{ottaviano2023cheshirealightw}}, we find the \gls{dmac} area to be less than \SI{10}{\percent} of the core's area while achieving similar clock speeds, confirming the scalability of our controller.

\begin{table}
    \caption{\gls{fpga} resource requirements of the \gls{dmac} at 200~MHz.}
    \vspace{-3.5mm}
    \begin{center}
    \renewcommand{\arraystretch}{1.1} %
    \resizebox{0.58\linewidth}{!}{
    \begin{tabular}{@{}ccc@{}}
    \hline
    \textbf{Configuration} & \textbf{LUTs} & \textbf{FFs} \\
    \hline
    \emph{base}          & 2610 & 3090 \\
    \emph{speculation}   & 2480 & 3935 \\
    \emph{scaled}        & 6764 & 11353 \\
    \emph{LogiCORE IP DMA}~\cite{xilinx2022axidmav71logico} & \emph{2784} & \emph{5133} \\
    \hline
    \end{tabular}
    \vspace{-5mm}
    }
    \label{tab:util_soc_fpga}
    \end{center}
\end{table}

\subsubsection{In-System Results}
\label{sec:soc-results}

To evaluate the required resources on a \gls{fpga}, we synthesized {\cheshire} with the various configurations of our \gls{dmac} integrated. %
Synthesis was done using \emph{Vivado 2019.2} targeting the \emph{Genesys 2} board, which features a \emph{Kintex 7} \gls{fpga} from \emph{Xilinx}. %
In the \emph{base} configuration, the footprint of the \gls{dmac} is \teb{\SI{2610}{\lut}\,\footnote{\teb{\emph{\si{\lut}}: lookup table}}} and \teb{\SI{3090}{\ff}\,\footnote{\teb{\emph{\si{\ff}}: flip-flop}}}, while the entire \gls{soc} occupies \teb{\SI{79142}{\lut}} and \teb{\SI{58086}{\ff}}, see \Cref{tab:util_soc_fpga}. %
This puts the base configuration at \SI{3.3}{\percent} of total \teb{\si{\lut}} usage and \SI{5.3}{\percent} of total \teb{\si{\ff}} usage, and is a reduction of \SI{6.25}{\percent} \teb{\si{\lut}} and \SI{39.8}{\percent} \teb{\si{\ff}} utilization compared to the \emph{LogiCORE IP DMA}. %
Compared to the \emph{base} configuration, the \emph{speculation} configuration uses \SI{27}{\percent} more \teb{\si{\ff}}, but reduces the number of \teb{\si{\lut}} by \SI{5}{\percent}. %
The scaled configuration increases resource utilization further, requiring \SI{2.59}{\x} as many \teb{\si{\lut}} and \SI{3.67}{\x} as many \teb{\si{\ff}} as the \emph{base} configuration. %

\begin{table}
    \caption{\gls{dmac} latencies between various events and memory systems for the \emph{scaled} configuration}
    \vspace{-2mm}
    \begin{center}
    \renewcommand{\arraystretch}{1.1} %
    \resizebox{0.9\linewidth}{!}{
    \begin{tabular}{@{}lccccc@{}}
        \hline
        \textbf{Metric} && \emph{LogiCORE IP \gls{dma}}~\cite{xilinx2022axidmav71logico} & \emph{scaled} \\
        \hline
        \texttt{i-rf} && 10  & 3 \\
        \texttt{rf-rb} &
        1 cycle latency & 22 & 8 \\
        &13 cycles latency & 48 & 32 \\
        &100 cycles latency & 222 & 206 \\
        \texttt{r-w}  && 1 & 1  \\
        \hline
    \end{tabular}
    }
    \label{tab:latencies}
    \end{center}
\end{table}

We use our latency-configurable memory system presented in \Cref{chap:dmaext:linux:results:ooc}, which we integrate into the upstream CVA6-\gls{soc} to measure the following three different latencies:
\begin{itemize}
    \item \emph{i-rf:} the \gls{cpu} issuing a write to the \teb{\idma} \teb{\fe} issuing a read request %
    \item \emph{rf-rb:} between the issue of the read request from \teb{\idma} \teb{front-} and \teb{\be} %
    \item \emph{r-w:} the latency between \teb{our \gls{dmac}} reading and writing the same data %
\end{itemize}

As can be seen in \Cref{tab:latencies}, we achieve three cycles of latency for \emph{i-rf}, an improvement of \SI{3.33}{\x} over the \emph{LogiCORE IP DMA}. %
For \emph{rf-rb}, we achieve a latency of eight cycles in ideal memory, 32 cycles with a memory latency of 13, and 206 cycles in the case of 100 cycles of latency. %
This results in an improvement of \SI{2.75}{\x}\!, \SI{1.5}{\x}\!, and \SI{1.08}{\x}\!, respectively. %
Latencies for \emph{r-w} are equal at one cycle for both our \gls{dmac} and the \emph{LogiCORE IP DMA}. %

\newpage
\section{Virtual Memory Capabilities}
\label{chap:dmaext:vm}

\subsection{Introduction}

With the advent of multi-user- and multi-application-capable computing systems in the second half of the last century, the necessity of an elegant and efficient way to implement address space isolation, program modularity, resource sharing, and machine independence has become an undeniable truth~\cite{denning1970virtualmemory}.
\Glsf{vm} has established itself as the solution to achieve these objectives by distinguishing the \emph{virtual} address regions an application uses to refer to its information and the \emph{physical} addresses used by the memory system to identify hardware storage devices.

With all the benefits introduced, \gls{vm} also incurs challenges to be solved when designing a system's memory architecture.
This is especially true in heterogeneous systems, where \gls{vm}-enabled hosts are paired with physically addressed accelerator subsystems~\cite{rodriguez2023opensourceriscv, koenig2025evaluatingiommu, garofalo2025areliabletimepr, kurth2017heroheterogeneo, kurth2022herov2fullstack}.
If designed \teb{improperly}, the required translation layer between the accelerator and the host\teb{,} leads to a substantial loss in performance transferring data between the regions~\cite{kurth2018scalableandeffi}.

Traditional approaches handle physically addressed accelerators coupled to \gls{vm}-enabled application-class hosts either \teb{by placing} an \gls{iotlb}~\cite{kurth2017heroheterogeneo, kurth2018scalableandeffi} or a full \gls{iommu}~\cite{rodriguez2023opensourceriscv, koenig2025evaluatingiommu} between the domains.

Having a dedicated unit handling the translation process higher up in the memory hierarchy leads to both a programming overhead and a slow reaction to unexpected situations.
Setting up a dedicated \gls{iotlb} or \gls{iommu} requires configuration setup through dedicated memory-mapped registers.
Depending on the exact implementation, periodic reconfiguration is required to ensure all of the accelerator's memory accesses are mapped and can be properly translated.
In the case of a non-mapped access, \teb{a} \emph{page table fault}, either an interrupt is raised, requiring around one hundred cycles of latency to be communicated to the application-class core~\cite{zelioli2024vclictowardsfas}, or an error response is propagated through the interconnect back to the initiator device, where it is then handled (e.g. \Cref{chap:idma:architecture:errorhandling}).
Both communication methods\teb{,} in the case of these loosely coupled translation devices\teb{,} introduce a substantial latency to even communicate the fault to the core in charge of defining the translation table.

Currently used industry-grade on-chip interconnects, see \Cref{chap:introduction:protocols}, do not feature a mechanism for communicating streaming intent, and thus specifying its access patterns.
An upstream translation unit thus can only reactively handle memory transfers~\cite{koenig2025evaluatingiommu}.
Ahead-of-arrival translation can thus be only accomplished by employing predictive schemes identifying access patterns and preparing the required \glspl{pte} to be present in the unit once the memory access arrives.
With {\axipack}~\cite{zhang2023axipacknearmemo}, we tackled the challenge of transferring a streaming intent over \gls{axi4} interconnects by introducing additional signals.
Whilst communicating the nature and shape of the stream, the latency between the stream initiator and the \gls{mmu} still prevails, limiting the agility of such an approach.

We thus propose our \glsf{smmu}; specifically designed to translate memory streams at high efficiency.
Placed close to the stream initiator, e.g., an {\idmae}, \gls{smmu} can use the stream source to perfectly fetch page table \glspl{pte} into its internal \gls{tlb} unit, hiding translation latency and achieving a high bus utilization, achieving over \SI{95}{\percent} of the ideal bandwidth at a transaction granularity of \SI{256}{\byte}.

\subsection{Architecture}

\begin{figure}[t]
    \centering%
    \includegraphics[width=1\textwidth]{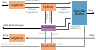}%
    \caption{%
        Architecture of our \gls{smmu} implemented as a {\me}.
        It is placed between a 1-D {\fe} and an {\idma} {\be}.
    }%
    \label{fig:arch:smmu}%
\end{figure}

We present an overview of the architecture in \Cref{fig:arch:smmu}.
The stream request enters the \gls{smmu}, where it is legalized into 4-\si{\kilo\byte}-sized and page-aligned fragments.
The fragments are then enumerated and sent to the translation unit, consisting of the \gls{tlb} and the \gls{ptw}.
Fragments hitting in the \gls{tlb} will directly be translated in-cycle, whereas misses are handed to the \gls{ptw} to be fetched from the page table.
The translation latency of the \gls{ptw} depends on the system's memory system and the number of levels on non-leaf \glspl{pte} stored in the \gls{tlb}.
To increase throughput, the individual 4-\si{\kilo\byte}-sized fragments get translated and forwarded \gls{ooo}.

\Gls{ooo}-completion of individual fragments of a memory stream is not an issue, as most \gls{api} abstractions, e.g., \emph{memcpy}, only operate on the granularity of an individual transfer.
To keep consistency, we just need to ensure that a transfer response, marking the transfer as completed, is only issued once all fragments of a transfer are completed, and that fragments of different transfers are not mixed.
A small \gls{rob} is used to ensure this consistency.

Our \gls{smmu} checks \alc{ the \gls{tlb}} access permissions and prevents the corresponding transfer fragments from being forwarded should an access violation be detected.
\alc{Should a transfer violate access permissions, a fault is issued.}
In this case, the \gls{smmu} waits for the outstanding (and valid) fragments of the transaction to complete and then flushes all metadata information of the transfer that caused the violation.
The violating transfer is then terminated and an error response is issued.
A similar abort mechanism is employed should a page table fault be detected.
In this case, one or multiple fragments of the transfer are not properly mapped in the translation table, making them impossible to complete.
Thanks to the tight coupling between \teb{our} \gls{smmu} and the device initializing the stream, these error responses can be forwarded to the issuing manager within a few cycles, rendering this approach very agile.

\begin{figure}[t]
    \centering%
    \includegraphics[width=0.8\textwidth]{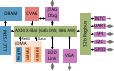}%
    \caption{%
        Our \gls{smmu} integrated into {\cheshire}'s \gls{soc}-level {\idmae}.
    }%
    \label{fig:arch:smmu_integrated}%
\end{figure}

\subsection{Case Study: {\cheshire}}

We integrated our \gls{smmu} into {\cheshire}'s \gls{soc}-level {\idmae}, see \Cref{fig:arch:smmu_integrated} and implemented the system on a \emph{Diligent Genesys 2} \gls{fpga}~\cite{genesys2referen}.
Our \gls{smmu} is implemented as an {\idma} {\me} translating both source and destination addresses using \emph{SV48}~\cite{riscv2025theriscvinstruc}.
The \gls{ptw} accesses {\cheshire}'s memory system through an additional \gls{axi4} manager port, giving the system's \gls{dma} now one subordinate port for configuration and two manager ports for fetching \glspl{pte} and for transferring the data.

Implemented on the \emph{Diligent Genesys 2}, \gls{smmu} requires \SI{8.1}{\kilo\ff} and \SI{9.0}{\kilo\lut} whilst not impacting the critical path of the \gls{soc}.
Copying transfers of varying length, \gls{smmu} can achieve in excess of \SI{99}{\percent} of the non-translated performance when issuing 4-\si{\kilo\byte}-sized transfers.
At at transfer size of \SI{256}{\byte} and \SI{64}{\byte}, a performance of over \SI{95}{\percent} and over \SI{80}{\percent}, respectively can be achieved allowing \gls{smmu} to achieve good performance even for fine-granular transfer accesses.

\newpage
\section[Tensor Extension]{Register {\Fe} and N-D Tensor Extension}
\label{chap:dmaext:tensor}

\subsection{Introduction}

Operations on matrices or \teb{\gls{nd}} tensors are the fundamentals of scientific computing~\cite {golub2014scientificcompu} and machine learning~\cite{golub2013matrixcomputati}.
\Gls{blas}~\cite{gates2024blasquickrefere} implements the fundamental set of both matrix and vector operations and is one of the most optimized and widespread libraries to date.
\Gls{blas}-based benchmark suits, \emph{LINPACK}/\emph{LAPACK}, are an integral part of the \emph{Top500} workload sets.
With such a widespread use of matrix operations, many \gls{dma} engines feature support for \teb{2-D} data transfers~\cite{fjeldtvedt2019cubedmaoptimizi, rossi2014ultralowlatency}.
However, \gls{soa} units only implement two or three dimensions.
Higher-level transfers are either implemented via descriptor chaining (\Cref{sec:arch:descr}) or by programmatically launching multiple lower-dimensional transfers.

\subsection{Architecture}

\paragraph*{Register-Based {\Fe}}

\begin{figure}[t]
    \centering%
    \includegraphics[width=\textwidth]{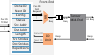}%
    \caption{%
        Architecture of the register-based {\fe} and its connection to our tensor {\me}. %
        Each \gls{pe} has its private register file containing the transfer shape.
    }%
    \label{fig:arch:reg-frontend}%
\end{figure}

Core-private register-based configuration interfaces are the simplest class of {\fes}, their core architecture is shown in \Cref{fig:arch:reg-frontend}. %
Each \gls{pe} uses its own dedicated configuration space to eliminate race conditions and ensure transfer-atomicity while programming the attached {\dmaes}~\cite{rossi2014ultralowlatency}.
We employ different memory-mapped register layouts depending on the host system's word width and whether our \emph{\teb{\gls{nd}} tensor \me} is present. %
The \emph{src\_address}, \emph{dst\_address}, \emph{transfer\_length}, \emph{status}, \emph{configuration}, and \emph{transfer\_id} registers are shared between all variants. %
In the case of a \teb{\gls{nd}} configuration, every tensor dimension introduces three additional fields: \emph{src\_stride}, \emph{dst\_stride}, and \emph{num\_repetitions}. %
After configuring the shape of a transfer, it is launched by reading from  \emph{transfer\_id}, which returns an incrementing unique \gls{jid}. %
The last completed \gls{jid} may be read from the \emph{status} register, enabling agile transfer-level synchronization. %

\paragraph*{N-D Tensor {\Me}}

\begin{figure}[t]
    \centering%
    \includegraphics[width=\textwidth]{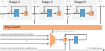}%
    \caption{%
        Architecture of our tensor {\me}.
        Each stage is activated by the previous; the strides of the highest stage are added to the base address.
    }%
    \label{fig:arch:nd-midend}%
\end{figure}

The tensor {\me} consists of a compile-time-configurable number of stage counters, which handle the number of times each dimension is repeated.
On transfer launch, each stage is loaded with its corresponding number of repetitions (\emph{num\_repetitions}).
Once the lower-dimensional transfer completes, this could be the linear transfer executed by the backend, or when the lower stage counter arrives at zero, the stage counter is decremented.
While the transfer is still active, once a stage counter is at zero, it is reloaded with its respective \emph{num\_repetitions} register value.
The \teb{\gls{nd}} transfer is completed once the highest-stage counter is at zero.

Once stage counters expire, their corresponding stride has to be added to the transfer's base address.
Should more than one stage conclude in one cycle, only he highest stage's stride is added, simplifying the hardware substantially by ensuring each cycle at most one stride is added to the base address.
As a direct consequence of this hardware simplification, the first two dimensions of transfers feature inclusive, whereas higher dimensions feature exclusive strides.
This mix of stride definitions does not complicate software and thus is a valid hardware optimization.

\subsection{Case Study: \teb{\pulpopen}}
\label{chap:dmaext:tensor:pulpopen}

{\pulpopen} is a \gls{ulp} edge compute platform consisting of a 32-\si{\bit} {\riscv} microcontroller host and a parallel compute cluster~\cite{pullini2019mrwolfanenergyp}. %
The compute cluster comprises eight 32-\si{\bit} {\riscv} cores with custom \gls{isa} extensions to accelerate \gls{dsp} and \glsu{ml} workloads, enabling energy-efficient \gls{ml} inference in extreme-edge AI nodes. %
These cores are connected to an \gls{sram}-based \gls{tcdm} with single-cycle access latency, providing the processing cores with fast access to shared data. %
While the \gls{tcdm} is fast, it is very limited in size;
the platform thus features a \gls{l2} on-chip and \gls{l3} off-chip HyperBus RAM~\cite{semiconductor2019hyperbusspec}. %
To allow the cluster fast access to these larger memories, a {\dma} unit is embedded, specialized for transferring data from and to the \gls{l1} memory. %

\paragraph*{\textbf{\teb{\iDmaE} Integration}}

\begin{figure}
    \centering%
    \begin{subcaptionblock}{\linewidth}%
        \centering%
        \includegraphics[width=\linewidth]{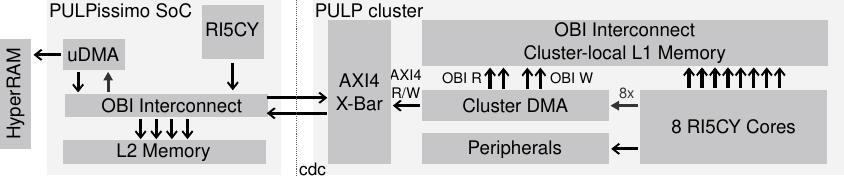}%
        \caption{System architecture.}%
        \vspace{3mm}
        \label{fig:case:pulp:bd}%

        \centering%
        \includegraphics[width=\linewidth]{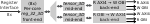}%
        \caption{{\Fe}, {\Me}, and {\Be} configuration.}%
        \vspace{3mm}
        \label{fig:case:pulp:dma}%
    \end{subcaptionblock}
    \caption{%
        (a) Block diagram of the {\pulpopen} system. %
        (b) Configuration of the cluster {\idmae}. %
    }%
    \label{fig:case:pulp:arch}%
\end{figure}

In the {\pulpopen} system, our {\idmae} is integrated into the processing cluster with a 64-\si{\bit} \emph{AXI4} interface to the host platform and an \emph{OBI} connection to the \gls{tcdm}, see \Cref{fig:case:pulp:arch}. %
The multi-protocol {\be} is fed by a \emph{tensor{\tus}ND} {\me}, configured to support three dimensions, allowing for fast transfer of 3-D data structures common in \gls{ml} workloads. %
At the same time, higher-dimensional transfers are handled in software. %
The back- and {\me} are configured through per-core \emph{reg\_32\_3d} {\fes} and two additional {\fes}, allowing the host processor to configure the {\idmae}. %
Round-robin arbitration is implemented through a round-robin arbitration {\me} connecting the {\fes} to the \emph{tensor{\tus}ND} {\me}.%
Multiple per-core {\fes} ensure atomic {\dma} access and prevent interference between the cores launching transactions.

\paragraph*{\textbf{Benchmarks}}

To evaluate {\idmae} performance in a realistic application, we use Dory~\cite{burrello2021doryautomaticen} to implement {MobileNetV1} inference on {\pulpopen}. %
This workload relies heavily on the {\idmae} to transfer the data for each layer stored in \gls{l2} or off-chip in \gls{l3} into the cluster's \gls{tcdm} in parallel with cluster core computation. %
\teb{2-D}, \teb{3-D}, and very small transfers are frequently required for this workload. %

In previous versions of {\pulpopen}, MCHAN~\cite{rossi2014ultralowlatency} was used to transfer data between the host \gls{l2} and the cluster's \gls{tcdm}. We assume this as a baseline for our evaluation. %

\paragraph*{\textbf{Results}}

In {\pulpopen}, {\idmae} can almost fully utilize the bandwidth to the \gls{l2} and \gls{tcdm} in both directions: %
measuring with the on-board timer, a transfer of \SI{8}{\kibi\byte} from the cluster's \gls{tcdm} to \gls{l2} requires \emph{1107} cycles, of which \emph{1024} cycles are required to transfer the data using a 64-\si{\bit} data bus. The minimal overhead is caused by configuration, system latency, and contention with other ongoing memory accesses. %
During {MobileNetV1} inference, individual cores frequently require short transfers, incurring a potentially high configuration overhead. %
With its improved \emph{tensor\_3D} {\me}, {\idma} improves the cores' utilization and throughput for the network over MCHAN, achieving an average of \SI{8.3}{MAC\per cycle} compared to the previously measured \SI{7.9}{MAC\per cycle}. %
Furthermore, configured with similar queue depths as MCHAN, {\idmae} with its \emph{reg\_32\_3d} achieves a \SI{10}{\percent} reduction in the utilized area within a \teb{\pulp} cluster. %

\newpage
\section{Multichannel Operation}
\label{chap:dmaext:multichannel}

\subsection{Introduction}

As mentioned in \Cref{sec:idma:architecture:backend}, we design our {\idma} {\be} to handle one data stream as efficiently as possible.
Should an architecture be capable of handling multiple data streams at once, either because it features multiple compute units working on different threads or data sets~\cite{riedel2023mempoolascalabl, scheffler2025occamya432cored}, a compute unit accesses multiple \gls{numa} endpoints, or a mixture of both.

Multiple channels, or {\idma} {\bes}, can easily be integrated into a system's memory architecture, as each {\be} features a doubly handshaked request and response port.
The true challenge arises in coordinating the different channels, minimizing interference of the channels in the process, and providing a simple programming abstraction for the user to launch and synchronize on transfers.

To facilitate transfer scheduling in a single-cluster manycore architecture, we device two {\mes} dividing and distributing transfers over multiple channels or {\bes}, efficiently transferring data from/into physically separated cluster-local memory endpoints, while giving the user a single-\gls{dma} memory abstraction supporting a regular \emph{memcpy} abstraction.

\subsection{Architecture}

\begin{figure}
    \centering
    \begin{subcaptionblock}{0.45\linewidth}%
        \centering%
        \includegraphics[width=\textwidth]{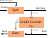}%
        \vspace{0em}%
        \caption{%
            \emph{mp\_split}
        }%
        \label{fig:arch:mp-midend-split}%
    \end{subcaptionblock}
    \hspace{0.5cm}
    \begin{subcaptionblock}{0.45\linewidth}%
        \centering%
        \includegraphics[width=\textwidth]{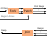}%
        \vspace{0em}%
        \caption{%
            \emph{mp\_dist}
        }%
        \label{fig:arch:mp-midend-dist}%
    \end{subcaptionblock}
    \caption{%
        The architecture of our multichannel \emph{\mempool} (\emph{mp}) {\mes}.
    }%
    \label{fig:arch:mp-midends}%
\end{figure}

To distribute work among {\bes}, we create two specialized {\mes} called \emph{mp{\tus}split} and \emph{mp{\tus}dist}. %
Our \emph{mp{\tus}split}, seen in \Cref{fig:arch:mp-midends}a, splits a single linear transfer into multiple transfers aligned to a parametric address boundary, guaranteeing that no resulting transfer crosses specific address boundaries. %
This is required when sending distributed transfers to multiple {\bes}. %
When accepting a transfer, the length of the transfer is checked against a maximum support transfer length.
Larger transfers will be split and emitted as multiple shorter transfers.
Responses are coalesced to keep this action transparent to the {\fe}.

\emph{Mp{\tus}dist}, presented in \Cref{fig:arch:mp-midends}b, then distributes the split transfers over $N$ parallel downstream mid- or {\bes}.
Transfers are distributed with their addresses modified, so that the $N$ parallel {\bes} each complete a non-overlapping interleaved block.
Responses are collected, and the transfer will be marked as completed if all {\bes} working on an interleaved block range are completed.

\subsection{Case Study: Mempool}

\mempool{}~\cite{riedel2023mempoolascalabl} is a flexible and scalable single-cluster manycore architecture featuring 256 32-\si{\bit} {\riscv} cores that share \SI{1}{\mebi\byte} of low-latency \gls{l1} \gls{spm} distributed over 1024 banks. %
All cores are individually programmable, making \mempool{} well-suited for massively parallel regular workloads like \alc{computer vision, deep learning, and machine learning} and irregular workloads like graph processing. %
The large shared \gls{l1} memory simplifies the programming model as all cores can directly communicate via shared memory without explicit dataflow management. %
The \gls{l1} banks are connected to the cores via a pipelined, hierarchical interconnect. %
Cores can access banks close to them within a single cycle, while banks further away have a latency of three or five cycles. %
In addition to the \gls{l1} interconnect, the cores have access to a hierarchical \glsunset{axi4}\gls{axi4}~\cite{arm2023ambaaxiandacepr} interconnect connecting to the \gls{soc}. %

\paragraph*{\textbf{\teb{\iDmaE} Integration}}

\mempool{}'s large scale and distributed \gls{l1} memory make a monolithic \dmae{} incredibly expensive, as it would require a dedicated interconnect, spanning the whole \mempool{} architecture, connecting all 1024 memory banks. %
The existing interconnect between cores and \gls{l1} memory is built for narrow, single-word accesses; thus unsuitable for wide, burst-based transfers. %

To minimize interconnect overhead to \gls{l1} memory, multiple {\bes} are introduced into \mempool{} placed close to a group of banks. %
To connect to the \gls{soc}, it can share the existing \gls{axi4} interconnect used to fetch instructions. %

These distributed \idmae{} {\bes}, each controlling exclusive regions of the \gls{l1} memory, greatly facilitate physical implementation. %
However, individually controlling all \bes{} would burden the programmer and massively increase overhead due to transfer synchronization of the individual {\dmaes}. %
Instead, our \idmae{}'s modular design allows for hiding this complexity in hardware by using a single \fe{} to program all the distributed \bes{}. %

\begin{figure}
    \centering%
    \includegraphics[width=\columnwidth]{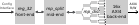}%
    \caption{%
        Our distributed {\idmae} is implemented in \mempool{}. %
        \emph{Mp{\tus}split} splits the transfers along their \gls{l1} boundaries, and a tree of \emph{mp{\tus}dist} \mes{} distribute the transfer. %
    }%
    \label{fig:case:mempool}%
\end{figure}

As seen in \Cref{fig:case:mempool}, the \emph{mp{\tus}split} {\me} splits a single \dma{} request along lines of \mempool{}'s \gls{l1} memory's address boundaries, and a binary tree of \emph{mp{\tus}dist} {\mes} distributes the resulting requests to all \bes{}. %

\paragraph*{\textbf{Benchmarks}}

We evaluate \mempool{}'s \idmae{} by comparing the performance of various kernels compared to a baseline without {\dma}. %
Since \mempool{} requires our modular \idmae{} to implement a distributed \dma{}, a comparison with another \dma{} unit is not feasible here. %
First, we compare the performance copying \SI{512}{\kibi\byte} from \gls{l2} to \gls{l1} memory. %
Without a \dma{}, the cores can only utilize one-sixteenth of the wide \gls{axi4} interconnect. %
The \idmae{} utilizes \SI{99}{\percent} and speeds up memory transfers by a factor of \SI{15.8}{\x} while incurring an area overhead of less than \SI{1}{\percent}. %

The performance improvement for kernels is evaluated by comparing a double-buffered implementation supported by our \idmae{} to the cores copying data in and out before and after the computation. %
Even for heavily compute-bound kernels like matrix multiplication, {\idmae} provides a speedup of \SI{1.4}{\x}\!. %
Less compute-intensive kernels like the convolution or discrete cosine transformation benefit even more from the \idmae{} with speedups of \SI{9.5}{\x} and \SI{7.2}{\x}\!, respectively. %
Finally, memory-bound kernels like vector addition and the dot product are dominated by the data transfers and reach speedups of \SI{15.7}{\x} and \SI{15.8}{\x}\!. %

\newpage
\section{I/O\tebsr{-}DMA}
\label{chap:dmaext:iodma}

\subsection{Introduction}

\Gls{soc} \glspl{mcu} and general-purpose systems feature a wide range of peripherals, including {UART}, {SPI}, {I2C}, {I2S}, among many~\cite{pullini2017textmudmaanauto, sauter2025basiliska34mm²e}.
Although different in their exact operation, these peripherals share a common trait; they are low-bandwidth and feature a high latency compared to on-chip communications and memories.
It is further common to connect the different peripheral devices over a low-performance on-chip interconnect to reduce the area overhead~\cite{pullini2017textmudmaanauto}.
These factors combined make accessing peripheral data an intensive task requiring a high number of cycles if done exclusively through the \gls{cpu}.

Most systems feature dedicated and complex peripheral subsystems to autonomously move data~\cite{pullini2017textmudmaanauto, pullini2019mrwolfanenergyp}.
While effective, having a monolithic peripheral subsystem combining both data movement and peripheral configuration through a single subsystem, complicates the design of \glspl{mcu}, limits its extendability, and complicates \gls{pnr}~\cite{solt2020aflexibleperiph}.
This is especially true if the interfaces between the individual peripherals and the \gls{io} subsystems are not pipelinable~\cite{solt2020aflexibleperiph}.

It is overall simpler and more flexible to couple individual peripherals with highly area-optimized {\dmaes}.
Such engines can be instantiated using our {\idma} architecture~\cite{solt2020aflexibleperiph}; see \Cref{chap:idma}.

\subsection{A General Template for I/O Data Movement}

\begin{figure}[t]
    \centering%
    \includegraphics[width=0.85\textwidth]{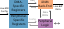}%
    \caption{%
        I/O-\gls{dma} template.
    }%
    \label{fig:arch:io-dma-template}%
\end{figure}

Thanks to {\idma}'s multi-protocol capabilities and its flexible system-binding interface, integrating {\idmae}s into a peripheral device is facilitated and straightforward.
An integration \teb{overview} can be seen in \Cref{fig:arch:io-dma-template}.
The {\idma} {\be} first needs to be customized to the peripherals' and the system's needs.
This includes selecting between unidirectional and bidirectional operation, deciding on the manager interface towards the system, and parameterizing it for the targeted use case.

For most architectures, \gls{axi4}-Lite is the preferred protocol to select for the system-facing manager interface as it is latency-tolerant and pipelinable.
Peripherals usually deliver data at a low rate compared to the throughput of the on-chip interconnect; thus, one outstanding transfer is sufficient to meet the peak bandwidth requirements.
Scaled-down, a minimal {\idmae} only incurs an area overhead of less than \SI{2}{\kilo\gateeq}.

To configure the \gls{io}-\gls{dma}, an extension of the peripheral's registers can be used and directly connected to the {\be} using the request and response interfaces.

The preferred way of interaction between the {\be} and the peripheral is done using the \emph{Fifo/Init} protocol, see \Cref{sec:arch:protocol-managers}, for its simplicity.
The {\be} exposes one stream \gls{fifo} containing valid read data and one where data to be written is placed, behaving like a memory-mapped \gls{fifo} device.

\subsection{Case Study: Neopixel}

Neopixel~\cite{burgess2013adafruitneopixe} describes both a type of light-emitting diode modules and a protocol to control them.
Neopixel modules are chainable to up to 5050 pixels, whilst still remaining individually RGB-addressable.

We implemented a Neopixel host controller as part of MLEM~\cite{wuethrich2024mlemapulpsocfor, sauter2025crocanendtoendo}\teb{, a} pipecleaner \gls{asic} to silicon-harden \emph{Croc}, our educational \gls{soc} platform~\cite{contributors2024croc, sauter2025crocanendtoendo}.

\begin{figure}
    \centering%
    \includegraphics[width=0.81\textwidth]{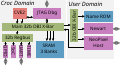}%
    \caption{%
        Top-level architecture of MLEM, including our Neopixel host controller.
    }%
    \label{fig:arch:mlem}%
\end{figure}

It is connected to MLEM through an \gls{obi} subordinate device exposing its configuration registers as well as a read-only manager port to fetch data from the \gls{soc}'s memory system.
We provide a block diagram of MLEM in \Cref{fig:arch:mlem}.

\begin{figure}
    \centering%
    \includegraphics[width=1\textwidth]{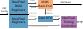}%
    \caption{%
        Architecture of our Neopixel host controller including an {\idma} {\be}.
    }%
    \label{fig:arch:neopixel}%
\end{figure}

Internally, the Neopixel host controller consists of a configuration register controlling both the \gls{io}-\gls{dma} and the configuration of the Neopixel protocol \gls{fsm}, a pixel \gls{fifo}, and the \gls{io}-\gls{dma} engine; as can be seen in \Cref{fig:arch:neopixel}.
The pixel \gls{fifo} can be filled both through explicit \gls{obi} writes of individual pixel values, called \emph{direct mode}, or by configuring the {\idma} to fetch a stream of pixels from main memory, called \emph{\gls{dma} mode}.

\paragraph*{Direct Mode}

Direct mode is ideal to supply a huge amount of procedurally generated pixel data to the device without storing it into \gls{sram} beforehand.
It needs to be ensured that the \gls{cpu} provides the data fast enough for the Neopixel protocol \gls{fsm} to only emit valid pixel data to the devices.

\paragraph*{\gls{dma} Mode}

Once configured, our {\idma} engine fetches the pixel stream over the host controllers \gls{obi} manager port and delivers it over its \emph{Fifo/Init} interface to the pixel \gls{fifo} without requiring any external protocol conversion, facilitating the integration.
The {\dmae} can be configured to fetch as many pixel values as Neopixel devices in the chain, updating the entire array in one operation.
Repeated \gls{dma} transaction allows the MLEM to display animations stored in \gls{sram} over Neopixel.

\newpage

Our Neopixel host controller was implemented on {MLEM}~\cite{contributors2024croc, sauter2025crocanendtoendo} using {IHP}'s \SI{130}{\nano\metre} open \gls{pdk}.
The entire Neopixel host controller requires \SI{16.4}{\kilo\gateeq}, with the \gls{io}-\gls{dma} occupying \SI{5.0}{\kilo\gateeq}.
The \gls{io}-\gls{dma} instantiation is around \SI{2}{\x} larger than expected (see \Cref{tab:res:soac}) due to an \gls{obi} \emph{rready} converter required by croc.
This converter adds a two-deep stream \gls{fifo} between the \gls{dma} and the interconnect.
The \gls{io}-\gls{dma} can achieve asymptotically \SI{100}{\percent} bus utilization, requiring four cycles of latency until the first data arrives at the output \gls{fifo} ports, ensuring agile and efficient data transport to our Neopixel host controller.

\newpage
\section{Real-Time Sensor Scheduling}
\label{chap:dmaext:realtime}

\subsection{Introduction}

Cyber-physical systems must continuously sense and actuate their environment with a high degree of predictability and meeting all deadlines imposed by their real-time schedule.
This, in particular, includes reading a multitude of sensor values from memory-mapped \gls{io} devices~\cite{ottaviano2025controlpulpleta}.

\Gls{io} devices are usually attached over a slow, low-bandwidth, and high-latency interface, keeping the initiator waiting until the sensor data has been retrieved.
To complicate things, distributed sensor arrays usually feature complex address maps~\cite{ottaviano2024controlpulparis,ottaviano2024controlpulpariscvpowercontrollerforhpcprocessorswithparallelcontrollawcomputationacceleration, ottaviano2025controlpulpleta}.
To keep the \gls{cpu} free to complete useful workloads, we introduce our, \emph{\gls{sdma} engine}, specialized to retrieve sensor data \alc{autonomously on a schedule, ensuring critcal information of the outside world is always present on time and the real-time core is freed of this data movement burden}.
We configure our \gls{sdma} unit to be light-weight, only supporting one outstanding transaction in line with the sensor interfaces, keeping the area overhead minimal.

\subsection{Architecture}

\begin{figure}
    \centering%
    \includegraphics[width=\textwidth]{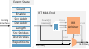}%
    \caption{%
        Architecture of the real-time {\me}.
        The event state, holding the frequency and the shape of the transfer, can be part of the {\fe}.
    }%
    \label{fig:arch:rt-fe-me}%
\end{figure}

Our real-time extension adds additional fields to the {\fe} specifying the length, shape, and periodicity of the sensor accesses.
A compile-time \gls{sv} parameter specifies the maximum number of events supported.
A bypass mechanism allows the core to dispatch unrelated transfers using the same front- and {\be}. %

We present a block diagram of our real-time {\me} in \Cref{fig:arch:rt-fe-me}.
It introduces a counter for each event tracked.
Once the corresponding event is enabled, the counter decrements from the specified cycle count.
An event is triggered once the counter reaches zero.
A stream arbiter selects between all triggered events and the bypass port and relays the selected transfer through a \teb{\gls{nd}} {\me} to the {\be} to execute the transfer and gather the sensor data.

\subsection{Case Study: \teb{\controlpulp}}

{\controlpulp}~\cite{ottaviano2024controlpulparis, ottaviano2024controlpulpariscvpowercontrollerforhpcprocessorswithparallelcontrollawcomputationacceleration, ottaviano2025controlpulpleta} is an on-chip parallel \gls{mcu} employed as a \gls{pcs} for manycore \gls{hpc} processors. %
It comprises a single 32-\si{\bit} {\riscv} \emph{manager domain} with \SI{512}{\kibi\byte} of \gls{l2} scratchpad memory and a programmable accelerator (\emph{cluster domain}) hosting eight 32-\si{\bit} {\riscv} cores and \SI{128}{\kibi\byte} of \gls{tcdm}. %

A \gls{pcf} running on FreeRTOS implements a reactive power management policy. %
{\controlpulp} receives (i) \gls{dvfs} directives such as frequency target and power budget from high-level controllers and (ii) temperature from \gls{pvt} sensors and power consumption from \glspl{vrm}, and is tasked to meet its constraints. %
The \gls{pcf} consists of two periodic tasks, \gls{pfct} (low priority) and \gls{pvct} (high priority) that handle the power management policy. %

{\controlpulp} requires an efficient scheme to collect sensor data at each periodic step without adding overhead to the computation part of the power management algorithm. %

\paragraph*{\textbf{\teb{\iDmaE} Integration}}

As presented by Ottaviano~\etal~\cite{ottaviano2024controlpulparis, ottaviano2024controlpulpariscvpowercontrollerforhpcprocessorswithparallelcontrollawcomputationacceleration}, the \emph{manager domain} offloads the computation of the control action to the \emph{cluster domain}, which independently collects the relevant data from \gls{pvt} sensors and \glspl{vrm}. %
We redesign {\controlpulp}'s data movement paradigm by integrating a second dedicated {\idmae}, called \emph{\gls{sdma} engine}, in the \emph{manager domain} to simplify the programming model and redirect non-computational, high-latency data movement functions to the \emph{manager domain}, similar to {IBM's} \emph{Pstate} and \emph{Stop} engines~\cite{rosedahl2017powerperformanc}. %
Our \gls{sdma} engine is enhanced with \emph{rt\tus3D}, a {\me} capable of autonomously launching repeated 3-D transactions. %
{\controlpulp}'s architecture is heavily inspired by {\pulpopen}; {\idmae} integration can thus be seen in \Cref{fig:case:pulp:arch}. %
The goal of the extension is \alc{on one hand} to further reduce software overhead for the data movement phase, which is beneficial to the controller's \emph{slack} within the control hyperperiod~\cite{ripoll2013periodselection}, %
\alc{and to ensure sensor data is always present on time.}
The \gls{sdma} engine supports several interface protocols, thus allowing the same underlying hardware to handle multiple scenarios. %

\paragraph*{\textbf{Benchmarks and \teb{R}esults}}

We evaluate the performance of the enhanced \gls{sdma} engine by executing the \gls{pcf} on top of FreeRTOS within an \gls{fpga}-based (Xilinx Zynq UltraScale+) hardware-in-the-loop framework that couples the programmable logic implementing the \gls{pcs} with a power, thermal, and performance model of the plant running on top of the ARM-based Processing System~\cite{ottaviano2024controlpulparis, ottaviano2024controlpulpariscvpowercontrollerforhpcprocessorswithparallelcontrollawcomputationacceleration}. %

Data movement handled by \emph{rt\tus3D}, which allows repeated 3-D transactions to be launched,  brings several benefits to the application scenario under analysis. %
First, it decouples the main core in the \emph{processing domain} from the \gls{sdma} engine in the \emph{I/O domain}. %
The \gls{sdma} engine autonomously realizes periodic external data accesses in hardware, minimizing the context switching and response latency suffered by the manager core in a pure software-centric approach. %
We consider a \gls{pfct} running at \SI{500}{\micro\second} and the \gls{pvct} at \SI{50}{\micro\second}, meaning at least ten task preemptions during one \gls{pfct} step with FreeRTOS preemptive scheduling policy. %
The measured task context switch time in FreeRTOS for {\controlpulp} is about 120 clock cycles~\cite{ottaviano2024controlpulparis, ottaviano2024controlpulpariscvpowercontrollerforhpcprocessorswithparallelcontrollawcomputationacceleration}, while {\idmae} programming overhead for reading and applying the computed voltages is about 100 clock cycles. %
From \gls{fpga} profiling runs we find that the use of \gls{sdma} engine saves about 2200 execution cycles every scheduling period, thus increasing the slack of the \gls{pvct} task. %
\alc{As it is no longer the core's burden to fetch the data, it reduces the variability of the \gls{pvct} task by isolating computation from data retrieval over slow and unpredictable peripheral interconnects.}
\alc{Further, sensor data of physical qualities usually are continuous in values, thus most control algorithms are robust against spuriously outdated sensor data.}
Autonomous and intelligent data access from the \emph{I/O domain} is beneficial as it allows the two subsystems to reside in independent power and clock domains that could be put to sleep and woken up when needed, reducing the \emph{uncore} domain's power consumption. %

Our changes add minimal area overhead to the system. %
In the case of eight events and sixteen outstanding transactions, the \gls{sdma} engine is about \SI{11}{\kGE} in size, accounting for an area increase of only \SI{0.001}{\percent} of to the original {\controlpulp}'s area. %
The overhead imposed by \gls{sdma} engine is negligible when {\controlpulp} is used as an on-chip power manager for a large \gls{hpc} processors.
It has been shown~\cite{ottaviano2024controlpulparis, ottaviano2024controlpulpariscvpowercontrollerforhpcprocessorswithparallelcontrollawcomputationacceleration} that the entire {\controlpulp} only occupies a small area of around \SI{0.1}{\percent} on a modern \gls{hpc} CPU dies. %

\newpage
\section{{\riscv} Instruction {\Fe}}
\label{chap:dmaext:instrction}

To enable closer coupling between the core and the {\idmae}, we design and implement a {\fe}, called \emph{inst\tus64}, encoding transfers directly as {\riscv}-compliant instructions.
Our \emph{inst\tus64} allows connected cores to launch \teb{1-D} \gls{dma} transactions within only three cycles, enabling highly agile transfers and reducing the setup cost to a minimum.

The architecture of \emph{inst\tus64}, its integration into the Snitch cluster~\cite{zaruba2021snitchatinypseu}, and a scale-out study will be detailed in \Cref{chap:comcpu}.

\newpage

\section{Summary and Conclusion}
\label{chap:dmaext:conclusion}

In this chapter, we present how {\idma} is integrated in a wide range of systems and how we enhance our architecture with multiple extension further accelerating \gls{dma} transfers and increasing their efficiency.

In \Cref{chap:dmaext:linux} we present a scalable, platform-independent, synthesizable, transfer-descriptor-based \gls{dmac} for fast and efficient data transfers in Linux-capable \gls{axi4}-based systems. %
Integrated in {\cheshire}, our Linux-capable \gls{soc} platform~\cite{ottaviano2023cheshirealightw}, we achieve \SI{1.66}{\x} less latency, increasing bus utilization by up to \SI{2.5}{\x} in an ideal memory system with 64-\si{\byte} transfers, overall requiring \SI{11}{\percent} fewer \si{\lut} and \SI{23}{\percent} fewer \si{\ff} without requiring any block {RAM}s compared to Arm's {LogiCore} \gls{dma} \gls{ip}~\cite{xilinx2022axidmav71logico}. %
In deep memory systems, we show an even more significant increase in the utilization of \SI{3.6}{\x} with 64-\si{\byte} transfers.

\Cref{chap:dmaext:vm} presents our very capable \gls{smmu} architecture and its integration into {\idma} enhancing our {\dmaa} with \gls{vm} support.
Implemented on the \emph{Diligent Genesys 2}, \gls{smmu} requires \SI{8.1}{\kilo\ff} and \SI{9.0}{\kilo\lut} whilst not impacting the critical path of the \gls{soc}.
Copying transfers of varying length, \gls{smmu} can achieve in excess of \SI{99}{\percent} of the non-translated performance when issuing 4-\si{\kilo\byte}-sized transfers.

Equipped with our \alc{\gls{nd}} tensor {\me} and our light-weight register-based {\fe} presented in \Cref{chap:dmaext:tensor}, {\idmae} almost fully utilizes the bandwidth to the \gls{l2} and \gls{tcdm} in both directions in {\pulpopen}: %
measuring with the on-board timer, a transfer of \SI{8}{\kibi\byte} from the cluster's \gls{tcdm} to \gls{l2} requires 1107 cycles, of which 1024 cycles are required to transfer the data using a 64-\si{\bit} data bus.
During {MobileNetV1} inference, with its improved \emph{tensor\_3D} {\me}, {\idma} improves the cores' utilization and throughput for the network over MCHAN, achieving an average of \SI{8.3}{MAC\per cycle} compared to the previously measured \SI{7.9}{MAC\per cycle}. %
Configured with similar queue depths as MCHAN, {\idmae} with its \emph{reg\_32\_3d} achieves a \SI{10}{\percent} reduction in the utilized area within a \teb{\pulp} cluster. %

Our multi-channel extensions presented in \Cref{chap:dmaext:multichannel}, show perfromance improvements in {\mempool}.
Even for heavily compute-bound kernels like matrix multiplication, {\idmae} provides a speedup of \SI{1.4}{\x}\!. %
Less compute-intensive kernels like the convolution or discrete cosine transformation benefit even more from the {\idmae} with speedups of \SI{9.5}{\x} and \SI{7.2}{\x}\!, respectively. %
Finally, memory-bound kernels like vector addition and the dot product are dominated by the data transfers and reach speedups of \SI{15.7}{\x} and \SI{15.8}{\x}\!.

We show in \Cref{chap:dmaext:iodma}, how small configurations of {\idma} can be used to enhance peripherals with fast and standalone memory accesses.
On the example of our Neopixel host controller, we show that integrating an \gls{io}-\gls{dma} incurs only an area overhead of around \SI{5.0}{\kilo\gateeq}.
With an even more efficient memory system implementation, \gls{io}-\glspl{dma} requiring less than \SI{2.0}{\kilo\gateeq} can be configured whilst still achieving near-ideal bus utilization.

\teb{Finally, we} present our real-time {\me} extension and the corresponding integration into {\controlpulp} in \Cref{chap:dmaext:realtime}.
From \gls{fpga} profiling runs we find that the use of our \gls{sdma} unit saves about 2200 execution cycles every scheduling period, thus increasing the slack of the \gls{pvct} task. %
Our changes add minimal area overhead to the system. %
In the case of eight events and sixteen outstanding transactions, the \gls{sdma} engine is about \SI{11}{\kilo\gateeq} in size, accounting for an area increase of only \SI{0.001}{\percent} of to the original {\controlpulp}'s area. %

\alc{To summarize; {\idma} brings clear benefits in systems spanning the entire compute continuum, thus proofing the versatility and flexibility of our architecture.}

\chapter{Communication Processor}
\label{chap:comcpu}

\section{Introduction}
\label{chap:comcpu:introduction}

With the \emph{Memory Wall} or the \emph{Processor-Memory Gap} still prevailing~\cite{burger1996memorybandwidth, wilkes2001thememorygapand, hennessy2011computerarchite} and with the advent of \gls{ai} further demanding ever more compute~\cite{villalobos2022machinelearning}, pressure on the memory system is tightening.
Classical computing architectures employ a tiered cache architecture trying to mitigate the performance loss when accessing slow and low-bandwidth off-chip memory at the cost of a high area and power consumption~\cite{hennessy2011computerarchite}.
An other approach is to use explicitly managed and tightly coupled scratchpad memory, called \gls{tcdm}, located close to the \glspl{pe}.
\Gls{dma} engines are used to copy data from and to \gls{tcdm} traditionally employing long bursts to hide the latency into last-level storage~\cite{pullini2019mrwolfanenergyp}.
With \teb{\pulp}, and thus with this thesis, we use the latter approach and thus focus our discussion on explicitly managed storage architectures.

\Glsf{hbm} is the dominating \gls{dram} organization for low-latency and high-bandwidth applications.
Its immense bandwidth is provided through a combination of a massively parallel \gls{io} connection into the \gls{dram} array and 3D-stacked multi-channel packages~\cite{jun2017hbmhighbandwidt}.
Recently, systems have evolved to use multi-site \gls{hbm} configurations~\cite{corporation2024nvidiah100tenso, nassif2022sapphirerapidst, biswas2021sapphirerapids} to further increase the available bandwidth into main memory.
Using such multi-site configurations with systems employing explicitly managed \gls{tcdm} leads to a set of challenges;
to get the prospective bandwidth advantage, \gls{dma} transfers need to be optimized to access multiple sites \teb{concurrently}, as datasets might be interleaved over the channels,
at the same time, long bursts need to be split to equalize the access intensity of the sites and channels,
and the {\dmaes} need to be tolerant towards non-uniform site access latencies.

Compressing the data set size is another way of overcoming the \emph{Memory Wall}~\cite{dave2021hardwareacceler}.
Leveraging data sparsity poses a very elegant compression strategy, especially as data sets from a wide range of fields, like image and signal processing, computer vision, and pattern recognition, are inherently sparse~\cite{zhang2015asurveyofsparse}.
Moving and processing these sparse data sets fundamentally incurs indirect data accesses and data-dependent program flows~\cite{dave2021hardwareacceler, qureshi2020tearingdownthem}, which are incompatible with traditionally employed coarse-grained data transfers used to hide last-level memory access latencies~\cite{qureshi2020tearingdownthem}.

{\dmaes} need to evolve to suit these application scenarios by designing them to be latency-tolerant, capable of handling short and irregular transfers.
In \Cref{chap:idma}, with {\idma}, we propose exactly such a configurable, agile, and latency-tolerant {\dmaa}.
\Cref{chap:dmaext} extends the architecture's capabilities with regular strided tensor accesses and describes loosely coupled register- and transfer-descriptor-based programming models.

To maximize agility, we tightly couple our {\idmae} directly to a core inside of our compute cluster, allowing it to launch \gls{dma} transactions in as little as three \teb{instructions} and have the first data moved in as little as four \teb{cycles}.
Compared to the prevailing approach of using custom \glspl{isa}~\cite{stmicroelectronicsdmaengineovervi, arm2012corelinkdma330d}, our instruction-based {\fe} extends the {\riscv} \gls{isa}, allowing data-dependent and irregular \gls{dma} movement patterns to be easily calculated alongside the program execution with minimal synchronization and communication overhead.

We present the general architecture of the instruction-based {\fe}, \emph{inst\tus64}, in \Cref{chap:comcpu:frontend:arch}, followed by our {\riscv} \gls{dma} instruction extension in \Cref{chap:comcpu:frontend:encoding}, and its programming model in \Cref{chap:comcpu:frontend:prog}.
We then present the integration of our {\fe} into the Snitch cluster~\cite{zaruba2021snitchatinypseu} in \Cref{chap:comcpu:integration} and a case study of a scaled-out system based on our instruction-based {\idma} integration in \Cref{chap:comcpu:occamy}.
\alc{Tightly coupling our inst\tus64 and our {\idma} {\be} with a Snitch core creates a \emph{communication processor} in-charge of orchestrating data transfers and synchronizing with other \glspl{pe} doing useful compute.}
Finally, we end the chapter with a summary and conclusion.

\newpage
\section{Instruction-Based {\Fe}}
\label{chap:comcpu:frontend}

In the following, we will describe the architecture, the instruction encoding, and the programming model of the tightly coupled instruction-based \emph{inst\_64} {\idma} {\fe}.

\subsection{{\DmaE} Architecture}
\label{chap:comcpu:frontend:arch}

\begin{figure}
    \centering%
    \includegraphics[width=\textwidth]{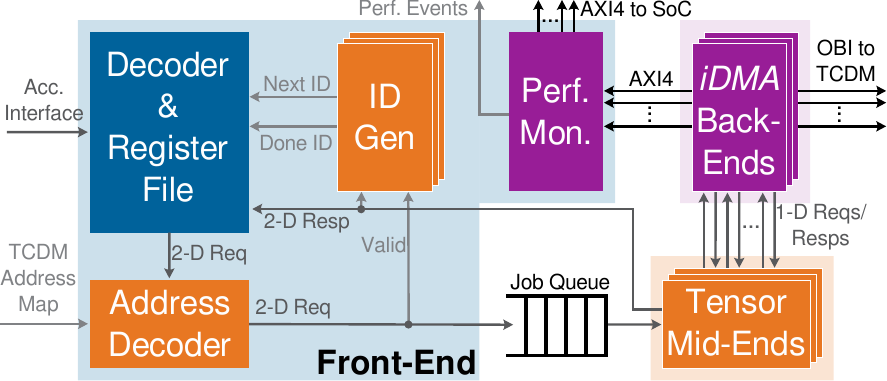}%
    \caption{%
        Architecture of the tightly coupled \emph{Snitch} {\dmae}. %
        It combines the \emph{inst\_64} {\fe} with a configurable number of \emph{tensor{\tus}ND} {\mes} and {\idma} {\bes}.
    }%
    \label{fig:comcpu:frontend:snitch_dma}%
\end{figure}

\begin{table}[t]
    \centering
    \caption{%
        Events emitted by the tightly coupled Snitch \gls{dma}.%
    }%
    \label{tab:dma_events}
    \renewcommand*{\arraystretch}{1.15}
    \resizebox{\linewidth}{!}{%
    \begin{threeparttable}
        \begin{tabular}{@{}lll@{}}
            \toprule
            & \textbf{Event} & \textbf{Description} \\
            \midrule
            \multirow{3}{*}{\rot{AW}} & \emph{done}   & \emph{AW} handshake successfully completed \\
                                      & \emph{stall}  & {\idma} {\be} has a valid \emph{AW} beat, but memory is not ready \\
                                      & \emph{issued} & Number of bytes~\tnote{a}~~requested during successful \emph{AW} handshake \\
            \midrule
            \multirow{3}{*}{\rot{W}}  & \emph{done}   & \emph{W} handshake successfully completed \\
                                      & \emph{stall}  & {\idma} {\be} has a valid \emph{W} beat, but memory is not ready \\
                                      & \emph{bytes}  & Number of bytes~\tnote{b}~~transferred during successful write operation \\
            \midrule
            \multirow{2}{*}{\rot{B}}  & \emph{done}   & \emph{B} handshake successfully completed \\
                                      & \emph{stall}  & {\idma} {\be} can accept \emph{B} beat, but memory is not providing it \\
            \midrule
            \multirow{2}{*}{\rot{Buffer}} & \emph{in stall}  & Buffer is full, put back-pressure on read transfers \\
                                          & \emph{out stall} & Buffer is empty, cannot write \\
            \midrule
            \multirow{3}{*}{\rot{AR}} & \emph{done}   & \emph{AR} handshake successfully completed \\
                                      & \emph{stall}  & {\idma} {\be} has a valid \emph{AR} beat, but memory is not ready \\
                                      & \emph{issued} & Number of bytes~\tnote{a}~~requested during successful \emph{AR} handshake \\
            \midrule
            \multirow{2}{*}{\rot{R }} & \emph{done}  & \emph{R} handshake successfully completed \\
                                      & \emph{stall} & {\idma} {\be} can accept \emph{R} beat, but memory is not providing read data \\
            \bottomrule
        \end{tabular}
        \begin{tablenotes}[para, flushleft]
            \item[a] \teb{a}ccording to $(AX_{len} + 1) << AX_{size}$
            \item[b] \teb{o}nly \emph{strobed} bytes are counted
        \end{tablenotes}
    \end{threeparttable}
    }
\end{table}

\Cref{fig:comcpu:frontend:snitch_dma} shows the architecture of our \emph{inst\_64} {\fe} coupled with a configurable number of \emph{tensor{\tus}ND} {\mes} and {\idma} {\bes}.

Our \emph{inst\_64} {\fe} is connected to a core through its \emph{accelerator offloading interface}~\cite{zaruba2021snitchatinypseu}, allowing our {\fe} to receive and decode the \emph{dm} subset, see \Cref{chap:comcpu:frontend:encoding}, of the core's issued {\riscv} instructions.
A \emph{decoder} handles the \emph{dm} subset and interfaces the {\dmae} by either returning status information, updating the internal register file holding the job currently being constructed to be dispatched, or launching a job.

Transfers are identified through the software via \emph{\glsplf{jid}}.
On launch, the {\fe} returns a monotonically increasing \gls{jid} identifying the job just launched.
As each {\be} handles transfers strictly in-order, see \Cref{sec:idma:architecture:backend}, the \gls{jid} are marked as completed in the same monotonically increasing order as they were launched.
If the last completed \gls{jid} is larger than or equal to the \gls{jid} of a job previously launched, the {\dmae} has successfully completed said job.

In multi-channel {\dmae} configurations, we currently select the active channel through the configuration instruction.
Automatic channel assignment can be implemented if required; our \emph{in-network accelerator} works~\cite{di2021ariscvinnetwork, di2021pspinahighperfo, khalilov2024osmosisenabling} employ multi-channel-selection using \gls{rr} arbitration.
As the channels are independent of each other, we have one \emph{\gls{jid} generator} for each channel present in the system.

The {\bes} used are configured to have both \gls{axi4}, connecting to the \gls{soc}, and \gls{obi}, for efficiently accessing \gls{tcdm}, read and write ports.
An \emph{address decoder} determines the port combination used for the current job.

A \emph{job queue} is used to allow the core to launch multiple jobs while the {\dmae} is busy; once the queue is full, the \emph{dmcpy} instruction will be blocking until the queue can accept a next job.

Finally, all \gls{axi4} interfaces are channeled through a \emph{performance event unit}, which emits the \gls{dma} events described in \Cref{tab:dma_events}.

\afterpage{%
\begin{sidetab}
    \centering
    \renewcommand{\arraystretch}{1.25}
    \caption{Our {\riscv} \emph{dm} instruction\alc{s} introduced by \emph{inst\_64}.}%
    \label{tab:comcpu:frontend:instr}
    \resizebox{\columnwidth}{!}{
        \begin{threeparttable}
            \begin{tabular}{@{}lllllllll@{}}
                \toprule
                \textbf{Type} & \textbf{Mnemonic}          & \textbf{Immediate}      & \textbf{Funct 7}
                              & \textbf{Source 2}          & \textbf{Source 1}       & \textbf{Funct 3}
                              & \textbf{Destination}       & \textbf{Opcode}         \\
                \midrule
                \emph{R}      & \emph{dmsrc}               & \nad                    & \emph{0}
                              & \emph{address\_high}       & \emph{address\_low}     & \emph{0}
                              & \nad                       & \emph{0x2b}             \\
                \emph{R}      & \emph{dmdst}               & \nad                    & \emph{1}
                              & \emph{address\_high}       & \emph{address\_low}     & \emph{0}
                              & \nad                       & \emph{0x2b}             \\
                \emph{I}      & \emph{dmcpyi}~\tnote{a}    & \emph{config}~\tnote{b} & \nad
                              & \nad                       & \emph{num\_bytes}       & \emph{1}
                              & \emph{\gls{jid}}           & \emph{0x2b}             \\
                \emph{R}      & \emph{dmcpy}~\tnote{a}     & \nad                    & \emph{2}
                              & \emph{config}~\tnote{b}    & \emph{num\_bytes}       & \emph{0}
                              & \emph{\gls{jid}}           & \emph{0x2b}             \\
                \emph{I}      & \emph{dmstati}             & \emph{metric}           & \nad
                              & \nad                       & \nad                    & \emph{2}
                              & \emph{information}         & \emph{0x2b}             \\
                \emph{R}      & \emph{dmstat}              & \nad                    & \emph{3}
                              & \nad                       & \emph{metric}           & \emph{0}
                              & \emph{information}         & \emph{0x2b}             \\
                \emph{R}      & \emph{dmstr}               & \nad                    & \emph{4}
                              & \emph{dst\_stride}         & \emph{src\_stride}      & \emph{0}
                              & \nad                       & \emph{0x2b}             \\
                \emph{R}      & \emph{dmrep}               & \nad                    & \emph{5}
                              & \nad                       & \emph{num\_repetitions} & \emph{0}
                              & \nad                       & \emph{0x2b}             \\
                \emph{R}      & \emph{dmuser}~\tnote{c}    & \nad                    & \emph{6}
                              & \emph{user\_high}          & \emph{user\_low}        & \emph{0}
                              & \nad                       & \emph{0x2b}             \\
                \emph{I}      & \emph{dminiti}~\tnote{a,d} & \emph{config}~\tnote{b} & \nad
                              & \nad                       & \emph{num\_bytes}       & \emph{3}
                              & \emph{\gls{jid}}           & \emph{0x2b}             \\
                \bottomrule
            \end{tabular}
            \begin{tablenotes}[para, flushleft]
              \item[a] \teb{l}aunches \gls{dma} job on execution
              \item[\teb{b}] \teb{u}p to eight channels selectable
              \item[c] \teb{p}rimarily used as \emph{dmcast} for multi-cast operations~\cite{colagrande2025tamingoffloadov}
              \item[d] \teb{b}yte to be written is specified using \emph{dmsrc}
            \end{tablenotes}
        \end{threeparttable}
    }
\end{sidetab}
}

\afterpage{%
\begin{sidefig}
    \centering%
    \begin{subcaptionblock}{\linewidth}
        \centering%
        \includegraphics[width=\linewidth]{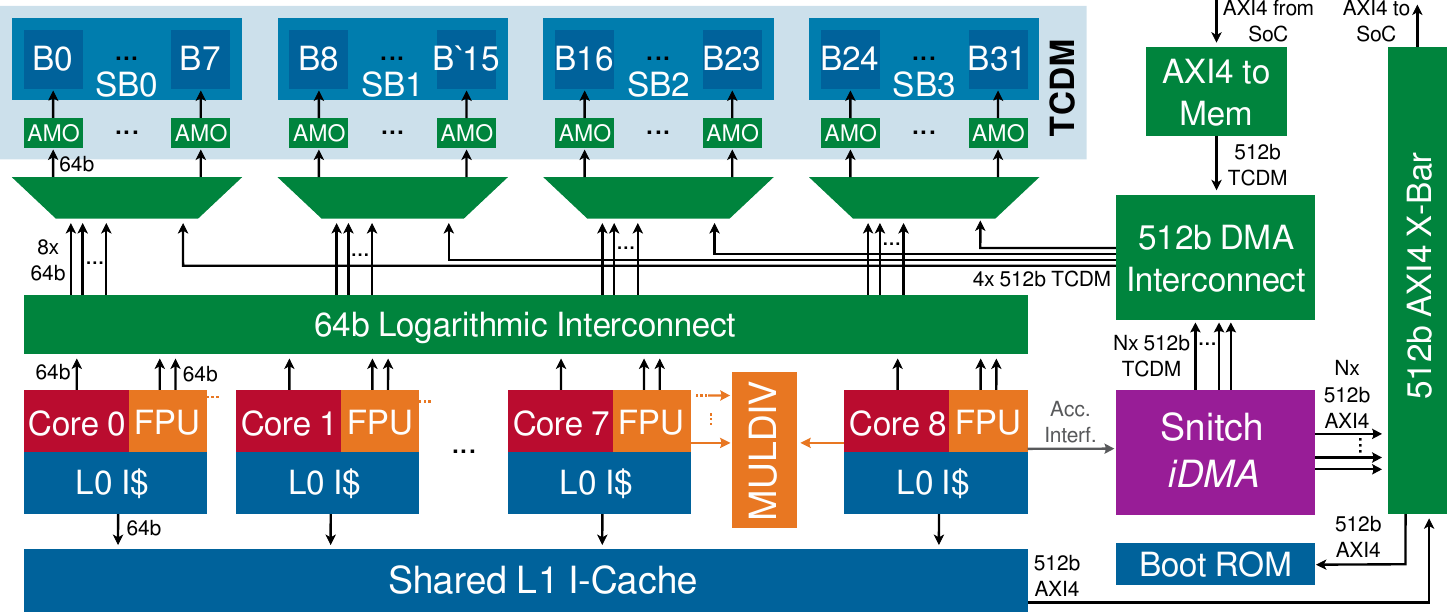}%
    \end{subcaptionblock}\hfill
    \caption{\teb{Our {\idma} integration into a Snitch compute cluster.}}
    \label{fig:cluster_integration}
\end{sidefig}
}

\subsection{Instruction Encoding}
\label{chap:comcpu:frontend:encoding}

We extend the {\riscv} \gls{isa} with our ten \emph{dm} instructions presented in \Cref{tab:comcpu:frontend:instr}.
Our instructions use \emph{0x2b} as \emph{opcode}, partially occupying one of four custom blocks, \emph{0x0A} of quadrant \emph{3}.

Our \emph{dmsrc}, \emph{dmdst}, \emph{dmstr}, and \emph{dmrep} are used to configure the \emph{shape} of the transfer.
They specify the source and destination addresses as well as the strides and number of repetitions in the case of a 2-D transfer.
The status instructions, \emph{dmstat} and \emph{dmstati}, can be used to retrieve the current state of the \gls{dma}, including the last completed \gls{jid}, the next \gls{jid}, and the busy signals of the mid- and {\bes}.
Jobs are launched using \emph{dmcpy} and \emph{dmcpyi}, as well as \emph{dminiti}; the former two launch regular \gls{dma} transfers where data is copied according to the defined shape, the latter initializes a memory region located at \emph{dmdst} with a specified byte in \emph{dmsrc}.
The length of the transfer, as well as the \teb{\be} configuration and the chosen \gls{dma} channel, \teb{are} given by the launching instruction, while the \gls{jid} is placed in the specified destination register.

The \emph{dmuser} instruction populates \gls{axi4} user signals with a value specified in the source registers.
This functionality is currently used to issue multicast operations through {\idma}~\cite{colagrande2025tamingoffloadov}.

\newpage
\section{Cluster Integration}
\label{chap:comcpu:integration}

We present the integration of our tightly coupled {\dmae} into the Snitch cluster~\cite{zaruba2021snitchatinypseu} in \Cref{fig:cluster_integration}.
Compared to the original architecture~\cite{zaruba2021snitchatinypseu}, we introduce an \emph{(N+1)}th Snitch core, dedicated to orchestration and data movement operations, we add a 512-\si{bit}-wide \gls{dma} interconnect dedicated to move data, and we group the \gls{tcdm} into \emph{superbanks \teb{(SB\tebsr{s})}}, \alc{each \SI{512}{bit} wide}.

We present \teb{a case study integrating {\idma}} into \teb{\emph{\occamy}}, a \teb{scaled-out} Snitch system in \Cref{sec:occamy:mem_arch}.

\paragraph*{\textbf{Data Movement Core}}

Our tightly coupled {\dmae} is connected to a ninth Snitch core\alc{, constituting our system's communication processor,} purposefully introduced to perform initialization, \teb{synchronization,} orchestration, and data management operations, hence we call it \emph{data movement core}.
The eight regular \emph{worker cores} constitute the \emph{worker team}.

To allow the data movement core to compute \emph{more complex} data movement \teb{patterns} for \teb{\gls{nd}} tensor transfers, it has access to the clusters' shared \emph{mul/div} unit over its \emph{accelerator offloading interface}.
The data movement core can be coupled to a \emph{simplified} \gls{fpu} subsystem, should all nine cores be required to execute floating-point workloads, but for most control and \emph{data} scheduling operations, this is not required.

\paragraph*{\textbf{Superbanks}}

We group the 32 64-\si{bit}-wide \gls{tcdm} banks into four 512-\si{bit}-wide superbanks.
Each superbank has a priority-arbited multiplexer which connects either the eight individual banks to the cluster's logarithmic interconnect or as one superbank to the \alc{512-\si{\bit}-wide} \gls{dma} interconnect.
The \gls{dma} connection has priority over fine-grained \gls{tcdm} accesses by the cores and accelerator subsystems.
This increased priority of the \gls{dma} accesses reduces back-pressure in the global \gls{soc} interconnect.
In the \alc{default} configuration with one \gls{dma} channel present in the system, having four superbanks gives the Snitch cores \SI{75}{\percent} and \SI{50}{\percent} of the undisturbed \gls{tcdm} bandwidth with only {\idma} and {\idma} and an external manager \teb{being} active, respectively.

A minimum of four superbanks is required to provide the Snitch cores and their tightly coupled accelerator subsystems enough bandwidth into \gls{tcdm}, whilst still maintaining the priority-based arbitration mechanism.

\paragraph*{\textbf{Wide DMA Interconnect}}

To move data between the cluster and the \gls{soc}, we introduce a second, 512-\si{\bit}-wide \gls{axi4} interconnect next to the already existing 64-\si{\bit}-wide bus.
After our modifications, the \emph{narrow} interconnect handles only synchronization traffic between the clusters and the host, with all the bulk data transfers and instruction traffic being moved to the wide interconnect.
A width of \SI{512}{\bit} directly corresponds to the size of a cache line in most host systems, and it is the default bus width of \gls{hbm} memory controllers, see \Cref{sec:arch_chiplet}.

As mentioned in \Cref{chap:comcpu:frontend:arch}, the {\idmae} included into the Snitch cluster is configured to feature both \gls{axi4} and \gls{obi} ports, allowing it to be a bridge between the wide \gls{axi4} interconnect towards the \gls{soc} and the \SI{512}{\bit} \gls{dma} interconnect.
External accesses are directly connected to the \SI{512}{\bit} \gls{dma} interconnect through an \emph{axi\_to\_mem} adapter.
Having both an {\idmae} connected to and an external path into the the cluster-local \gls{tcdm}, allows two clusters to bidirectionally exchange data among themselves while all involved Snitch cores are still executing useful workloads.

\subsection{Programming Model}
\label{chap:comcpu:frontend:prog}

For primarily data-oblivious~\cite{goldreich1996softwareprotect} and tileable workloads, we use the data movement core to double-buffer the tiled workload data into the cluster-local \gls{tcdm} scratchpad.
Once the data is ready, the \alc{worker team} can then be launched to do the \teb{computation} on the tiles present in \gls{tcdm}.
Communicating through semaphores in \gls{tcdm}, the data movement core can then relaunch the team on the next tiles.
\alc{So far\teb{,} we have not observed significant interference in the instruction cache between the worker team and the data movement core.}

For data-depended workloads running on Snitch systems~\cite{scheffler2024sarisaccelerati, scheffler2023sparsestreamsem, scheffler2025occamya432cored}, the split between data movement core and worker team has been proven to be a fruitful architectural decision~\alc{\cite{scheffler2024sarisaccelerati}}.

\newpage
\section{Case Study: Occamy}
\label{chap:comcpu:occamy}

\subsection{Introduction}

Sparse \gls{ml} and high-performance computing applications in fields like multiphysics simulation and graph analytics often rely on sparse \gls{la}, stencil codes, and graph pattern matching~\cite{zhang2015asurveyofsparse}.
These workloads achieve low FPU utilization on modern CPUs and GPUs because of their sparse, irregular memory accesses and complex, indirection-based address computation~\cite{alappat2020performancemode, alappat2023levelbasedblock, niu2021tilespmvatileda}.

This \teb{section} presents \teb{the integration of {\idma} into} {\occamy}, a flexible, general-purpose, dual-chiplet system with two \SI{16}{\gibi\byte} \gls{hbm2e} stacks optimized for a wide range of irregular-memory-access workloads.
\teb{We present {\occamy}} as an industry-scale use case \teb{study} for our tightly coupled instruction-based \teb{Snitch} {\idmae} \teb{introduced in \cref{chap:comcpu:integration}}.

{\occamy} demonstrates in silicon three innovations: \textbf{(A)} efficient multi-precision compute cores with sparse \glspl{su} supporting indirection, intersection, and union operations to accelerate general sparse computations, \textbf{(B)} a scalable, latency-tolerant, hierarchical architecture with separate data and control interconnects and distributed {\dmaes} for agile on-die and \gls{d2d} traffic, and \textbf{(C)} an innovative system-in-package 2.5D integration for two compute chiplets with two \SI{16}{\gibi\byte} \gls{hbm2e} stacks.

\subsection{Memory Architecture}
\label{sec:occamy:mem_arch}

\begin{figure}[ht!]
    \begin{subcaptionblock}{0.62\linewidth}%
        \centering%
        \includegraphics[width=\linewidth]{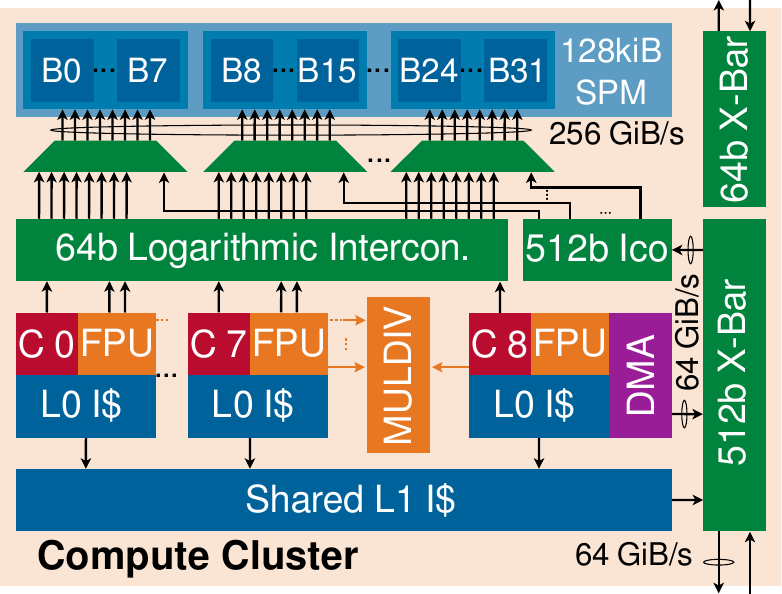}%
        \vspace{-0.4em}%
        \caption{Compute cluster}%
        \vspace{0.25em}%
        \label{fig:arch_cluster}%
        \vfill%
        \centering%
        \includegraphics[width=\linewidth]{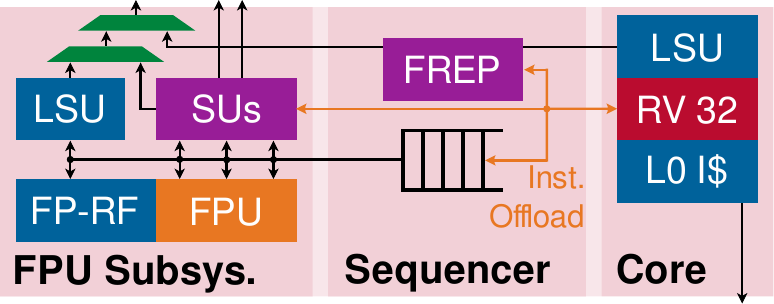}%
        \vspace{-0.3em}%
        \caption{Worker core}%
        \label{fig:arch_cc}%
    \end{subcaptionblock}\hfill
    \begin{subcaptionblock}{0.36\linewidth}
        \centering%
        \includegraphics[width=\linewidth]{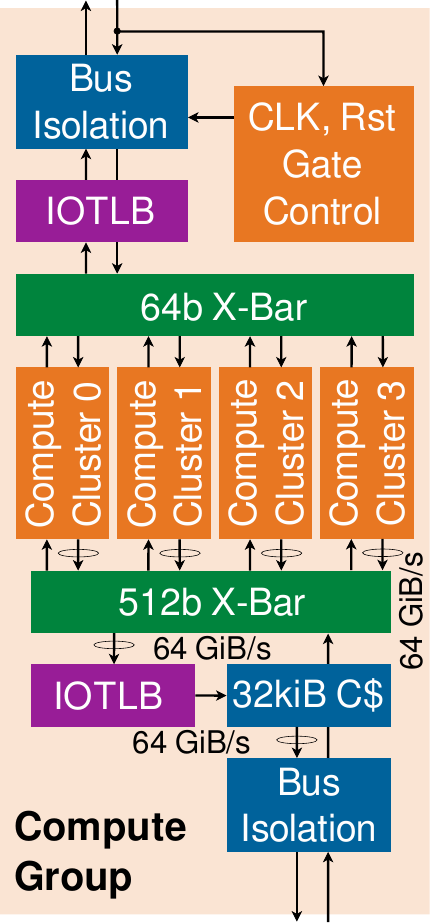}%
        \vspace{-0.3em}%
        \caption{Compute group}%
        \label{fig:arch_group}%
    \end{subcaptionblock}
    \caption{Architecture of the Occamy compute clutser, the worker core, and the compute group.}
    \label{fig:arch_parts}
\end{figure}

Occamy combines two 216-core compute chiplets on a passive interposer, each paired with an eight-device, \SI{16}{\gibi\byte} \gls{hbm2e} stack and connected through a fully digital, fault-tolerant \gls{d2d} link.
\Cref{fig:arch_parts} and \Cref{fig:arch_chiplet} show the hierarchical compute chiplet architecture:
at the lowest level, nine {\riscv} \teb{Snitch} cores are organized into clusters (\Cref{sec:arch_cluster}, four such clusters form a group (\Cref{sec:arch_group}), and each chiplet contains six such groups (\Cref{sec:arch_chiplet})
, which we describe in a bottom-up fashion in the following sections.

\subsection{Compute Cluster}
\label{sec:arch_cluster}

Occamy's compute cores are based on the RV32G \gls{isa} and are organized into compute \emph{clusters}~\cite{zaruba2021snitchatinypseu} shown in \Cref{fig:arch_cluster}.
Within each cluster, eight worker cores and one \teb{\emph{communication processor}} share a \SI{128}{\kibi\byte} 32-bank \gls{spm} through a single-cycle logarithmic interconnect with double-word interleaving.
The clusters also feature \SI{8}{\kibi\byte} of shared \gls{l1} instruction cache, a shared integer multiply-divide unit, \teb{\emph{mul/div}}, a local hardware synchronization barrier, and \teb{sixteen} retargetable performance counters
capable of tracking various per-core and cluster-wide events.

Each worker core, shown in \Cref{fig:arch_cc}, features a 64-bit-wide \glsunset{simd}\gls{simd} \gls{fpu} supporting FP64, FP32, FP16, FP16alt (8,7), FP8, and FP8alt (4,3) formats.
In addition to \gls{fma} instructions, the \gls{fpu} supports widening sum-dot-product and three-addend summation instructions for all FP8 and FP16 formats~\cite{bertaccini2024minifloatsonris}.
Two worker-core \gls{isa} extensions maximize the \gls{fpu} utilization for both regular and irregular workloads: a hardware loop buffer~\cite{zaruba2021snitchatinypseu} and three sparsity-capable \glspl{su}~\cite{scheffler2023sparsestreamsem}.

\begin{figure}[t]
\centering
\includegraphics[width=0.75\linewidth]{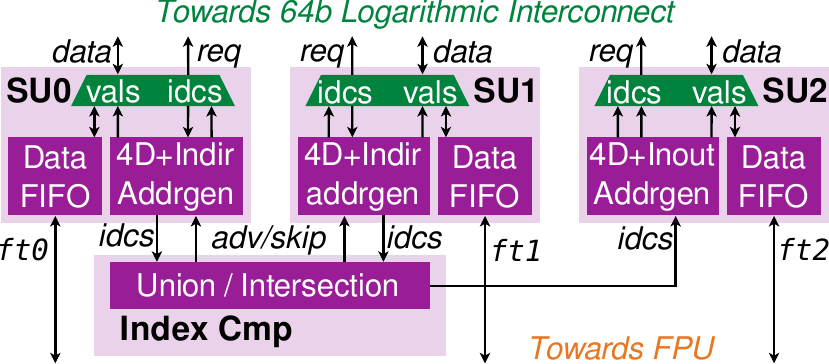}
\caption{Architecture and interconnection of the three cooperating sparsity-capable SUs in each worker core.}
\label{fig:arch_su}
\end{figure}

The \glspl{su}, shown in \Cref{fig:arch_su}, map buffered streams of \gls{spm} accesses directly to floating-point registers, generating the necessary addresses in hardware.
All three \glspl{su} in each core support up to \teb{4-D} strided accesses to accelerate dense tensor workloads; for instance, a regular \gls{gemm} utilizes all four \gls{su} loop levels to cover the algorithm's three nested loops and an inner unroll for performance.
Two \glspl{su} additionally support indirect streams with \mbox{8-,} 16-, or 32-bit indices to accelerate the scatter-gather accesses underlying irregular-access applications such as sparse-dense \gls{la} and stencil codes.
Finally, these indirect \glspl{su} can also compare their indices to accelerate the sparse tensor intersection and union underlying sparse-sparse \gls{la} or graph matching; the third \gls{su} can optionally write out the joint indices for sparse result tensors.
Our \glspl{su} can handle any sparse tensor format whose major axis is given by a value-index array pair, which includes the widespread and scalable CSR, CSC, CSF, and many of their variations,
without restrictive assumptions on operand structure or density.
In general, our \glspl{su} enable a sustained per-core bandwidth of up to three double-words per cycle into the shared \gls{spm}.

The cluster \gls{spm} is dimensioned to balance throughput, interconnect complexity, and area.
32 banks are chosen to reduce the probability of banking conflicts between \glspl{su} (24 per cluster) \teb{and the \gls{dma}} while keeping the logarithmic interconnect physically implementable.
A capacity of \SI{128}{\kibi\byte} achieves a reasonable \glsunset{sram}\gls{sram} bit density (avoiding significant drops that would appear for $\leq$\SI{4}{\kibi\byte} banks) while keeping area prevalently allocated to compute logic.

The \teb{communication processor} in each cluster features a tightly coupled 512-bit \teb{\idmae}, described in \Cref{chap:idma}, programmed through \emph{inst\tus64} (\Cref{chap:comcpu:frontend}), enabling asynchronous $\leq$\teb{2-D} transfers between external memory (other clusters, \gls{hbm2e}, or other chiplet) and the local \gls{spm}.
This core coordinates the computation of worker cores and their fine-grained, low-latency accesses to the \gls{spm} with the latency-tolerant \gls{dma} transfers of large, double-buffered data tiles to and from the \gls{spm}\teb{, as presented in \Cref{chap:comcpu:frontend:prog}.}
The \teb{\idmae} accesses blocks of eight \gls{spm} banks (\emph{superbanks}, see \Cref{chap:comcpu:integration}) at once through a secondary interconnect, transferring up to \SI{64}{\byte} per cycle or \SI{64}{\gibi\byte\per\second};
this way, it maximizes throughput without significantly slowing down ongoing memory-intensive computations, which can access up to 24 \gls{spm} banks at once using all \glspl{su}.
To reduce backpressure in the group- and chiplet-level interconnect, the \gls{dma} has priority accessing the \gls{spm} through the secondary interconnect over the cores and \glspl{su}.

\subsection{Compute Group}
\label{sec:arch_group}

Four clusters together form the next compute hierarchy level, a \emph{group}, shown in \Cref{fig:arch_group}.
Clusters within a group have full-bandwidth access to each other through two fully connected \gls{axi4} crossbars: a 512-bit crossbar used by \gls{dma} engines and instruction caches for bulk transfers and an atomics-capable 64-bit crossbar used by cores for global synchronization and message passing.
Thus, groups allow their clusters to locally share data at higher bandwidths than at the global level and constitute a replicable multi-cluster design that significantly simplifies top-level chiplet implementation.

On each crossbar, a group has one outgoing and one incoming port, providing a shared bandwidth to and from the chiplet interconnect of \SI{64}{\gibi\byte\per\second} for bulk transfers and \SI{8}{\gibi\byte\per\second} for message passing.
Providing one outgoing port per group best matches the \gls{hbm2e} bandwidth available at the chiplet level and keeps the chiplet-level interconnect implementable.
\glspl{iotlb} on the outgoing ports allow for per-group address remapping and access control at page granularity.
A remappable \SI{32}{\kibi\byte} constant cache on the outgoing 512-bit port can be used to cache program code and other immutable data.
Finally, the groups can be individually clock-gated, reset, and isolated from their interconnect ports to the chiplet level through memory-mapped registers for online power management.

\begin{figure}[t]
    \begin{subcaptionblock}{\linewidth}
        \centering%
        \includegraphics[width=\linewidth]{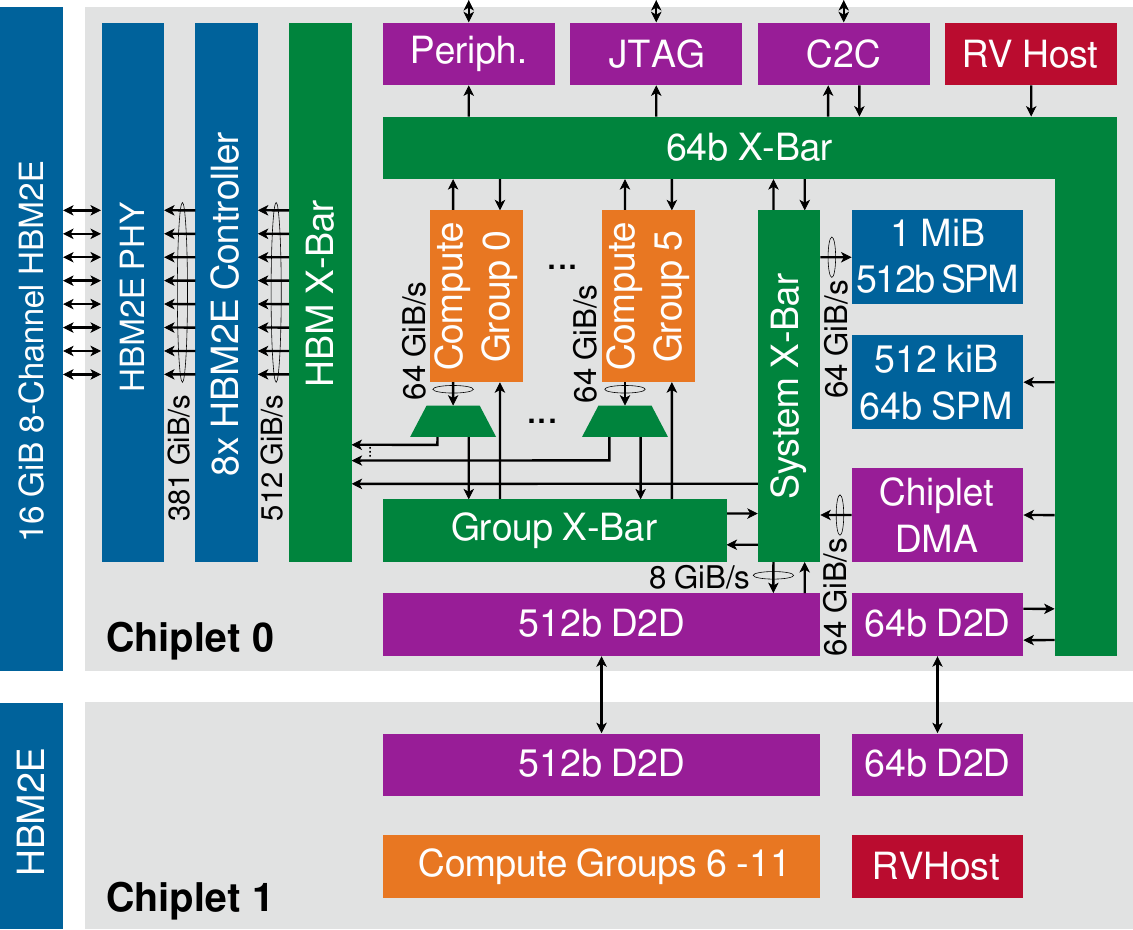}%
    \end{subcaptionblock}
    \caption{%
    Architecture of the two Occamy chiplets. %
    \alc{Occamy features only one address region; narrow managers can access wide subordinates and vice versa through the \emph{system} crossbar}. %
    }
    \label{fig:arch_chiplet}
\end{figure}

\subsection{Occamy Chiplets}
\label{sec:arch_chiplet}

\begin{figure}[t]
\centering
\includegraphics[width=0.97\linewidth]{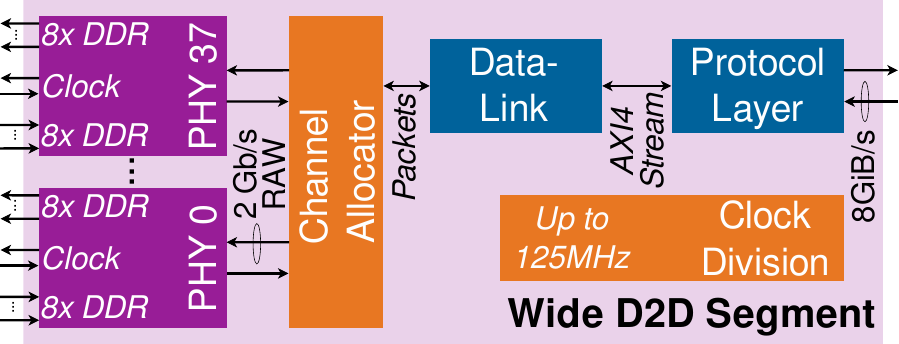}
\caption{Architecture of the wide D2D segment with its 38 source-synchronous double-data-rate PHYs, carrying up to \SI{2}{\gibi\byte\per\second} of raw data at a clock speed of \SI{125}{\mega\hertz}.}
\label{fig:arch_d2d}
\end{figure}

\Cref{fig:arch_chiplet} shows Occamy's top-level chiplet architecture.
Each chiplet features six groups, totaling 216 cores, and a single Linux-capable 64-bit \riscv~host processor managing the groups and all other on-chip resources.
Like the groups, the chiplets feature a hierarchical 512-bit \gls{axi4} network for bulk data transfers and an atomics-capable 64-bit \gls{axi4} network for synchronization, message passing, and management.
The \gls{d2d} link, which serializes cross-chiplet transactions, comprises a \emph{narrow} and a \emph{wide} segment carrying 64-bit and 512-bit transactions, respectively.

The 512-bit network is composed of three fully connected crossbars.
Each group's \SI{64}{\gibi\byte\per\second} outgoing port provides access to an \emph{HBM} crossbar and a \emph{group} crossbar, interconnecting the six on-chip groups.
The \emph{HBM} crossbar provides access to the eight on-chip \SI{47.68}{\gibi\byte\per\second} \gls{hbm2e} controllers (\SI{381.47}{\gibi\byte\per\second} in total).
It can be configured at runtime to interleave the \gls{hbm2e} channels at page granularity, facilitating load balancing and data reuse.
A \emph{system} crossbar connects the group and HBM crossbars to the wide D2D segment and 64-bit network, providing all actors across both chiplets access to the entire memory space.
It also connects to a \SI{1}{\mebi\byte} \gls{spm} used for low-latency on-chip storage of shared data and a chiplet-level \gls{dma} engine used by the host for fast, explicit data movement and memory initialization.
As a chiplet-level memory, the \SI{1}{\mebi\byte} \gls{spm} was dimensioned to be significantly larger than the \SI{128}{\kibi\byte} cluster \glspl{spm}, but small enough to fit into the area available to chiplet-level logic and routing.

The 64-bit network consists of a single crossbar. It connects the Linux-capable host and the 64-bit ports of the groups to the narrow \gls{d2d} segment, the 512-bit network, and various peripherals.
It also features a \SI{512}{\kibi\byte} \gls{spm} used by the host for management tasks.
The peripherals include UART, I2C, QSPI, GPIOs, a \SI{1.33}{\giga\bit\per\second} off-interposer \gls{c2c} link, and a JTAG test access point for live host processor debugging.
They also include {\riscv}-compliant timers and platform-level interrupt controllers providing interrupts for both the host processor and all on-chip compute cores.

The main benefit of our hierarchical crossbar-based interconnect over a ring or mesh topology is its \emph{symmetry}: the memory topology looks identical to all compute cores and clusters, meaning that the architectural bandwidth and latency of accesses to each hierarchy level (cluster, group, chiplet) are \emph{constant}.
This greatly simplifies programming, as network performance is homogeneous and code can be written in a \emph{cluster-agnostic} way without sacrificing performance.
However, crossbars pose challenges in physical implementation, requiring effort to provision sufficient bandwidth while managing area costs.
While this was feasible for our design, larger systems may face scalability limitations.
In this case, alternative \gls{noc} topologies such as mesh and torus could be more suitable.

The \gls{d2d} link enables seamless communication across chiplets, allowing the system to scale to 432 cores and two \gls{hbm2e} stacks.
This approach improves overall performance while avoiding the yield challenges and high manufacturing costs associated with large monolithic dies~\cite{naffziger202022amdchipletarc}.
The \gls{d2d} link consists of a narrow segment, optimized for synchronization and message passing, and a wide segment, designed for high-throughput bulk data transfers.

The wide segment is a scaled-up version of the narrow segment with 38 \glspl{phy} to increase bandwidth; its architecture is shown in \Cref{fig:arch_d2d}.
The \emph{protocol layer}, shown in \Cref{fig:arch_d2d}, arbitrates between \gls{axi4} requests and responses and converts them to \gls{axi4}-Stream payloads.
It handles the five independent channels of the \gls{axi4} protocol (AW, AR, W, R, B) and applies backpressure to the \gls{axi4} interface to prevent protocol-level deadlocks.
Each \teb{\gls{axi4}-Stream} payload includes a header that contains information about the packet type and credits required for flow control.
The \emph{data-link} layer further packetizes the payloads based on the available number of off-chip lanes;
it also handles credit-based flow-control to ensure that no packets are lost, as well as synchronization and alignment of packets if multiple \glspl{phy} are configured.
Additionally, the data-link layer features debugging capabilities, including a \emph{Raw Mode} that allows the link to operate independently of the \gls{axi4} interface by sending patterns over specific PHY channels for fault detection.
Each \gls{phy} features an all-digital and source-synchronous interface with eight \gls{ddr} lanes in each direction.
On the transmitter side, the PHY operates with a forwarded clock derived from the system clock. On the receiver side, the transmitted packets are synchronized with the system clock and reassembled into the original payload.
The narrow and wide segments can achieve effective duplex bandwidths of up to \SI{1.33}{\giga\bit\per\second} and \SI{64}{\giga\bit\per\second}, respectively.
The wide segment additionally features a \emph{channel allocator} to enable fault tolerance.
An initial calibration detects faulty \glspl{phy}, which can individually be disabled;
the channel allocator then reshuffles packets among the functional \glspl{phy} with only linear bandwidth degradation.
This fault tolerance mechanism ensures reliable communication across chiplets, even in the presence of manufacturing defects.

The {\occamy} chiplet relies on \glspl{fll} to generate on-chip clocks for each of its three clock domains: the \emph{compute} domain, the \emph{peripheral} domain, and the \emph{\gls{hbm2e} PHY} domain.
The \emph{compute} domain includes the compute groups, the 64-bit host, the chiplet-level interconnect, the \gls{d2d} link, and the \gls{hbm2e} controller.
Like the groups, the \gls{d2d} link and the \gls{hbm2e} subsystem can be clock-gated through memory-mapped configuration registers.

\subsection{Physical Design}

\begin{figure}
    \centering%
    \includegraphics[width=\linewidth]{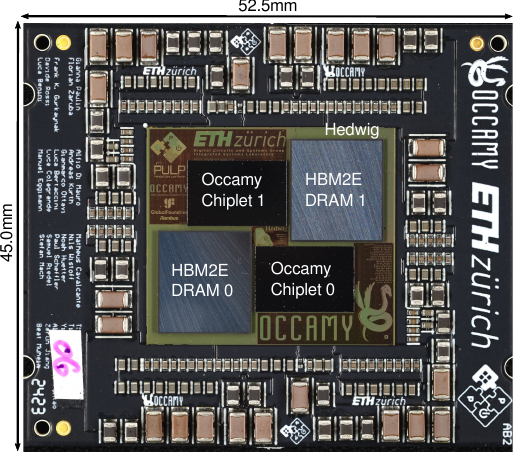}%
    \vspace{-0.5mm}%
    \caption{2.5D Occamy assembly mounted on carrier PCB.}%
    \label{fig:module}%
    \vspace{3.4mm}%
\end{figure}

The full Occamy 2.5D system was implemented and fabricated along with a dedicated carrier board, resulting in the compute module shown in \Cref{fig:module}.
The \SI{73}{\milli\meter^2} chiplets were fabricated in {\gfs} \SI{12}{\nano\meter} LP+ FinFET node using a 13-metal stack.
The two compute dies with their respective Micron \emph{MT54A16G808A00AC-32} \gls{hbm2e} stacks were mounted on \emph{Hedwig}, a passive, 4-metal-stack \SI{65}{\nano\meter} PKG-25SI interposer from \gf, shown in \Cref{fig:hedwig:render}.

\afterpage{%
\begin{sidefig}
    \centering%
    \hspace{0.5cm}
    \begin{subcaptionblock}{0.395\linewidth}
        \centering%
        \includegraphics[width=\linewidth]{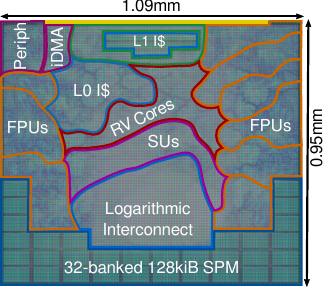}%
        \vspace{-1.3mm}%
        \caption{Compute cluster}%
        \vspace{0.5mm}%
        \label{fig:impl_cluster}%
    \end{subcaptionblock}\hspace{1.4cm}
    \begin{subcaptionblock}{0.355\linewidth}
        \centering%
        \includegraphics[width=\linewidth]{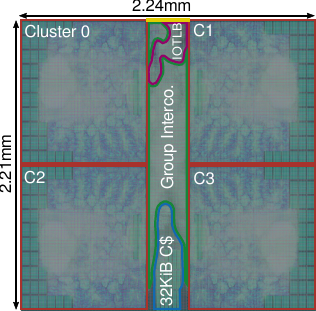}%
        \vspace{-1.3mm}%
        \caption{Compute group}%
        \vspace{0.5mm}%
        \label{fig:impl_group}%
    \end{subcaptionblock}\hfill

    \begin{subcaptionblock}{\linewidth}
        \centering%
        \includegraphics[width=0.94\linewidth]{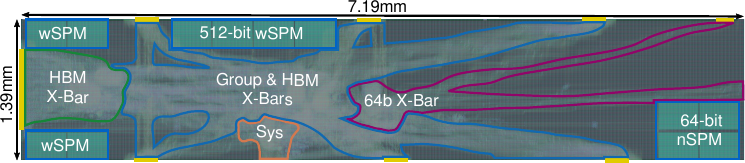}%
        \vspace{-1.3mm}%
        \caption{Chiplet-level interconnect}%
        \vspace{-2mm}%
        \label{fig:impl_interco}%
        \vfill%
    \end{subcaptionblock}\hfill
    \caption{Annotated physical layouts of the {\occamy} chiplet's hierarchical components. }
    \label{fig:impl_chip}
\end{sidefig}
}

\afterpage{%
\begin{sidefig}
    \centering%
    \begin{subcaptionblock}{\linewidth}
        \centering%
        \includegraphics[width=\linewidth]{fig-60.pdf}%
    \end{subcaptionblock}\hfill
    \caption{Annotated physical layouts of the {\occamy} chiplet (block IO is shown in yellow). }
    \label{fig:impl_chiplet}
\end{sidefig}
}

\begin{figure}[t]
    \includegraphics[width=\linewidth]{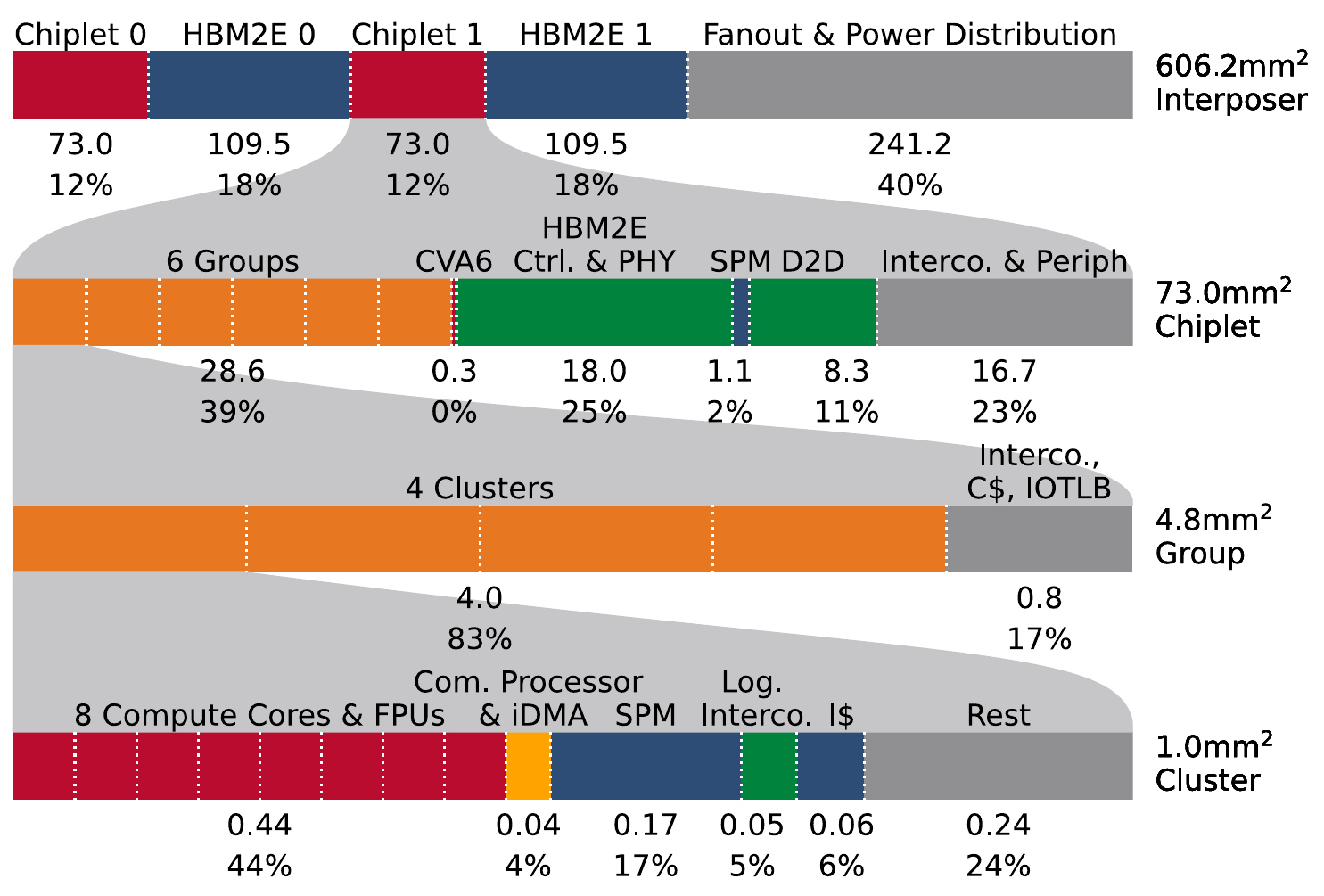}
    \caption{Hierarchical area breakdown of the Occamy system.}
    \label{fig:impl_area}
\end{figure}

\subsection{Occamy Chiplets}

The compute chiplets were synthesized, placed, and routed hierarchically using Synopsys' \emph{Fusion Compiler} 2022.3.
We used 7.5-track standard cell libraries from Arm and \gls{hbm2e} \glsunset{ip}\glspl{ip} (controller and \gls{phy}) provided by Rambus.
We targeted a nominal compute domain clock of \SI{1}{\giga\hertz} under typical conditions (\SI{0.8}{\volt}, \SI{25}{\celsius}) with a worst-case \teb{(\SI{0.72}{\volt}, \SI{125}{\celsius})} constraint of \SI{950}{\mega\hertz}.
The peripheral domain was constrained to \SI{500}{\mega\hertz} under worst-case conditions.
The \gls{hbm2e} controllers and \gls{phy} were constrained as specified to match the DRAM's peak \SI{3.2}{\giga\bit\per\second\per pin} throughput.

\Cref{fig:impl_area} presents a hierarchical area breakdown of an entire {\occamy} assembly.
\Cref{fig:impl_chip} \teb {and \Cref{fig:impl_chiplet} show} the resulting hierarchical chiplet layout; we describe our implementation in a bottom-up fashion.
In accordance with our findings in \cite{paulin2022softtilescaptur}, we arranged the compute cluster's IO ports and L1 instruction cache on one side and its \gls{spm} \glsunset{sram}\glspl{sram} in a U shape on the opposite side, resulting in the \SI{1.0}{\milli\meter^2} cluster layout shown in \Cref{fig:impl_cluster}.
The cluster is area-dominated by the nine {\riscv} compute cores with extended FPU functionality (\SI{44}{\percent}) and \gls{spm} (\SI{17}{\percent}).
The group layout, shown in \Cref{fig:impl_group}, is almost entirely comprised of its four cluster macros (\SI{83}{\percent}) and funnels its shared interconnect ports to a narrow interval on its north edge.
The chiplet interconnect is shown in \Cref{fig:impl_interco}; the global \gls{spm} \glspl{sram} were placed to avoid obstructing the shortest path for crossbar connections.
Further,  we implemented the host processor, the eight \gls{hbm2e} controllers, and the \gls{d2d} link as hierarchical macros to simplify integration.

The top-level chiplet layout is shown in \Cref{fig:impl_chiplet};
it is area-dominated by the six compute groups (\SI{39}{\percent}), \gls{hbm2e} interface (\SI{25}{\percent}), and \gls{d2d} link (\SI{11}{\percent}).
In accordance with the interposer arrangement, the west and south chiplet beachfronts are fully reserved for the \gls{hbm2e} and \gls{d2d} interfaces, respectively, while the remaining off-interposer IO drivers are kept on the north and west edges.
Designing a chiplet for interposers brings additional constraints to the floorplan.
First of all, in scaled technologies, complex \teb{\gls{ip}} like the \gls{hbm2e} physical interface are sometimes designed specifically for one orientation (either East-West or North-South).
Second is the arranmgement of dies on the interposer to satisfy both short signal propagation delays, stay within routing limitations of the interposer structure and consider mechanical stability of the assembled module (bending).
Occamy has been designed as a reserach vehicle, and therefore does not feature high-bandwidth external \teb{\gls{io}} reducing the constraints on off-module communication.
The central chiplet bumps are reserved for power delivery.

\subsubsection{Hedwig Interposer}
\label{sec:hedwig}

\begin{figure}
        \centering%
        \includegraphics[width=\linewidth]{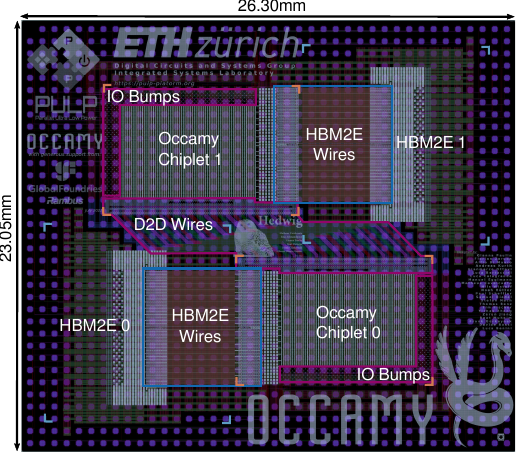}%
        \vspace{-0.5mm}%
        \caption{Layout of the Hedwig interposer.}%
        \label{fig:hedwig:render}%
        \vspace{3.4mm}%
\end{figure}

\begin{figure}
        \hspace{0.35cm}%
        \includegraphics[width=\linewidth]{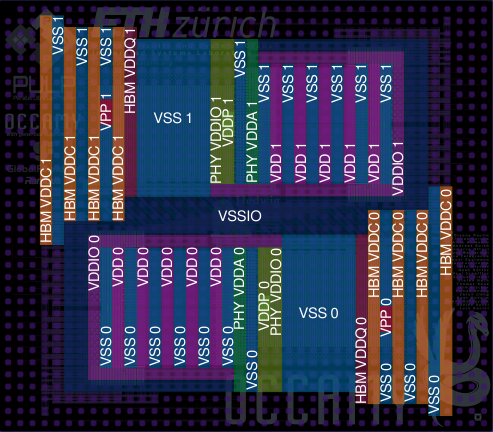}%
        \vspace{-0.5mm}%
        \caption{C4 bump map of the Hedwig interposer.}%
        \label{fig:hedwig:c4}%
\end{figure}

The Hedwig interposer is at the center of the 2.5D system, connecting the Occamy chiplets to each other, to their \gls{hbm2e} stacks, and to the carrier board.
On its top side, it exposes \SI{45}{\micro\metre}-wide \emph{microbump} openings accepting connections from the Occamy chiplets and \gls{hbm2e} stacks through \SI{15}{\micro\metre}-wide copper micropillars.
On its bottom side, {1399} custom-designed hexagonal \emph{C4 pads} with \SI{320}{\micro\metre} openings and a \SI{650}{\micro\metre} pitch allow us to solder the 2.5D assembly onto the carrier PCB using low-temperature bismuth-based \emph{C4 bumps}.

The connections between the Occamy chiplets and \gls{hbm2e} stacks were routed on the odd metal layers with a wire width of \SI{2.5}{\micro\metre} and a pitch of \SI{4.1}{\micro\metre}.
The \gls{hbm2e} wire length was kept below \SI{4.9}{\milli\metre}, and ground shields were introduced on the even layers to ensure signal integrity.
\Cref{fig:hedwig:render} highlights the \gls{hbm2e} wires; the routing density inside the rectangles approaches \SI{100}{\percent}.
The 39 \gls{d2d} link channels were length-matched to \SI{8.8}{\milli\metre} and combined into 24 wire bundles located at the center of Hedwig as shown in \Cref{fig:hedwig:render}.
The bundles were routed on the three upper metal layers in an alternating pattern, leaving the lowest layer for a shared ground connection.
A routing width of \SI{3.2}{\micro\metre} with a pitch of \SI{6.4}{\micro\metre} was chosen.

The digital IO of the Occamy chiplets and \gls{hbm2e} stacks was routed to Hedwig's edge using \SI{2.5}{\micro\metre} traces on the three uppermost layers;
placing IO-related C4 pads at the edge of Hedwig facilitates the fan-out to the carrier PCB.
\Cref{fig:hedwig:c4} shows Hedwig's C4 pad map, highlighting the IO-related connections as well as the sixteen power and three ground domains.
The ground pads of IO drivers of the two chiplets are connected through the shielding of the \gls{d2d} link, resulting in a shared IO ground \emph{VSSIO}.
For the rest of the power domains, each chiplet-\gls{hbm2e} pair has its own ground \emph{VSSx}.
The compute fabric, host, and \gls{hbm2e} controller on each chiplet are supplied through the core power net \emph{VDDx};
the remaining power domains supply the \gls{hbm2e} PHYs and memory stacks.
We designed Hedwig to keep the maximum current below \SI{540}{\milli\ampere} and \SI{34}{\milli\ampere} for each C4 pad and through-silicon via, respectively.

\subsubsection{Carrier PCB}
\label{sec:carrier-pcb}

The carrier PCB completes the Occamy assembly at the lowest level.
It serves three main purposes: giving the assembly mechanical stability, implementing the fan-out of the IO signals, and stabilizing the individual power domains.
The 12-layer stack-up with \SI{70}{\micro\metre}-thick copper foil and \emph{ROGERS RO4350B} high-stability, low-CTE laminate ensures high current delivery capabilities while providing sufficient mechanical strength.
Decoupling capacitors are placed close to Hedwig on the carrier to comply with the power delivery requirements of the \gls{hbm2e} stacks.

The carrier PCB implements an industry-standard \emph{LGA 2011-3} CPU socket interface compatible with off-the-shelf mainboard sockets, facilitating the creation of application boards.
Most of the 2011 pins are used for power and ground connections, with 877 exposed IOs.
The carrier is designed to be strong enough to protect the fragile, \SI{120}{\micro\metre}-thick Hedwig interposer from mechanical stress while being placed into and removed from the test socket.

\subsection{Bringup and Silicon Measurement Setup}

\begin{figure}
    \begin{subcaptionblock}{\linewidth}
        \centering%
        \includegraphics[width=0.69\linewidth]{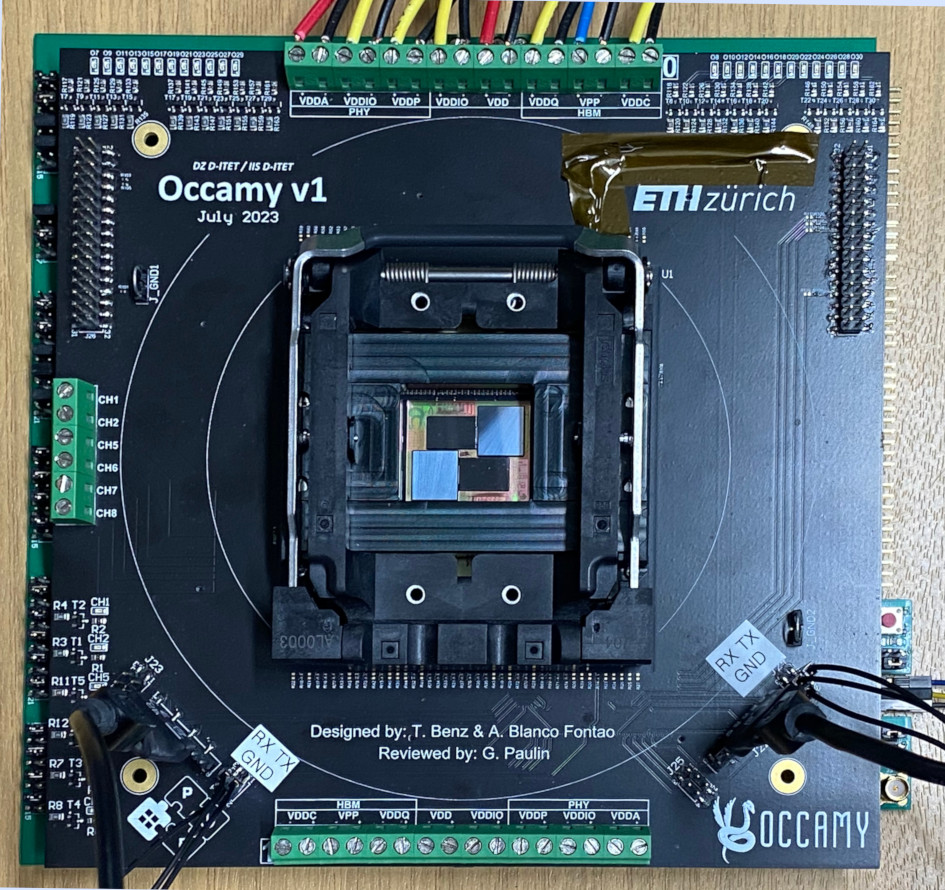}%
        \vspace{-1.7mm}%
        \caption{}%
        \vspace{3mm}
        \label{fig:tester-pcb}%
    \end{subcaptionblock}

    \begin{subcaptionblock}{\linewidth}
        \centering%
        \includegraphics[width=0.69\linewidth]{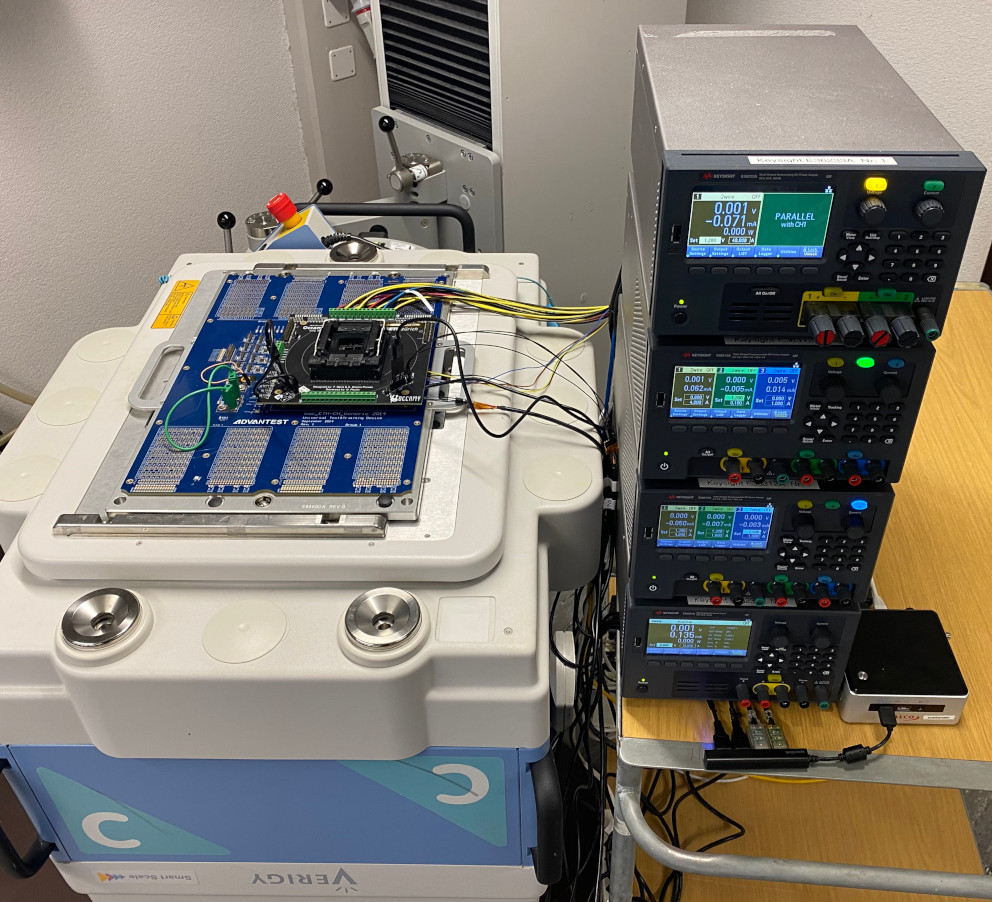}%
        \vspace{-1.7mm}%
        \caption{}%
        \label{fig:setup}%
    \end{subcaptionblock}\hfill
\caption{(a) Bringup board enabling both testing on an %
V93000 ATE and standalone operation, and (b) measurement setup.}
\label{fig:impl_board}
\end{figure}

\Cref{fig:tester-pcb} shows {\occamy}'s \emph{bringup} board, which consists of a stack of two individual PCBs.
The upper PCB holds an {LGA 2011} \gls{zif} socket for the {\occamy} system, connectors for the 16 power domains, an interface to a \emph{V93000} \gls{ate} system, as well as JTAG, UART, and GPIO headers.
For standalone operation, the lower PCB provides clocking resources, reset circuitry, configuration headers, and an SD card slot for each chiplet.

To evaluate {\occamy}, we connect the bringup board to an \gls{ate} system,
ensuring a stable, low-jitter clock delivery.
In all experiments, the temperature of the {\occamy} system is kept at 25°C through an active temperature forcing system.
We use Keysight \emph{E36200} series power supplies for power delivery and current measurements.
Two Digilent \emph{JTAG-HS2} programmers are used to program the {\occamy} system through {JTAG}, and two {FTDI} chips relay {UART} data to a testing workstation using \emph{DUTCTL} (\Cref{chap:dutctl}).

\newpage
\section{Experimental Results}
\label{sec:occamy:results}

\begin{figure}[t]
    \centering%
    \includegraphics[width=\columnwidth]{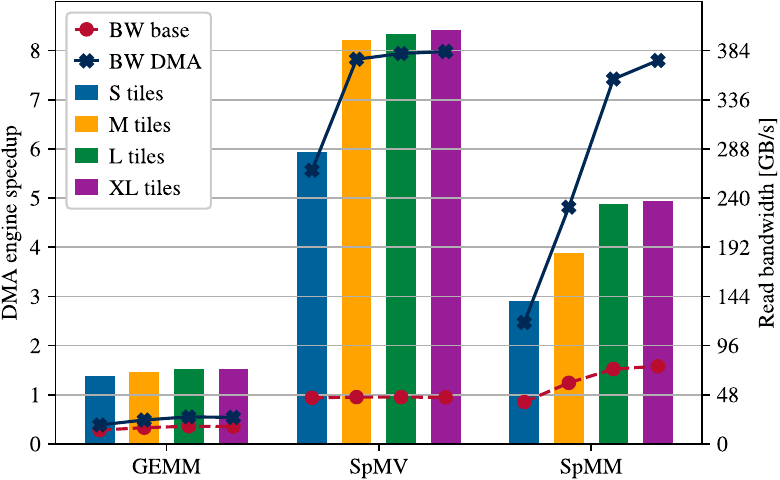}%
    \caption{%
        {\occamy} chiplet bandwidths and speedups enabled by {\idma} on workloads with varying tile sizes. %
    }%
    \label{fig:case:occamy:results}%
\end{figure}

We evaluate \gls{gemm}, \gls{spmv}, and \gls{spmm} on {\occamy} with and without the use of cluster {\dmaes}. %
We run double-precision tiles and use these results to compute the performance of a single chiplet, taking into account bandwidth bottlenecks and assuming all reused data is ideally cached. %

Each workload is evaluated with four cluster tile sizes \emph{S}, \emph{M}, \emph{L}, and \emph{XL}; for \gls{gemm}, these are square tiles of size 24, 32, 48, and 64, while the two sparse workloads use the matrices of increasing density \emph{diag}, \emph{cz2548}, \emph{bcsstk13}, and \emph{raefsky1} from the SuiteSparse matrix collection~\cite{davis2011theuniversityof} as tiles. %

As Snitch~\cite{zaruba2021snitchatinypseu} originally does not include a {\dmae}, we compare {\occamy} to an architecture where the data movement is handled by the worker cores.
\alc{In this base architecture, we assume the main interconnect to be bandwidth limited to \SI{64}{\bit} per cycle, corresponding to the \gls{axi4} bandwidth of a worker core, but we do not assume any limitation in the maximum number of outstanding transfers.}

\Cref{fig:case:occamy:results} shows our results. %
For the highly compute-bound \gls{gemm}, our {\idmaes} enable moderate, but significant speedups of \SI{1.37}{\x} to \SI{1.52}{\x}\!. %
Since all tile sizes enable ample cluster-internal data reuse, we see only small benefits as tiles grow. %
Nevertheless, the cluster engines still increase peak HBM read bandwidth from 17 to \SI{26}{\giga\byte\per\second}. %

\gls{spmv} performance, on the other hand, is notoriously data-dependent and memory-bound due to a lack of data reuse. %
Unable to leverage on-chip caches fed by a wider network, the baseline nearly saturates its narrow interconnect for all tile sizes at \SI{48}{\giga\byte\per\second}. %
The {\idmaes} only become memory-bound past \emph{M}-sized tiles, but then approach the wide interconnect peak throughput of \SI{384}{\giga\byte\per\second}. %
Overall, the engines enable significant speedups of \SI{5.9}{\x} to \SI{8.4}{\x}\!. %

\gls{spmm} is similar to \gls{spmv}, but enables on-chip matrix data reuse, becoming compute-bound for both the baseline and {\idmaes}. %
Since data caching is now beneficial, the baseline overcomes the \SI{48}{\giga\byte\per\second} bottleneck and speedups decrease to \SI{2.9}{\x} to \SI{4.9}{\x}\!. %
Still, {\idmaes} unlock the full compute of clusters on sparse workloads while \alc{asymptotically} approaching the  \SI{384}{\giga\byte\per\second} peak throughput only for \emph{XL} tiles. %

\newpage
\section{Summary and Conclusion}
\label{chap:comcpu:conclusion}

In this chapter, we present tightly coupled instruction-based {\fe}, called \emph{inst\_64}.
Coupled with our 2-D tensor {\me} and {\idma} {\be}, we create an agile, tightly coupled \emph{data movement accelerator} to be used in Snitch-based systems.

With our \emph{dm} instructions, we define a light-weight {\riscv} \gls{isa} extension for {\dmaes}, allowing to launch linear transfers in as few as three instructions and 2-D tensor transfers in as few as 5 instructions.
Coupled with our very agile {\idma} architecture, we can thus launch \gls{dma} jobs in as low as five cycles from the issue of the first \emph{dm} instruction to the first data element issued on the \gls{axi4} interface.
We further provide additional instructions for memory initialization and user-defined extensions, including issuing multicast operations.

We integrate our tightly coupled {\dmae} into the Snitch cluster by introducing a dedicated \teb{\emph{communication processor}} to orchestrate the \emph{worker cores} and manage the data movement operations on a high level.
We further optimize Snitch's memory architecture by introducing the concept of wide superbanks and prioritized access via remote units and cluster-local {\idmaes} into cluster-local \gls{tcdm} while still employing a physically implementable single-cycle logarithmic interconnect.

We present the feasibility of our approach by integrating our engine into {\occamy}, a 2.5D multi-chiplet 432-core {\riscv} system featuring \gls{hbm} memory endpoints.
We optimize our {\idmaes} and the corresponding wide cluster, group, and chiplet-level interconnect hierarchies to maximize cluster-to-cluster and cluster-to-\gls{hbm} throughput.

Running both dense and sparse workloads on Occamy, we can report speedups of up to \SI{8.4}{\x}\! whilst only increasing cluster area by less than \SI{5}{\percent}.

\chapter{Real-time Interconnect Extensions}
\label{chap:realm}

\section{Introduction}
\label{chap:realm:introduction}

The current trend in industrial domains such as automotive, robotics, and aerospace is towards autonomy, connectivity, and electrification, significantly increasing the demand for onboard computing power and communication infrastructure, thus driving a paradigm shift in their design~\cite{srcmaptmicroelectr, burkackygettingreadyfor, mutschlerautomotiveoemsf, jiang2023towardshardreal, kasarapu2025performanceande, fletcherthecaseforanend}.

A clear example is the automotive domain, where the traditional approach --- relying on hundreds of embedded real-time \glspl{ecu} distributed throughout the vehicle --- cannot meet the growing compute demands and complicates cable harness management, impacting \gls{swapc}~\cite{fletcherthecaseforanend, lim2025aframeworkforde, pinto2019virtualizationo}.
This paradigm cannot support the rapid shift toward \gls{aces}, which is laying the foundation of \gls{adas} and \glspl{sdv}~\cite{burkackygettingreadyfor}.
Hence, integrated, interconnected \emph{zonal} and \emph{domain} architectures are becoming the preferred replacements for discrete \glspl{ecu}, as they deliver the flexibility and compute capability required for \gls{aces} mobility and the \gls{swapc} problem~\cite{jang2023designofahybrid, pinto2019virtualizationo}.

These architectures are heterogeneous \glspl{mcs}~\cite{esper2018anindustrialvie, burns2017asurveyofresear}.
They comprise general-purpose and domain-specific sub-systems with diverse real-time and specialized computing requirements that execute concurrently on the same silicon die, sharing communication, storage, and micro-architectural resources~\cite{majumder2020partaaarealtime, sá2022afirstlookatris}.
Some subsystems handle hard safety- and time-critical workloads, such as engine, brake, and cruise control~\cite{fletcherthecaseforanend, pagani2019abandwidthreser, gray2023axiicrttowardsa}, while others run less time-critical but computationally demanding tasks like perception pipelines, infotainment, and commodity applications~\cite{fletcherthecaseforanend}.

Time- and safety-critical tasks require strict real-time guarantees, ensured through time-predictable run-time mechanisms, composable timing analysis, and safety assessments~\cite{wilhelm2008theworstcaseexe}.
However, in heterogeneous \glspl{mcs}, this process is complicated by the increased interference generated by multiple domains contending for shared hardware resources on the same platform~\cite{pagani2019abandwidthreser}.
This additional contention may introduce unpredictable behavior during the system's execution, causing possible deadline misses for time- and safety-critical tasks~\cite{restuccia2020axihyperconnect, certificationpositionpaperca}.
To preserve the timing behavior of the system under known and predictable bounds, techniques such as spatial and temporal isolation become a prerequisite, as they enhance the \emph{observability} and \emph{controllability} of shared \gls{hw} resources~\cite{rehm2021theroadtowardsp, jiang2023towardshardreal}.
The \emph{interconnect} in modern \glspl{soc} is of particular concern; several previous works have highlighted that interference in accessing shared resources regulated by bus arbiters and interconnects is a major source of unpredictability~\cite{pagani2019abandwidthreser, restuccia2019isyourbusarbite, restuccia2020axihyperconnect, gray2023axiicrttowardsa}.

In this chapter, we present {\axirealm}, an \glsu{axi4}-based, interconnect extension that improves real-time and predictability behavior of \glspl{mcs} by monitoring and controlling both the \emph{ingress} and \emph{egress} data streams.
The architecture is split into two sub-systems: \emph{\irealm}, tasked to guard and regulate the \emph{ingress} stream (requests and data issued by managers) with a \emph{budget and time-slicing} approach; \emph{\erealm}, tasked to supervise the \emph{egress} stream (data and responses issued by subordinates) with bandwidth and latency statistics, and eventually protect the system from malfunctioning subordinate devices.

In the following sections, we provide a detailed description of the internal \gls{hw} components,
we include a sequence diagram presenting an example write transaction passing through the various sub-units,
combine the ingress~\cite{benz2024axirealmalightw, benz2025axirealmsafemod} with the egress units~\cite{liang2025towardsreliable} to a unified {\axirealm} system,
capable of shaping the ingress traffic and protecting the interconnect and the system from malfunctioning subordinate devices (\Cref{chap:realm:architecture}),
we provide an extensive evaluation at the system-level by integrating {\axirealm} in an open-source \gls{mcs} characterizing functional, energy, and power performance (\Cref{chap:realm:carfield}), we give an \glsu{ip}-level evaluation (\Cref{chap:realm:archres}), and we provide a detailed \gls{soa} comparison for the unified system (\Cref{chap:realm:relwork}).
Finally, \Cref{chap:realm:conclusion} concludes \lb{this chapter} by summarizing its key contributions and achievements.
In more detail, this thesis provides the following contributions to \gls{soa}:

\begin{itemize}

    \item \textbf{\axirealm:} We present a scheme to enforce predictable behavior compatible with any \gls{axi4}-based interconnect which relies on \emph{observing} and \emph{controlling} both its \emph{ingress} and \emph{egress} data streams using ad-hoc \gls{hw} methodologies.
    The resulting architecture demands minimal additional hardware resources and no internal modifications to the baseline interconnect, enabling portability across diverse \gls{soc} targets.

    \item \textbf{\Gls{hw}-driven traffic controllability:} {\axirealm}'s {\irealm} unit implements a configurable number of subordinate \emph{regions} per manager.
    Each \emph{region} is runtime-programmable with address range, transfer fragmentation size, transfer budget, and reservation period to control the bus traffic through a \emph{time slicing} approach.

    \item \textbf{\Gls{hw}-driven traffic observability:} {\axirealm}'s {\irealm} and {\erealm} units include modules that observe and track per-manager access and interference statistics, such as transaction latency, bandwidth, and interference with each other manager.
    With bandwidth-based observability, {\axirealm} can perform per-manager bandwidth throttling, modulating back-pressure.

    \item \textbf{\Gls{hw}-driven safety measures for malfunctioning subordinates:} We include a \gls{hw} mechanism~\cite{liang2025towardsreliable} to isolate and reset malfunctioning subordinates individually, taking advantage of {\axirealm}'s {\erealm} latency tracking capabilities to identify response timeouts, mismatching transactions, and invalid handshakes.

    \item  \textbf{IP-level characterization:} We extensively characterize {\axirealm} in a \SI{12}{\nano\metre} technology, presenting an area model as well as timing and latency information.

    \item \textbf{In-system implementation assessment:} We evaluate {\axirealm} in an open-source heterogeneous \gls{mcs} research platform\,\footnote{\url{github.com/pulp-platform/carfield}}.
    We demonstrate the versatility of the proposed approach under interference scenarios between critical and non-critical managers of the system,
    achieving at least \SI{68}{\percent} of the single-initiator case (over \SI{95}{\percent} when distributing the budget in favor of the critical manager).
    Further, the proposed transfer fragmentation reduces the access latency of the critical manager by \SI{255}{cycles} from 266 to \SI{11}{cycles}.
    {\axirealm} incurs an area overhead of less than \SI{2}{\percent} in the presented \gls{mcs}.

\end{itemize}

The synthesizable and silicon-proven register transfer level description of {\axirealm} and its integration into the presented real-time system are available open-source under a libre Apache-based license\,\footnote{\url{github.com/pulp-platform/axi_rt}}.

\section{Background}
\label{sec:bkgrnd}

\subsection{The AMBA AXI4 On-Chip Interconnect}

\Gls{axi4} is an industry-standard protocol for high-bandwidth, non-coherent, on-chip communication.
It defines five separate channels for read and write requests (\emph{AR}, \emph{AW}, \emph{W}) and responses (\emph{R}, \emph{B}).
An \gls{axi4} \emph{beat} is the communication in one cycle on an \gls{axi4} channel~\cite{arm2023ambaaxiandacepr}.
An \gls{axi4} \emph{transaction} is the number of beats a manager requires to communicate to a subordinate.
The manager initiates a transaction by emitting an \emph{address and control beat} containing the meta information (address, attributes, and length in beats, ...) over either the \emph{AR} or the \emph{AW} channel.
The \emph{burst} attribute defines the increment mechanism of the write or read addresses during a transaction.

Each \gls{axi4} transaction carries a \emph{\gls{tid}}.
All beats in a transaction must have the same \gls{tid}.
The subordinate completes a transaction by sending a response over the \emph{B} channel in the write case or by returning the last read response over the \emph{R} channel read case.
The protocol also supports multiple \emph{outstanding} transactions, i.e., initiated by the same manager and simultaneously in progress with the same \gls{tid}.

Based on \gls{tid}, the protocol dictates three \emph{ordering rules} for write and read transactions.
We recall them in the following:
\circnum{1} for different \glspl{tid}, write data on the \emph{W} channel must follow the same order as the address and control beats on the \emph{AW} channel, as \emph{W} beats do not have a \gls{tid} field;
\circnum{2} transactions with different \glspl{tid} can be completed in any order; on the \emph{R} channel, the read data can be interleaved;
\circnum{3} a manager can have multiple outstanding transactions with the same \gls{tid}, but they must be performed and completed in the order they were requested, for both writes and reads.

This work uses an open-source and silicon-proven implementation of \gls{axi4} network elements\,\footnote{\url{github.com/pulp-platform/axi}}.
We refer to a \emph{crossbar} as the main point-to-point network junction between managers and subordinates in the systems (\Cref{sec:arch:ico} and \Cref{chap:realm:carfield}).

\subsection{MCS Terminology: Essential Insights}

\emph{Criticality} designates \emph{"the level of rigor required to develop safety-critical functions so that the risk of failure can be brought to an acceptable level"}~\cite{esper2018anindustrialvie}.
An \gls{mcs} involves applications with different criticality requirements deployed on the same platform.
Safety functions in \glspl{mcs} are treated as belonging to the highest safety integrity level unless \emph{independence} between them is guaranteed, i.e., applications achieve freedom from interference with each other.
This implies demonstrating that (i) independence is achieved in both spatial and temporal domains, and (ii) violation of independence can be controlled (see \cite{iec615083functi}, Sect. 7.4.2.9).

A way to achieve independence is through \emph{isolation}, or partitioning, of \gls{hw} resources and \gls{sw} components.
Isolation allows the segregation of faults, improves predictability by providing bounds on resource access times~\cite{cilku2013towardstemporal}, and reduces the \gls{sw} \gls{vv} effort~\cite{esper2018anindustrialvie}.
\emph{Physical} isolation relies on federated \gls{hw} for each \gls{sw} component.
Hence, resources at all levels are physically decoupled.
\emph{Virtual} isolation establishes partitioned \gls{hw} provisions that allow multiple \gls{sw} components to run on the same \gls{hw} platform, namely, an \gls{mcs}~\cite{burns2017asurveyofresear}.

Within \emph{virtual isolation}, we distinguish between \emph{spatial} and \emph{temporal isolation}.
Spatial isolation means that an application shall not change data used by another application; it can be achieved with virtualization techniques through a \gls{mmu}~\cite{rehm2021theroadtowardsp}.
However, on an integrated \gls{hw} platform, virtual partitions still share resources such as caches, interconnects, and memory endpoints, making their temporal behavior inter-dependent~\cite{cilku2013towardstemporal, burns2017asurveyofresear}.
Temporal isolation ensures that one application will not cause malfunction of another application by blocking a shared resource over time or consuming another resource execution time.
This can be achieved with adequate scheduling methods in \gls{sw}, or \gls{hw}/\gls{sw} time slicing and fencing~\cite{esper2018anindustrialvie}.

Current industrial and academic activities around \glspl{mcs} do not share a holistic view of fully exploiting \gls{hw}'s potential to isolate executing application layers.
However, some initiatives are being developed and studied~\cite{arm2023armarchitecture}.
A promising direction is that \gls{hw} can cooperate with \gls{sw} by enabling fine-grained \emph{observability} and \emph{controllability} of individual application behavior~\cite{rehm2021theroadtowardsp}.
Proceeding from this premise and terminology background, this work aims to improve the observability and controllability of shared interconnect buses by leveraging time slicing, a temporal isolation technique.

\newpage
\section{Architecture}
\label{chap:realm:architecture}

\begin{figure}[ht!]
    \centering
    \includegraphics[width=0.9\linewidth]{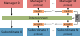}
    \caption{%
        Overview of a generic system extended with {\axirealm}.
        The {\irealm} units monitor and control data from the managers and {\erealm} units guard the subordinate devices.
    }
    \label{fig:realm-system}
\end{figure}

An overview of a generic system extended with {\axirealm} is provided in~\Cref{fig:realm-system}.
At the \emph{ingress} of the interconnect, \emph{\irealm} units~\cite{benz2024axirealmalightw, benz2025axirealmsafemod} monitor and shape traffic injected by managers, enforcing fairness and reducing congestion within the network as well as at the target devices.
The {\irealm} unit tracks the bandwidth and budget on the granularity of a \emph{region}.
Each region can encompass a subordinate space, combine multiple subordinates, or only cover a fraction thereof.
The number of supported regions can be set through a \gls{sv} parameter at design time, and the address space covered by each region through \gls{sw} at runtime.
This is explicitly designed to be independent of the addressing of the interconnect.

At the \emph{egress}, \emph{{\erealm}} units~\cite{liang2025towardsreliable} guard the subordinate devices. They protect the interconnect and prevent deadline misses of real-time tasks in the case of protocol failures and subordinate regions/devices that extensively delay their responses. %
Our {\erealm} unit provides two messaging options to inform the core of the unresponsive subordinates: interrupts and \gls{axi4} protocol responses.
Moreover, it integrates a framework to isolate, reset, and reinitialize malfunctioning devices within a single cycle~\cite{benz2021a10coresocwith2}.
\begin{sidefig}
    \centering
    \includegraphics[width=0.99\linewidth]{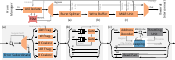}
    \caption{%
        Internals of the {\irealm} unit: (a) \emph{granular burst splitter}, (b) \emph{write buffer}, and (c) \emph{management and regulation unit}.
    }
    \label{fig:irealm-unit}
\end{sidefig}
\FloatBarrier

\begin{figure}[H]
    \centering
    \includegraphics[width=0.85\linewidth]{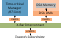}
    \caption{%
        \alc{Block diagram of an exploratory system featuring two regulated managers, a crossbar-based main interconnect using \gls{rr} arbitration and a guarded subordinate device.
        Configuration interfaces are not shown.}
    }
    \label{fig:seqbd}
\end{figure}

\subsection{The {\irealm} Unit and Architecture}
\label{sec:arch:irealm}

The {\irealm} unit comprises three main submodules, shown in~\Cref{fig:irealm-unit}: the \emph{burst splitter} (a), the \emph{write buffer} (b), and \gls{mtunit} (c).
At the {\irealm} unit's input, an \gls{axi4} isolation block isolates the manager during dynamic reconfiguration of the unit, once the manager's assigned budget expires, or when commanded through \gls{sw}.

\alc{\Cref{fig:seqbd} shows an exemplary system equipped with {\axirealm} featuring two regulated managers, a crossbar using \gls{rr} arbitration and a guarded subordinate device.}
\Cref{fig:seq} shows the function of the {\irealm} \alc{and {\erealm}} units at two exemplary write transactions\alc{; one} from a time-critical manager and \alc{one from} a \gls{dma} engine (\Cref{fig:seqiso}).
\alc{Only the three write-related channels --- see \Cref{sec:bkgrnd} --- of the \gls{axi4} buses are shown.}
\alc{The \gls{dma} requires five cycles to read the data from \teb{a memory internally to the} \gls{dsa} before issuing its first write beat, see \Cref{fig:seqiso}.}
\begin{figure}[H]
    \centering%
    \begin{subcaptionblock}{0.95\linewidth}%
        \centering%
        \includegraphics[width=\linewidth]{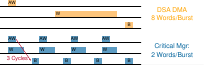}%
        \caption{Isolated waveforms after interconnect.}%
        \vspace{3mm}
        \label{fig:seqiso}%
        \vfill%
        \centering%
        \includegraphics[width=\linewidth]{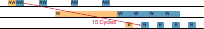}%
        \caption{Unregulated waveform after the interconnect.}%
        \vspace{3mm}
        \label{fig:sequnreg}%
        \vfill%
        \centering%
        \includegraphics[width=\linewidth]{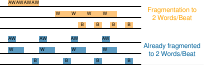}%
        \caption{Regulated waves inside of {\irealm} after the \emph{burst splitter}.}%
        \vspace{3mm}
        \label{fig:seqsplitter}%
        \vfill%
        \centering%
        \includegraphics[width=\linewidth]{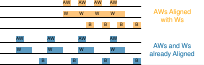}%
        \caption{Regulated waves inside of {\irealm} after the \emph{write buffer}.}%
        \vspace{3mm}
        \label{fig:seqbuffer}%
    \end{subcaptionblock}
    \caption{Write transaction passing through our {\irealm} unit. (continued on next page)}
    \label{fig:seq}
\end{figure}
\begin{figure}[H]
    \ContinuedFloat
    \centering%
    \begin{subcaptionblock}{\linewidth}%
        \centering%
        \includegraphics[width=\linewidth]{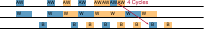}%
        \caption{Regulated waveform after the interconnect.}%
        \vspace{1mm}
        \label{fig:seqrealm}%
        \vfill%
        \centering%
        \includegraphics[width=\linewidth]{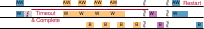}%
        \caption{Incomplete transaction gets completed by {\erealm} and is restarted.}%
        \vspace{1mm}
        \label{fig:seqerror}%
    \end{subcaptionblock}
    \caption{Write transaction passing through our {\irealm} unit. (continued from previous page)}
    \label{fig:seq_cont}
\end{figure}
In the \emph{unregulated} case, \Cref{fig:sequnreg}, the critical transaction experiences a completion latency of up to fifteen cycles.
With our {\irealm} unit activated, the \gls{dma} beats are fragmented after the burst splitter, \Cref{fig:seqsplitter}, and bandwidth reservation is mitigated, \Cref{fig:seqbuffer}, by stalling \emph{AWs} until their corresponding \emph{W} beats arrive.
In this \emph{regulated} case, \Cref{fig:seqrealm}, the transaction latency of the critical manager is at most four cycles.
\Cref{fig:seqerror} shows the case of a critical subordinate device timing out and how {\erealm} completes and restarts this pending transaction.

This section explains the architecture and functionality of the {\irealm} unit, detailing how it addresses unfairness from unregulated burst-based communication in \gls{rr}\teb{-}arbitrated interconnect systems.
It describes how the unit ensures execution predictability using time slicing through static or dynamic budget and period assignments to the managers.

\subsubsection{Granular Burst Splitter}
\label{sec:arch:split}

On-chip interconnects can employ burst-based transactions to increase the efficiency of non-coherent interconnect architectures.
Such transactions increase bus utilization and decrease the addressing overhead.
In heterogeneous \glspl{soc}, transactions of different granularities, e.g., short, cache-line-sized transactions issued by a core and a long burst requested by a \gls{dsa}, are common transaction patterns~\cite{restuccia2020axihyperconnect}.
Classic and fair \gls{rr} arbitration on individual transactions affects bandwidth distribution fairness by increasing the completion latency of short, fine-granular transactions in the presence of long bursts.

As shown in \Cref{fig:irealm-unit}b, the \emph{burst splitter} accepts incoming burst transactions and splits them to a runtime-configurable granularity, from one to 256 beats, according to the \gls{axi4} specification~\cite{arm2023ambaaxiandacepr}.
Any transaction not supported by the burst-splitter is rejected and handled by an \emph{error subordinate}.
For instance, atomic bursts and \emph{non-modifiable} transactions of length sixteen or smaller cannot be fragmented~\cite{arm2023ambaaxiandacepr}.
We store a burst transaction's meta information (address, transaction size, \gls{axi4} flags), emit the corresponding fragmented transactions, and update the address information.
Write responses of the fragmented bursts are coalesced transparently.
Read responses are passed through, except for the \emph{R.last} signal, which is gated according to the length of the original transaction. %
A large granularity requires the write buffer module following the burst-splitter to be large enough to hold a single fragmented write burst.
If a manager only emits single-word transactions, the granular burst splitter can be disabled from the {\irealm} unit to reduce the area footprint. %

\subsubsection{Write Buffer}
\label{sec:arch:buf}

The meta information of a transaction is inherently tied to the data being written.
\gls{axi4} physically decouples meta information from the payload to increase bus efficiency~\cite{arm2023ambaaxiandacepr}.
However, the write data beats and meta information are not fully decoupled as the write channel does not have a \gls{tid} field.
Most interconnect architectures reserve the bandwidth for an entire write transaction on the \emph{W} channel once the corresponding \emph{AW} is received~\cite{arm2023ambaaxiandacepr}.
Additionally, according to the \gls{axi4} standard, the \emph{W} channel remains indefinitely allocated to the request's issuer once the request has been propagated through the interconnect.
The standard does not specify a maximum delay between the propagation of the request and the provisioning of the corresponding data.
A manager device can reserve a large transaction by holding the \emph{W} channel, potentially stalling the interconnect by delaying data injection.
In practice, this mechanism is observed with slow manager devices or \gls{dma} units copying data from high-latency or bandwidth-limited endpoints, which cause interference in the downstream memory system, as discussed in~\cite{restuccia2022cutandforwardsa}.
We prevent this behavior by storing the fragmented write burst in a \emph{write buffer}, \Cref{fig:irealm-unit}b.
The buffer forwards the \emph{AW} request and the \emph{W} burst only if the write data is fully contained within the buffer.
The transaction buffer is configured to hold two \emph{AWs} and one fragmented write burst.

\subsubsection{Monitoring and Regulation Unit}
\label{sec:arch:mr}

In contrast to safety- and time-critical tasks executed on general-purpose processors, \glspl{dsa} often work independently~\cite{pagani2019abandwidthreser} and employ double buffering using their large internal memories; this results in memory-intensive phases followed by compute-intensive phases.
The asynchronous nature of \glspl{dsa} accessing the system's memory coupled with coarse-grained synchronization results in unpredictable memory access patterns, increasing the timing uncertainty of critical tasks.

The \gls{mtunit}, presented in \Cref{fig:irealm-unit}c, uses a \teb{period-based hardware-implemented} bandwidth limiting mechanism to prevent managers from injecting more bandwidth for each subordinate region into the network than allowed.
The \gls{mtunit} is symmetrically designed with identical read and write components.
Transactions first pass through the address decoder, which maps them to their respective subordinate region.
A \emph{bus probe} measures the transaction bandwidth and latency, providing this data to the \emph{bookkeeping} unit, which is responsible for budget checks and monitoring bandwidth and latency.
Each {\irealm} unit can track the data sent to each region.

A different time period and budget can be specified for each {\irealm} unit and each subordinate region.
Once activated, this specified budget amount is available and is reduced by every beat passing the unit.
Once one subordinate region's budget exceeds the allocated amount, the number of outstanding transactions is reduced.
The corresponding manager is completely isolated once one budget is depleted using the {\irealm} unit's isolation cell, see \Cref{fig:irealm-unit}.
The budget is automatically renewed once the time period expires.

If the total budget assigned to each manager's {\irealm} unit is less or equal to what the interconnect and the subordinate devices can handle within a period, the {\axirealm} system ensures each manager can use its assigned budget.
The budget distribution should be set according to the real-time task running on the general-purpose cores~\cite{pagani2019abandwidthreser}.
The time period, budget checking, and budget renewal are tracked and handled entirely in hardware, allowing the system to react with clock-cycle accuracy.
This allows the {\irealm} unit to set very short periods, ensuring agile regulation and fair bandwidth sharing in the presence of a \gls{dsa} manager.

Furthermore, we extend this monitor to track the average transaction latency issued through the {\irealm} unit.
After a simple profiling run measuring the latency of each manager to each subordinate region in isolation during \gls{vv}, average completion latency can be used to reveal inter-manager interference within the network and its subordinate regions or devices.
Thus, online performance data can help fine-tune the budget and period settings for each manager and assess how well {\irealm} ensures time-critical tasks meet their deadlines.

\subsection{Interconnect layer}
\label{sec:arch:ico}

We design {\axirealm} to be independent of the system's memory architecture, except for fundamental properties.
Our approach expects the interconnect to ensure a progress guarantee and route transactions using \gls{rr} arbitration, which is the most common policy in commercial interconnects~\cite{restuccia2020axihyperconnect, gray2023axiicrttowardsa, serranocases2021leveraginghardw}.
{\axirealm} is primarily intended for mixed-critical systems where \glspl{dsa} require a high-performance interconnect to satisfy their data demand, and critical actors rely on real-time guarantees.
We specifically choose an \gls{rr}-arbitration mechanism over more classical real-time distribution patterns, like \gls{tdma}~\cite{poletti2003performanceanal}, to maintain the high-performance memory access required by the \glspl{dsa} in today's emerging heterogeneous systems.
We aim to enhance the determinism and fairness of a classic \gls{rr}-arbitrated interconnect by minimally intruding on its design, using lightweight helper modules at its boundaries.

Thanks to the independence from instance-specific assumptions on the interconnect architecture and to the compliance with the \gls{axi4} specification of our {\axirealm} architecture, verification and maintenance are facilitated by allowing the use of unit-level standalone verification infrastructure.

As mentioned in~\Cref{sec:bkgrnd}, the in-system case study and \gls{ip}-level evaluation presented in this work use a point-to-point-based interconnect constructed from \gls{axi4} crossbars.
{\axirealm} can handle hierarchical point-to-point interconnects thanks to the concept of subordinate regions.
While outside of the scope of this work, we also ensured compatibility of {\axirealm} with \gls{axi4}-based \gls{noc} architectures, such as presented in~\cite{fischer2025floonoca645gbsl}.

\begin{figure}[t]
    \centering
    \includegraphics[width=0.85\linewidth]{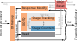}
    \caption{%
        The internal architecture of the {\erealm} unit. The read and write parts are constructed equally.
    }
    \label{fig:erealm-unit}
\end{figure}

\subsection{The {\erealm} Unit and Architecture}
\label{sec:arch:erealm}

The {\erealm} unit uses a \gls{hw}-based approach to monitor the latency of transactions sent to a subordinate device or region, responding promptly to unexpected misbehavior or malfunctions that disrupt the predictability of real-time tasks.
This unit is essential because the assumption of perfect behavior by subordinate devices, often made in \gls{soa}~\cite{restuccia2019isyourbusarbite, restuccia2020axihyperconnect}, does not hold in real-world scenarios.
Additionally, {\erealm} proactively ensures protocol compliance for all transactions without impacting system throughput or latency.

When responses from a subordinate device exceed user-programmable timeouts or when a protocol violation is detected, {\erealm} completes the outstanding transactions and communicates the cause of the issue with the core either using interrupts or the \gls{axi4} response channel.
Error information, including \gls{tid}, address, and the \emph{specific transaction stage} in which the error occurred, are logged into registers.
The unit can reset the connected subordinates through an agile reset controller~\cite{benz2021a10coresocwith2} either within one clock cycle upon fault detection, or when commanded by the core as part of the fault handling.
Overall, {\erealm} guards subordinate devices, guaranteeing responses within user-defined time frames, preventing the interconnect from locking up.

The architecture of {\erealm} is shown in \Cref{fig:erealm-unit}.
An ID remapper at the unit's input compacts the typically sparsely used \gls{tid} space, requiring fewer \gls{tid} bits to track all transactions.
The data path is then split into a similarly constructed \emph{Write} and a \emph{Read module}, presented in more detail below.

\subsubsection{Dynamic Outstanding Transaction Queue (DOTQ)}

{\axirealm} supports multiple outstanding,  multi-id transactions commonly occurring when accessing high-performance subordinate devices; requiring dynamic tracking of multiple data streams, each with several outstanding transactions.

The {\erealm} unit manages this through a dynamic queue consisting of three linked tables: an \emph{ID Head-Tail (HT)} table, a \emph{Transaction Linked Data (LD)} table, and a \emph{Write (W)} or \emph{Read (R)} table present in the write and read path, respectively.
The HT table keeps track of active \glspl{tid} and enforces ordering for transactions with the same \gls{tid} and supports efficient \gls{tid} lookups without scanning through all transactions in the LD table.
The LD table stores metadata such as \gls{tid}, address, state, latency, and timeout, allowing detailed tracking of each outstanding transaction.
Finally, the W/R table ensures that write data maintains the correct sequence with address beats, aligning data properly even when \glspl{tid} are not explicitly available on the write data channel.
The tracking capacity is defined by two design parameters: the maximum number of unique \glspl{tid} and the maximum number of transactions each unique \gls{tid} can support simultaneously.

\subsubsection{Stage-Level Tracking for each Transaction}
\label{sec:arch:erealm:stage}

\begin{figure}
    \centering
    \includegraphics[width=\linewidth]{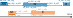}
    \caption{%
        Tracked stages in the {\erealm} unit.
    }
    \label{fig:erealm-stage}
\end{figure}

\gls{axi4} transactions occur in multiple \emph{stages}: address and meta information is sent first, followed by the data beats, and finally, the response stage.
As shown in \Cref{fig:erealm-stage}, the counter-based tracking logic monitors \emph{six} and \emph{four} stages for write and read transactions, respectively.
The first stage is the initial handshake transferring address and meta information \emph{(ax\_valid to ax\_ready)} to confirm the transaction acceptance.
For the write case, this is followed by the transition from address acceptance to data availability \emph{(aw\_ready to w\_valid)} to ensure that data beats readiness promptly follows.
Monitoring continues with the acceptance of the first data beat \emph{(w\_valid to w\_ready)} and in a subsequent stage, the data beat from the first to the last beat \emph{(w\_\teb{first} to w\_last)} to guarantee continuous and correct data flow.
After the last write data beat, the monitoring tracks from \emph{(w\_last to b\_valid)}, which is to confirm that the subordinate device sends the write response in time.
Finally, the transition from write response valid to response ready \emph{(b\_valid to b\_ready)} checks that the acknowledgment is properly issued by the manager device, which marks the end of the transaction.
For read operations, the transition from address acceptance to data availability \emph{(rw\_ready to r\_valid)} is monitored, along with the data beats \emph{(r\_\teb{first} to r\_last)} and the timely delivery of the read response.

\subsubsection{Stage Budget Allocation}

In \gls{axi4} systems, transactions of the same \gls{tid} are processed sequentially.
The stage budget for each transaction is dynamically calculated by multiplying the per-word stage budget, obtained by profiling the system during \gls{vv}, with the burst length, granting longer bursts more time to complete.

For write transactions, this impacts the time budget from the \emph{aw\_ready to w\_first} stage, while the time from \emph{w\_first to w\_last} is counted only when the transaction begins servicing, thus excluding prior transaction latency.
A similar dynamic stage budgeting method is applied for read transactions.
This ensures that transactions, particularly those with large data beats, have sufficient time to complete.

\subsection{Configuration Interface}
\label{sec:arch:prog}

In its basic configuration, {\irealm} and {\erealm} units are configured through a shared set of memory-mapped registers, as shown in \Cref{fig:realm-system}.
The shared configuration register file can be physically decoupled to increase the scalability of the {\axirealm} architecture in larger designs.
Configurations with dedicated configuration register files for each {\irealm} and {\erealm} unit are supported.

The register values are reset to a default configuration on startup, the {\erealm} units are deactivated, and the {\irealm} are bypassed.
In this reset state, the {\axirealm} system is inert, interconnect accesses are unregulated, and no additional latency is introduced.
One privileged manager, e.g., the booting core from a secure domain, programs the {\axirealm} system.
The system's memory protection or {\axirealm}'s \emph{bus guard} restricts configuration space access to privileged managers.
{\axirealm} can be dynamically reconfigured during the system's runtime.

\begin{figure}
    \centering
    \includegraphics[width=0.85\linewidth]{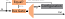}
    \caption{%
        The architecture of the \emph{bus guard}.
    }
    \label{fig:bus-guard}
\end{figure}

\subsection{Bus Guard}
\label{sec:arch:bus-guard}

{\axirealm} 's configuration space must be protected against malicious or erroneous accesses.
Most systems use physical memory protection units or virtual memory space, e.g., through \glspl{mmu}, to isolate critical configuration spaces.

Even in systems with no such protection device, our minimal \emph{bus guard} unit, presented in \Cref{fig:bus-guard}, restricts unwanted access to the configuration interface. %
After a system reset, a trusted manager must claim ownership of the configuration space by writing to a \emph{guard register} within the \emph{bus guard}.
In the unclaimed state, every access to the configuration space except for the \emph{guard register} returns an error. %
Once a manager has claimed the address space, it can perform a \emph{handover} operation to transfer the exclusive read/write ownership to any other manager in the system.
The \emph{bus guard} differentiates between managers using their unique \gls{tid}. %

\newpage
\section{Architectural Results}
\label{chap:realm:archres}

\begin{table}[t]
    \centering
        \centering
        \caption{%
            Area contribution \emph{weights} of {\axirealm}'s building blocks as a function of their parameters. %
            All numbers are in \si{\GE}, at \SI{1}{\giga\hertz} using typical conditions.%
        }%
        \label{tab:ooc-area}
        \centering
        \resizebox{1\columnwidth}{!}{%
            \begin{threeparttable}
                \renewcommand{\arraystretch}{1.1}
                \renewcommand{\tabcolsep}{2pt}
                \begin{tabular}{@{}lllccccccccc@{}}
                    \toprule
                    \multirow{7}{*}{\rot{Config. Registers}} &
                    \multirow{3}{*}{PUR} &
                    (i) Status &
                    \noc &
                    \noc &
                    \noc &
                    \noc &
                    \noc &
                    \noc &
                    \noc &
                    \noc &
                    24.6 \\
                    &
                    &
                    (i) Budget/Period &
                    \noc &
                    \noc &
                    \noc &
                    \noc &
                    \noc &
                    \noc &
                    \noc &
                    \noc &
                    1320 \\
                    &
                    &
                    (i) Region Bound. &
                    20.6 &
                    \noc &
                    \noc &
                    \noc &
                    \noc &
                    \noc &
                    \noc &
                    \noc &
                    \noc \\
                    \arrayrulecolor{ieee-dark-black-40}\cdashline{2-12}
                    &
                    \multirow{3}{*}{PU} &
                    (i) Config &
                    \noc &
                    \noc &
                    \noc &
                    \noc &
                    \noc &
                    \noc &
                    \noc &
                    \noc &
                    83.5 \\
                    &
                    &
                    (e) Status/Config &
                    \noc &
                    \noc &
                    \noc &
                    \noc &
                    \noc &
                    \noc &
                    \noc &
                    \noc &
                    9.7 \\
                    &
                    &
                    (e) R/W Budget &
                    \noc &
                    \noc &
                    \noc &
                    \noc &
                    \noc &
                    \noc &
                    770 &
                    \noc &
                    \noc \\
                    \arrayrulecolor{ieee-dark-black-40}\cdashline{2-12}
                    &
                    \multirow{1}{*}{PS} &
                    Bus Guard &
                    \noc &
                    \noc &
                    \noc &
                    \noc &
                    \noc &
                    \noc &
                    \noc &
                    \noc &
                    261 \\
                    \arrayrulecolor{ieee-dark-black-100}\midrule
                    \teb{\multirow{6}{*}{\rot{\irealm}}} &
                    \multirow{2}{*}{PUR} &
                    Tracking Cnts. &
                    \noc &
                    \noc &
                    \noc &
                    \noc &
                    \noc &
                    \noc &
                    \noc &
                    \noc &
                    1930 \\
                    &
                    &
                    Region Decoders &
                    20.8 &
                    \noc &
                    \noc &
                    \noc &
                    \noc &
                    \noc &
                    \noc &
                    \noc &
                    \noc \\
                    \arrayrulecolor{ieee-dark-black-40}\cdashline{2-12}
                    &
                    \multirow{4}{*}{PU} &
                    Isolate/Throttle &
                    3.5 &
                    2.7 &
                    9.0 &
                    \noc &
                    \noc &
                    \noc &
                    \noc &
                    \noc &
                    267 \\
                    &
                    &
                    Burst Splitter &
                    49.3 &
                    1.5 &
                    729 &
                    \noc &
                    \noc &
                    \noc &
                    \noc &
                    \noc &
                    4840 \\
                    &
                    &
                    Meta Buffer &
                    38.1 &
                    \noc &
                    \noc &
                    \noc &
                    \noc &
                    \noc &
                    \noc &
                    \noc &
                    1310 \\
                    &
                    &
                    Write Buffer &
                    \noc &
                    \noc &
                    \noc &
                    \noc &
                    \noc &
                    264 &
                    \noc &
                    \noc &
                    11.4 \\
                    \arrayrulecolor{ieee-dark-black-100}\midrule
                    \multirow{6}{*}{\rot{\erealm}} &
                    \multirow{6}{*}{PU} &
                    ID Remap. &
                    \noc &
                    \noc &
                    \noc &
                    \noc &
                    \noc &
                    \noc &
                    \noc &
                    \noc &
                    \noc \\
                    &
                    &
                    Stage Cnts. &
                    \noc &
                    \noc &
                    \noc &
                    \noc &
                    \noc &
                    \noc &
                    \noc &
                    129 &
                    735 \\
                    &
                    &
                    HT Table &
                    \noc &
                    \noc &
                    201 &
                    \noc &
                    \noc &
                    \noc &
                    \noc &
                    \noc &
                    \noc \\
                    &
                    &
                    LD Table &
                    \noc &
                    \noc &
                    \noc &
                    51 &
                    \noc &
                    \noc &
                    \noc &
                    \noc &
                    \noc \\
                    &
                    &
                    R/W Table/Ctrl. &
                    \noc &
                    \noc &
                    \noc &
                    \noc &
                    \noc &
                    \noc &
                    329 &
                    \noc &
                    356 \\
                    &
                    &
                    Reset Ctrl. &
                    \noc &
                    \noc &
                    \noc &
                    \noc &
                    \noc &
                    \noc &
                    \noc &
                    \noc &
                    1270 \\
                    \arrayrulecolor{ieee-dark-black-100}\midrule
                    &
                    &
                    &
                    \tilt{Addr Width~\tnote{a}~\tnote{b}}             &
                    \tilt{Data Width~\tnote{a}~\tnote{b}}             &
                    \tilt{Num Pending~\tnote{c}}                      &
                    \tilt{Num \glspl{tid}}                            &
                    \tilt{Buffer Depth~\tnote{c}}                     &
                    \tilt{Storage Size~\tnote{a}~\tnote{d}~\tnote{e}} &
                    \tilt{Num Counters~\teb{\tnote{f}}}               &
                    \tilt{Counter Storage~\teb{\tnote{g}}}            &
                    \tilt{Constant~\teb{\tnote{h}}}                   \\
                    \arrayrulecolor{ieee-dark-black-100}\bottomrule
                \end{tabular}

                \begin{tablenotes}[para, flushleft]
                    \fontsize{7.7pt}{7.7pt}\selectfont
                    \item[a] \teb{i}n [bit]
                    \item[b] \teb{e}valuated \SIrange{32}{64}{\bit}
                    \item[c] \teb{e}valuated \SIrange{2}{16}{elements}
                    \item[d] \teb{p}roduct of \emph{Buffer Depth} and \emph{Data Width}
                    \item[e] \teb{e}valuated \SIrange{256}{8192}{\bit}
                    \item[f] \teb{p}roduct of \emph{Num Pending} and\emph{Num \glspl{tid}}
                    \item[g] \teb{p}roduct of \emph{Num Counters} and the counter width (\SIrange{10}{32}{\bit})
                    \item[h] \teb{b}ase area independent of params
                \end{tablenotes}
            \end{threeparttable}
        }
\end{table}

This section provides an extensive area, timing, and latency model to enable quick design-space exploration and promotes fair comparison with other works.
For gate-level assessment, we use {\gfs} {\gftech} node with a 13-metal stack and 7.5-track standard cell library in the typical corner. %
We synthesize the designs using {\dc} in topological mode to account for place-and-route constraints, congestion, and physical phenomena.
We provide all area results in \emph{gate equivalent (\si{\GE})}, a technology-independent circuit complexity metric, allowing comparisons among technology nodes.
A \si{\GE} represents the area of a two-input, minimum-strength {NAND} gate.

\subsection{Area Model}
\label{sec:ooc:area}

Our linear area model, given in \Cref{tab:ooc-area}, allows us to estimate a system's {\axirealm} configuration given the number of {\irealm} and {\erealm} units with their respective parameters.
We synthesize our {\axirealm} system to construct the model, sweeping the parameter space.
The resulting areas are correlated with the input parameters, and a linear model is fitted.
The model is divided into three categories: \emph{Configuration Registers}, \emph{\irealm}, and \emph{\erealm}.
Each category is grouped into the sub-categories: \emph{\gls{ps}}, \emph{\gls{pu}}, and \emph{\gls{pur}}.
To estimate the area of an {\axirealm} system or its components, the system's desired configuration is determined.
This includes the number of {\irealm} and {\erealm} units, configuration register files, and regions, as well as the \gls{ip}'s desired \gls{sv} parameters.
The area of the individual sub-units is given by a linear function of the \gls{ip} parameters with the coefficient in \Cref{tab:ooc-area}.
The total {\axirealm} area can be obtained by summing over the individual sub-unit's contributions.

We use the {\axirealm} \gls{mcs}'s configuration (\Cref{tab:params}) presented in \Cref{chap:realm:carfield} as an example of how to use our model.
The \emph{bus guard} is the only \gls{ps} item with a constant contribution of \SI{261}{\GE}.
For each \gls{pu} and \gls{pur} elements we evaluate
\begin{equation*}
A_{contrib} = \sum_{i}{param_i}*{weight_i} + constant
\end{equation*}
to obtain their respective area contribution.
E.g., for the \emph{write buffer}, we multiply \SI{264}{\GE} with the storage size of 256 and add \SI{11.4}{\GE}.
Our example features three {\irealm} units, bringing the total area contribution of all \emph{write buffers} in the {\axirealm} system to \SI{203}{\kGE}.
The total modeled area, presented in \Cref{tab:params}, is calculated by summing all \gls{ps}, \gls{pu}, and \gls{pur} contributions together. %

Our {\axirealm} architecture does not have inherent limitations in terms of throughput, the supported amount of \glspl{tid}, and the number of outstanding transfers as long as the units are tuned to the encompassing system and its use cases.

\begin{table}
    \centering
    \caption{Parametrization, the resulting modeled, and actual area of the {\irealm} and {\erealm} units in {\carfield} (\Cref{chap:realm:carfield}).}
    \resizebox{\columnwidth}{!}{%
        \begin{threeparttable}
            \renewcommand{\arraystretch}{1.2}
            \renewcommand{\tabcolsep}{3pt}
            \begin{tabular}{@{}llllccccccccc@{}}
                \toprule
                \dll{\textbf{Sub-}}{\textbf{system}} &
                \dll{\textbf{Num.}}{\textbf{Units}} &
                \dll{\textbf{Num.}}{\textbf{Regions}} &
                \multicolumn{8}{l}{\dll{\textbf{SystemVerilog}}{\textbf{Parameters}}} &
                \tll{\textbf{Model}}{\textbf{Area}}{\textbf{[\si{\kGE}]}} &
                \tll{\textbf{Design}}{\textbf{Area}}{\textbf{[\si{\kGE}]}} \\
                \midrule
                \emph{\irealm} &
                3 &
                2 &
                \SI{48}{\bit} &
                \SI{64}{\bit} &
                16 &
                \nad &
                4 &
                \SI{256}{\bit} &
                \nad &
                \nad &
                330 &
                328 \\
                \emph{\erealm} &
                1 &
                \nad &
                \teb{\SI{48}{\bit}} &
                \teb{\SI{64}{\bit}} &
                2 &
                2 &
                \nad &
                \nad &
                20 &
                \SI{200}{\bit} &
                50 &
                45 \\
                \midrule
                &
                &
                &
                \tilt{Addr Width}           &
                \tilt{Data Width}           &
                \tilt{Num Pending}          &
                \tilt{Num \glspl{tid}}      &
                \tilt{Buffer Depth}         &
                \tilt{Storage Size}         &
                \tilt{Num Counters}         &
                \tilt{Counter Storage}      &
                \\
                \bottomrule
            \end{tabular}
        \end{threeparttable}
        \label{tab:params}
    }
\end{table}

\subsection{Timing and Latency}
\label{sec:ooc:timing}

The {\axirealm} architecture and its units are designed to achieve clock speeds exceeding \SI{1.5}{\giga\hertz} (corresponding to \SI{25}{logic} levels) in {\gftech} when combined with optimized \gls{axi4} \glspl{ip} for \glspl{asic}.
The achieved frequency can be further increased at the cost of additional latency by either adding \gls{axi4} cuts around the {\axirealm} units or introducing pipelining into the {\axirealm} units.
The {\irealm} unit adds no additional cycle of latency when bypassed and introduces one cycle through the write buffer (\Cref{sec:arch:buf}) when active.
The {\erealm} unit adds no additional latency, whether bypassed or active.

\newpage
\section{Case Study: Automotive MCS}
\label{chap:realm:carfield}

\begin{figure}
    \centering
    \includegraphics[width=\columnwidth]{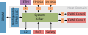}
    \caption{%
        Architectural block diagram of the {\carfield} platform.
    }
    \label{fig:carfield-bd}
\end{figure}

\subsection{Introduction}

\emph{{\carfield}} establishes a heterogeneous platform for mixed-criticality systems and application across domains like automotive, space, and cyber-physical embedded systems.
At the core of {\carfield}, the \emph{host} domain consists of a Linux-capable dual-core \emph{CVA6} system enhanced with virtualization extensions, namely {\riscv}'s \emph{H-extension} and virtualized fast interrupts.
The platform is complemented by a \emph{safety} and a \emph{security} domain, allowing for reliable operation and secure boot, respectively.
{\carfield} computational capabilities are enhanced through general-purpose \glspl{dsa}.
We instantiate two \glspl{dsa}: one specialized for integer and one for floating-point workloads.
Each accelerator features an internal \gls{dma} engine to copy data between its private \gls{spm} and the \gls{mcs}'s main storage.
{\carfield} features two memory endpoints, a \SI{512}{\kibi\byte} banked \gls{l2} memory and an off-chip {DRAM} accessed through a \gls{llc}.
Each rank of the platform's \gls{llc} can be configured either as software-managed \gls{spm} or cache for the {DRAM}.
All five domains are connected through a 64-bit point-to-point \gls{axi4} crossbar.%

We integrate {\axirealm} into {\carfield} by adding four {\irealm} units at the crossbar's ingress of all time-critical managers the two \glspl{dsa} and the two CVA6 cores.
An {\erealm} unit ensures real-time responses from the platform's \gls{eth} controller, which accesses time-critical sensor data.
Characterization in {\carfield} shows \gls{eth} to be prone to data loss, and it further exhibits strong fluctuations in timing behavior, thus being a representative candidate for a subordinate guarded by {\erealm}.
\Cref{fig:carfield-bd} shows the enhanced architectural block diagram enhanced with our {\axirealm} units.

\begin{figure}
    \centering
    \includegraphics[width=\linewidth]{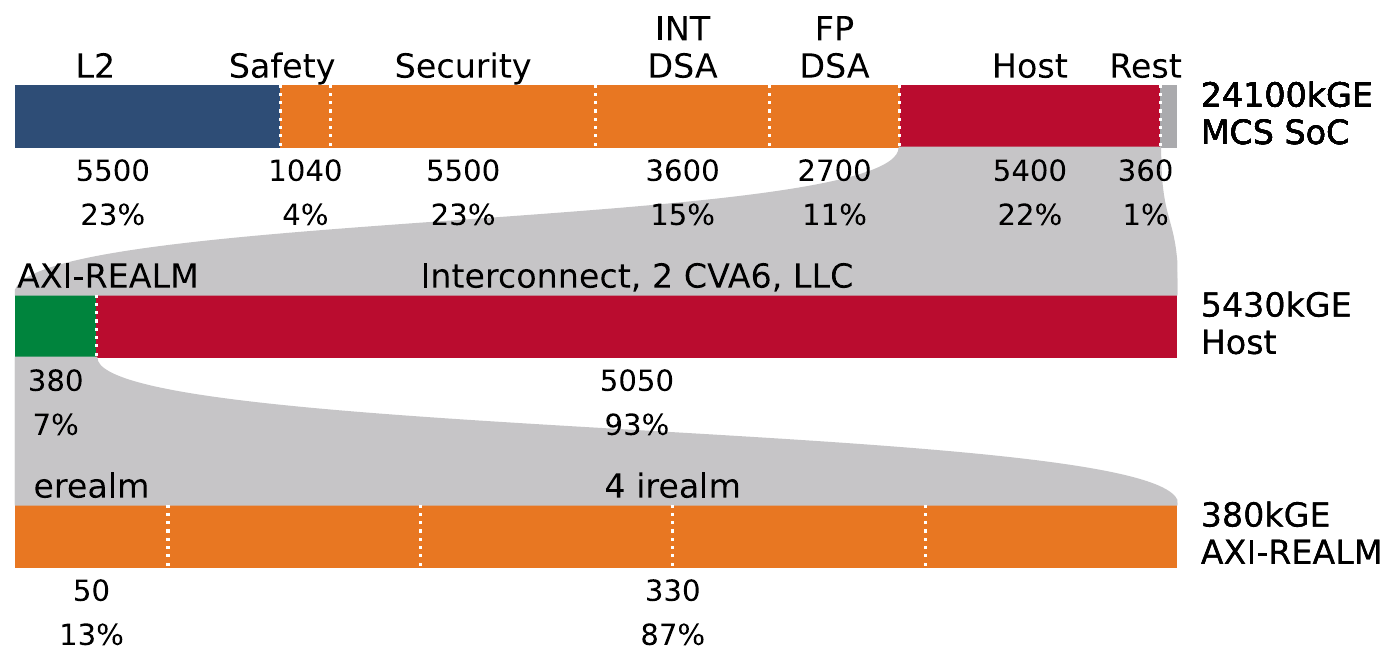}
    \vspace{-5mm}
    \caption{%
        {\carfield}'s hierarchical area including {\axirealm}.
    }
    \label{fig:carfield-area}
\end{figure}

\subsection{Area Impact}
\label{sec:case:area}

We synthesize {\carfield} with the {\axirealm} extensions in {\gfs} \SI{12}{\nano\metre} node using typical timing corners.
\Cref{fig:carfield-area} presents the total \gls{soc} area of \SI{24}{\mega\GE} and the hierarchical area contributions of our units introduced.
The {\irealm} units incur a total overhead of \SI{330}{\kilo\GE}, contributing \SI{1.4}{\percent} to the total area.
The {\erealm} unit uses \SI{50}{\kilo\GE} (\SI{0.21}{\percent}) of the \gls{soc}'s area.
The parameterization of the {\axirealm} units implemented in {\carfield} is given in \Cref{tab:params}, \Cref{chap:realm:archres}.

\subsection{Synthetic Performance Analysis of {\irealm}}
\label{sec:case:synth-eval}

We first evaluate the functional performance of the {\irealm} architecture using a memory-bound synthetic benchmark, which emulates the real-time-critical task, to maximize the effects of interconnect and subordinate interference between the processor cores and the platform's \glspl{dsa}.
This synthetic benchmark uses {CVA6} to copy data between different memory locations while a \gls{dma} engine in one of the \glspl{dsa} performs data transfer operations.
The default configuration is to copy \SI{1}{\kibi\byte} of data with the core from {\carfield}'s hybrid \gls{llc}, configured as a \gls{spm}, to \gls{l2} memory while the \gls{dsa} causes interference in the \gls{spm} using long bursts of 256 beats.
Large and equal budget periods as well as a fragmentation size of one are used if nothing else is specified.
Results from application benchmarks are presented in \Cref{sec:case:bmk-eval}.

\begin{figure}
    \centering
    \includegraphics[width=0.9855\linewidth]{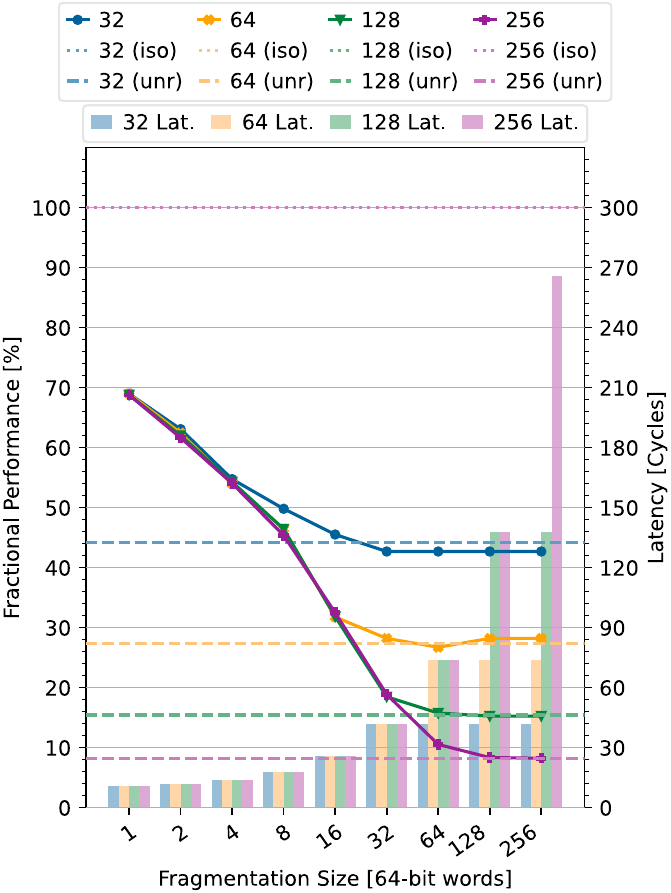}
    \caption{%
        Performance results of CVA6 copying data between \gls{spm} and \gls{l2} with the \gls{dsa} accessing \gls{spm} at various granularities in 64-\si{bit} \si{beats}.
        \emph{iso} denotes the isolated\alc{ --- \SI{100}{\percent} fractional performance}, \emph{unr} the unregulated performance.
    }
    \label{fig:result:memcpy-spm}
\end{figure}

\begin{figure}
    \centering
    \includegraphics[width=\linewidth]{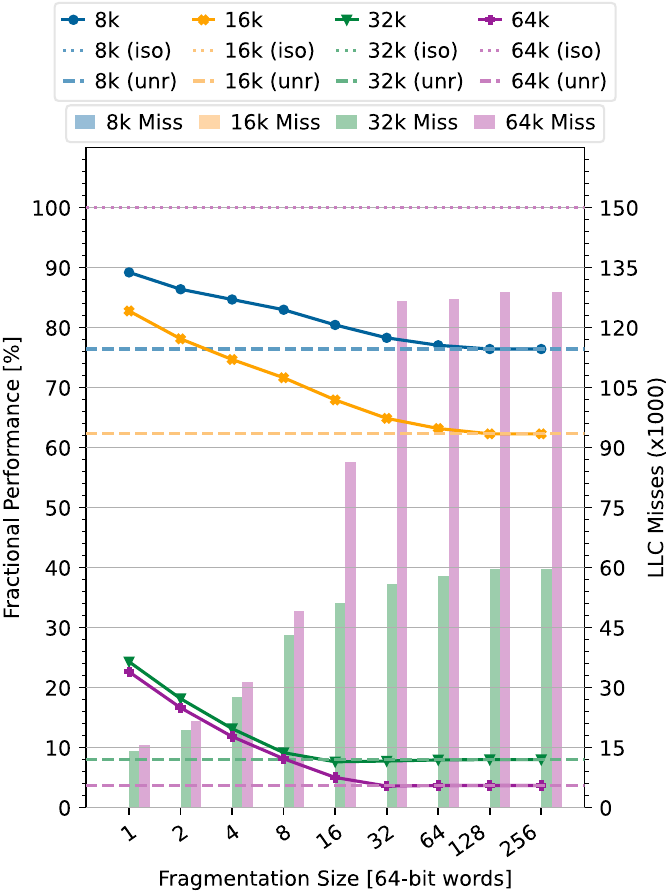}
    \caption{%
        {DRAM} and \gls{l2} with the \gls{dsa} accessing {DRAM} at various problem sizes in \si{byte}.
        \emph{iso} denotes the isolated\alc{ --- \SI{100}{\percent} fractional performance}, \emph{unr} the unregulated performance.
    }
    \label{fig:result:memcpy-llc}
\end{figure}

\begin{figure}
    \centering%
    \begin{subcaptionblock}{\linewidth}%
        \centering
        \includegraphics[width=\linewidth]{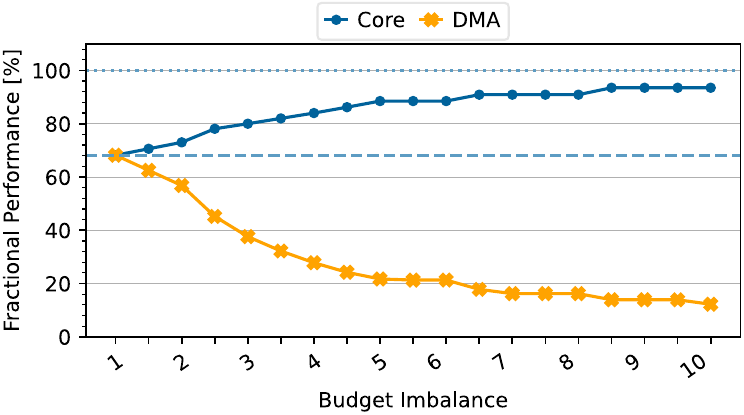}
        \caption{Fractional performance at different budget imbalances.}
        \label{fig:period-bmk-perf-budget}
        \vspace{0.25cm}
    \end{subcaptionblock}

    \begin{subcaptionblock}{\linewidth}%
        \centering
        \includegraphics[width=\linewidth]{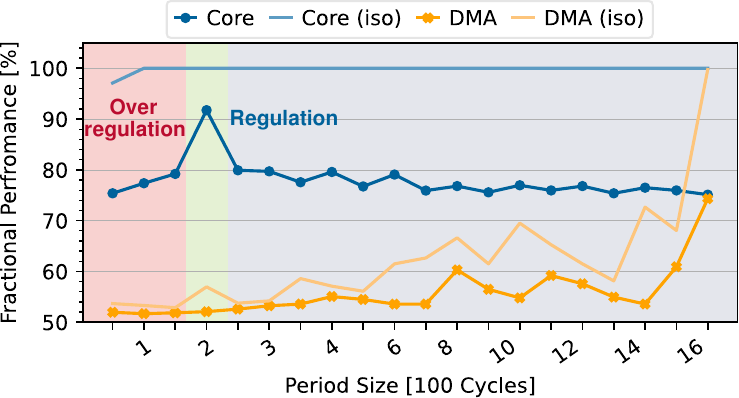}
        \caption{Fractional performance at different period sizes.}
        \label{fig:period-bmk-perf-period}
    \end{subcaptionblock}
    \caption{%
        (a) Fractional performance at different budget imbalances favoring the critical manager assuming fragmentation one and (b) fractional performance at different period sizes assuming fragmentation size one and equal budget.
    }
    \label{fig:period-bmk-perf}
\end{figure}

\subsubsection{Controlling Fairness: Burst Fragmentation}
\label{sec:case:frag}

We configure the {\irealm} units of {CVA6} and one of the \gls{dsa} to fragment transactions at different granularities without any budget limitation.
For this synthetic assessment, the write buffer is disabled, as both managers are under our control, eliminating the possibility of bandwidth stealing.
To simulate different \gls{dsa} workloads, we vary the \gls{dma} transaction length in \Cref{fig:result:memcpy-spm} from 32 to 256 64-\si{bit} words.
Fragmenting all beats to single-world granularity results in the best performance independent of the nature of the \gls{dma} transfer; in this setting, {CVA6} achieves \SI{68}{\percent} of its isolated performance.
The worst slowdown can be observed in the presence of 256-word-long bursts without fragmentation activated, which represents the unregulated case: {CVA6} only achieves \SI{1}{\percent} of the isolated performance.
The latency of each core access increases from \SI{11}{cycles} to \SI{266}{cycles}, as core transfers are interleaved by the 256-\si{cycle}-long \gls{dsa} transfers.

In a second experiment, \Cref{fig:result:memcpy-llc}, we copy data between the system's external DRAM cached by the \gls{llc} and the \gls{l2} memory.
The \gls{dsa} \gls{dma} is configured to emit 256-word-long bursts at varying data set sizes.
Due to conflict misses between the \gls{dsa} and the core, and capacity misses when the \gls{dsa} transfer size surpasses the \gls{llc} capacity, the fraction of isolated performance drops from \SI{80}{\percent} (\SI{16}{\kilo\byte}) to \SI{23}{\percent} (\SI{64}{\kilo\byte}) at single-word fragmentation.
{\axirealm} can help mitigate interference in accessing a shared cached memory location, allowing {\carfield} to achieve up to \SI{23}{\percent} of the isolated performance as opposed to \SI{4}{\percent} without regulation.
To improve performance further, complementary regulation strategies must be put in place for the shared \gls{llc}.
For example, cache coloring or partitioning can mitigate conflict misses or \gls{dma} cache bypassing to eliminate capacity misses.

\subsubsection{Period and Budget Considerations}
\label{sec:case:per-bud}

\begin{figure}
    \centering
    \includegraphics[width=\columnwidth]{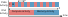}
        \caption{%
            Schedule of the periodic (\emph{p}) real-time-critical transactions running on {CVA6} and the data copy operation (period \emph{P}) by the \gls{dsa} \gls{dma}.
        }
    \label{fig:period-bmk-plot}
\end{figure}

This section assesses {\axirealm}'s period and budget functionality at fixed fragmentation.
In particular, we demonstrate that tuning each manager's period and budget can prioritize traffic of certain managers over others and even increase fairness~\cite{pagani2019abandwidthreser} further compared to solely acting on the fragmentation size.
We observe that the budget imbalance favoring the real-time task restores performance up to \SI{95}{\percent} of the isolated case, at a performance detriment to the \gls{dsa} \gls{dma}, see \Cref{fig:period-bmk-perf-budget}.
Similarly, the period can be used as a knob for online traffic regulation~\cite{pagani2019abandwidthreser}.
{\axirealm} does not limit the managers in how to spend the budget within each period.
Larger periods introduce less regulation overhead but allow \gls{dsa} managers to cause more interference.
We assume the periodic execution schedule for the critical manager and the \gls{dma} \gls{dsa} given in \Cref{fig:period-bmk-plot}.
The first has a period \emph{p}, \SI{200}{cycles}, and the latter a period \emph{P}, set to \SI{1600}{cycles}.
Both managers utilize the interconnect \SI{50}{\percent} within their period; for the critical manager this corresponds to \SI{800}{\byte} in \SI{200}{cycles}, for the \gls{dma} \SI{6400}{\byte} in \SI{1600}{cycles}.
The corresponding {\irealm} units are configured equally to a fragmentation size of one.
The regulation time period is swept from 50 to \SI{1600}{cycles} with the budget set to half the maximum transfer size possible during the regulation period.
E.g., the budget for the critical manager and the \gls{dma} are set to \SI{6400}{\byte} each when selecting a period of \SI{1600}{cycles}.

\Cref{fig:period-bmk-perf-period} shows the fractional performance of the \gls{dma} and the core given selected period sizes.
The performance of the \gls{dma} decreases when the {\irealm} period falls below the \gls{dma}'s period of \emph{P}, regardless of whether the critical manager is active.
This happens due to \emph{overregulation} as every \gls{dma} transaction no longer fits a period, interrupting it at least once. %

The critical manager nearly matches isolated memory performance --- over \SI{93}{\percent} of the isolated case --- when the {\irealm} period aligns with the core's task period of \emph{p}.
Below this critical period, the core's transfer is again overregulated.

\subsubsection{Power and Energy Efficiency Analysis}
\label{sec:case:power}

\begin{figure}
    \centering
    \includegraphics[width=\columnwidth]{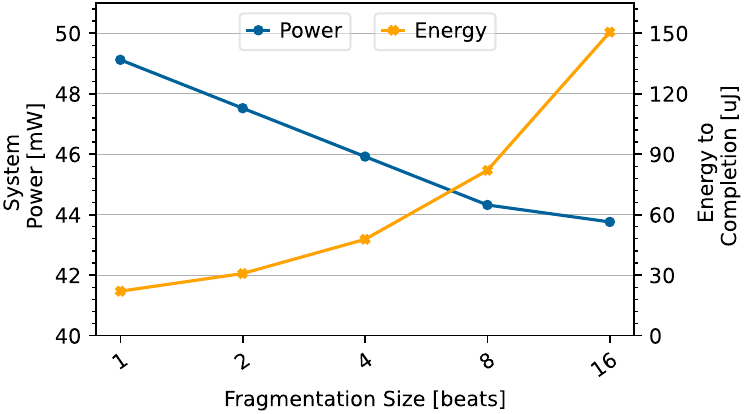}
    \caption{%
        Power and energy of CVA6 copying data between \gls{spm} and \gls{l2} with the \gls{dsa} accessing \gls{spm} given different fragmentation sizes.%
    }
    \label{fig:carfield-power}
\end{figure}

\Cref{sec:case:frag} establishes full fragmentation as the configuration achieving the best performance in the presence of \gls{dsa} interference.
However, fragmenting transfers to word-level accesses increases the switching activity (e.g., the address changes on every access) and, thus, the power consumed by the interconnect and the subordinate devices.
We evaluate the energy and power consumption of our synthetic benchmark running on {\carfield} using timing-annotated switching activity on a post-layout netlist using {\pt}.
We vary the fragmentation size of the {\irealm} units from one to sixteen beats.
The peak power consumed by the host domain, which includes the host, the \gls{mcs}'s main interconnect, and the \gls{spm} memory, is linearly increasing with decreasing fragmentation size.
When evaluating {CVA6}'s energy spent to copy \SI{1}{\kibi\byte} of data, the energy required is minimal at a fragmentation size of one.
\alc{Increasing the burst granularity increases the interference in the memory system prolonging the execution time and thus increasing energy.}
Even though the activity of fragmenting transactions is increasing the power consumption of the benchmark, the reduction in execution time outweighs the increase in activity, as seen in \Cref{fig:carfield-power}.

\subsection{Synthetic Performance Analysis of {\erealm}}
\label{sec:case:synth-eval-erealm}

To evaluate the functional performance of {\erealm} connected in front of the \gls{eth} peripheral in {\carfield}, we inject \gls{axi4} transaction faults within the \gls{eth} peripheral by either delaying to accept requests or stalling responses.
This behavior may occur due to a full transmission buffer or the \gls{eth} device failing to send the requested data.

We define the \gls{wcdt} as the time between the occurrence of a fault and the {\erealm} unit detecting it.
Most protocol errors, e.g., a wrong \gls{tid} or a superfluous handshake, are detected instantaneously.
To derive the \gls{wcdt}, we thus consider a timeout event.
As explained in \Cref{sec:arch:erealm}, an \gls{axi4} transaction is tracked in different stages, each having a dedicated time budget assigned.
The \gls{wcdt} is equal to the time of the largest budget configured for all stages.
For most transactions, the longest stage monitors the read or write data beats (\emph{r/w\_first to r/w\_last}), see \Cref{fig:erealm-stage}.
A timeout happens when a peripheral sends the first read or write \emph{r/w\_first} but never continues to issue any more beats; the {\erealm} unit detects this after the stage's budget is depleted.

In {\carfield}, we set the budgets for all stages, but the \emph{r/w\_first to r/w\_last} to \SI{20}{cycles}.
As the \gls{eth} \gls{ip} supports bursts up to \SI{256}{beats} in length, we set the budget for the read and write monitoring to \SI{300}{cycles}, corresponding to the \gls{wcdt} of the \gls{eth} \gls{ip}.
{\carfield} implements fast virtualized interrupt support through a core-local interrupt controller, enabling best-in-class interrupt responses of~\SI{100}{cycles} on the {CVA6} host core, ensuring quick and agile system reactions to disturbances.
The worst-case latency, from the fault occurring to the core reacting, is thus at most \SI{400}{cycles}.
The {\erealm} unit can be configured to automatically reset the faulty subordinate device within two cycles from fault detection, preparing it to resume operation immediately once the core is informed.

Thanks to the stage-level tracking of {\erealm} and the fast interrupt support of {\carfield}, we can inform the core and reset the subordinate in as low as \SI{100}{cycles} with a \gls{wcdt} of \SI{400}{cycles} for the \gls{eth} \gls{ip} in {\carfield}.
This response time is substantially lower than waiting for a deadline miss or a system-wide watchdog reset due to an interconnect stall from a faulty subordinate.

\begin{figure}
    \centering%
    \begin{subcaptionblock}{\linewidth}%
        \centering%
        \includegraphics[width=\linewidth]{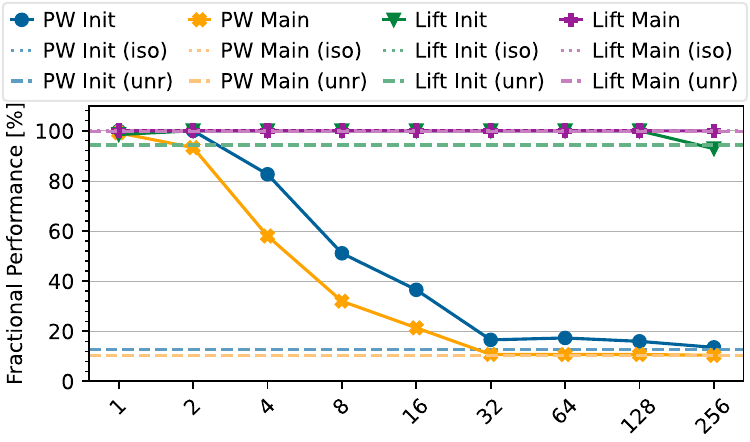}%
        \caption{Single-\glspl{dsa} interference.}%
        \vspace{2.5mm}
        \label{fig:taclebench-sa}%
        \vfill%
        \centering%
        \includegraphics[width=\linewidth]{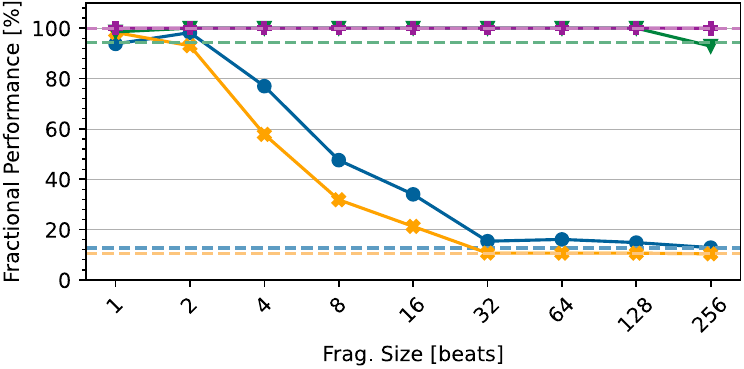}%
        \caption{Dual-\glspl{dsa} interference.}%
        \vspace{1mm}
        \label{fig:taclebench-da}%
    \end{subcaptionblock}
    \caption{%
        TACLeBench~\cite{falk2016taclebenchabenc} performance of {CVA6}
        with (a) one and (b) two \glspl{dsa} interfering.
        A budget imbalance of 1:1 and a period longer than the application's runtime is selected.
        We vary the fragmentation granularity.
        The applications contain an initialization, \emph{init}, and a \emph{main} phase.
        \emph{iso} denotes the isolated, \emph{unr} the unregulated performance.
    }
    \label{fig:taclebench}
\end{figure}

\subsection{Case Study on {\carfield}}
\label{sec:case:bmk-eval}

We evaluate the performance of {\axirealm} using heterogeneous benchmark applications on {\carfield}, \alc{as synthetic benchmarks alone do not suffice to fully characterize memory interference~\cite{carletti2025takingacloserlo}}.
We combine a \gls{ml} inference application running on either one or both of {\carfield}'s machine-learning-optimized \glspl{dsa} while executing time-critical applications from \emph{TACLeBench}~\cite{falk2016taclebenchabenc} on the platform's host {\riscv} cores.
To simulate a truly heterogeneous application including communication and coarse-grained synchronization, both the \gls{dsa} and the time-critical tasks access {\carfield}'s shared \gls{l2} \gls{spm} memory.
{TACLeBench} provides two real-world applications: \emph{lift} and \emph{powerwindow}; the first mimics the controller of an industrial elevator, the latter one of the four car windows controlled by the driver and the passenger~\cite{falk2016taclebenchabenc}.

We use {\axirealm} to control and monitor the data streams injected by the \glspl{dsa} and the {\riscv} cores to restore the performance of the {TACLeBench} applications.
We keep the budget between the actors equal and large enough to restrict neither the accelerator's \glspl{dma} nor the cores from accessing \gls{l2}.
We chose a reasonably small period \alc{of \SI{100}{cycles}} to mitigate any imbalance issues presented in \Cref{sec:case:per-bud}.
The fragmentation is swept between \emph{one}, fairest, and \emph{255} (no fragmentation).

For compute-bound applications (\emph{lift}), we observe no interference of the \glspl{dsa} with the execution of the critical program.
{\axirealm} does not introduce any measurable overheads in these cases.
\alc{If security is not a main concern,} {\axirealm} can be rapidly disabled \alc{by the \gls{os}}, see \Cref{sec:arch:prog}, should a non-memory-intensive task be executed on the platform, fully eliminating any dynamic power overhead of the {\axirealm} units.
The \emph{powerwindow} task faces up to \SI{9.7}{\x} interference from the DSAs in shared memory. (\Cref{fig:taclebench}).
{\axirealm} reduces this delay by achieving in both cases up to \SI{99}{\percent} of isolated performance with full fragmentation.

\newpage
\section{Related Work}
\label{chap:realm:relwork}

We structure {\axirealm}'s related work into two parts: \Cref{sec:relwrk:irealm} compares the {\irealm} unit to real-time interconnects and regulation modules and \Cref{sec:relwrk:erealm} compares the {\erealm} unit to \gls{soa} transaction monitoring units.

\subsection{Real-time Extensions: \irealm}
\label{sec:relwrk:irealm}

The \gls{soa} of real-time interconnect extensions can be divided into two fundamental design strategies: \emph{drop-in regulation modules}, which are integrated between the managers and the interconnect itself, or intrusive \emph{interconnect architecture customizations}.
The latter strategy profoundly changes the interconnect's internal structure, intertwining the enhancements with a given memory system architecture and, thus, with a given system~\cite{restuccia2020axihyperconnect, gray2023axiicrttowardsa, serranocases2021leveraginghardw}.

\subsubsection{Drop-in Regulation Modules}
\label{subsec:relwork_helpers}

Credit-based mechanisms are commonly introduced at the boundary of existing interconnect configurations to impose spatial and temporal bounds on non-coherent, on-chip interconnect networks.

Pagani~{\etal} and Restuccia~{\etal} analyze and address the problem of multiple \glspl{dsa} either competing for bandwidth or causing interference in heterogeneous, \gls{axi4}-based \gls{fpga} \glspl{soc}.
They propose three units to mitigate contention.
The \emph{\gls{axi4} budgeting unit} (ABU)~\cite{pagani2019abandwidthreser} extends the concept of inter-core memory reservation established in \glspl{mpsoc} to heterogeneous \glspl{soc}.
The {ABU} uses counter-based budgets and periods assigned for each manager in the system, reserving a given bandwidth to each manager.
The \emph{\gls{axi4} burst equalizer} (ABE)~\cite{restuccia2019isyourbusarbite} tackles unfair arbitration by limiting the nominal burst size and a maximum number of outstanding transactions for each manager.
The \emph{Cut and Forward} (C\&F)~\cite{restuccia2022cutandforwardsa} unit prevents ahead-of-time bandwidth reservations by holding back transactions until they can certainly be issued.
Our {\axirealm} architecture tackles these challenges while optimizing the design to be suitable for high-performance systems and use cases.
{\axirealm} adds only one cycle of latency (\Cref{chap:realm:architecture}) and extremely low area overhead (\Cref{chap:realm:archres}).

\alc{Similarly to {ABU}, Valente at al.~\cite{valente2025finegrainedqosc} present \emph{Runtime Bandwidth Regulator} (RBR) implementing a bandwidth equalization mechanism using counter-based budgets and periods, targeting primarily \gls{fpga} accelerators.}
\alc{Our {\axirealm} provides a similar functionality, whilst additionally preventing bandwidth reservations and automatically equalizing burst lengths among all memory actors.}

Farshchi~\teb{\etal}~\cite{farshchi2020brubandwidthreg} propose the \emph{Bandwidth Regulation Unit} (BRU), a \gls{hw} module aimed at reducing the regulation overheads of \gls{sw} approaches.
Designed for coherent multicore \glspl{soc}, BRU manages memory traffic per core.
Akin to {\axirealm}, it employs a time-slicing approach, albeit with one global period shared by all domains.
The design can only regulate the maximum bandwidth, whose size is fixed to the dimension of a cache line, while the number of memory access transactions is user-configurable.
Implemented in a \SI{7}{\nano\meter} node, BRU adds minimal logic overhead ($<$\teb{\SI{0.3}{\percent}}) and reduces the maximum achievable frequency by $<$\teb{\SI{2}{\percent}}.
Furthermore, BRU can independently control the write-back traffic to the main memory, mitigating write-read imbalances.
Unlike \gls{sw} techniques~\cite{bechtel2019denialofservice}, this functionality requires significant \gls{hw} modifications to the cache hierarchy.
We argue that it would favor a cache-centric partitioning strategy over an auxiliary module for the system bus.

A key aspect of achieving temporal isolation for shared resources involves extracting significant data from functional units during \gls{vv}.
This step is essential for determining an optimal upper limit for resource usage during operation.
Cabo~{\etal}~\cite{cabo2022safesu2asafesta, andreu2025expandingsafesu} propose \emph{SafeSU}, a minimally invasive statistics unit.
SafeSU tracks inter-core interference in \glspl{mpsoc} using dedicated counters.
Instead of limiting the number of transferred bytes, the maximum-contention control unit (MCCU) allocates timing interference quotas to each manager core in clock cycles.
Whenever the allocated quota is exceeded, an interrupt is raised.
SafeSU uses temporal information, including contention, request duration, and interference quota, as knobs to enhance traffic observability and enforce controllability.
Furthermore, the mechanism addresses interference exclusively in symmetric, general-purpose, multicore systems.
{\axirealm} leverages spatial and temporal information for traffic regulation (i.e., bandwidth reservation and time slicing) and extends the monitoring capabilities to heterogeneous \glspl{mcs} comprising of real-time-critical, general-purpose, and high-performance domain-specific managers.

\subsubsection{Interconnect Customization}
\label{subsec:relwork_bus_redesign}

Restuccia~{\etal}~\cite{restuccia2020axihyperconnect} propose \emph{HyperConnect}, a custom \gls{axi4}-based functional unit block for virtualized \gls{fpga}-\glspl{soc}.
While being the closest \gls{soa} to {\axirealm}, HyperConnect does not tackle ahead-of-time bandwidth reservation issues caused by a slow manager stalling the interconnect (\Cref{sec:arch:buf}).

Recently, Jiang~{\etal} introduced \emph{AXI-IC$^{RT}$}~\cite{gray2023axiicrttowardsa}, one of the first end-to-end \gls{axi4} microarchitectures tailored for real-time use cases.
AXI-IC$^{RT}$ leverages the \gls{axi4} user signal to assign priorities and introduces a dual-layer scheduling algorithm for the dynamic allocation of budget and period to each manager during runtime.
To prevent request starvation on low-priority managers, {\axirealm} does not depend on the concept of priority, but rather on a credit-based mechanism and a \emph{burst splitter} to distribute the bandwidth according to the real-time guarantee of the \gls{soc}.
While AXI-IC$^{RT}$ supports several budget reservation strategies, it limits the assessment to managers with equal credit (bandwidth).
Finally, from an implementation angle, the design strategy followed by AXI-IC$^{RT}$ adds extensive buffering to the microarchitecture to create an observation window for early service of incoming transactions based on priorities.
Overall, HyperConnect and AXI-IC$^{RT}$ lack monitoring capabilities to track traffic statistics.

In industry, Arm's \emph{CoreLink QoS-400} is widely integrated into modern \gls{fpga}-\glspl{soc} to manage contention using the QoS signal defined in the \gls{axi4} and AXI5 specifications.
However, QoS-400 has several limitations, as analyzed in~\cite{serranocases2021leveraginghardw}.
One significant drawback is its intrusiveness; for instance, in a {Zynq Ultrascale+} \gls{fpga}, the authors report the need to coordinate over 30 {QoS} points to effectively control traffic~\cite{serranocases2021leveraginghardw}.

\afterpage{%
\begin{sidetab}
    \centering
        \centering
        \caption{%
            \Gls{soa} comparison of {\axirealm}.%
        }%
        \label{tab:soa}
        \centering
        \resizebox{0.978\linewidth}{!}{%
            \begin{threeparttable}
                \renewcommand{\arraystretch}{1.2}
                \begin{tabular}{@{}lllccccccccccc@{}}
                    \toprule
                    &
                    &
                    &
                    \tlb{Budget/}{Time}{Slicing} &
                    \textbf{Granularity} &
                    \dlb{Fair}{Arbitration} &
                    \tlb{\teb{B}andwidth}{Reservation}{Prevention} &
                    \dlb{Manager}{Isolation} &
                    \textbf{Statistics} &
                    \dlb{Protocol}{Checking} &
                    \dlb{Fault}{Handling} &
                    \dlb{Multiple}{Outstanding} &
                    \dlb{Target}{Technology} &
                    \dlb{Area}{Overhead} \\
                    \arrayrulecolor{ieee-dark-black-100}\midrule
                    \multirow{9}{*}{\vspace{-2.5cm}\rot{\textbf{\irealm}}} &
                    \multirow{5}{*}{\vspace{-2cm}\rot{\dlb{Regulation}{Helpers}}} &
                    ABU~\cite{pagani2019abandwidthreser} &
                    \dl{\teb{P}eriod-}{\teb{B}ased~\tnote{a}} &
                    \tl{One}{\teb{S}ubordinate}{\teb{O}nly} &
                    \xmark &
                    \xmark &
                    \xmark &
                    \xmark &
                    \na &
                    \tl{Ideal}{\teb{S}ubordinate}{\teb{A}ssumed} &
                    \xmark &
                    \dl{Xilinx}{FPGA} &
                    \dl{\teb{\SI{715}{\lut}}}{\teb{\SI{908}{\ff}}}~\tnote{b} \\
                    &
                    &
                    ABE~\cite{restuccia2019isyourbusarbite} &
                    \xmark &
                    \na &
                    \dl{Transfer}{\teb{F}ragmentation} &
                    \xmark &
                    \xmark &
                    \xmark &
                    \na &
                    \tl{Ideal}{\teb{S}ubordinate}{\teb{A}ssumed} &
                    \cmark &
                    \dl{Xilinx}{FPGA} &
                    \dl{\teb{\SI{1130}{\lut}}}{\teb{\SI{582}{\ff}}}~\tnote{b} \\
                    &
                    &
                    C\&F~\cite{restuccia2022cutandforwardsa} &
                    \xmark &
                    \na &
                    \xmark &
                    \dl{Write}{\teb{B}uffering} &
                    \xmark &
                    \xmark &
                    \na &
                    \tl{Ideal}{\teb{S}ubordinate}{\teb{A}ssumed} &
                    \na &
                    \dl{Xilinx}{FPGA} &
                    \dl{\teb{\SI{3088}{\lut}}}{\teb{\SI{1467}{\ff}}}~\tnote{c,d} \\
                    &
                    &
                    \alc{RBR~\cite{valente2025finegrainedqosc}} &
                    \alc{\dl{\teb{P}eriod-}{\teb{B}ased~\tnote{a}}} &
                    \alc{\dl{\gls{fpga}}{Accelerator}}  &
                    \alc{\xmark} &
                    \alc{\xmark} &
                    \alc{\xmark} &
                    \alc{\xmark} &
                    \alc{\na} &
                    \alc{\tl{Ideal}{\teb{S}ubordinate}{\teb{A}ssumed}} &
                    \alc{\xmark} &
                    \alc{\dl{Xilinx}{FPGA}} &
                    \alc{\dl{\teb{\SI{654}{\lut}}}{\teb{\SI{607}{\ff}}}~\tnote{e}} \\
                    &
                    &
                    SafeSU~\cite{cabo2022safesu2asafesta,andreu2025expandingsafesu} &
                    SW &
                    SW &
                    \na &
                    \xmark &
                    \xmark &
                    \tl{\teb{B}andwidth,}{\teb{L}atency,}{\teb{I}nterference} &
                    \na &
                    \xmark &
                    \xmark &
                    \dl{Tech.-}{\teb{I}ndependent} &
                    \dl{\SI{83.1}{\kGE}~\tnote{f, g}}{\SI{223}{\kGE}~\tnote{f, h}} \\ %
                    &
                    &
                    BRU~\cite{farshchi2020brubandwidthreg} &
                    \tl{Global}{\teb{P}eriod-}{\teb{B}ased~\tnote{a}} &
                    \dl{Shared}{\teb{M}emory} &
                    \xmark &
                    \xmark &
                    \xmark &
                    \xmark &
                    \na &
                    \tl{Ideal}{\teb{S}ubordinate}{\teb{A}ssumed} &
                    \xmark &
                    \dl{Tech.-}{\teb{I}ndependent} &
                    \SI{57.2}{\kGE}~\tnote{i} \\ %
                    \arrayrulecolor{ieee-dark-black-40}\cdashline{2-14}
                    &
                    \multirow{4}{*}{\rot{\dlb{Interconn.}{Cust.}}} &
                    Hyperconnect~\cite{restuccia2020axihyperconnect} &
                    \dl{\teb{P}eriod-}{\teb{B}ased~\tnote{a}} &
                    \dl{Per}{\teb{S}ubordinate} &
                    \dl{Transfer}{\teb{F}ragmentation} &
                    \dl{Write}{\teb{B}uffering} &
                    \xmark &
                    \xmark &
                    \na &
                    \tl{Ideal}{\teb{S}ubordinate}{\teb{A}ssumed} &
                    \cmark &
                    \dl{Xilinx}{FPGA} &
                    \dl{\teb{\SI{3020}{\lut}}}{\teb{\SI{1289}{\ff}}}~\tnote{b} \\
                    &
                    &
                    AXI-IC$^{RT}$~\cite{gray2023axiicrttowardsa} &
                    \dl{\teb{P}eriod-}{\teb{B}ased~\tnote{a}} &
                    \dl{Per}{\teb{S}ubordinate} &
                    \dl{Transfer}{\teb{B}uffering} &
                    Buffering &
                    \xmark &
                    \xmark &
                    \na &
                    \xmark &
                    \cmark &
                    FPGA &
                    \dl{\teb{\SI{4745}{\lut}}}{\teb{\SI{4184}{\ff}}}~\tnote{j} \\
                    &
                    &
                    QOS-400~\cite{serranocases2021leveraginghardw} &
                    \dl{Prio.-}{\teb{B}ased~\tnote{a}} &
                    \dl{Per}{\teb{S}ubordinate} &
                    \xmark &
                    \xmark &
                    \xmark &
                    \xmark &
                    \na &
                    \na &
                    \cmark &
                    \dl{Tech.-}{\teb{I}ndependent} &
                    \na \\
                    \arrayrulecolor{ieee-dark-black-40}\midrule
                    \multirow{6}{*}{\vspace{-1.6cm}\rot{\textbf{\erealm}}} &
                    \multirow{6}{*}{\vspace{-1.5cm}\rot{\dlb{\teb{S}ubordinate}{Guarding}}} &
                    SP805~\cite{arm2024armwatchdogmodu} &
                    \na &
                    \dl{Entire}{\teb{S}ystem} &
                    \na &
                    \na &
                    \xmark &
                    \cmark &
                    \xmark &
                    \tl{IRQ,}{Global}{\teb{R}eset}&
                    \xmark &
                    \dl{Tech.-}{\teb{I}ndependent} &
                    \na \\
                    &
                    &
                    Synopsys~\cite{synopsys2023enhancingarmsoc} &
                    \na &
                    \dl{Per}{\teb{S}ubordinate} &
                    \na &
                    \na &
                    \xmark &
                    \cmark &
                    \xmark &
                    \na &
                    \xmark &
                    \dl{Tech.-}{\teb{I}ndependent} &
                    \na \\
                    &
                    &
                    AMD~\cite{amd2017axiperformancem} &
                    \na &
                    \dl{Per}{\teb{S}ubordinate} &
                    \na &
                    \na &
                    \xmark &
                    \cmark &
                    \xmark &
                    \na &
                    \xmark &
                    \dl{Tech.-}{\teb{I}ndependent} &
                    \na \\
                    &
                    &
                    Ravi~\etal~\cite{ravi2014designofabusmon} &
                    \na &
                    \dl{Per}{\teb{S}ubordinate} &
                    \na &
                    \na &
                    \xmark &
                    \cmark &
                    \xmark &
                    \na &
                    \xmark &
                    \dl{Tech.-}{\teb{I}ndependent} &
                    \SI{8.86}{\kGE}~\tnote{k} \\ %
                    &
                    &
                    Kyung~\etal~\cite{kyung2007performancemoni} &
                    \na &
                    \dl{Per}{\teb{S}ubordinate} &
                    \na &
                    \na &
                    \xmark &
                    \cmark &
                    \xmark &
                    \na &
                    \xmark &
                    \dl{FPGA}{\teb{P}latform} &
                    \na \\
                    &
                    &
                    Lee~\etal~\cite{lee2014reconfigurableb} &
                    \na &
                    \dl{Per}{\teb{S}ubordinate} &
                    \na &
                    \na &
                    \xmark &
                    \cmark &
                    \cmark &
                    \dl{Logging-}{\teb{O}nly} &
                    \xmark &
                    \dl{FPGA}{\teb{P}latform} &
                    \na \\
                    &
                    &
                    AXIChecker~\cite{chen2010asynthesizablea} &
                    \na &
                    \dl{Per}{\teb{S}ubordinate} &
                    \na &
                    \na &
                    \na &
                    \xmark &
                    Performance &
                    \xmark &
                    \xmark &
                    \dl{Tech.-}{\teb{I}ndependent} &
                    \SI{70.7}{\kGE} \\
                    \arrayrulecolor{ieee-dark-black-40}\midrule
                    \multicolumn{2}{c}{\textbf{\irealm}} &
                    \multirow{2}{*}{\emph{\axirealm~[Ours]}} &
                    \dl{\teb{P}eriod-}{\teb{B}ased~\tnote{a}} &
                    \tlb{Configurable}{\teb{S}ubordinate}{\teb{R}egions} &
                    \dl{Transfer}{\teb{F}ragmentation} &
                    \dl{Write}{\teb{B}uffering} &
                    \dlb{Per}{\teb{M}anager} &
                    \tlb{Per-\teb{R}egion}{\teb{B}andwidth,}{\teb{L}atency} &
                    \na &
                    \multirow{2}{*}{\tlb{IRQ,}{\teb{P}er-\teb{S}ubordinate}{\teb{R}eset}} &
                    \dlb{Throttling}{\teb{M}echanism} &
                    \multirow{2}{*}{\dl{Tech.-}{\teb{I}ndependent}} &
                    \dlb{As \teb{L}ow \teb{A}s}{\SI{5}{\kGE}} \\
                    \multicolumn{2}{c}{\textbf{\erealm}} &
                    &
                    \na &
                    \dl{Per}{\teb{S}ubordinate} &
                    \na &
                    \na &
                    \dlb{Per}{\teb{S}ubordinate} &
                    \dlb{\teb{L}atency}{\teb{F}ine-\teb{G}ranular} &
                    \cmark &
                    &
                    \cmark &
                    &
                    \dlb{As \teb{L}ow \teb{A}s}{\SI{15}{\kGE}} \\
                    \arrayrulecolor{ieee-dark-black-100}\bottomrule
                \end{tabular}

                \begin{tablenotes}[para, flushleft]
                    \teb{\item[a]} in hardware
                    \teb{\item[b]} Xilinx Zynq-7020
                    \teb{\item[c]} Xilinx ZCU102
                    \teb{\item[d]} C=4
                    \teb{\item[e]} \alc{Xilinx XCZU9EG}
                    \teb{\item[f]} assuming \SI{1.28}{\micro\metre\squared} for \SI{1}{\GE}
                    \teb{\item[g]} SafeSU
                    \teb{\item[h]} SafeSU-2
                    \teb{\item[i]} assuming \SI{0.08748}{\micro\metre\squared} for \SI{1}{\GE} (\emph{NAND2x1\_ASAP7\_75t\_R})
                    \teb{\item[j]} Xilinx VC709
                    \teb{\item[k]} assuming \SI{0.718}{\micro\metre\squared} for \SI{1}{\GE}

                \end{tablenotes}
            \end{threeparttable}
        }
\end{sidetab}
}

\subsection{Subordinate Guarding: \erealm}
\label{sec:relwrk:erealm}

All the works described in \Cref{sec:relwrk:irealm} assume \emph{perfect} subordinate behavior, providing a response within a bounded time, and focus on the manager side without considering malfunctioning or misbehaving subordinates.
Our {\erealm} unit tackles these challenges by monitoring and guarding subordinate devices.
Transaction monitoring and guarding are crucial in studying security, performance analysis, fault detection, and system reliability.
With the {\erealm} unit, we use these established concepts in real-time memory interconnect systems.

Arm's \emph{SP805 Watchdog}~\cite{arm2024armwatchdogmodu} is primarily designed for fault detection and system protection by safeguarding the \glspl{soc} against \gls{sw} malfunctions due to unresponsive or runaway processes.
The operating system has to reset an internal counter regularly; if it becomes unresponsive, SP805 can either emit an interrupt or reset the entire system.
In contrast, {\erealm} provides a \gls{hw} solution that monitors every subordinate access, reducing fault detection latency.
We allow dynamic time budgeting of each subordinate device's transaction phases and thus support tight latency bounds for each device individually.
Unlike SP805, our approach allows the selective reset of the non-responsive subordinate device within a single cycle, leaving the rest of the system operational.

We identify multiple units specialized in monitoring subordinate devices.
With Synopsys' \emph{Smart Monitor}~\cite{synopsys2023enhancingarmsoc} and AMD's \emph{\gls{axi4} Performance Monitor}~\cite{amd2017axiperformancem}, industry provides performance monitoring solutions for \gls{axi4} buses and subordinate devices.
These units monitor bus traffic and compute key performance metrics, such as data byte count, throughput, and latency.
In academia, Ravi~{\etal} present a \emph{Bus Monitor}~\cite{ravi2014designofabusmon} and Kyung~{\etal} describe their \emph{Performance Monitoring Unit} (PMU)~\cite{kyung2007performancemoni} to capture key performance metrics such as transaction count, transfer size, and latency distributions for \gls{axi4} transactions through \gls{hw} counters.
Compared to {\erealm}, neither support multiple outstanding transactions nor provide detailed, \emph{stage-specific} transaction insights.
This limits their use in heterogeneous \glspl{soc}, where high-performance \glspl{dsa} emit complex transactions, and detailed performance reports of individual transactions are required.
Delayed or missing responses are not the only critical fault a subordinate device can experience.
Lee~{\etal}~\cite{lee2014reconfigurableb} describe a \emph{Reconfigurable Bus Monitor Tool Suite} for on-chip monitoring of \glspl{soc}.
The suite offers a \emph{Bus Monitor IP} designed to monitor the device's performance and check key protocol properties.
For the latter, it verifies simple specification-compliance, but unlike {\erealm}, it does not offer any protection in multi-{ID} scenarios with multiple outstanding transactions, e.g., \gls{tid} mismatch.
Chen~{\etal} developed \emph{AXIChecker}~\cite{chen2010asynthesizablea}, a rule-based, synthesizable protocol checker enforcing 44 rules ensuring managers and subordinates operate protocol-compliant.
Compared to {\erealm}, it can log protocol issues but lacks performance monitoring and reaction capabilities.

\subsection{Final Remarks: \axirealm}
\label{sec:relwrk:axirealm}

A distinctive aspect of \emph{{\axirealm}} is in its modular design.
It seamlessly combines ingress monitoring and throttling to ensure real-time behavior across managers with egress monitoring and guarding to guarantee timely responses from subordinate devices.
Its transparent and modular design requires minimal changes to the system, and its compatibility with many well-tested, silicon-proven crossbars and interconnects eases integration and verification.
Most \gls{soa} solutions explicitly restrict the design and evaluation on \gls{fpga} platforms, lacking support for \glspl{asic}.
Our technology-independent approach provides in-system and \gls{ip}-level gate-level characterization in a modern technology node, facilitating \gls{soa} comparisons.

\newpage
\section{Conclusion and Summary}
\label{chap:realm:conclusion}

In this chapter, we present {\axirealm} a lightweight, minimally invasive, architecture-independent, open-source interconnect extension to enable real-time behavior in high-performance interconnects used in heterogeneous systems.

The {\irealm} unit offers an effective solution for monitoring and moderating manager traffic during interference scenarios on a shared interconnect.
It provisions isolation and enforces real-time guarantees to managers executing critical tasks in heterogeneous systems.
Integrated into {\carfield}, an open-source \gls{mcs} research platform, we achieve \SI{68}{\percent} of the ideal performance in memory-bound applications, massively reducing the memory access latency by \SI{24}{\x}\!, while incurring less than \SI{2}{\percent} of additional area.
When distributing the budget in favor of the core, we achieve over \SI{95}{\percent} of the isolated performance.
Running applications from {TACLeBench}, we achieve over \SI{98}{\percent} of the isolated performance.

With the {\erealm} unit, we include an effective \gls{hw}-based solution to gracefully handle malfunctioning subordinates individually without stalling or locking the rest of the interconnect or the system.
The unit monitors the transaction latency and protocol correctness of each guarded subordinate, being able to inform the application-class core in as low as \SI{100}{cycles}, handshake open transaction, and reset the device should a transaction be overly delayed or the subordinate malfunctioning ensuring timely responses and real-time guarantees.

\chapter{Conclusions and Future Directions}
\label{chap:conclusion}

\section{Summary and Main Results}
\label{chap:realm:summary}

In my thesis titled \emph{\thetitle}, we firstly tackle efficient, agile, and high-performance data movement in heterogeneous systems by presenting a modular and highly customizable \gls{dma} architecture, \emph{\idma}, applicable across the entire compute continuum and serving the diverse needs of today's heterogeneous platforms.
Secondly, we elaborate how we integrate our {\dmaa} in a wide range of systems and how we enhance our architecture with multiple extensions, further accelerating \gls{dma} transfers and increasing their efficiency.
In multiple case studies, we show the applicability and the benefits of {\idma} in silicon-implemented systems and across the entire compute continuum, including an application-grade Linux-capable \gls{soc}, an energy-efficient multicore compute cluster, and an automotive-grade \gls{mcs}, each introducing its own {\idma} extension tailored to the system's use case.
We then explored the integration and use of {\idma} as an agile, tightly coupled data movement accelerator in a scaled-out manycore system optimized for massive floating\teb{-}point compute.
Finally, we introduce \emph{\axirealm}, a modular interconnect extension, tackling the predictability problem arising in heterogeneous \gls{mcs} running real-time-critical applications in the presence of \gls{dma} transfers originating from domain-specific accelerators.

We summarize the key achievements below:

\paragraph*{\textbf{Modular DMA Architecture}}
We first design a modular and highly versatile \gls{dma} architecture, called {\idma}, which can be used in applications ranging from simple and dedicated \glsu{io}-\glspl{dma} controlling autonomous data transfer for peripherals to high-performance multi-channel engines providing massive throughput at near-ideal bus utilization with minimal latency overheads of only one to two cycles.
With the key insight in mind, that most commonly used on-chip interconnect protocols share a byte-addressed data stream, {\idma} is designed to support five industry-grade protocols, translating seamlessly between them, facilitating {\idma}'s integration in heterogeneous \glspl{soc}.
Our architecture enables the creation of both ultra-small {\idmaes} incurring less than \SI{2}{\kilo\gateeq}, as well as large high-performance {\idmaes} running at over \SI{1}{\giga\hertz} on a \SI{12}{\nano\metre} node.
Flexibility and parameterization allow us to create configurations that achieve asymptotically full bus utilization and can fully hide latency in arbitrary deep memory systems while incurring less than \SI{400}{\gateeq} per trackable outstanding transfer. %
In a \teb{32-\si{\bit}} system, our {\idmaes} achieve almost perfect bus utilization for \SI{16}{\byte}-long transfers when accessing an endpoint with 100 cycles of latency.

\paragraph*{\textbf{Architectural DMA Extensions}}
We integrate {\idma} in a wide range of systems, ranging from \gls{ulp} to \gls{hpc} use cases, covering a large set of application scenarios and proving the applicability of the architecture.
As part of these integration efforts, we develop three different system bindings for {\idma}:
a simple configuration-register-based interface, a Linux-capable transfer-descriptor-based scheme, and an agile {\riscv}-compliant instruction-based {\fe}.
Integrated in {\cheshire}, our Linux-capable \gls{soc} platform~\cite{ottaviano2023cheshirealightw}, we achieve \SI{1.66}{\x} less latency, increasing bus utilization by up to \SI{2.5}{\x}\! in an ideal memory system with 64-\si{\byte} transfers, overall requiring \SI{11}{\percent} fewer \teb{\si{\lut}} and \SI{23}{\percent} fewer \teb{\si{\ff}} without requiring any block {RAM}s compared to Arm's {LogiCore} \gls{dma} \gls{ip}~\cite{xilinx2022axidmav71logico}. %
In deep memory systems, we show an even more significant increase in the utilization of \SI{3.6}{\x}\! with 64-\si{\byte} transfers.

To accelerate commonly occurring transfer patterns, we develop a set of \gls{dma} extensions facilitating data orchestration by handling \alc{\gls{nd}} tensor transfers, and scheduling of real-time transfers directly in hardware. We further include a standardized interface in {\idma}'s data path to integrate in-stream operations.
Finally, we develop a stream-optimized \gls{mmu} to be integrated into {\idma}, allowing the \gls{dma} to autonomously handle \gls{vm} in conjunction with an application-grade host.
Equipped with our \alc{\gls{nd}} tensor {\me} and our light-weight register-based {\fe}, our {\idmae} almost fully utilizes the bandwidth to the \gls{l2} and \gls{tcdm} in both directions in {\pulpopen}: %
measuring with the on-board timer, a transfer of \SI{8}{\kibi\byte} from the cluster's \gls{tcdm} to \gls{l2} requires 1107 cycles, of which 1024 cycles are required to transfer the data using a 64-\si{\bit} data bus.
Our multi-channel extensions show performance improvements in memory-bound kernels like vector addition and the dot product, which are dominated by the data transfers and reach speedups of \SI{15.7}{\x} and \SI{15.8}{\x}\!.
From \gls{fpga} profiling runs, we find that our real-time \glsf{sdma} engine saves about 2200 execution cycles every scheduling period, thus increasing the slack of the \gls{pvct} task.
In the case of eight events and sixteen outstanding transactions, our \gls{sdma} unit is about \SI{11}{\kilo\gateeq} in size, accounting for an area increase of only \SI{0.001}{\percent} of the original {\controlpulp}'s area. %
Enhancing our {\dmaa} with \gls{vm} support with our \gls{smmu} requires \SI{8.1}{\kilo\ff} and \SI{9.0}{\kilo\lut} whilst not impacting the critical path of the \gls{soc}.
Copying transfers of varying length, \gls{smmu} can achieve in excess of \SI{99}{\percent} of the non-translated performance when issuing 4-\si{\kilo\byte}-sized transfers.

\paragraph*{\textbf{Communication Processor}}
We combine {\idma} with a tiny \SI{20}{\kilo\gateeq} {\riscv} processor tightly coupled through our instruction-based system binding, creating an agile communication processor capable of complex data scheduling and orchestration.
We integrate this communication processor into a compute-cluster-based accelerator architecture with a focus on an optimal connection between the {\dma} engine and the cluster-local scratchpad memory.
This cluster architecture is then scaled out to a 432-core dual-chiplet manycore system featuring fourteen {\idma} engines and \SI{16}{\gibi\byte} of HBM2E memory connected through a hierarchical crossbar-based point-to-point interconnect.
For the highly compute-bound \gls{gemm}, our {\idmaes} enable moderate, but significant speedups of \SI{1.37}{\x} to \SI{1.52}{\x} increasing peak \gls{hbm} read bandwidth from 17 to \SI{26}{\giga\byte\per\second}.
For \gls{spmv} performance, our cluster-level {\idmaes} approach the wide interconnect peak throughput of \SI{384}{\giga\byte\per\second}, overall enabling significant speedups of \SI{5.9}{\x} to \SI{8.4}{\x}\!.
\gls{spmm} is similar to \gls{spmv}, but enables on-chip matrix data reuse, becoming compute-bound for both the baseline and {\idmaes}, still achieving speedups of \SI{2.9}{\x} to \SI{4.9}{\x}\!.

\paragraph*{\textbf{Real-time Interconnect Extensions}}
With an efficient \gls{dma} engine developed and its integration into multiple systems studied, we shift focus towards the host by investigating real-time data transfers in heterogeneous \glspl{mcs}, which become increasingly important with the advent of \gls{ai}-driven applications in automotive and aeronautical use cases.
We develop {\axirealm}, a lightweight and interconnect-agnostic helper-module-based architecture to modulate the amount of data managers can inject into the shared interconnect.
This is especially important, as {\idma}-equipped accelerators are tuned for maximal bandwidth efficiency through long data bursts, which can, if kept unregulated, lead to adverse performance of time-critical applications running on the host cores.
Integrated into {\carfield}, an open-source \gls{mcs} research platform, we achieve \SI{68}{\percent} of the ideal performance in memory-bound applications, massively reducing the memory access latency by \SI{24}{\x}\!, while incurring less than \SI{2}{\percent} of additional area.
When distributing the budget in favor of the core, we achieve over \SI{95}{\percent} of the isolated performance.
Running applications from {TACLeBench}, we achieve over \SI{98}{\percent} of the isolated performance.
introducing the {\erealm} units to monitor the transaction latency and protocol correctness of each guarded subordinate, being able to inform the application-class core in as low as \SI{100}{cycles}, handshake open transaction, and reset the device should a transaction be overly delayed or the subordinate malfunctioning, we can ensure timely responses and real-time guarantees.

These contributions address pressing challenges in designing energy-efficient and high-performance real-time memory systems for heterogeneous \glspl{mcs} required to carry the next generation of cyber-physical systems.

A large portion of this work has been contributed to the \gls{epi} project\,\footnote{\url{https://www.european-processor-initiative.eu}}.
An even larger portion of this work has been silicon-proven in a wide range of tapeouts, see \Cref{chap:chip_gallery}, providing an energy-efficient and high-performance data movement solution.
To allow anyone to make use of this work, all design files and many \gls{asic} implementations are freely available under a libre license\,\footnote{\url{https://github.com/pulp-platform}}.

\newpage
\section{Outlook}
\label{chap:realm:outlook}

This thesis lays the groundwork for multiple future research directions.
Thanks to the open-source nature of this work and our detailed \gls{ooc} models of our architectural components, we provide a strong baseline for broader adaptations of our ideas.

\paragraph*{\textbf{Reliable Data Movement}}

With \Cref{chap:realm}, we present a light-weight interconnect extension to enable high-performance real-time interconnect architectures in \glspl{mcs}, which is a fundamental requirement to adopt zonal and domain architectures in next-generation \gls{aces}-enabled \glspl{sdv}.
To expand our systems to include aerospace applications, tackling reliable transfer becomes a core necessity.
Our {\idma} architecture can be expanded to not only feature error handling, but also complete data integrity checks and perform, if required, retransmission of erroneous data transfers.
With our ability to support multiple read and write ports with {\idma}, we could read data from multiple sources, and add an in-stream accelerator to check the data's integrity and add redundant data before sending it over our high-performance, real-time-enabled interconnect.

\paragraph*{\textbf{Extend the Concept of an Intelligent {\DmaE}}}

Integrated in our Snitch cluster and coupled with our data movement core, our {\idmae} becomes a data movement accelerator allowing us to efficiently perform complex and even data-dependent transfer patterns.
Still coupled to a Snitch core, but outside a compute-cluster, {\idma} could truly become an intelligent data mover.
Coupled with in-stream accelerators, e.g., matrix transposition units, such a unit could be programmed using high-level \gls{api} calls and complete complex \gls{dma} transfers autonomously as a service.

\paragraph*{\textbf{Scaling-out of an {\axirealm} System}}

\Cref{chap:realm} focuses enhancing high-performance on crossbar-based point-to-point interconnects with real-time guarantees.
Future work could tackle a fully scaled-out \gls{mcs} using a \gls{noc}-based interconnect.
{\axirealm} is designed to fundamentally be compatible with \gls{axi4}-based \glspl{noc}, but work is needed to evaluate our ideas in such distributed contexts.

\appendix

\clearpage
\chapter{Chip Gallery}
\label{chap:chip_gallery}

This appendix lists all fabricated chips the author has contributed to\teb{, in both} the form of technical work or supervision.
A complete, up-to-date list can be found online at:
\url{http://asic.ethz.ch/authors/Thomas_Benz.html}.

\newpage

\section{Scarabaeus}
\begin{figure}[H]
    \centering
    \includegraphics[width=1\textwidth]{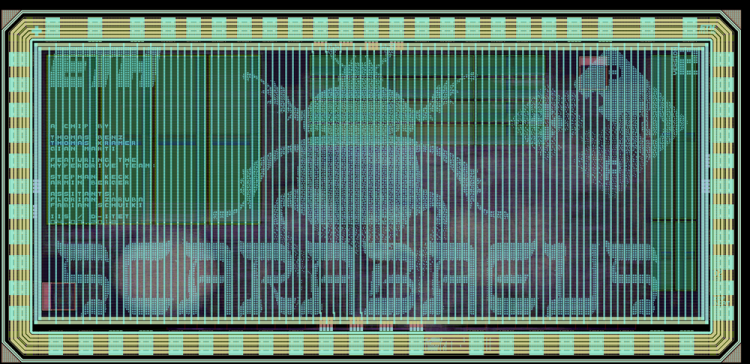}   \\
\end{figure}

\begin{figure}[H]
    \centering
    \includegraphics[width=1\textwidth]{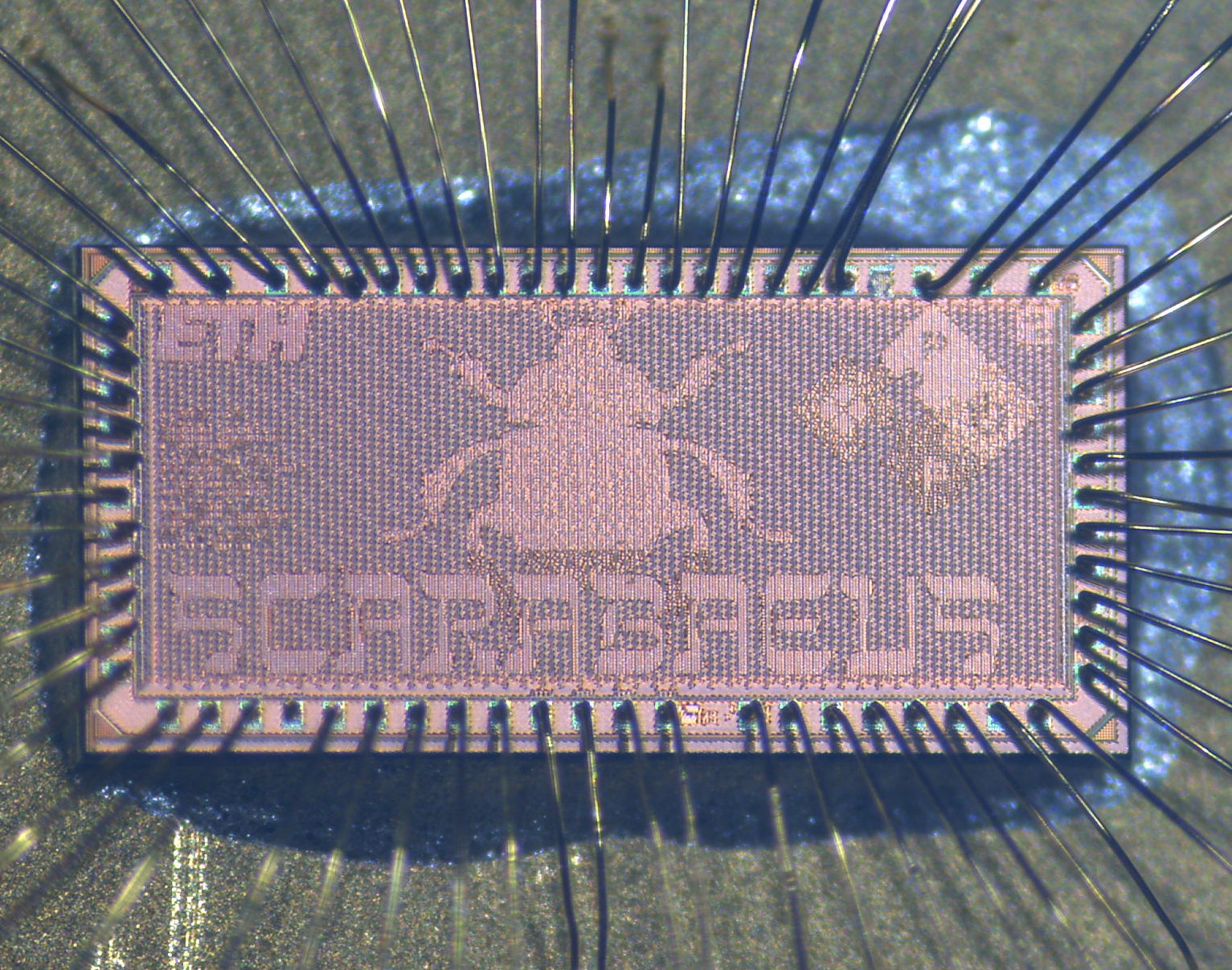}   \\
\end{figure}

\begin{table}[H]
    \centering
    \small
    \begin{tabularx}{\linewidth}{@{}ll@{}}
        \toprule
        Name                      & Scarabaeus                            \\
        \midrule
        Designers                 & \textbf{Thomas Benz}, Thomas Krammer, \\
                                  & Gian Marti, Stephan Keck,             \\
                                  & Armin Berger, Fabian Schuiki,         \\
                                  & Florian Zaruba                        \\
        Application / Publication & \teb{{\pulp} Student \gls{mcu}}       \\
        Technology / Package      & UMC65 / QFN56                         \\
        Dimensions                & \SI{2.626}{\milli\meter}~$\times$~\SI{1.252}{\milli\meter} \\
        Circuit Complexity        & \SI{1200}{\kilo\GE}                   \\
        Voltage                   & \SI{1.2}{\volt}                       \\
        Power                     & \SI{45.97}{\milli\watt} @ \SI{1.2}{\volt} and \SI{200}{\mega\hertz} \\
        Clock \teb{Frequency}     & \SI{200}{\mega\hertz}                 \\
        \bottomrule
    \end{tabularx}
\end{table}

\Gls{scarabaeus}, affectionately named after a \emph{Dung Beetle}, implements an Ariane-based~\teb{\cite{zaruba2019thecostofapplic}} RISC-V (RV64IMC) \gls{soc} including \teb{a} \SI{64}{\kibi\byte} \gls{l2} memory, \teb{and a} \SI{4}{\kibi\byte} data and instruction cache.
In this project, \teb{we extend} the existing \teb{{\pulp}} modules by a \gls{plic} as specified by the {\riscv} standard, and by a \gls{dma} controller that is capable of transferring arbitrarily shaped, aligned and strided data of up to \teb{4-D} on a 64-\si{\bit} \gls{axi4} bus.

The chip also includes a \emph{HyperBus} controller, \teb{rewritten from scratch}, and \teb{supporting} \emph{Cypress HyperRAM}~\cite{semiconductor2019hyperbusspec} \teb{chips}.
This interface contains \teb{thirteen} \teb{\gls{io}} pads occupies the lower side of the chip.
The control logic has been placed between the pads and occupies less than \SI{10}{\kilo\GE} including clock synchronization between two independent clock domains (\teb{\emph{HyperBus}} \teb{runs} at \SI{166}{\mega\hertz} and \teb{the \gls{soc} at} \SI{200}{\mega\hertz}).
When operating with long \teb{data} bursts, \teb{we achieve an} effective bandwidth \teb{of} \SI{327}{\mega\byte\per\second}.

The chip \teb{further} includes several peripeharls, a JTAG \teb{debug module} for programming, \teb{eight} GPIOs, and an UART.
Finally the chip contains an 8-\si{bit} RISC processor called \gls{fireore}.
\Gls{fireore} uses the \gls{aox-01} instruction set architecture, which specifies a four-class reduced instruction set with three 8-\si{bit} special-purpose registers.

\newpage

\section{Thestral}
\begin{figure}[H]
    \centering
    \includegraphics[width=0.65\textwidth]{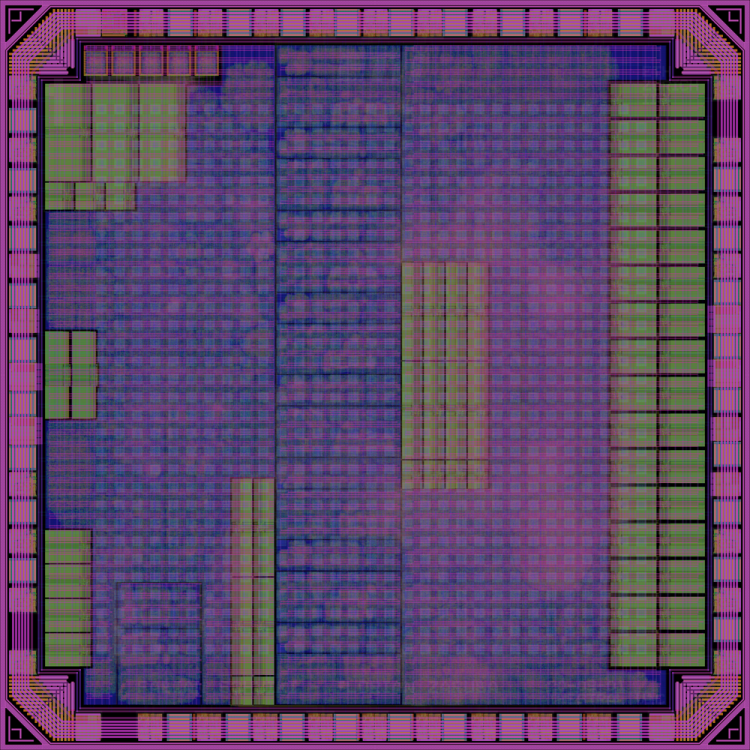}   \\
\end{figure}

\begin{figure}[H]
    \centering
    \includegraphics[width=0.65\textwidth]{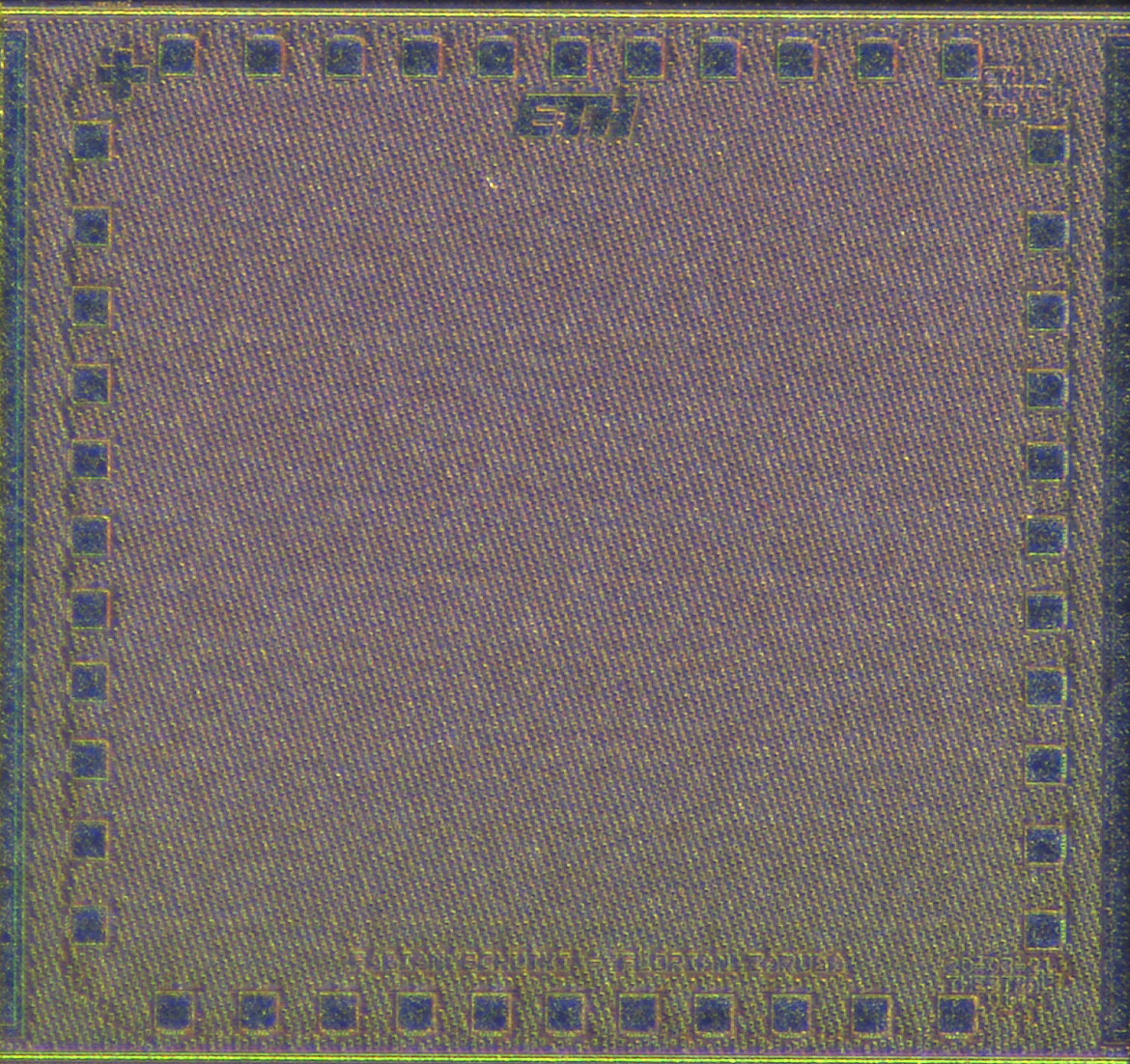}   \\
\end{figure}

\begin{table}[H]
    \centering
    \small
    \begin{tabularx}{\linewidth}{@{}ll@{}}
        \toprule
        Name                      & Thestral                              \\
        \midrule
        Designers                 & Fabian Schuiki, Florian Zaruba,       \\
                                  & \textbf{Thomas Benz}, Paul Scheffler, \\
                                  & Wolfgang Roenninger                   \\
        Application / Publication & \teb{{\pulp} \gls{soc} / Benz~\etal~\cite{benz2021a10coresocwith2}} \\
        Technology / Package      & GF22 / QFN40                          \\
        Dimensions                & \SI{1.25}{\milli\meter}~$\times$~\SI{1.25}{\milli\meter} \\
        Circuit Complexity        & \SI{6}{\mega\GE}                      \\
        Voltage                   & \SI{0.8}{\volt}                       \\
        Clock \teb{Frequency}     & \SI{650}{\mega\hertz}                 \\
        \bottomrule
    \end{tabularx}
\end{table}

Thestral is a small test chip \teb{implementing our } Snitch compute cluster \teb{architecture~\cite{zaruba2021snitchatinypseu}}.
Snitch \teb{is centered around a} small 32-\si{\bit} {\riscv} (RV32IMAF) core \teb{coupled to a capable 64-\si{\bit}} \gls{fpu} \teb{subsystem optimized} for stream processing.

The chip contains our next \emph{Snitcholution} (\emph{Snitch evolution}) with architectural improvements and \teb{fine-grain} power gating infrastructure.
Notable changes compared to the system in \emph{Baikonur}~\teb{\cite{zaruba2021manticorea4096c}} are:

\begin{itemize}
    \item Thestral features 20 different power domains. Each \gls{fpu}/\gls{ipu}, as well as the cluster, can be individually \teb{power gated}.
    \item Double-pumped \glspl{fpu} and \glspl{ipu}.
    The compute units (and the \gls{tcdm}) can be operated at twice the integer core speed, allowing for higher peak throughput.
    \item SV32 \teb{\gls{vm}} support
    \item \teb{More accurate \gls{l1} instruction cache prefetching}
    \item \teb{Thestral is the first silicon demonstrator including {\idma}}
\end{itemize}

\teb{Thestral's architecture} features one governor core with \teb{it's private} \gls{fpu}, \gls{ipu}, \gls{tcdm}, and \teb{\gls{l1}} instruction cache.
The governor manages the \gls{soc}, \teb{by controlling the system {\idma} and the chip's fine-grain} power management.

The primary compute is provided by a Snitch cluster consisting of eight cores, \teb{each coupled to a} \gls{fpu} \teb{and a \gls{ipu} subsystem} accessing a \SI{64}{\kibi\byte} \gls{tcdm}.

Additional peripherals such as \emph{HyperBus}, a \teb{fully digital} serial link, \emph{USB device}, and \emph{I2C} provide enough infrastructure to operate the chip stand-alone.
\SI{24}{\kibi\byte} \gls{llc} are used to reduce expensive off-chip accesses to the \emph{HyperBus} device.

Thestral is the third chip in our Snitch-based systems, following \emph{Billywig} and \emph{Baikonur}~\teb{\cite{zaruba2021manticorea4096c}}, \teb{the latter includes three} clusters and \teb{one} governor.
The naming continues \teb{our} tradition \teb{naming chips after beasts from the} \emph{Harry Potter} universe.
As described in the \emph{Harry Potter Fandom wiki}, a \emph{Thestral} can only be seen by people who've seen death, \teb{or in our case, dead chips}.

\cleardoublepage
\newpage
\ %
\newpage

\section{Dogeram}

\begin{figure}[H]
    \centering
    \includegraphics[width=1\textwidth]{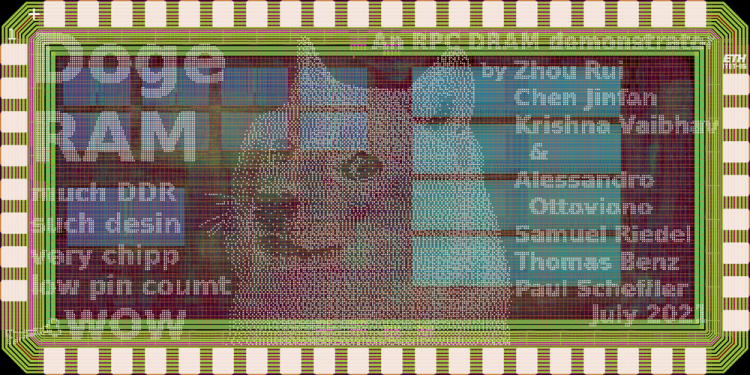}   \\
\end{figure}

\vspace{0.25cm}

\begin{figure}[H]
    \centering
    \includegraphics[width=\textwidth]{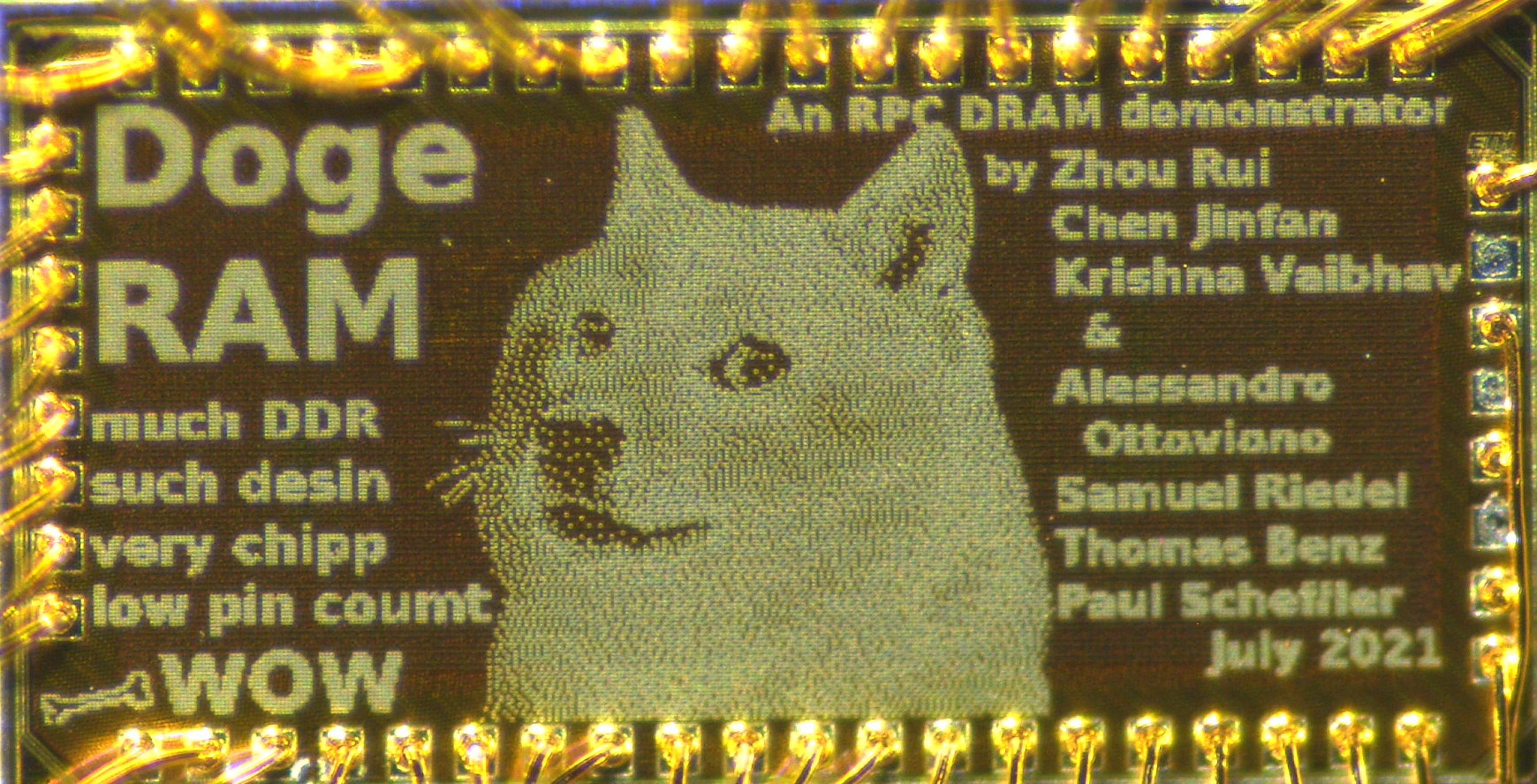}   \\
\end{figure}

\vfill

\newpage

\begin{table}[H]
    \centering
    \small
    \begin{tabularx}{\linewidth}{@{}ll@{}}
        \toprule
        Name                      & Dogeram                                            \\
        \midrule
        Designers                 & Rui Zhou, Jinfan Chan,                             \\
                                  & Vaibhav Krishna, Alessandro Ottaviano,             \\
                                  & Samuel Riedel, \textbf{Thomas Benz},               \\
                                  & Paul Scheffler                                     \\
        Application / Publication & \teb{{\pulp} Student \gls{mcu}} / \teb{Ottaviano~\etal~\cite{ottaviano2023cheshirealightw}} \\
        Technology / Package      & TMSC65 / QFN56                                     \\
        Dimensions                & \SI{2}{\milli\meter}~$\times$~\SI{1}{\milli\meter} \\
        Voltage                   & \SI{1.2}{\volt}                                    \\
        Clock \teb{Frequency}     & \SI{200}{\mega\hertz}                              \\
        \bottomrule
    \end{tabularx}
\end{table}

DogeRAM is a chip to test \teb{our newly developed} \gls{rpc} \gls{dram} interface\teb{, which is connected over the system's} \gls{axi4} crossbar to a minimal \teb{two-core} Snitch-based \teb{{\pulp} \gls{soc}}.
At \SI{200}{\mega\hertz} (at worst case conditions), the chip achieves up to \SI{751.5}{\mega\byte\per\second} throughput over the \gls{rpc} \gls{dram} interface.
The name captures a bit the Zeitgeist and is a play on the meme culture.

\newpage

\section{Zest}

\begin{figure}[H]
    \centering
    \includegraphics[width=0.85\textwidth]{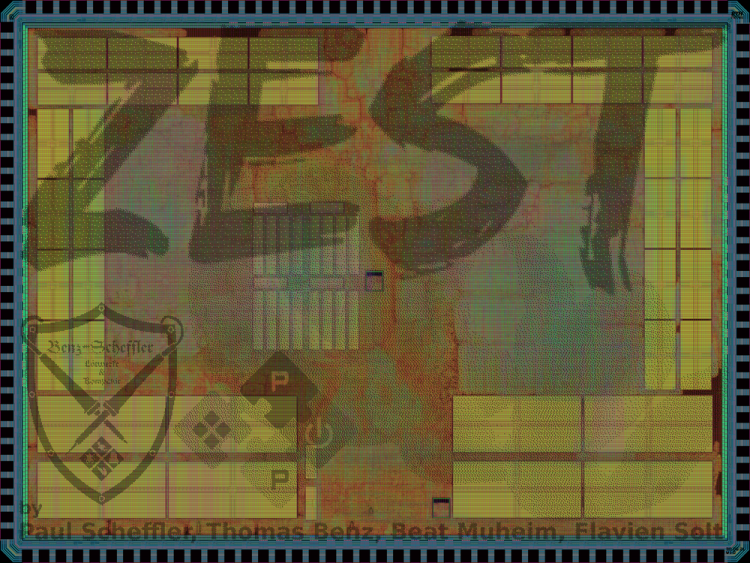}   \\
\end{figure}

\begin{figure}[H]
    \centering
    \includegraphics[width=0.85\textwidth]{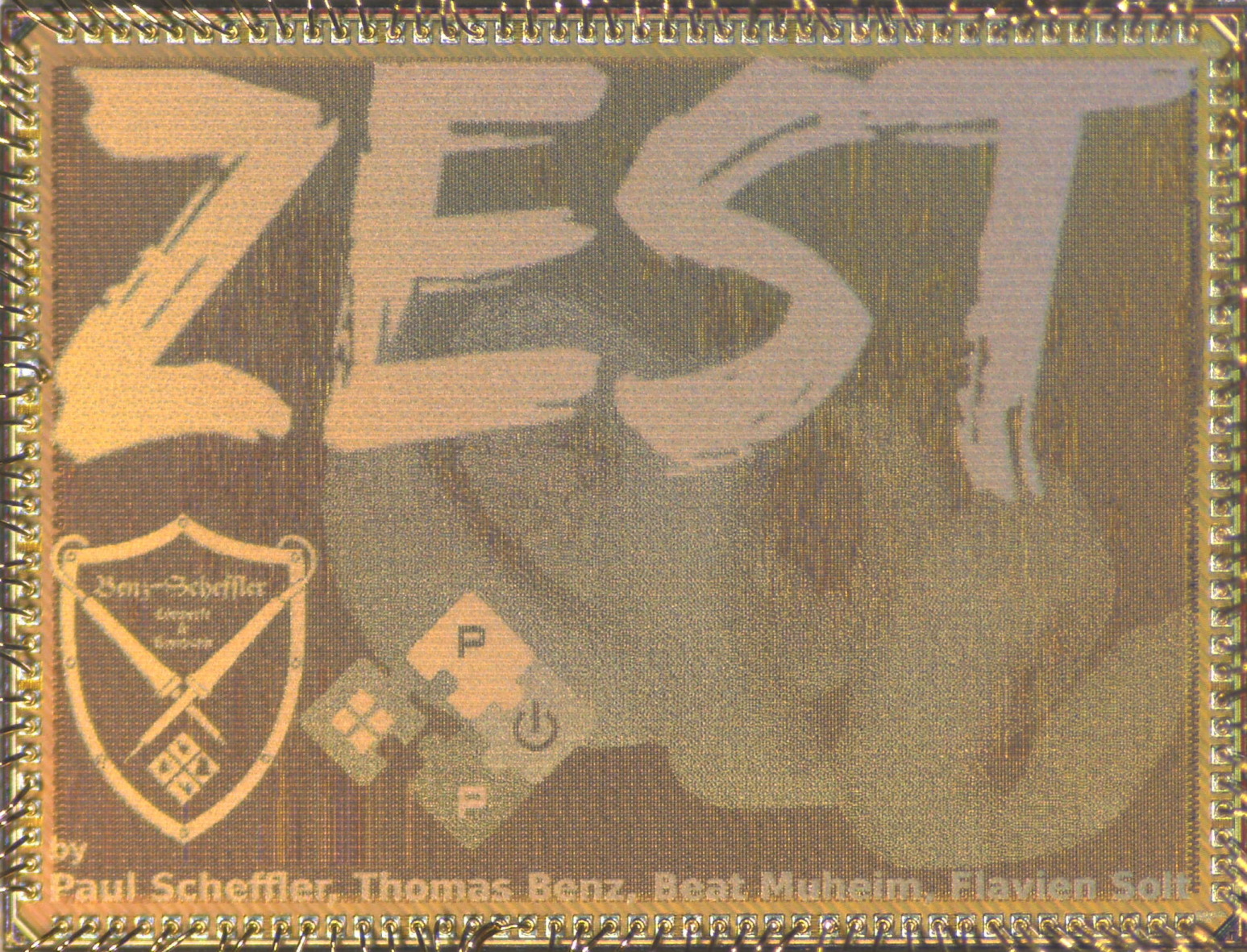}   \\
\end{figure}

\begin{table}[H]
    \centering
    \small
    \begin{tabularx}{\linewidth}{@{}ll@{}}
        \toprule
        Name                      & Zest                                              \\
        \midrule
        Designers                 & Flavien Solt, \textbf{Thomas Benz},                \\
                                  & Paul Scheffler, Beat Muheim                        \\
        Application / Publication & \teb{\pulp} Student \gls{soc}                      \\
        Technology / Package      & TMSC65 / QFN88                                     \\
        Dimensions                & \SI{4}{\milli\meter}~$\times$~\SI{3}{\milli\meter} \\
        Circuit Complexity        & \SI{6}{\mega\GE}                                   \\
        Voltage                   & \SI{1.2}{\volt}                                    \\
        Clock \teb{Frequency}     & \SI{230}{\mega\hertz}                              \\
        \bottomrule
    \end{tabularx}
\end{table}

Zest is the first \gls{soc} of our group \teb{featuring a} Snitch and \teb{ a {\pulp}} clusters in \teb{one \gls{soc}}.
It also demonstrates the \emph{Fenrir peripheral system}\teb{, a predecessor to the \gls{io}-\gls{dma}, presented in \Cref{chap:dmaext:iodma}}, supporting \emph{UART}, \emph{I2C}, \emph{QSPI}, \emph{Camera interface}, and \emph{DVSI event camera link}.
Zest additionally features a \emph{HyperBus} \teb{controller} and a custom double data rate serial link.
As a result of all these additions, it uses a rather large QFN88 package.

In total Zest has eight \emph{RI5CY} cores connected to a \SI{128}{\kibi\byte} \gls{tcdm} memory, \emph{4+1} Snitch cores in their own cluster with \SI{128}{\kibi\byte} \gls{tcdm}, and an additional \SI{256}{\kibi\byte} of shared memory for both clusters.

There is an orange peel on the logo as a play on the name Zest.

\newpage

\section{Neo}

\begin{figure}[H]
    \centering
    \includegraphics[width=1\textwidth]{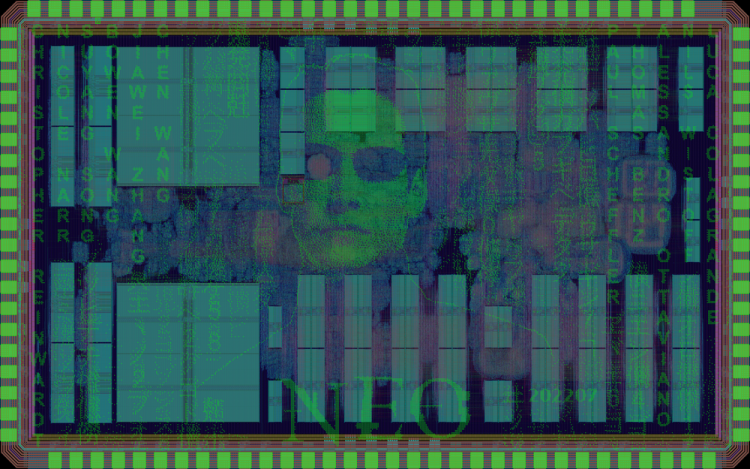}   \\
\end{figure}

\begin{figure}[H]
    \centering
    \includegraphics[width=1\textwidth]{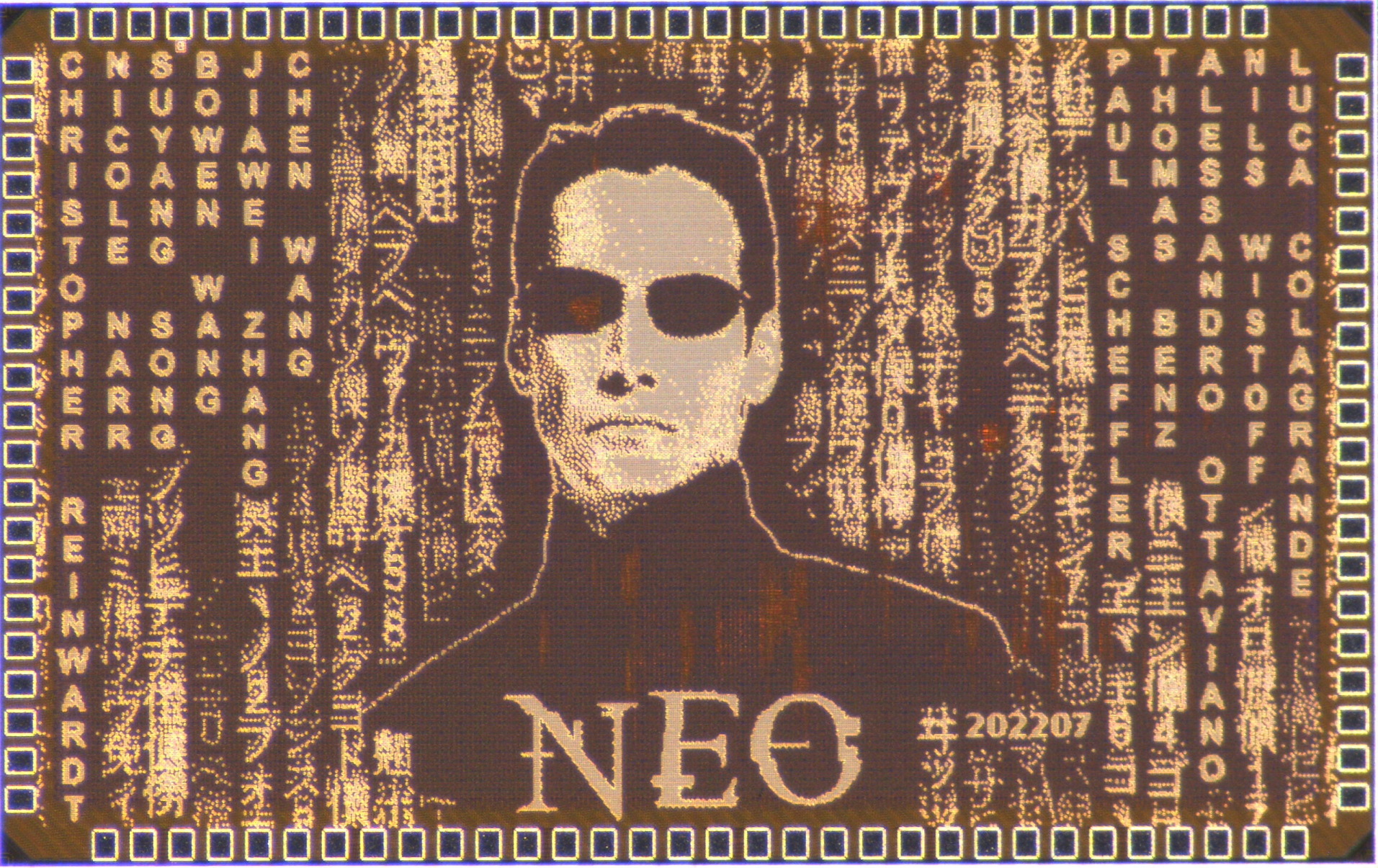}   \\
\end{figure}

\begin{table}[H]
    \centering
    \small
    \begin{tabularx}{\linewidth}{@{}ll@{}}
        \toprule
        Name                      & Neo                                                   \\
        \midrule
        Designers                 & Christopher Reinwardt, Nicole Narr,                   \\
                                  & Suyong Song, Bowen Wang,                              \\
                                  & Jiawei Zhang, Chen Wang,                              \\
                                  & Paul Scheffler, \textbf{Thomas Benz},                 \\
                                  & Alessandro Ottaviano, Nils Wistoff,                   \\
                                  & Luca Colagrande                                       \\
        Application / Publication & \teb{\pulp} Student \gls{soc} / \teb{Ottaviano~\etal~\cite{ottaviano2023cheshirealightw}} \\
        Technology / Package      & TMSC65 / QFN64                                        \\
        Dimensions                & \SI{3.2}{\milli\meter}~$\times$~\SI{2}{\milli\meter}  \\
        Circuit Complexity        & \SI{2500}{\kilo\GE}                                   \\
        Voltage                   & \SI{1.2}{\volt}                                       \\
        Power                     & \SI{65}{\milli\watt} @ \SI{1.2}{\volt} and \SI{200}{\mega\hertz} \\
        Clock \teb{Frequency}     & \SI{200}{\mega\hertz}                                 \\
        \bottomrule
    \end{tabularx}
\end{table}

Neo is the first design that integrates a Linux-capable {\riscv} core - the well-known Ariane (CVA6) - with an enhanced \gls{rpc} \gls{dram} memory controller for off-chip communication and peripherals to form an \gls{soc} capable of booting Linux.

In addition, a new VGA controller allows to drive VGA displays, and its functionality has been proven by emulating the design on an \gls{fpga} before chip tapeout.
The \gls{rpc} controller is fully \gls{axi4}-compliant, which makes it stand as \teb{a} reusable \gls{ip}, that allows off-chip communication with \SI{700}{\mebi\byte\per\second} bandwidth while preventing an explosion in the number of off-chip memory pins.

The name of the chip was supposed to be \emph{Neo-Scarabaeus} as a follow up chip to \emph{Scarabaeus} which had a similar goal.
The name was just a bit too long, and we ended up keeping the Neo part.
Once the name stuck like that, the logo followed as well.

\newpage

\section{Occamy}

\begin{figure}[H]
    \centering
    \includegraphics[width=1\textwidth]{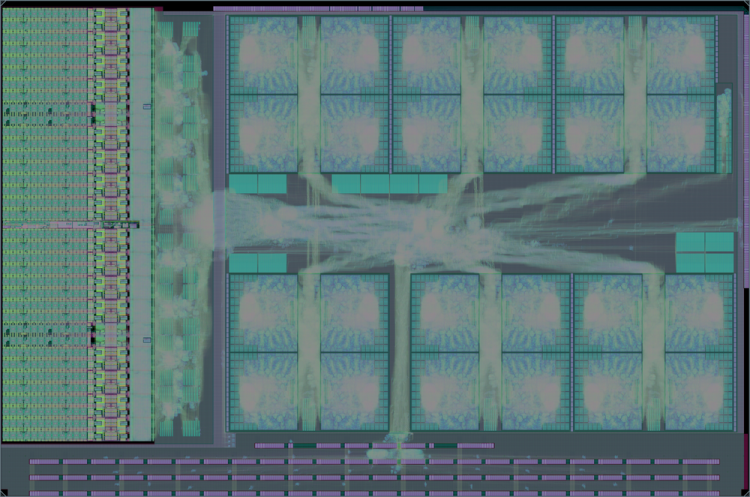}   \\
\end{figure}

\begin{figure}[H]
    \centering
    \includegraphics[width=1\textwidth]{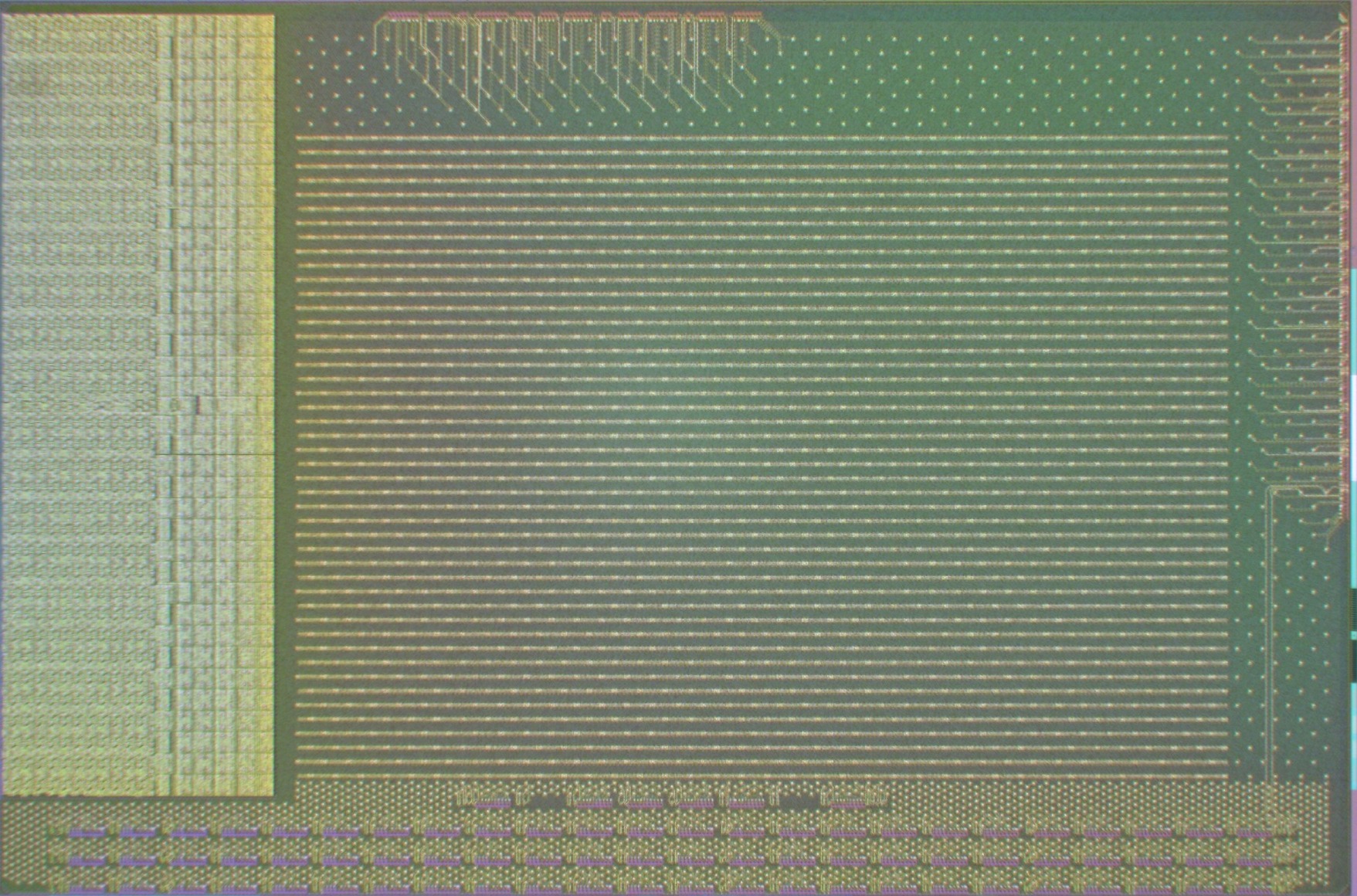}   \\
\end{figure}

\begin{table}[H]
    \centering
    \small
    \begin{tabularx}{\linewidth}{@{}ll@{}}
        \toprule
        Name                      & Occamy                                                \\
        \midrule
        Designers                 & Gianna Paulin, Florian Zaruba,                        \\
                                  & Fabian Schuiki, Stefan Mach,                          \\
                                  & Manuel Eggimann, Matheus Cavalcante,                  \\
                                  & Paul Scheffler, Yichao Zhang,                         \\
                                  & Tim Fischer, Nils Wistoff,                            \\
                                  & Luca Bertaccini, \textbf{Thomas Benz},                \\
                                  & Luca Colagrande, Alfio Di Mauro,                      \\
                                  & Andreas Kurth, Samuel Riedel,                         \\
                                  & Noah Huetter, Gianmarco Ottavi,                       \\
                                  & Zerun Jiang, Beat Muheim,                             \\
                                  & Frank K. Gurkaynak, Davide Rossi,                     \\
                                  & Luca Benini                                           \\
        Application / Publication & \teb{\pulp} \gls{hpc} / \teb{Scheffler~\etal~\cite{scheffler2025occamya432cored}} \\
        Technology / Package      & GF12 / LGA2011                                        \\
        Dimensions                & \SI{10.5}{\milli\meter}~$\times$~\SI{6.95}{\milli\meter}  \\
        Circuit Complexity        & \SI{600}{\mega\GE}                                    \\
        Voltage                   & \SI{0.8}{\volt}                                       \\
        Power                     & \SI{10}{\watt} @ \SI{0.8}{\volt} and \SI{1}{\giga\hertz} \\
        Clock \teb{Frequency}     & \SI{1}{\giga\hertz}                                   \\
        \bottomrule
    \end{tabularx}
\end{table}

\teb{\occamy} is a research prototype to demonstrate and explore the scalability, performance, and efficiency of our {\riscv}-based architecture in a 2.5D- integrated chiplet system showcasing {\gf}' technologies and \teb{their} \gls{ip} ecosystem, as well as Rambus' and Micron's \teb{\gls{hbm2e} \glspl{ip}}.

The \teb{\occamy} project started as a serendipitous outcome of the \emph{Manticore}\teb{~\cite{zaruba2021manticorea4096c}} high-performance architecture concept we presented at the Hot Chips symposium in 2020.
After Hot Chips 2020, the PULP platform team was approached by {\gf} with an exciting proposal to turn a concept architecture into a real silicon design.
The project was made possible by the generous contribution and strong support of {\gf} (technology access, expert advice, ecosystem enablement, and silicon budget), Rambus (\gls{hbm2e} controller \gls{ip} and integration support), Micron (\gls{hbm2e} \gls{dram} supply and integration support), Synopsys (\gls{eda} tool licenses and support) and Avery (\gls{hbm2e} \gls{dram} verification model).
We kick-started the \teb{\occamy} project on the 20th of April 2021 and taped out the compute chiplet in {\gf}' \SI{12}{\nano\metre} FinFet technology in July 2022 after less than \SI{15}{months} of hard work with a team of less than \SI{25}{people}.

In this work, we combine a small and super-efficient, in-order, 32-\si{\bit} {\riscv} integer core called Snitch with a large multi-precision-capable \gls{fpu} \teb{subsystem} enhanced with \gls{simd} capabilities.
In addition to the standard {\riscv} \gls{fma} instructions, the two 8-\si{\bit} and two 16-\si{\bit} \gls{fp} formats have the new expanding sum-dot-product and three-addend summation (\emph{exsdotp}, \emph{exvsum}, and \emph{vsum}) instructions.

To achieve ultra-efficient computation on data-parallel \emph{FP} workloads, two architectural extensions are exploited: data-prefetchable register file entries and repetition buffers.
The corresponding {\riscv} \gls{isa} extensions \glspl{ssr} and \gls{frep} enable the Snitch core to achieve \gls{fpu} utilization higher than \SI{90}{\percent} for compute-bound kernels.

Each \teb{\occamy} chiplet contains 216 Snitch cores organized in \teb{six} groups of four compute clusters.
Each cluster shares a \gls{tcdm} among eight compute cores and a high-bandwidth (512-\si{\bit}) \gls{dma}-enhanced core orchestrating the data flow\teb{, see \Cref{chap:comcpu}}.
An \gls{axi4}-based wide, multi-stage interconnect and dedicated {\idmaes} help manage the massive on-chip bandwidth.
A CVA6 Linux-capable {\riscv} core manages all compute clusters and system peripherals.
Each chiplet has a private \SI{16}{\gibi\byte} \teb{\gls{hbm2e}} and can communicate with a neighboring chiplet over a \SI{8}{\giga\byte\per\second} wide, source-synchronous technology-independent die-to-die DDR link.
The dual-chiplet \teb{\occamy} system achieves a peak performances of \SI{0.768}{\tera\flop\per\second} for FP64, \SI{1.536}{\tera\flop\per\second} for FP32, \SI{3.072}{\tera\flop\per\second} for FP16/FP16alt, and \SI{6.144}{\tera\flop\per\second} for FP8/FP8alt.

\cleardoublepage
\newpage
\ %
\newpage

\section{Carfield}

\begin{figure}[H]
    \centering
    \includegraphics[width=1\textwidth]{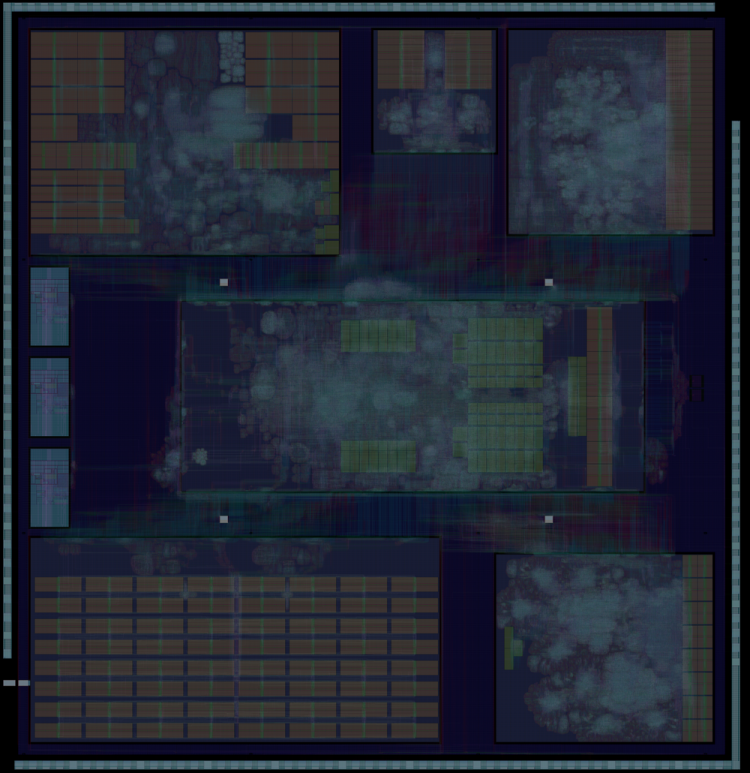}   \\
\end{figure}

\begin{figure}[H]
    \centering
    \includegraphics[width=1\textwidth]{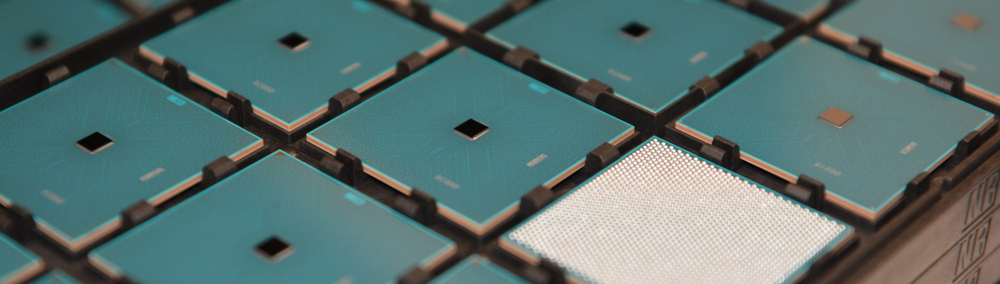}   \\
\end{figure}
\vfill
\newpage

\begin{table}[H]
    \centering
    \small
    \begin{tabularx}{\linewidth}{@{}ll@{}}
        \toprule
        Name                      & Carfield                                              \\
        \midrule
        Designers                 & Angelo Garofalo, Alessandro Ottaviano,                \\
                                  & Matteo Perotti, Robert Balas,                         \\
                                  & Yvan Tortorella, Michael Rogenmoser,                  \\
                                  & Chi Zhang, Luca Bertaccini,                           \\
                                  & Nils Wistoff, Maicol Ciani,                           \\
                                  & Thomas Benz, Cyril Koenig,                            \\
                                  & Luca Valente, Mattia Sinigaglia,                      \\
                                  & Paul Scheffler, Manuel Eggimann,                      \\
                                  & Matheus Cavalcante, Beat Muheim,                      \\
                                  & Zerun Jiang, Davide Rossi,                            \\
                                  & Frank K. Gurkaynak, Luca Benini                       \\
        Application / Publication & \teb{\pulp} \gls{mcs} / \teb{Garofalo~\etal~\cite{garofalo2025areliabletimepr}} \\
        Technology / Package      & Intel16 / BGA1733                                     \\
        Dimensions                & \SI{4.085}{\milli\meter}~$\times$~\SI{3.922}{\milli\meter}  \\
        Circuit Complexity        & \SI{70}{\mega\GE}                                     \\
        Voltage                   & \SI{0.8}{\volt}                                       \\
        Power                     & \SI{450}{\milli\watt} @ \SI{0.8}{\volt} and \SI{600}{\mega\hertz} \\
        Clock \teb{Frequency}     & \SI{600}{\mega\hertz}                                 \\
        \bottomrule
    \end{tabularx}
\end{table}

\teb{\carfield} is our first prototype in Intel's \SI{16}{\nano\metre} FinFet technology of our open-research platform for safety, resilient and time-predictable systems.

The rapid evolution of \gls{ai} algorithms, the massive amount of sensed data and the pervasive influence of \gls{ai}-enhanced applications across application-domains like automotive, space and cyber-physical embedded systems, call for a paradigm shift from simple micro-controllers towards powerful and heterogeneous edge computers for the design of next-generation mixed-criticality systems.
These must not only deliver outstanding performance and energy efficiency but also ensure steadfast safety, resilience, and security.

The \teb{\carfield} \gls{soc} aims to tackle these architectural challenges establishing itself as a pre-competitive heterogeneous research platform for mixed-criticality systems, underpinned by fully open-source \glspl{ip}.
The \gls{soc} showcases pioneering hardware solutions, addressing challenges related to time-predictable on/off-chip communication, robust fault recovery mechanisms, secure boot processes, cryptographic acceleration services, hardware-assisted virtualization, and accelerated computation for both floating-point and integer workloads.
\teb{\carfield' architecture features three domains and two \glspl{dsa}}.

The host domain integrates two memory-consistent 64-\si{\bit} CVA6 cores with support for virtualization at different levels.
Leveraging H-extensions within the {\riscv} \gls{isa}, the system can accommodate the execution of multiple \glspl{os}, including rich, Unix-like \glspl{os} and real-time \glspl{os}, via an intermediate Hypervisor layer.
The host domain provides a flexible interrupt subsystem, allowing for fast and virtualized interrupts through the {\riscv} \gls{clic}, and ensures hardware-based spatial isolation of cache resources to comply with freedom from interference requirements in modern \gls{mcs}.
\teb{\carfield}'s host domain builds on top of \emph{Cheshire} by enhancing it with fully configurable real-time features in \gls{hw}.

The safety island~\cite{rogenmoser2024sentrycorearisc}, a simple MCU-like domain, comprises three 32-\si{bit} physically isolated cores operating in triple-lockstep mode.
These cores, enhanced with the \gls{clic} and optimized for fast interrupt handling and context switch, run \gls{rtos} and safety-critical applications, embodying a core tenet of the platform reliability.

The secure domain, based on the \emph{OpenTitan} project, serves as the \gls{hwrot}, handling secure boot procedures, system integrity monitoring, and cryptographic acceleration services.

To augment computational capabilities, \teb{\carfield} incorporates two general-purpose accelerators: the \emph{Spatz} cluster, which handles vectorizable multi-format floating-point workloads (down to FP8), and the \emph{HMR} 12-core integer cluster, specialized in executing \gls{qnn} operations, exploiting the \emph{HMR} technique for rapid fault recovery and integer arithmetic support in the \gls{isa} of the {\riscv} cores from 32-\si{\bit} down to 2-\si{\bit}.

The system's interconnect relies upon a 64-\si{\bit} \gls{axi4} crossbar, enriched with \teb{{\axirealm}, see \Cref{chap:realm}}.

A full set of peripherals, including \emph{SPI}, \emph{I2C}, \emph{Serial Link}, \emph{CAN FD}, \emph{HyperBUS}, and timers, including watchdogs, complete the features of the \gls{soc}.

\cleardoublepage
\newpage
\ %
\newpage

\section{Iguana}

\begin{figure}[H]
    \centering
    \includegraphics[width=0.65\textwidth]{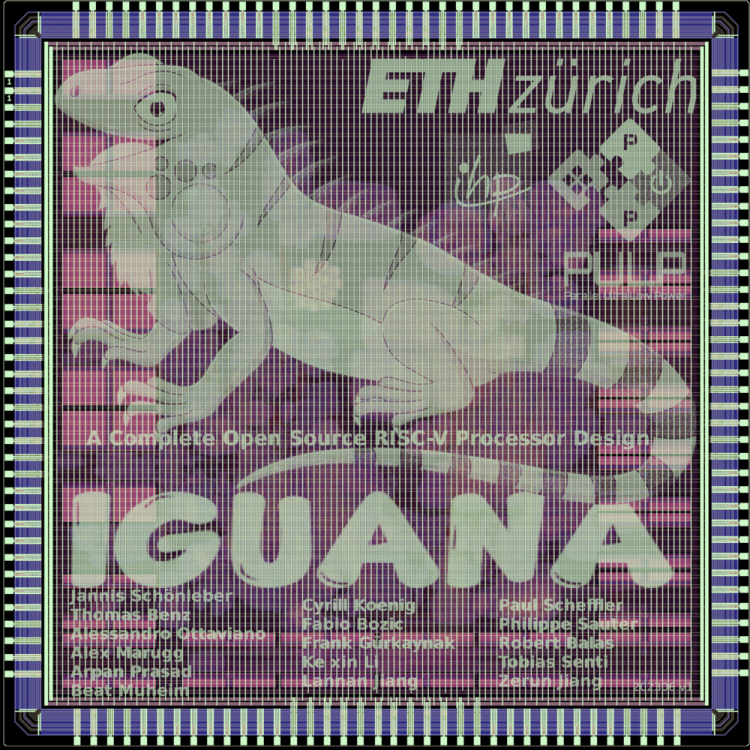}   \\
\end{figure}

\begin{figure}[H]
    \centering
    \includegraphics[width=0.65\textwidth]{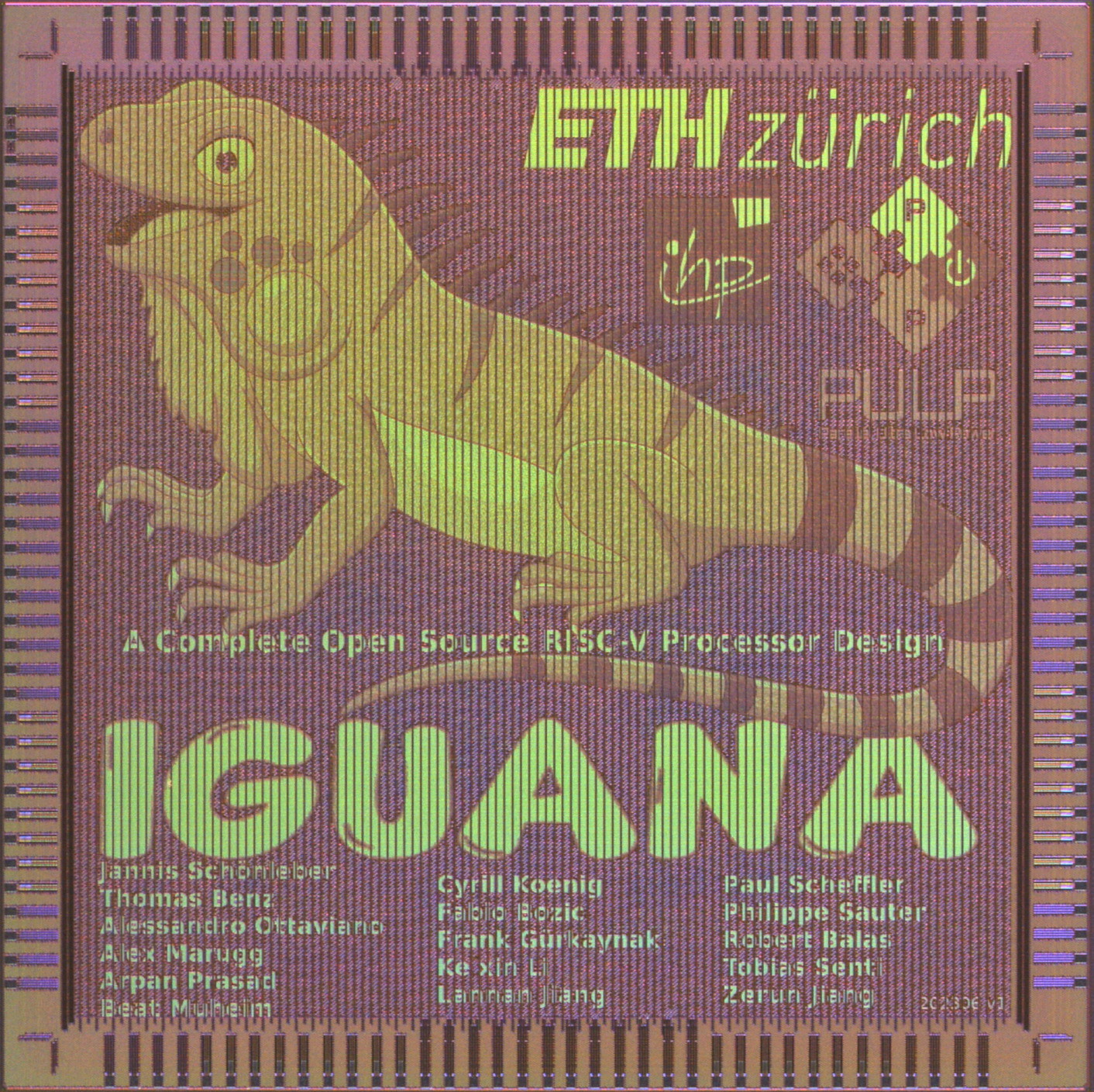}   \\
\end{figure}

\begin{table}[H]
    \centering
    \small
    \begin{tabularx}{\linewidth}{@{}ll@{}}
        \toprule
        Name                      & Iguana                                                \\
        \midrule
        Designers                 & Jannis Schoenleber, \textbf{Thomas Benz},             \\
                                  & Alessandro Ottaviano, Alex Marugg,                    \\
                                  & Arpan Prasad, Beat Muheim,                            \\
                                  & Cyril Koenig, Fabio Bozic,                            \\
                                  & Frank K. Gurkaynak, Kexin Li,                         \\
                                  & Lannan Jiang, Paul Scheffler,                         \\
                                  & Philippe Sauter, Robert Balas,                        \\
                                  & Tobias Senti, Zerun Jiang                             \\
        Application / Publication & \teb{\pulp} \gls{soc}-130-o / \teb{Benz~\etal~\cite{benz2023iguanaanendtoen}} \\
        Technology / Package      & IHP130 / QFN88                                        \\
        Dimensions                & \SI{6.264}{\milli\meter}~$\times$~\SI{6.264}{\milli\meter}  \\
        Circuit Complexity        & \SI{3}{\mega\GE}                                      \\
        Voltage                   & \SI{1.2}{\volt}                                       \\
        Clock \teb{Frequency}     & \SI{60}{\mega\hertz}                                  \\
        \bottomrule
    \end{tabularx}
\end{table}

Iguana is our first attempt at using IHP's \SI{130}{\nano\metre} open \gls{pdk}.
\teb{The chip's architecture connects our Cheshire \gls{soc} platform to a HyperBus memory controller for off-chip data storage.}

The design was completed using only open-source standard cell libraries, and although an almost complete backend run was made with OpenROAD, a last minute issue resulted in a backup design, made with commercial \gls{eda} tools, to be taped.

\teb{We introduce the terminology of \emph{\gls{soc}-130-o} to describe an end-to-end open-source (SW, RTL, EDA, PDK, PHYs) \gls{soc}.}

\newpage

\section{Basilisk}

\begin{figure}[H]
    \centering
    \includegraphics[width=0.755\textwidth]{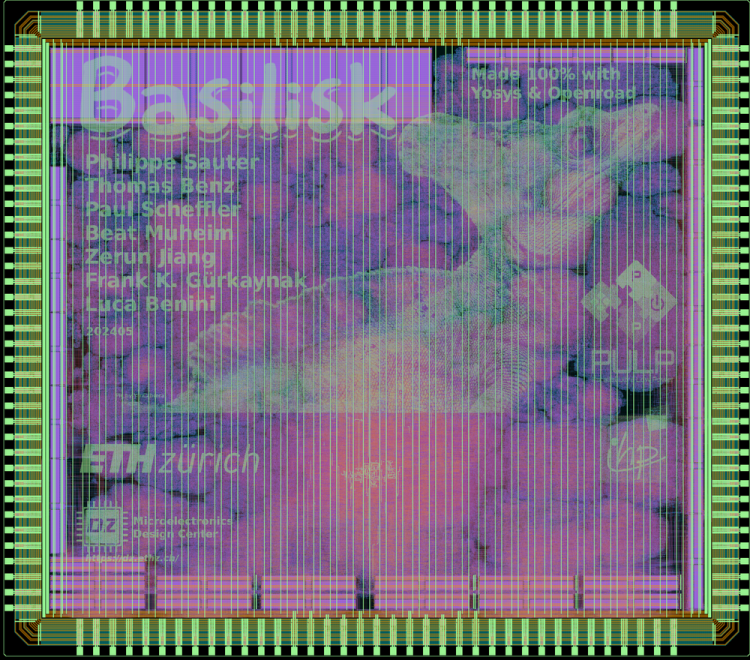}   \\
\end{figure}

\begin{figure}[H]
    \centering
    \includegraphics[width=0.755\textwidth]{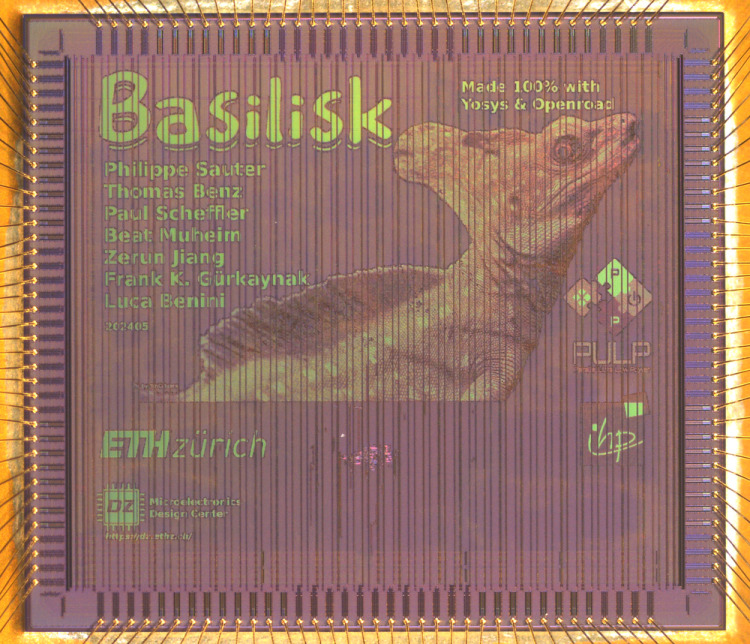}   \\
\end{figure}

\begin{table}[H]
    \centering
    \small
    \begin{tabularx}{\linewidth}{@{}ll@{}}
        \toprule
        Name                      & Basilisk                                              \\
        \midrule
        Designers                 & Philippe Sauter, \textbf{Thomas Benz},                \\
                                  & Paul Scheffler, Beat Muheim,                          \\
                                  & Zerun Jiang, Frank K. Gurkaynak,                      \\
                                  & Luca Benini                                           \\
        Application / Publication & \teb{\pulp} \gls{soc}-130-o / \teb{Sauter~\etal~\cite{sauter2025basiliska34mm²e}} \\
        Technology / Package      & IHP130 / QFN88                                        \\
        Dimensions                & \SI{6.264}{\milli\meter}~$\times$~\SI{5.498}{\milli\meter}  \\
        Circuit Complexity        & \SI{3}{\mega\GE}                                      \\
        Voltage                   & \SI{1.2}{\volt}                                       \\
        Clock \teb{Frequency}     & \SI{60}{\mega\hertz}                                  \\
        \bottomrule
    \end{tabularx}
\end{table}

Following \emph{Iguana}, Basilisk is our second attempt at using the IHP's \SI{130}{\nano\metre} open \gls{pdk}.
\teb{Basilisk features the same architecture as Iguana with some minor tweaks and improvements.}

This time around, a completely open flow, using Yosys for synthesis and OpenROAD for implementation, was successfully used.
Iguana also includes novel aging sensors developed by IHP. \tebsr{Basilisk successfully boots Linux.}

\newpage

\section{MLEM}

\begin{figure}[H]
    \centering
    \includegraphics[width=1\textwidth]{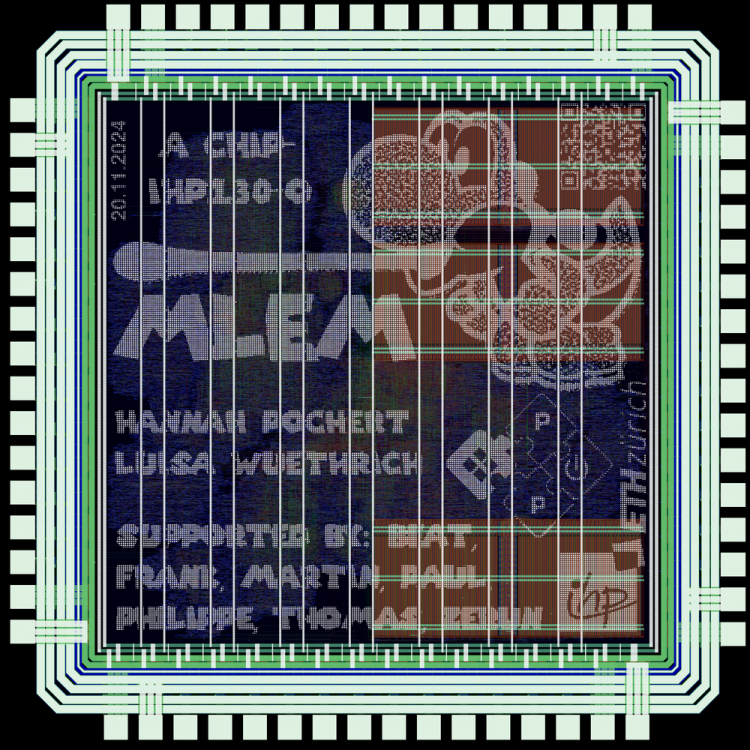}   \\
\end{figure}
\vfill
\newpage

\begin{table}[H]
    \centering
    \small
    \begin{tabularx}{\linewidth}{@{}ll@{}}
        \toprule
        Name                      & MLEM                                                  \\
        \midrule
        Designers                 & Hannah Pochert, Luisa Wuethrich,                      \\
                                  & Philippe Sauter, Paul Scheffler,                      \\
                                  & \textbf{Thomas Benz}                                  \\
        Application / Publication & \teb{\pulp} Student \gls{mcu}-130-o / \teb{Sauter~\etal~\cite{sauter2025crocanendtoendo}} \\
        Technology / Package      & IHP130 / QFN56                                        \\
        Dimensions                & \SI{2.235}{\milli\meter}~$\times$~\SI{2.235}{\milli\meter}  \\
        Circuit Complexity        & \SI{350}{\kilo\GE}                                    \\
        Voltage                   & \SI{1.2}{\volt}                                       \\
        Clock \teb{Frequency}     & \SI{80}{\mega\hertz}                                  \\
        \bottomrule
    \end{tabularx}
\end{table}

MLEM, named after the sound \emph{Yoshi} (from \emph{Super Mario}) makes when eating a tasty fruit, is the first tapeout based on the \emph{Croc} \gls{mcu} platform~\cite{sauter2025crocanendtoendo}.
MLEM was designed and prepared for tapeout by students using a complete open-source tool flow, including a new \gls{sv} frontend for Yosys developed by Martin Povišer.
This chip serves as a pilot for the new open VLSI II lecture starting in 2025, where students get the opportunity to prepare a Croc-based tapeout with any custom addition in IHP's \SI{130}{\nano\metre} open \gls{pdk}.

For MLEM, students contributed a \gls{sv} \emph{UART} peripheral to replace the existing VHDL version, a \emph{GPIO} peripheral, and a \emph{Neopixel} controller to drive RGB LEDs.

\newpage

\section{Koopa}

\begin{figure}[H]
    \centering
    \includegraphics[width=1\textwidth]{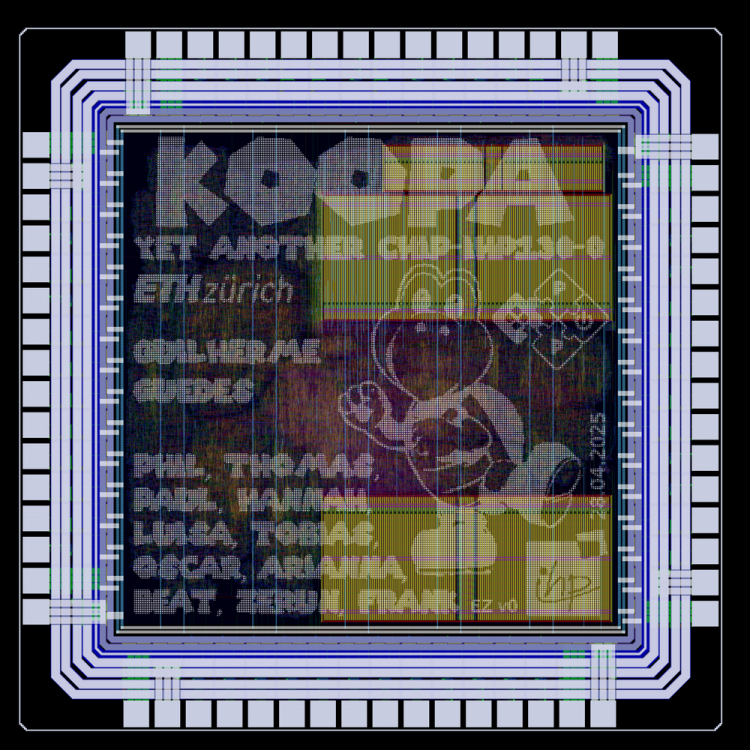}   \\
\end{figure}
\vfill
\newpage

\begin{table}[H]
    \centering
    \small
    \begin{tabularx}{\linewidth}{@{}ll@{}}
        \toprule
        Name                      & Koopa                                                  \\
        \midrule
        Designers                 & Guilherme Guedes, Philippe Sauter                     \\
                                  & \textbf{Thomas Benz}, Paul Scheffler,                 \\
                                  & Hannah Pochert, Luisa Wuethrich,                      \\
                                  & Tobias Senti, Oscar Castaneda,                        \\
                                  & Arianna Rubino, Beat Muheim,                          \\
                                  & Zerun Jiang, Frank K. Gurkaynak                       \\
        Application / Publication & \teb{\pulp} \gls{mcu}-130-o                           \\
        Technology / Package      & IHP130 / QFN56                                        \\
        Dimensions                & \SI{2.235}{\milli\meter}~$\times$~\SI{2.235}{\milli\meter}  \\
        Circuit Complexity        & \SI{250}{\kilo\GE}                                    \\
        Voltage                   & \SI{1.2}{\volt}                                       \\
        Clock \teb{Frequency}     & \SI{74}{\mega\hertz}                                  \\
        \bottomrule
    \end{tabularx}
\end{table}

Following \emph{MLEM}, Koopa is our second Croc-based \gls{mcu}.
Next to various minor updates \teb{to} the \glspl{ip} compared to MLEM, Koopa features a newly designed QSPI interface.
Koopa further uses a first version of \emph{EZ130}, a new open-source standard-cell library developed by and for the VLSI 5 course at ETH Zurich, led by Oscar Castañeda and Christoph Studer.

\newpage
\newpage

\section{Fluffy}

\begin{figure}[H]
    \centering
    \includegraphics[width=1\textwidth]{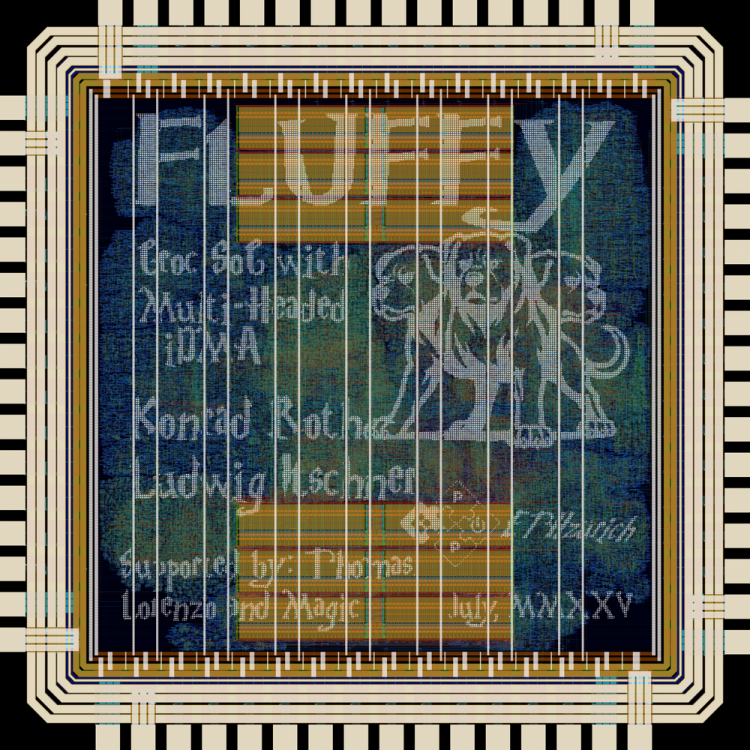}   \\
\end{figure}
\vfill
\newpage

\begin{table}[H]
    \centering
    \small
    \begin{tabularx}{\linewidth}{@{}ll@{}}
        \toprule
        Name                      & Fluffy                                                  \\
        \midrule
        Designers                 & Konrad Roth, Ludwig Itschner                            \\
                                  & \textbf{Thomas Benz}, Lorenzo Leone,                    \\
        Application / Publication & \teb{\pulp} Student \gls{mcu}-130-o                     \\
        Technology / Package      & IHP130 / QFN56                                          \\
        Dimensions                & \SI{2.235}{\milli\meter}~$\times$~\SI{2.235}{\milli\meter}  \\
        Circuit Complexity        & \SI{470}{\kilo\GE}                                      \\
        Voltage                   & \SI{1.2}{\volt}                                         \\
        Power                     & \SI{45}{\milli\watt} @ \SI{1.2}{\volt} and \SI{51}{\mega\hertz} \\
        Clock \teb{Frequency}     & \SI{51}{\mega\hertz}                                    \\
        \bottomrule
    \end{tabularx}
\end{table}

Fluffy integrates a multi-head-capable {\idmae} into a Croc-based \gls{mcu}.
This newly developed engine can copy two data streams simultaneously from Croc's {SRAM} banks, perform an element-wise arithmetic operation on them, and store the resulting vector back in {SRAM}.
Executing a vector add using our multi-head {\idmae} results in a speedup of \SI{9.14}{\x} compared to the baseline execution on Croc's CVE2 core.

Fluffy was designed as part of a mini project alongside our new VLSI II course.
It is one of five student designs chosen for a tapeout in 2025 thanks to its novel results, well-executed implementation, and outstanding report.

\newpage

\section{Skoll}

\begin{figure}[H]
    \centering
    \includegraphics[width=0.91\textwidth]{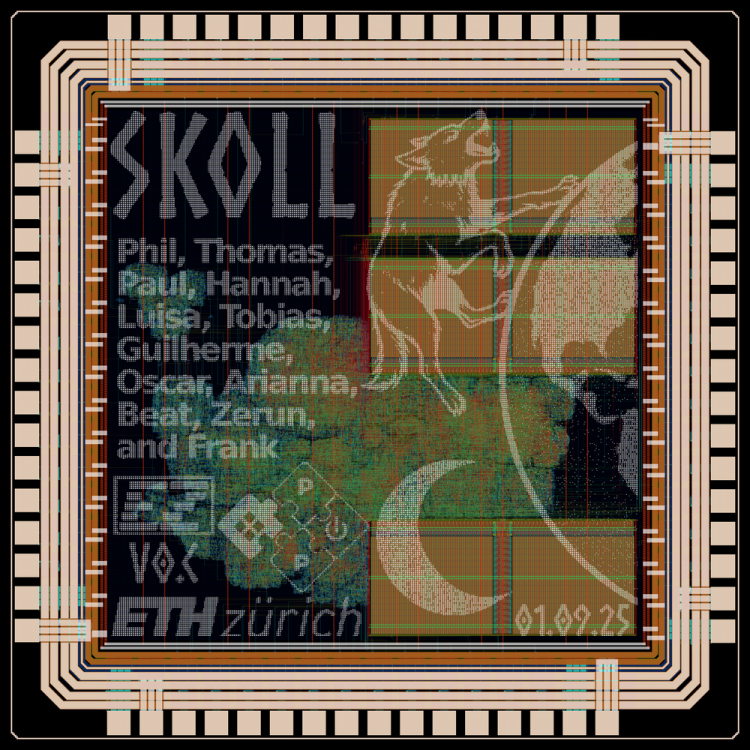}   \\
\end{figure}

\vspace{-0.3cm}

\begin{figure}[H]
    \centering
    \includegraphics[width=0.91\textwidth]{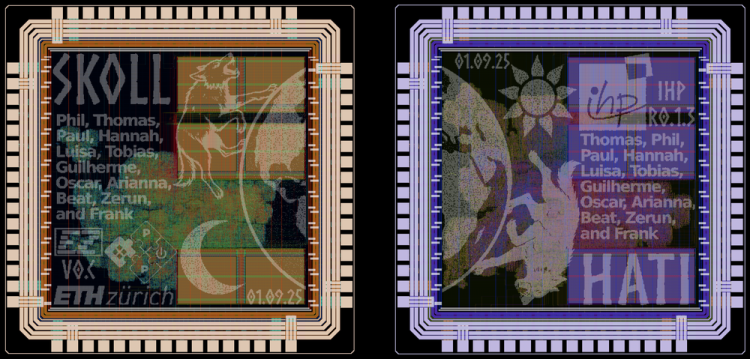}   \\
\end{figure}
\vfill

\newpage

\begin{table}[H]
    \centering
    \small
    \begin{tabularx}{\linewidth}{@{}ll@{}}
        \toprule
        Name                      & Skoll                                                  \\
        \midrule
        Designers                 & Philippe Sauter, \textbf{Thomas Benz}                 \\
                                  & Paul Scheffler, Hannah Pochert                        \\
                                  & Luisa Wuethrich, Tobias Senti,                        \\
                                  & Guilherme Guedes, Oscar Castaneda,                    \\
                                  & Arianna Rubino, Beat Muheim,                          \\
                                  & Zerun Jiang, Frank K. Gurkaynak                       \\
        Application / Publication & \teb{\pulp} \gls{mcu}-130-o                           \\
        Technology / Package      & IHP130 / QFN56                                        \\
        Dimensions                & \SI{2.235}{\milli\meter}~$\times$~\SI{2.235}{\milli\meter}  \\
        Circuit Complexity        & \SI{250}{\kilo\GE}                                    \\
        Voltage                   & \SI{1.2}{\volt}                                       \\
        Clock \teb{Frequency}     & \SI{80}{\mega\hertz}                                  \\
        \bottomrule
    \end{tabularx}
\end{table}

Hati and Skoll are two Croc-based \gls{mcu} featuring an identical \gls{rtl} code based on the architecture of Koopa.
They serve to silicon-harden v1.2.0 of Croc, evaluate the improvements of the open-source \gls{eda} tools since the tapeout of \emph{MLEM}, and to evaluate and compare the newest versions of the two standard cell libraries in IHP's \SI{130}{\nano\metre} open \gls{pdk}.

Skoll implements the \gls{mcu} using \emph{v0.c} of the EZ130 standard cells.

\newpage

\section{Hati}

\begin{figure}[H]
    \centering
    \includegraphics[width=0.91\textwidth]{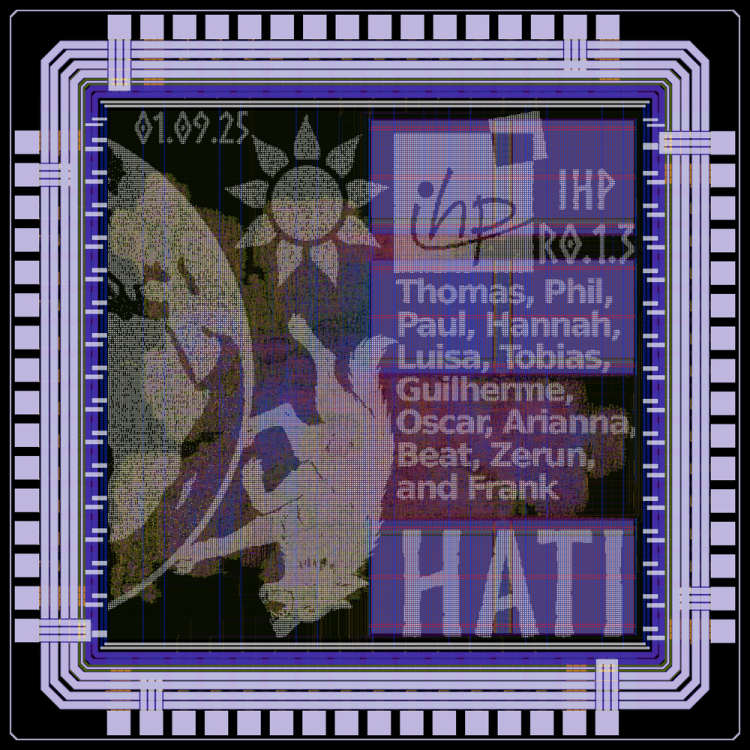}   \\
\end{figure}

\vspace{-0.3cm}

\begin{figure}[H]
    \centering
    \includegraphics[width=0.91\textwidth]{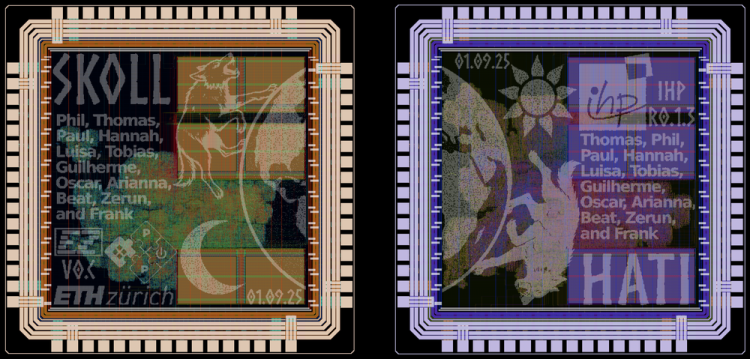}   \\
\end{figure}
\vfill

\newpage

\begin{table}[H]
    \centering
    \small
    \begin{tabularx}{\linewidth}{@{}ll@{}}
        \toprule
        Name                      & \teb{Hati}                                            \\
        \midrule
        Designers                 & \textbf{Thomas Benz}, Philippe Sauter                 \\
                                  & Paul Scheffler, Hannah Pochert                        \\
                                  & Luisa Wuethrich, Tobias Senti,                        \\
                                  & Guilherme Guedes, Oscar Castaneda,                    \\
                                  & Arianna Rubino, Beat Muheim,                          \\
                                  & Zerun Jiang, Frank K. Gurkaynak                       \\
        Application / Publication & \teb{\pulp} \gls{mcu}-130-o                           \\
        Technology / Package      & IHP130 / QFN56                                        \\
        Dimensions                & \SI{2.235}{\milli\meter}~$\times$~\SI{2.235}{\milli\meter}  \\
        Circuit Complexity        & \SI{250}{\kilo\GE}                                    \\
        Voltage                   & \SI{1.2}{\volt}                                       \\
        Clock \teb{Frequency}     & \SI{90}{\mega\hertz}                                  \\
        \bottomrule
    \end{tabularx}
\end{table}

Hati and Skoll are two Croc-based \gls{mcu} featuring an identical \gls{rtl} code based on the architecture of Koopa.
They serve to silicon-harden v1.2.0 of Croc, evaluate the improvements of the open-source \gls{eda} tools since the tapeout of \emph{MLEM}, and to evaluate and compare the newest versions of the two standard cell libraries in IHP's \SI{130}{\nano\metre} open \gls{pdk}.

Hati implements the \gls{mcu} using \emph{r0.1.3} of IHP's standard cells.

\newpage

\section{Flamingo}

\newpage
\null
\thispagestyle{empty}
\newpage
\thispagestyle{empty}
\cleardoublepage

\clearpage
\chapter{ArtistIC}
\label{chap:artistic}

\section{Introduction}

Recent advances in \gls{oseda} spawned countless \gls{asic} projects developed by industry, research, and hobbyists.
Even though the content and organization of these projects could not be more diverse, they share a common requirement to present their project using outreach material to gain attention, share resources and results, and secure funding.
Especially in the early design phases of projects, sharing the \gls{asic} under design can prove very difficult as no running prototype can be shown.

Methodologies have been established to render and visualize layout (\glsunset{gdsii}\gls{gdsii}) files, but they are limited either in scope or fidelity.
The \emph{TinyTapeout} project~\cite{venn2024tinytapeoutasha} uses a {3D}-\gls{gdsii} viewer to visualize their layouts directly within a web browser~\cite{mbalestrini2022tinytapeoutgdsv}.
Proving an excellent tool to visualize tiny designs on the standard-cell level in {3D}, it is not suited to render research or industry-grade chips on a poster scale.
{\klayout}~\cite{kofferlein2020klayout} has established itself as the default open-source \gls{gdsii} viewer featuring a powerful scripting \gls{api}, which can be used to export high-resolution ($\lesssim$ \SI{250}{\mega\pixel}) renders of the current view~\cite{kofferleinscreenshotwitho}.
Our experiments show that a much higher resolution is required to capture the intricate details of the lower-level metalization layers of research-scale~\cite{sauter2024insightsfrombas, scheffler2025occamya432cored} chips.
Furthermore, {\klayout} does not provide a transparency option for individual layers, leading to upper metal levels covering up lower-level details.

Adding top-metal artwork is a well-established way of branding the design files and the fabricated silicon~\cite{goldstein2002thesecretartofc}.
To our knowledge, no open-source tool exists to translate and embed artwork into \gls{gdsii} layouts.

To tackle these points,we propose {\artistic}, an open-source framework to translate and insert top-metal \gls{asic} art into \gls{gdsii} layout files and create artistic ultra-high-fidelity renders thereof.
In particular, we present the following contributions:
\begin{itemize}
    \item A \emph{Gdspy}-based~\cite{gdspysdocumenta} script translating and inserting \glsunset{drc}\gls{drc}-clean top-metal artworks into \gls{gdsii} files.
    \item A tile-based image rendering methodology supporting arbitrary high resolutions through tiling and capable of individual layer transparencies.
    \item A case study presenting insights from analyzing the layout renders of two {\riscv} \glspl{asic}~\cite{sauter2024insightsfrombas, scheffler2025occamya432cored}.
\end{itemize}

\section{Toolflow}

\begin{figure}
\centering%
    \includegraphics[width=1\linewidth]{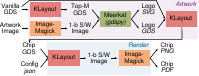}%
    \caption{The {\artistic} toolflow.}
    \label{fig:toolflow}
\end{figure}

{\artistic}'s toolflow, presented in \Cref{fig:toolflow}, accepts a \gls{drc}-clean \emph{vanilla} \gls{gdsii} file and a lossless artwork image as primary inputs.
In the first step, {\klayout} is used to export the top-metal layer from the \gls{gdsii} file and the logo is translated into a 1-\si{\bit} \emph{b/w} image.
\emph{Meerkat}, a \emph{Gdspy}-based script, converts the \emph{b/w} image to \emph{tetromino}-like shapes following \gls{drc} and density rules as well as flowing them around existing top-metal structures.
\emph{Meerkat} exports a vector graphic image and a \gls{gdsii} file containing the logo only; {\klayout} is then used to merge the vanilla and logo \gls{gdsii} files.

A \emph{json} configuration file specifies the resulting image size, render resolution, layer colors and transparencies, and the used layer stack.
Using a maximum individual tile size of around \SI{250}{\mega\pixel}, we use {\klayout} to export each layer of the \gls{asic} as \emph{b/w} image tiles.
The tiles can be rendered at a higher resolution than the final image size to keep fine layout details while ensuring a manageable image file size.
In the last step, \emph{\imagemagick} is used to color the layer tiles, merge the colored layers using the specified transparencies, and resize and merge the tiles to a complete image, which is then embedded into a \emph{PDF} container for printing.

\section{Results}

\begin{figure}
    \begin{subcaptionblock}{0.47\linewidth}%
        \centering%
        \includegraphics[width=\linewidth]{fig-116.png}%
        \vspace{-0.4em}%
        \caption{Klayout render}%
        \vspace{0.25em}%
        \label{fig:klayout_render}%
        \vfill%
        \centering%
        \includegraphics[width=\linewidth]{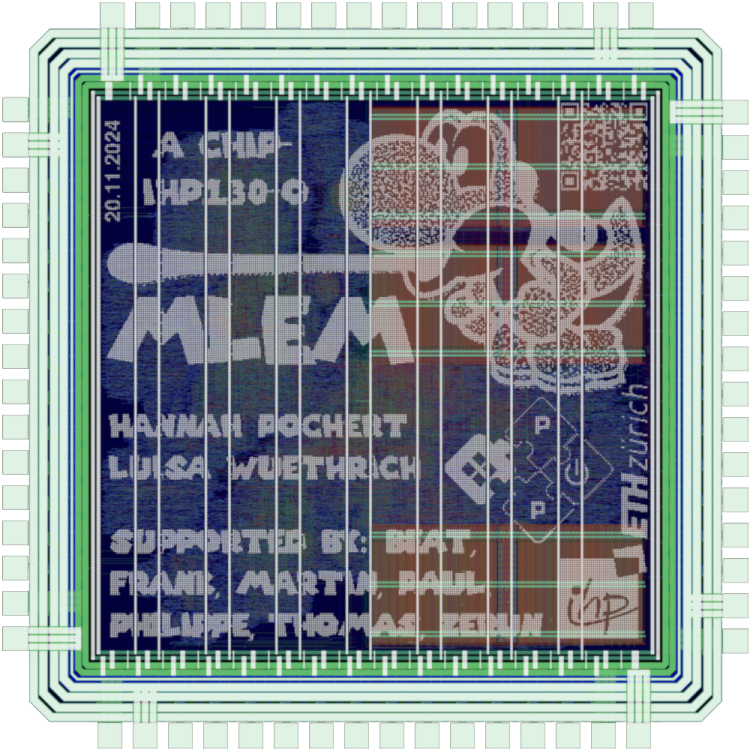}%
        \vspace{-0.3em}%
        \caption{{\artistic} render}%
        \label{fig:artistic_render}%
    \end{subcaptionblock}\hfill
    \begin{subcaptionblock}{0.52\linewidth}
        \centering%
        \includegraphics[width=\linewidth]{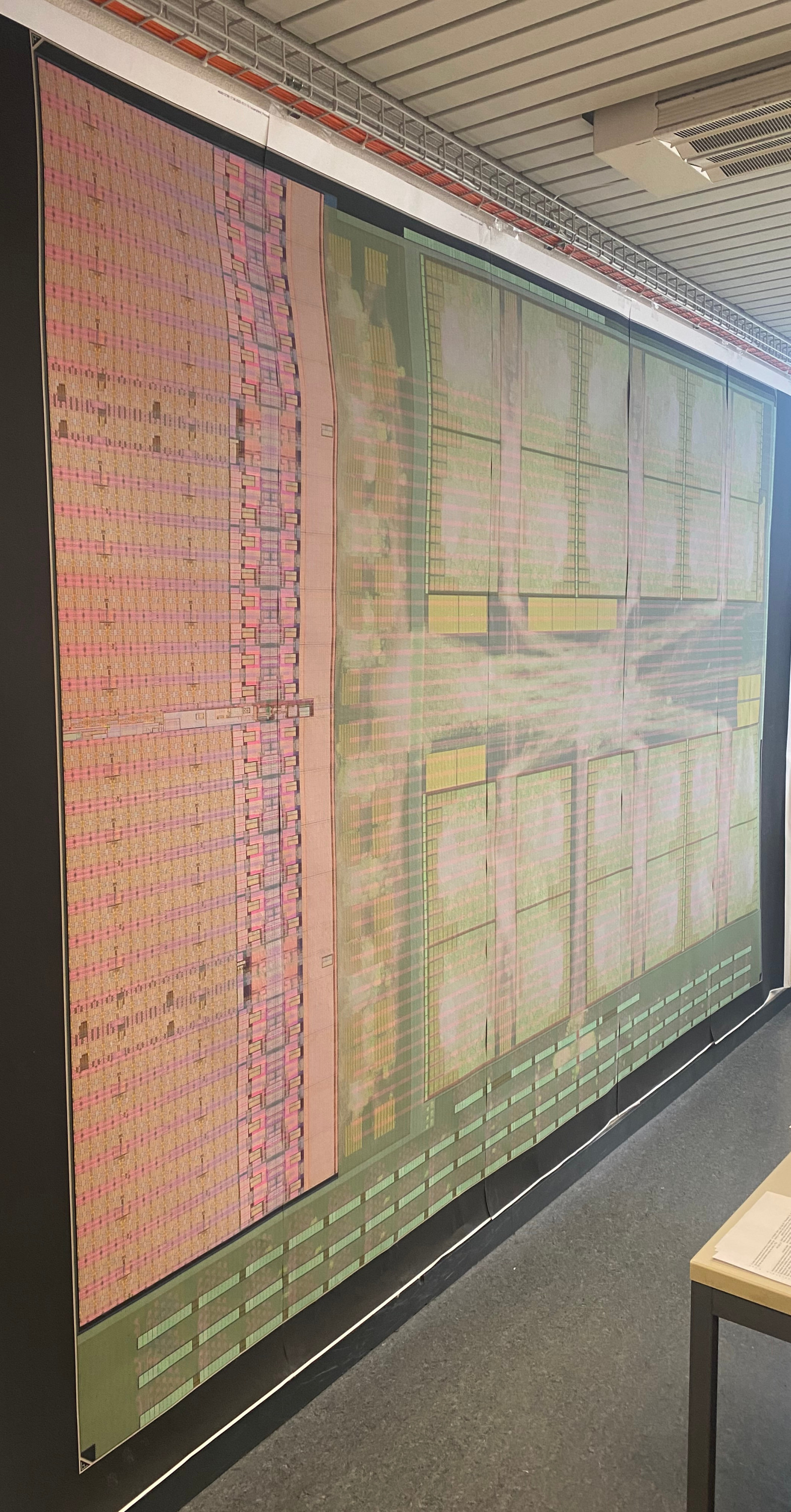}%
        \vspace{-0.3em}%
        \caption{Wall-spanning poster}%
        \label{fig:artistic_poster}%
    \end{subcaptionblock}\hfill

    \caption{\gls{gdsii} renders}
    \label{fig:artistic}
\end{figure}

\Cref{fig:klayout_render} shows the \emph{vanilla} \gls{gdsii} file~\cite{wuethrich2024mlemapulpsocfor, pochert2024mlemapulpsocfor, contributors2024croc} rendered with {\klayout}~\cite{kofferlein2020klayout, kofferleinscreenshotwitho} and \Cref{fig:artistic_render} with {\artistic} including a generated top-level metal artwork.
\Cref{tab:runtime} presents {\artistic}'s runtimes for two open-source \glspl{asic}.
\Cref{fig:artistic_poster} displays a 2.3 by \SI{4.2}{\metre} poster of one Occamy die~\cite{scheffler2025occamya432cored} rendered at \SI{57}{\giga\pixel} (\SI{44}{\nano\metre\per\pixel}) and printed at \SI{1600}{\dpi} demonstrating the scalability of {\artistic}.

\begin{table}
    \setlength{\tabcolsep}{4pt}
    \centering
    \scriptsize{%
        \centering
        \caption{%
            {\artistic} runtime of open-source {\riscv} \glspl{asic}.%
        }%
        \label{tab:runtime}
        \renewcommand*{\arraystretch}{0.95}
        \begin{threeparttable}
            \begin{tabular}{cccccc} \toprule
                \textbf{Chip} &
                \dl{\textbf{Chip}}{\textbf{Size}} &
                \dl{\textbf{Render}}{\textbf{Res.}} &
                \dl{\textbf{Print}}{\textbf{Res.}} &
                \dl{\textbf{Print}}{\textbf{Size}} &
                \dl{\textbf{Run-}}{\textbf{time}} \\

                \midrule

                \textit{Mlem/Croc~\cite{wuethrich2024mlemapulpsocfor, pochert2024mlemapulpsocfor, contributors2024croc}} &
                \SI{5}{\milli\metre\squared} &
                \SI{25}{\nano\metre\per\pixel} &
                \SI{2}{\giga\pixel} &
                A1~\tnote{a} &
                \SI{0.9}{\hour}~\tnote{b} \\

                \textit{Basilisk~\cite{sauter2024insightsfrombas}} &
                \SI{35}{\milli\metre\squared} &
                \SI{25}{\nano\metre\per\pixel} &
                \SI{55}{\giga\pixel} &
                2x\SI{2}{m}~\tnote{a} &
                \SI{6.1}{\hour}~\tnote{b} \\

                \bottomrule
                
            \end{tabular}

            \begin{tablenotes}[para, flushleft]
                \item[a] target: \SI{2400}{\dpi}
                \item[b] \SI{2.5}{\GHz} Xeon E5-2670
            \end{tablenotes}
        \end{threeparttable}
    }
\end{table}

\section{Case Study: RISC-V SoCs}

\begin{figure}
    \begin{subcaptionblock}{0.45\linewidth}
        \centering%
        \includegraphics[width=\linewidth]{fig-119.pdf}%
        \vspace{-0.3em}%
        \caption{Linux-capable SoC~\cite{sauter2024insightsfrombas}}%
        \label{fig:artistic_iguana}%
    \end{subcaptionblock}\hfill
    \begin{subcaptionblock}{0.508\linewidth}%
        \centering%
        \includegraphics[width=\linewidth]{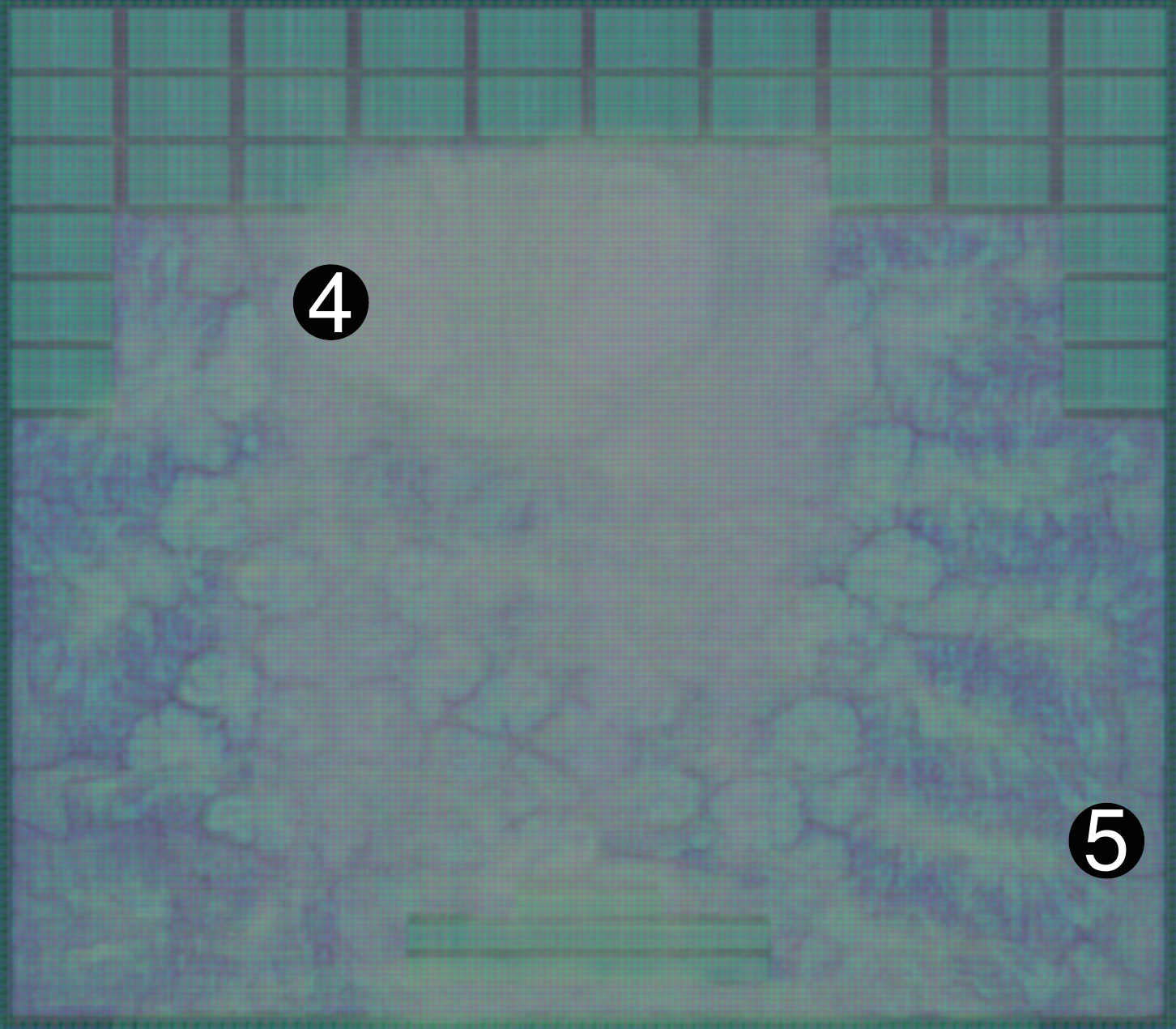}%
        \vspace{-0.3em}%
        \caption{Occamy cluster~\cite{scheffler2025occamya432cored}}%
        \label{fig:artistic_cluster}%
    \end{subcaptionblock}\hfill
    \begin{subcaptionblock}{\linewidth}
        \centering%
        \includegraphics[width=\linewidth]{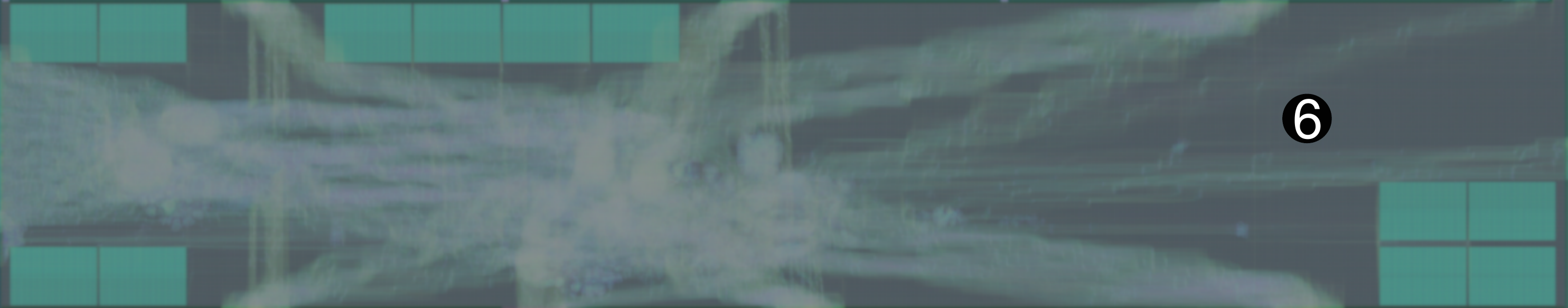}%
        \vspace{-0.3em}%
        \caption{Occamy main interconnect~\cite{scheffler2025occamya432cored}}%
        \label{fig:artistic_ico}%
    \end{subcaptionblock}\hfill

    \caption{Insights gained through {\artistic}.}
    \label{fig:case}
\end{figure}

{\artistic} can directly be used in scientific publications to analyze, evaluate, and compare \gls{asic} layouts.
\Cref{fig:case} presents three high-fidelity layout renders with annotations of exemplary insights gained; for a more detailed analysis, we refer to our peer-reviewed works on these systems~\cite{sauter2024insightsfrombas, scheffler2025occamya432cored}.
In \Cref{fig:artistic_iguana}, we see \circnum{1} the bootrom, which is closely interconnected and thus densely packed, \circnum{2} empty space, and \circnum{3} an area of high connectivity; the routing uses higher top metals (rendered in red).
In \Cref{fig:artistic_cluster}, \circnum{4} highlights the high routing effort of the \emph{logarithmic interconnect} and \circnum{5} the \emph{pipeline stages} of one of the computing units.
In \Cref{fig:artistic_ico}, \circnum{6} annotates one of the top-level interconnect buses with individual pipeline stages visible.

\section{Conclusion \& Outlook}

With {\artistic}, we present a framework to embed images as top-metal ASIC art into existing \gls{gdsii} layouts and to render them artistically with ultra-high fidelity.
{\artistic} cannot only be used to generate outreach material for \gls{asic} projects to gain attention, share results, and secure funding, but also to build the foundation for scientific discussions concerning chip layouts.
With {\artistic} available open source\,\footnote{\url{github.com/pulp-platform/artistic}}, we hope to see many high-resolution \gls{asic} renders and posters emerging.

\clearpage
\chapter{DUTCTL}
\label{chap:dutctl}

\section{Introduction}

Silicon testing and characterization, driven in industry by \gls{ate}~\cite{bushnell2013essentialsofele}, is an essential part of \gls{asic} production. %
Modern \gls{ate} systems enable comprehensive, calibrated, and high-resolution functional and physical evaluation of %
\glspl{asic};
however, they are expensive to purchase and maintain and require specialized training to program and operate.

With open-source hardware initiatives like {\riscv} on the rise, small organizations and research groups increasingly design their own prototype \glspl{soc}.
In such low-volume settings, purchasing and operating \gls{ate} usually cannot be justified.

Luckily, the bring-up and basic characterization of prototype \glspl{soc} can also be accomplished with commodity hardware. %
Unlike generic \glspl{asic}, \glspl{soc} are designed to be highly software-controllable;
{\riscv} \glspl{dbgm}~\cite{internationalriscvexternalde} enable full external control of on-chip harts through standard interfaces like JTAG.
Power and clocks can be provided by bench-top supplies and signal generators, respectively.
Most \gls{soc} \glspl{io} (e.g., UART, SPI, DRAM) adhere to established protocols and can be connected to off-the-shelf devices or adapters; debugging is possible with oscilloscopes and logic analyzers.

Nevertheless, this approach inherently lacks coordination and automation. To match \gls{ate} systems, a solution to externally control and coordinate power delivery, clocking, reset, and debugging is required. This, in turn, is what enables the sequencing, sweeping, and reproduction of tests that is required for automated, in-depth characterization across production runs.

We thus present {\dutctl}~\cite{contributorsdutctl, benz2024dutctlaflexible}, an open-source framework automating the rapid, \gls{ate}-less bring-up and characterization of {\riscv}-based \glspl{soc} by controlling and coordinating the necessary external devices. {\dutctl} provides the following features:

\begin{itemize}
\item It configures and controls any number of network-attached supplies, clock sources, and reset generators described in a customizable configuration.
\item It coordinates a full reset-and-power cycle, ensuring statelessness and reproducibility, and provides a fully scriptable GDB~\cite{freegdbthegnuprojec} debugging session.
\item It monitors and stores the \gls{soc}'s serial output; through control sequences, the \gls{soc} can communicate internal measurements (e.g. computed results, cycle counts) or trigger external measurements (e.g. supply power) with precise timing.
\item Through iterated sessions with different parameters and debugging payloads, it enables the design of full test flows and characterization sweeps.
\item It enables easy \emph{shared} and \emph{remote} access to limited engineering samples for time-efficient testing and software exploration.
\end{itemize}

\begin{figure}
    \centering
    \includegraphics[width=\columnwidth]{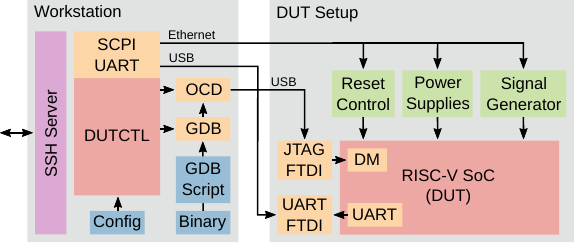}
    \caption{A generic {\dutctl} setup. Software running on a workstation orchestrates equipment around the \gls{dut}.}
    \label{fig:arch}
\end{figure}

\section{Architecture}

\Cref{fig:arch} shows a generic {\dutctl} setup composed of a software part running on a workstation and instruments and adapters connecting to the \gls{dut}.
Instruments are interfaced through an extensible {Python} library with generic bindings; in our case, they are controlled using \gls{scpi}~\cite{instrumentsdocumentationab} over {Ethernet}.
Adapters connect directly to the test workstation and leverage existing drivers and software; we currently support JTAG adapters for {\riscv} debugging and UART adapters for serial IO.

A {\dutctl} session is launched using a single command specifying an instrument configuration, a GDB script, and optionally a serial port.
{\dutctl} first connects to all instruments and configures them; this includes voltages and current limits for supply channels or frequencies and waveforms for signal generators.
After configuration, instrument outputs are enabled and any configured resets issued through instrument GPIOs.
{\dutctl} then launches a GDB session sourcing the passed script, which may preload, execute, and debug {\riscv} binaries.
In parallel, it monitors the \gls{dut}'s serial output for control sequences communicating internal results or triggering external instrument measurements with precise timing relative to binary execution.
The session ends when GDB exits, which may be triggered by the \gls{dut} using \texttt{ebreak}. Test failures are communicated through nonzero return codes. Instrument outputs are disabled to purge non-resettable state (e.g. SRAMs) within the DUT.

Launching multiple consecutive sessions with different parameters enables full test flows and characterization sweeps.
For example, a \emph{Schmoo} plot~\cite{womack1965schmooplotanaly} may be created by sweeping clock frequency and supply voltage and recording for each session whether a workload terminates with correct results.
With {\dutctl}, we can easily enhance this binary \emph{pass-or-fail} plot with execution times and power measurements.
By customizing instrument configurations and scripts, arbitrary test flows and sweeps may be assembled from {\dutctl} sessions, allowing custom \glspl{soc} to be extensively tested, characterized, and binned. %

\begin{figure}
    \centering
    \includegraphics[width=\columnwidth]{fig-123.pdf}
    \caption{Photograph of a running {\dutctl} setup.}
    \label{fig:dutctl-setup}
\end{figure}

\section{Results and Conclusion}

\begin{figure}
    \centering
    \includegraphics[width=\columnwidth]{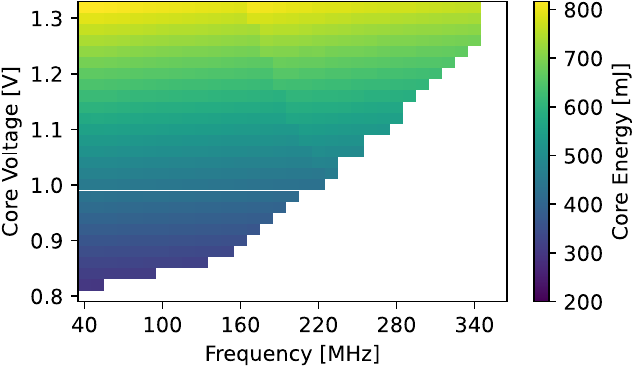}
    \caption{Voltage-frequency sweep running a matrix multiplication; failing operating points are denoted in \emph{white}, passing according to total core energy to completion.}
    \label{fig:schmoo}
\end{figure}

As shown in \Cref{fig:dutctl-setup}, we successfully used {\dutctl} to bring up, test, and characterize an open-source {\riscv} \gls{soc} using a testflow composed of a series of functional and sweep tests. %
Clock speed is an important figure of merit even for small-scale runs.
To create the enhanced {Schmoo} plot shown in \Cref{fig:schmoo}, hundreds of {\dutctl} sessions must be executed sequentially. %
We found the test runtime to be dominated by the payload's execution time, which requires a minimum duration to allow precise steady-state current measurements.

With {\dutctl}, we present and release an open-source framework automating the \gls{ate}-less bring-up and characterization of {\riscv}-based SoCs.
{\dutctl}'s modularity allows easy replacement of individual components and addition of test equipment.

\clearpage
\chapter{Student Theses Supervised}
\label{chap:theses}

\vspace{-3mm}
During my Ph.D. studies, I had the chance to supervise the following student theses.
I am very grateful for the opportunity to work with so many students, the chance to grow on the challenges tackled together, and to be able to improve my teaching and supervision skills.

\section{2020}

\begin{publications}
    \publication{huetter2020asnitchbasedcom}
    \publication{solt2020aflexibleperiph}
\end{publications}

\section{2021}

\begin{publications}
    \publication{bucheli2021asnitchbasedsoc}
    \publication{zhou2021anrpcdramimplem}
    \publication{eudine2021bringupandevalu}
    \publication{ho2021investigationof}
    \publication{hauser2021lcoreaminimalri}
    \publication{zhang2021extendingaxi4wi}
\end{publications}

\section{2022}

\begin{publications}
    \publication{schaerer2022enhancingourdma}
    \publication{senti2022extensionandeva}
    \publication{narr2022a64bitlinuxcapa}
    \publication{song2022anaxi4slaveinte}
    \publication{vanoni2022linuxsupportfor}
    \publication{brandl2022apowermeasureme}
    \publication{raeber2022developinganene}
    \publication{svelto2022towardsatechnol}
    \publication{zhang2022axipacknearmemo}
\end{publications}

\section{2023}

\begin{publications}
    \publication{werner2023aflexiblefpgaba}
    \publication{holborn2023towardsanafford}
    \publication{bozic2023verifyingiguana}
    \publication{kuenzli2023implementingvir}
    \publication{senti2023theopenroadtowa}
    \publication{jiang2023towardsopensour}
    \publication{sauter2023towardssystemve}
    \publication{guzenko2023implementingusb}
    \publication{leone2023areductioncapab}
    \publication{sauter2023basiliskopensou}
\end{publications}

\section{2024}

\begin{publications}
    \publication{gedik2024avideooutputper}
    \publication{frehner2024jtaggenerator}
    \publication{hauser2024integratingthes}
    \publication{dubochet2024fpgaverifcation}
    \publication{hirs2024fixingbringupre}
    \publication{gruenberg2024optimizingthepe}
    \publication{braun2024towardsafullver}
    \publication{dumik2024bufferlesstrans}
    \publication{hung2024designingascala}
    \publication{wuethrich2024mlemapulpsocfor}
    \publication{pochert2024mlemapulpsocfor}
    \publication{hauser2024creatingasystem}
    \publication{sommerhaeuser2024enablinglinuxdm}
    \publication{roth2024towardsavirtual}
    \publication{luzi2024multichannelmem}
\end{publications}

\section{2025}

\begin{publications}
    \publication{buchner2025towardsansdiope}
    \publication{itschner2025addingmultihead}
    \publication{koelbli2025commercialversu}
    \publication{schaerer2025exploringvirtua}
    \publication{rosetti2025performanceorie}
    \publication{vanoni2025acceleratingmmc}
\end{publications}

\chapter{Acronyms}
\label{chap:acronyms}

\renewcommand{\glossarysection}[2][]{}

\setglossarystyle{long}
\setlength{\glsdescwidth}{8.5cm}
\renewcommand{\glsnamefont}[1]{\textbf{#1}}
\renewcommand{\glspostdescription}{}         %
\renewcommand{\glsnumberformat}[1]{\hspace{1.5mm}(p.$\,$\glshypernumber{#1})}  %

\renewcommand{\glossarymark}[1]{\markright{\MakeUppercase{#1}}}  %

\printglossary[nonumberlist=true]

\listoffigures
\listoftables

\backmatter
%


%
\clearpage
\chapter{Curriculum Vitae}
\label{chap:cv}

\begin{wrapfigure}{l}{0.31\textwidth}
    \vspace{-0.45cm}
    \includegraphics[width=1.3in, height=1.4in, keepaspectratio]{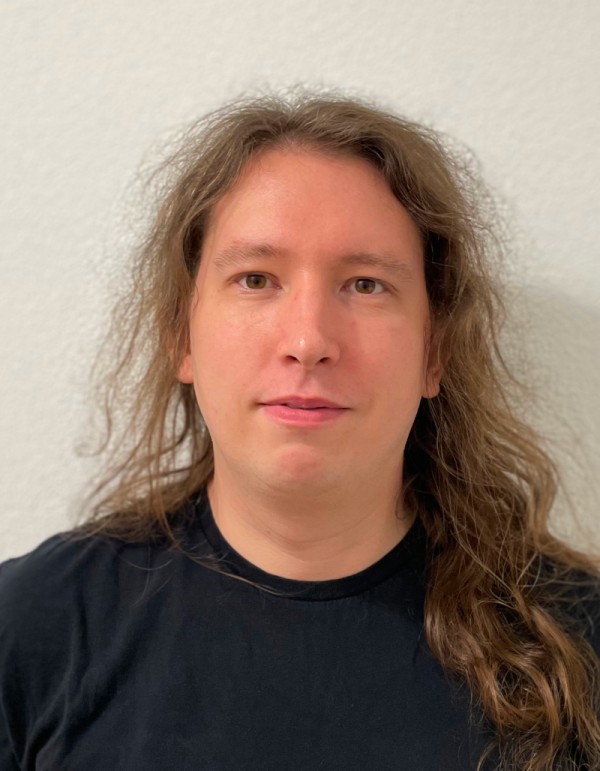}
\end{wrapfigure}

Thomas Benz was born in Z\"urich, Switzerland, in April 1994.
He received his B.Sc. and M.Sc. degrees in electrical engineering and information technology from ETH Zurich in 2018 and 2020, respectively.
\teb{He completed} his Ph.D. degree in the Digital Circuits and Systems group of Prof. Benini \teb{in 2025}.
His research interests include energy-efficient high-performance computer architectures, memory interconnects, data movement, \teb{heterogeneous mixed-criticality systems,} and the design of ASICs.
He \teb{received} the ETH Medal for outstanding Master's Theses in 2021 and won Best Student Poster at HotChips Symposium in 2025.
He has published more than 20 research articles in peer-reviewed journals and conference proceedings and contributed to over 15 ASIC tapeouts.
He has advised over 40 Bachelor's/Semester Theses and 6 Master's Theses.

\emptydoublepage
\
\null
\thispagestyle{empty}
\emptydoublepage

\end{document}